\pdfoutput=1
\documentclass[a4paper,11pt,chapterprefix,
cleardoubleempty,bibtotoc,idxtotoc,footexclude,normalheadings]{scrbook}

\pdfoutput=1

\usepackage{textcomp}
\let\endgraph\endgraf
\usepackage{longtable}
\usepackage{tensind}
\tensordelimiter{?}
\tensorformat{c}
\usepackage{feynmp}

\usepackage[utf8]{inputenc}
\usepackage{lmodern}
\usepackage[T1]{fontenc}
\usepackage{url}
\usepackage{graphicx}
\usepackage{epstopdf}
\let\Oldbar\bar
\usepackage{amsfonts,amssymb,amsmath}
\let\bar\Oldbar
\usepackage{amsmath}
\usepackage{cite}
\usepackage[bindingoffset=3mm,scale=0.76,paperwidth=180mm,paperheight=260mm,heightrounded]{geometry}
\usepackage{microtype}
\usepackage{makeidx}
\usepackage[catalan,english]{babel}

\addtokomafont{caption}{\small}
\setkomafont{captionlabel}{\slshape}
\setcapmargin{1.5em}
\setcapindent{0pt}

\newcommand{\longpage}[1][1]{\enlargethispage{#1\baselineskip}}
\newcommand{\shortpage}[1][1]{\enlargethispage{-#1\baselineskip}}
\newcommand{\fud}{\frac{1}{2}}
\newcommand{\D}{{\mathcal{D}}}
\DeclareMathOperator{\Tr}{Tr}%
\DeclareMathOperator{\sign}{sign}%
\DeclareMathOperator{\PV}{P}
\DeclareMathOperator{\csch}{csch}
\DeclareMathOperator{\Disc}{Disc}
\DeclareMathOperator{\Pf}{P}
\DeclareMathOperator{\diag}{diag}
\newcommand{\expp}[1]{ \mathop\mathit{e}\nolimits^{#1}}
\newcommand{\dert}[3][]{\frac{\mathrm{d}^{#1} #2}{\mathrm{d} #3^{#1}}}
\newcommand{\derp}[3][]{\frac{\partial^{#1} #2}{\partial #3^{#1}}}

\newcommand{\derf}[2]{\frac{\delta {#1}}{\delta {#2}}}
\newcommand{\derff}[3]{\frac{\delta^2 #1}{\delta #2 \, \delta #3}}
\newcommand{\av}[1]{\langle #1 \rangle}
\newcommand{\Av}[1]{\left\langle #1 \right\rangle}

\newcommand{\ud}[2][]{\textrm{d}^{#1}{#2}\,}
\newcommand{\uD}[1]{\D{#1}\,}
\newcommand{\vd}[2][]{\textrm{d}^{#1}{#2}}
\newcommand{\Eqref}[1]{eq.~\eqref{#1}}
\newcommand{\ie}{\emph{i.e.}}
\newcommand{\eg}{\emph{e.g.}}
\renewcommand{\Re}{\mathop\mathrm{Re}\nolimits}
\renewcommand{\Im}{\mathop\mathrm{Im}\nolimits}
\newcommand{\vect}{\mathbf}
\newcommand{\TO}{^{\scriptscriptstyle (T=0)}}
\newcommand{\udpi}[2][]{\frac{\textrm{d}^{#1}{#2}}{(2\pi)^{#1}}}
\newcommand{\Lp}{L_\mathrm{Pl}}
\newcommand{\Mp}{M_\mathrm{Pl}}
\newcommand{\Sint}{S_\mathrm{int}}
\newcommand{\mth}{m_T}
\newcommand{\GF}{G_\mathrm{F}}
\newcommand{\GD}{G_\mathrm{D}}
\newcommand{\SigmaR}{\Sigma_\mathrm R}
\newcommand{\SigmaA}{\Sigma_\mathrm A}
\newcommand{\Gret}{{G}_\mathrm R}
\newcommand{\Gadv}{{G}_\mathrm A}
\newcommand{\Ga}{G^{(1)}}
\newcommand{\Gc}{G}
\newcommand{\phim}{\phi_{m}}
\newcommand{\phiM}{\phi_{M}}
\newenvironment{widetext}{}{}

\newcommand{\dm}{\mathit{\Delta m}}
\newcommand{\GR}{\Gret}
\newcommand{\Mth}{M_T}
\newcommand{\PiR}{\Pi_\mathrm R}
\newcommand{\mph}{m_\mathrm{ph}}
\newcommand{\udpiE}[2]{\frac{\textrm{d}^{#1}{\vect #2}}{2E_\vect{#2}(2\pi)^{#1}}
  }
\newcommand{\SigmaN}{\Sigma^{(1)}}
\newcommand{\ti}{{t_\text{i}}}
\newcommand{\tf}{{t_\text{f}}}
\newcommand{\ssmall}{\scriptsize}
\newcommand{\GN}{{G^{(1)}}}
\newcommand{\GA}{G_\text{A}}
\newcommand{\pplus}{{\scriptscriptstyle (+)}}
\newcommand{\pminus}{{\scriptscriptstyle (-)}}
\newcommand{\meff}{m_\text{eff}}
\newcommand{\phir}{\bar\phi}
\newcommand{\pir}{\bar\pi}
\newcommand{\ar}{\bar{a}}
\newcommand{\Lie}{\mathcal L}
\newcommand{\SigmaImp}{\widetilde\Sigma_\text{R}}

\let\oldbibliography\thebibliography
\renewcommand{\thebibliography}[1]{%
  \small\oldbibliography{#1}%
  \setlength{\itemsep}{0pt}%
}

\providecommand{\abs}[1]{\lvert#1\rvert}
\providecommand{\norm}[1]{\lVert#1\rVert}

\newcommand{\fr}{^{(0)}}

\providecommand{\citep}{\cite}

\newcommand{\ret}{^\text{(ret)}}

\allowdisplaybreaks[3]

\makeindex

\begin{document}

\frontmatter

\newpage


\begin{titlepage}
	\linespread{1.05}
	\centering
	\sffamily
	\large
	Departament de Física Fonamental\\
	Grup de Gravitació i Cosmologia
	\vfill
	{\huge Particle propagation in non-trivial backgrounds:  a quantum field
theory approach\par}
	\vspace{2em}
	{\Large Daniel Arteaga Barriel\par}
	\vspace{2em}
	Memòria presentada per optar al títol de Doctor en Física\\
	Tesi dirigida pel Dr.~Enric Verdaguer Oms\\[1em]
	Abril de 2007\\
	\footnotesize{Segona impressió setembre de 2007}
	\vfill \vfill
	\includegraphics[width=7cm]{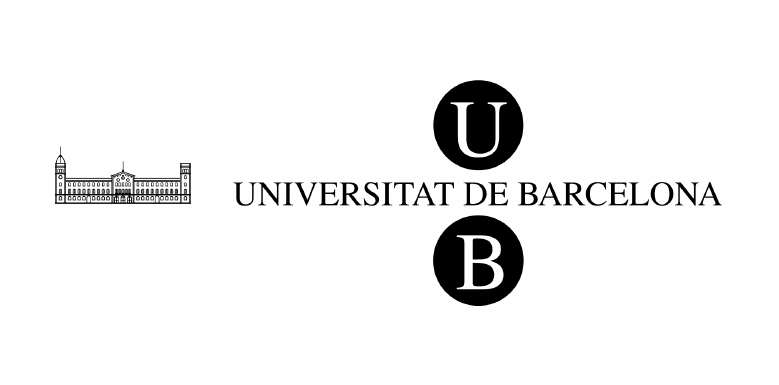}\\
	Programa de doctorat de Física Avan\c cada\\
	Bienni 2000--2001\\
	Universitat de Barcelona
\end{titlepage}

\cleardoublepage

\thispagestyle{empty}
\mbox{}\vfill\vfill

\begin{flushright}
	A Oriol
		
	\bigskip
	
	A tots els que heu cregut \\ 
	que era possible
\end{flushright}

\vfill\vfill\vfill\mbox{}
\cleardoublepage

\tableofcontents

	\selectlanguage{catalan}

\renewcommand*{\dictumwidth}{.50\textwidth} 

\setchapterpreamble[o]{\dictum[John Lennon]{Life is what happens to you while you're busy making other plans.}
\bigskip
\dictum[Antoine de Saint-Exupéry, \emph{Le petit prince}]{C'est le temps que tu as perdu pour ta rose qui fait ta rose si importante.
}}

\renewcommand{\chapterheadstartvskip}{\vspace*{8\baselineskip}} 
\chapter[Agra\"iments]{\emph{Agra\"iments}}
\renewcommand{\chapterheadstartvskip}{\vspace*{2\baselineskip}}

Hi ha tres persones que han marcat profundament el contingut d'aquesta tesi. L'Enric Verdaguer va acceptar dirigir-la i em va introduir dins del món de la física gravitatòria. Sempre ha tingut interès en mi i m'ha posat facilitats de tota mena. De l'Albert Roura he après bona part del que sé.  Vaig tenir el privilegi de compartir despatx amb ell durant els primers anys. Hem passat llargues estones discutint i xerrant, de física i de tot una mica.
Renaud Parentani a été mon collaborateur principal. Une partie importante de cette thèse est le résultat de notre travail ensemble. Nous avons partagé de longues discussions et échanges de
 courriels, mais aussi des vins, des fromages et des dîners. 


Aquesta tesi també està marcada per allò que he après parlant i discutint amb molts altres. El Diego Blas ha estat el meu company de despatx i fatigues aquests darrers anys: tots dos hem après física i català, i hem planejat grans projectes científics. El Guillem Pérez és la saba nova; les seves qüestions i observacions sempre m'han fet pensar. Ells dos han ajudat, amb les seves correccions i suggeriments, a fer aquesta tesi una mica millor. El Jaume Garriga ha resolt tots els meus dubtes amb una concisió remarcable. Esteban Calzetta me impresionó por su visión de la física y su agilidad mental; fue un placer colaborar con él.  Bei-Lok Hu gave me the opportunity of visiting Maryland, and his sharp questions have always been a motivation to me. And of course I should mention all the people from Peyresq: it has been a great opportunity to  go there  so many times. 

I naturalment també han influït en aquesta tesi els companys del departament i la facultat.  A banda del Diego i el Guillem, em vull recordar de l'Ernest, la Míriam, el Xavi, el Luca, el Sofyan, el Jan, el Lluís, el Toni, l'Àlex, l'Enrique, el Roman... Amb ells he compartit infinitat de dinars, cafès, seminaris i discussions, i algun amic invisible. I Peñíscola. I Granada. Per cert, també hi eren els de l'IFAE: l'Ester, el David, l'Olga\ldots I per suposat també haig d'esmentar l'Àngels, amb qui hem compartit cafès i alguna que altra preocupació pel final de la tesi.

En el terreny personal, en primer lloc vull recordar-me dels meus pares, el Jordi i la Montse. Ells m'han donat, a banda de l'estimació, tot allò que un fill també necessita per a créixer, que sovint no són coses materials. És gràcies a ells que sóc el que sóc. Tot seguit haig d'esmentar el meu germà Oriol. Ell sempre ha estat al meu costat. Ha tingut els ànims de seguir-me en el món de la física i del doctorat.  És per això que la tesi li està dedicada. Val a dir que a més a més m'ha ajudat amb les correccions. I naturalment no em puc oblidar de l'\emph{abuela} Maria, a qui estic convençut li farà molta i\l.lusió aquesta tesi, i la Maribel, que amb una mica de sort hi farà una contribució artística. Elles també m'han ajudat en tot. Voldria tenir un record molt especial per a dues persones que ja no són entre nosaltres, la \emph{iaia} Pepita, que em va cuidar i estimar com una mare, i l'\emph{abuelo} Vicente, que sempre va ser un referent per a mi. Tots han marcat profundament la meva manera de ser, i aquesta tesi no hagués estat sense ells.

Ara els hi toca  als meus amics i companys de pis, l'Estela, la Thais i el Vicen\c c. Amb ells he compartit durant aquests dos darrers anys moltes alegries i alguna que altra preocupació. I els viatges a París i Londres, i les discussions inacabables amb l'Estela, i el mercat, el dinar i la migdiada del dissabte, i les festes, i la multa, i els sopars, i els vins, i les cerveses, i mil coses més.  I tampoc em puc oblidar de les incorporacions de la nova temporada, el Sergio i la Pèrsida. Ni naturalment de les estrelles convidades, la Mireia i la Marta. 
 
Que passin ara els amics de la carrera: el Toni, la Mar, la Mari, la Míriam,  l'Estela, el Carles, la Nia... Quantes estones al passadís de química, estirant els descansos abans d'anar a classe... Com que no en vam quedar tots prou farts de la carrera, qui ho diria, doncs vam decidir continuar amb la tesi, alguns ben a prop ---la Míriam dos pisos per sobre, el Toni tres, i el Carles tot just aquí al costat--- i uns altres una mica més lluny ---per terres barcelonines, madrilenyes o de l'Ebre. I naturalment que passin també el Salva, la Cristina, l'Ernest, l'Ester...

Y como que lo mejor siempre se deja para el final, ahora les toca a mis amigos de Sant Andreu: el David, el Emilio, el Dani, la Eli, el Ábel, la Marta... y el Vicenç, naturalmente. Cuantos cosas me vienen a la cabeza: el JMJ,  la cerveza en los Romeros, l'Antic o el Versalles, el San Pedro, Italia, la masia, el pueblo del Emilio...

Gràcies a tots. Un trocet de la tesi us pertany.

\selectlanguage{english}

\bigskip

\noindent\emph{Nota afegida a la segona impressió.} I would also like to warmly thank the thoughtful readers Stefano Liberati and Albert Roura for providing corrections to the original printing.

\setchapterpreamble[o]{
\dictum[Bertrand Russell]{
Not to be absolutely certain is, I think, one of the essential things in rationality.}
\medskip
\dictum[Ernest Rutherford]{All of physics is either impossible or trivial. \\ It is impossible until you understand it, and then it becomes trivial. }}

\renewcommand{\chapterheadstartvskip}{\vspace*{8\baselineskip}} 
\chapter{Preface}
\renewcommand{\chapterheadstartvskip}{\vspace*{2\baselineskip}} 



If we had to summarize the goal of this thesis in one question, this could be: ``What can be learned about the propagation of interacting particles in non-trivial backgrounds through the analysis of the  two-point correlation functions?'' By ``non-trivial background'' we mean any background other than the Minkowski vacuum, \ie, either a non-vacuum state in Minkowski or a general state in curved spacetime.  While trying to address this question, we have found convenient to reanalyze some concepts, such as the particle and quasiparticle concept in general situations, and to deal with several complementary questions.

The main object of interest of this thesis are particles and quasiparticles, and the framework chosen to study them is relativistic field theory and, more in particular, the closed-time path approach to non-equilibrium field theory. 
We attempt, on the one hand, to elaborate on the appropriate theoretical framework suitable for analyzing the particle and quasiparticle properties and, on the other hand, to apply this framework to the analysis of several different physically motivated examples.  

We have tried to make an effort to make the thesis as self-contained as possible, by including review material as appropriate. We have introduced some concepts in simpler well-known situations (absence of interactions, Minkowski vacuum), facilitating the transition to the often more involved cases of interest. This  partially accounts for the length of some chapters of the thesis. 

We have attempted that the presentation of the different chapters of this thesis follows the structure which we believe is the most clear. However, this structure does not always correspond to the order in which the research project took place. Our original motivation was to investigate how interactions in a curved background could induce modifications to the dispersion relation in the context of the trans-Planckian problem. While trying to address this question, we found convenient to first address the corresponding flat spacetime case, and the feeling that some expressions were lacking the appropriate theoretical basis lead us to a more in depth analysis of the propagation of particles in general situations. The motivations will be further discussed in the introduction and the conclusions.

\subsection*{Organization of the thesis}

This thesis is divided into four parts, with parts II and III constituting the central body.

Part I is formed by the first two preliminary chapters. In chapter 1 we motivate the thesis and introduce in an informal way some concepts that are developed later on. In chapter 2 we review some aspects of the theory of linear open quantum systems, which are also needed later on.

Part II of the thesis is devoted to the study of the propagation of particles in flat spacetimes. It begins with chapter 3, which is one of the central chapters of the thesis. We study how the particle and quasiparticle concepts emerge from a field-theoretic point of view, and analyze how their properties can be extracted from the analysis of the two-point propagators. It must be noted that this is by far the more extended chapter of the thesis, in part because it contains review material. To keep the continuity of the exposition it did not seem convenient to divide it.
Chapters 4 and 5 are two different applications of the results of chapter 3 to two different physical systems.

Part III is dedicated to the analysis of the propagation in curved spacetime situations. In chapter 6 we extend the results of chapter 3 to a curved spacetimes, focusing on the cosmological propagation. It is also a rather lengthy chapter containing some review material. Chapter 7 is an application of the results of chapter 6 to the same physical model analyzed in chapter 5. We end with the conclusions (chapter 8), where we discuss of the main results of this thesis.

It must be noted that parts II and III follow a similar structure, with first a chapter developing the theoretical formalism (3 and 6), and then some chapters presenting applications (4, 5 and 7).

Part IV contains the appendices. They include the Catalan summary (appendix A), reference material (appendices B, C and D) and technical aspects (appendices E and F). Appendices constitute a significant part of the thesis, specially in the case of appendices B and C. 
 
See sect.~1.4 in the introduction for further details.

\subsection*{Publications}

The contents of this thesis have been partially published in several references:
\begin{itemize}
\item Chapter 2 basically consists of review material, although the presentation is mostly original. Section 2.8 is novel and not published anywhere as far as we are aware of.

\item The contents in chapter 3 remain unpublished at the moment of writing this preface, with the exception of subsect.~3.5.2, which appears in ref.~\cite{ArteagaEtAl04b}. As we have already mentioned, much review material is also present in this chapter (in sections 3.1 and 3.2). Novel results are mostly concentrated in sections 3.3 and 3.4.

\item Chapter 4 is based on ref.~\cite{ArteagaEtAl04a}, with some additions such as subsect.~4.3.3. The results of chapter 4 are also summarized in ref.~\cite{ArteagaEtAl04b}.

\item Chapter 5 is based on ref.~\cite{ArteagaEtAl05}, with some additions and modifications, such as sect.~5.4.

\item Chapter 6 remains partially unpublished. Preliminary physical insights were published in ref.~\cite{ArteagaEtAl04b}, and some cosmological results are summarized in refs.~\cite{ArteagaEtAl07a,Arteaga07a}. Sections 6.1 and 6.2 mostly (but not exclusively) consist of review material.

\item Preliminary results of chapter 7 are contained in ref.~\cite{ArteagaEtAl07a}, and are summarized in ref.~\cite{Arteaga07a}.
\end{itemize}
At the time of writing this preface, a publication including the results of chapters 6 and 7, and some contents from chapter 5, is in preparation \cite{ArteagaEtAlInPrep}. On the other hand, the contents of ref.~\cite{ArteagaEtAl03} have not been included in this thesis. See sect.~8.3. for a brief account of other work in progress.

\chapter{Notation and conventions}

\subsection*{Units}

We will use a system of natural units with $c = \hbar = k_\text{B}=1$. With this units Newton's constant $G$ has dimensions of squared length (or inverse squared energy). The Planck length and Planck mass are respectively given by $\Lp = \sqrt{G}$ and $\Mp = 1/\sqrt{G}$. It will be useful for us to define the gravitational coupling constant $\kappa = \sqrt{32\pi G} = \sqrt{32\pi} \Lp$.

\subsection*{Special and general relativity}

Four vectors  will be written in roman italic typeface. Three-vectors will be denoted by a bold symbol. Scalar products will be denoted by a dot ($\cdot$) or by explicit contraction of indices. We will assume the Einstein convention for summation of repeated indices. Greek indices ($\alpha, \beta, \gamma,\ldots,\mu,\nu,\ldots$) will be spacetime indices running from 0 to 3; Latin  indices ($i,j,k,\ldots$) will be space indices running from 1 to 3. As an exception, in chapter 6 we will use Latin indices ($a,b,c,\ldots,n,m,\ldots$) running from zero to three to indicate the four vectors of an orthonormal tetrad, and the components of the Riemann normal coordinates associated to that tetrad.


The signature of the metric will be $(-,+,+,+)$. Except where otherwise indicated, indices will be raised and lowered with the Minkowski metric $\vd s^2 = \eta_{\mu\nu} \vd x^\mu \vd x^\nu = -\vd t^2 + \vd x^2 + \vd y^2 + \vd z^2$.

Thus $x\cdot y
= x_\mu y^\mu=\eta_{\mu\nu}x^\mu y^\nu  = - x^0 y^0 + \vect x \cdot \vect y = - x^0 y^0 + \delta_{ij} x^i y^j = - x^0 y^0 + x^1 y^1 + x^2 y^2 +x^3 y^3$.

Notice that with this signature time-like vectors have negative norm. In particular this will be the case for the four-momentum: $\norm{p}^2=p^2 =p_\mu p^\mu= - m^2 \leq 0$.  

The sign conventions for the Riemann tensor will be those of Wald \cite{Wald} and Misner, Thorne and Wheeler \cite{MTW}.

\subsection*{Fourier transforms}

The Fourier transform $\tilde f(\omega)$ of a time-dependent function $f(t)$ is defined as
\begin{equation*}
	\tilde f(\omega) := \int^\infty_{-\infty} \ud t \expp{i t \omega} f(t).
\end{equation*}
The inverse Fourier transform is given by
\begin{equation*}
	f(t) = \int^\infty_{-\infty}  \frac{\vd \omega}{2\pi} \expp{-i \omega t} \tilde f(\omega).
\end{equation*}
Integration limits will be usually omitted when the integration domain is $\mathbf R^n$.

The Fourier transform $\tilde g(p)$ of a function $g(x)$ over the Minkowski spacetime  is defined as 
\begin{equation*}
	\tilde g(p) := \int \ud[4] x \expp{-i x \cdot p} g(x),
\end{equation*}
and the inverse function
\begin{equation*}
	g(x) = \int  \udpi[4]{p} \expp{i p \cdot x} \tilde g(p).
\end{equation*}

Tildes will be omitted whenever there is no danger of confusion, so the same symbol will be used for a quantity and its Fourier transform.


In general, all Fourier-transformed functions should be understood as generalized functions or distributions \cite{Schwartz} (in particular, tempered distributions), although little effort will be put in ensuring mathematical precision. In this context we will make use of the of the notion of principal part of an integral, which is a kind of generalization of the principal value, and the related concept of the principal part of a function, which is one of the possible ways to define distributions corresponding to functions such as $1/x^2$. Both the principal part of an integral and the principal part of a function will be noted by the symbol P.

\subsection*{Quantum mechanics and field theory}

Operators acting on a Hilbert space will be denoted with a hat ($\hat{\ }$); pure states will be denoted with the usual ket notation. We will often use the same letter for an Hermitian operator, its eigenstates and its eigenvalue. For instance, $\hat{q} |q\rangle = q |q\rangle$ means that $| q\rangle$ and $q$ are respectively an eigenstate and an eigenvalue of the operator $\hat q$. Time evolution operators will be denoted by $U(t,t')$ and will not carry a hat. 

Except where otherwise indicated, we will assume that time-independent (pure or mixed) states and time-dependent operators are expressed in the Heisenberg picture, and time-dependent states  and time-independent operators are expressed in the Schr\"odinger picture. Interaction picture will be always indicated explicitly.

Quantum mechanical averages will be indicated by angular brackets: $\av{\hat A(t)} := \Tr {[\hat \rho  \hat A(t)]} = \Tr {[\hat\rho(t) \hat A]}$, with $\hat\rho$ being the state of the system. 

Within field theory we will use the following normalization for the 1-particle states (which is not relativistic invariant):
\begin{equation*}
	\langle \vect p | \vect q \rangle =\delta_{\vect p\vect q}= \frac{1}{V} (2\pi)^3 \delta^{(3)} (\vect p- \vect q),
\end{equation*}
where $V$ is the volume of the spacetime, which is formally infinite. The conmutation relations for creation and annihilation operators [see eqs.~\eqref{ConmutRelat}] will be adjusted so that $|\vect p\rangle = \hat a^\dag _\vect p |0\rangle$ automatically verifies the above normalization. 

Notice that the spatial momentum label $\vect p$ will frequently appear as a subindex instead of being the argument of the function, except in chapter 5 (see the remark in page \pageref{NotationRemark}).

Within the closed time path (CTP) formalism (see appendix \ref{app:CTP}), Latin indices ($a,b,c\ldots$)  will take the values 1 or 2, to indicate the positive ($1$) or negative ($2$) time branches. An Einstein summation convention will be also used for repeated CTP indices. They will be raised or lowered with CTP ``metric'' $c_{ab} = c^{ab}= \textrm{diag}(1,-1)$.

\subsection*{Common symbols with their meaning}

\begin{longtable}{rp{0.85\textwidth}}
	* & Complex conjugate / excited state \\
	\dag & Hermitian conjugate \\  
	$\sim$ & Order of magnitude estimate\\
	$\approx$ & Approximately equal to\\
	$:=$ & Defined to be equal to\\ 
	$T$ & Time ordering \\
	$\widetilde T$ & Anti-time ordering \\
	\mbox{}$[\hat A, \hat B]$ & Conmutator ($=\hat A \hat B- \hat B \hat A$) \\
	$\{\hat A, \hat B\}$ & Anticonmutator  ($=\hat A \hat B+ \hat B \hat A$)\\
	$E_\vect p$ &	Energy of a free particle with momentum $\vect p$	($=\sqrt{m^2+ \vect p^2}$) \\
	$E_\vect p$ &	Energy of an asymptotic particle in the Minkowski vacuum with momentum $\vect p$	($= \sqrt{\smash[b]{\mph^2+ \vect p^2}}$)\\
	$E_\vect k$ &	Energy of an free particle with conformal momentum $\vect k$	($= \sqrt{m^2+ \vect k^2/a^2}$) \\
	$R_\vect p$ &	Energy of an interacting (quasi)particle with momentum $\vect p$ \\
	$R_\vect k$ &	Energy of an interacting (quasi)particle with conformal momentum $\vect k$\\
	$\Gamma_\vect p$ &	Decay rate of interacting (quasi)particle with momentum $\vect p$ in the laboratory rest frame\\
	$\gamma_\vect p$ &	Decay rate of interacting (quasi)particle with momentum $\vect p$ in the particle rest frame\\
	$\Gamma_\vect k$ &	Decay rate of interacting (quasi)particle with conformal momentum $\vect k$ in the comoving frame\\
	$\gamma_\vect k$ &	Decay rate of interacting (quasi)particle with conformal momentum $\vect k$ in the particle rest frame\\
\end{longtable}

\mainmatter

\part{Preliminaries}

\chapter{Introduction}


In this chapter we motivate this thesis, and present in an informal way some topics that will be discussed later on in greater depth. 

\section{Particles and quasiparticles in non-trivial backgrounds}

\index{Particle}
According to Weinberg \cite[p.~xxi]{WeinbergQFT}, ``the most immediate and certain consequences of relativity and quantum mechanics are the properties of particle states.'' Following this point of view, particles are associated with unitary representations of the Poincar\'e group in a Hilbert space, and they can be thought of as the main players in relativistic quantum mechanics (field theory). 

Notwithstanding its elegance and usefulness, this approach can be too restrictive in many circumstances. First, even in  Minkowski spacetime in the vacuum,  unstable particles cannot be associated to unitary representations of the Lorentz group \cite{Zwazinger63}. Therefore, either one can adopt the point of view that unstable particles are not true particles (the only true particles being their stable decay products), or enlarge the particle concept to embrace long-lived unstable excitations. 

\index{Particle!in curved spacetime}
Second, in spacetimes other than Minkowski the Poincaré group is not a global symmetry of the spacetime, and therefore the above approach is not of application. Several possibilities exist to extend the particle concept. If the spacetime is symmetric enough, one can attempt to associate particles to  representations of the symmetry group of the spacetime. However particles defined this way may bear little resemblance to the Minkowski particles, and moreover most spacetimes do not possess such high degree of symmetry. 
Alternatively, one can move to an operational definition of the particle concept, wherein the particle content of a given state is determined by the response of a particle detection device. A relevant fact is that the particle content measured by the detector depends on its own trajectory.\footnote{The lack of a universal particle concept seems to indicate that particles cannot be the fundamental objects in curved spacetimes. Compare the above above quotation by Weinberg with the following quotation by Wald \cite[p.~2]{WaldQFT}: ``Quantum field theory is a quantum theory of \emph{fields}, not particles. Although in appropriate circumstances a particle interpretation of the theory may be available, the notion of `particles' plays no fundamental role either in the formulation or interpretation of the theory'' (italics in the original).} We will further explore these possibilities in chapter 7. In any case, the equivalence principle (and the everyday experience with particle accelerators) suggests that for inertial observers the standard Minkowskian particle concept should be a valid approximation provided the de Broglie wavelength of the particle is much smaller than the typical curvature radius of the spacetime. This latter appoximation to the particle concept is the one we shall mostly refer to in this thesis.

Finally, one can consider particles propagating in media other than the vacuum, either in flat or curved spacetime. Particles  interact with the medium, and this affects their propagation, in a way that they can no longer be adequately described as a representation of the Poincaré group.  More generally, there are many physical systems possessing elementary excitations with some momentum and some energy, the quasiparticles. Quasiparticles  cannot be typically  associated to any fundamental symmetry group.\footnote{Although spontaneous symmetries may emerge in some situations; later on we will comment on condensed matter systems exhibiting approximate Lorentz invariance.}

\index{Quasiparticle}
For some purposes it is useful to think of the medium as an entity breaking (at least some sectors of) the global Poincaré invariance. Loosely speaking, particles become aware of the properties of the background through the interaction process, and in this way their propagation is no longer dictated by the Poincaré invariance. It may be also useful to imagine that something similar happens in curved backgrounds, where the background spacetime can be interpreted as a preferred reference frame. The particle properties are affected by the presence of the background spacetime. 

For the purpose of this thesis we will simply consider that particles and quasiparticles are  long-lived elementary excitations carrying some momentum and energy. By ``long lived'' we mean that their lifetime must be much larger than their inverse de Broglie frequency. More details will be given in chapter 3. We would like to provide a unified description of particles and quasiparticles, analyzing the analogies and differences arising between the propagation in a curved spacetime and the propagation in a non-trivial medium.
 
At least three different approaches to the study of particle propagation are possible. First, particles can be described from the point of view of second quantization, wherein particle properties can be extracted from the analysis of the mode corresponding to the momentum of the particle.  Another description is a first-quantized description, in which the particle dynamics can be studied through the time evolution of its first-quantized density matrix or its first-quantized Wigner function. This alternative description is more complete and more restrictive at the same time, since it requires the existence of a single particle excitation. Yet another possible approach is a statistical description, in which the dynamics of an ensemble of particles is studied.   In this thesis we will mostly adhere to the second-quantized description. The first-quantized and statistical descriptions will be commented in chapter 5.

\index{Dispersion relation}
Let us now briefly recall how the particle properties can be extracted in a second-quantized description through the analysis of the propagators. In flat spacetime in the vacuum, the propagation of a stable particle is fully characterized by the dispersion relation, \ie, by the relation between the energy and momentum of the particle. Because of the Poincaré symmetry, the dispersion relation is in turn fully determined by the value of the particle mass. If the particle is unstable, additionally the decay rate must be taken into account. In this case it is convenient to consider generalized complex dispersion relations,
\begin{equation}
	\mathcal E^2 = \mph^2 + \vect p^2 - i \mph \gamma,
\end{equation}
where $\mph$ is the physical mass of the particle and $\gamma$ is the decay rate in the particle rest frame. As mentioned before, the decay rate must be much smaller than the particle mass; otherwise one would speak of resonances rather than unstable particles. The generalized dispersion relation can be extracted from the location of the poles in the momentum representation of the Feynman propagator. Namely, the 2-point Feynman propagator can be approximated around the particle energy as: 
\begin{equation} \label{IntrodFeynman}
	\GF(E,\vect p) \approx \frac{-iZ}{-E^2 + \vect p^2 + \mph^2 -  i \mph \gamma}.
\end{equation} 

In a non-flat or non-empty spacetime, the propagation of a particle is no longer determined by its mass and its decay rate, since a much richer phenomenology appears in general (including, for instance, scattering, diffusion and decoherence). However one may think that at least some of the basic features of the particle propagation can be encoded in the form of a dispersion relation,
\begin{equation} \label{ModifiedEffective}
	\mathcal E^2 = R_\vect p^2 - i R_\vect p \Gamma_\vect p,
\end{equation}
with the particle energy $R_\vect p$ and the decay rate $\Gamma_\vect p$ being a function of the momentum. For convenience, the decay rate has been expressed in the observer rest frame. The momentum, energy and decay rate appearing in the above dispersion relation should be interpreted as those measured by a free-falling observer. 

One of the tasks we will address in this thesis is to analyze whether in general the dispersion relations can be obtained from the poles of a propagator. Several difficulties will be met. First, the Feynman propagator in a non-vacuum state does not exhibit the form \eqref{IntrodFeynman}, but it contains an additional term depending on the state.  Second, in curved spacetimes the momentum representation does not exist in general. Finally, even if a formal dispersion relation similar to \Eqref{ModifiedEffective} can be extracted from the poles of the propagators, it remains to be seen that the energy, momentum and decay rate which can be extracted from them correspond to physically observable quantities.  We will address these points in chapters 3 and 6.

\section{Modified dispersion relations} 

\index{Lorentz symmetry}

In recent years modified dispersion relations which break Lorentz
invariance have appeared in different contexts of gravitational physics. Let us briefly review some of them in the context of this work.

\subsection{The trans-Planckian problem}


\index{Dispersion relation!in the trans-Planckian problem}
On the one hand, modified dispersion relations appeared in several works which address the
\emph{trans-Planckian problem} \index{Trans-Planckian problem}
\cite{Jacobson91,Jacobson93,Jacobson99}. In black hole physics the trans-Planckian problem corresponds to the fact that 
 the modes responsible for the Hawking radiation at far distances reach arbitrarily high
energies near the black hole horizon when measured by a
free-falling observer ---in particular, they reach energies which are exponentially larger than the Planck mass. This observation has led some authors to
study the robustness of the prediction of Hawking radiation when modifying the dispersion relation near the Planck scale.

\index{Dumb hole}
\index{Black hole!sonic}
\index{Black hole}
The introduction of non-trivial dispersion relations has been originally suggested by Unruh's  sonic black hole analogy \cite{Unruh81}. In some inhomogeneous fluid flows, which are subsonic in some regions and supersonic in some others, a sonic horizon appears which has many resemblances to a black hole horizon.\footnote{Further analogies between black holes and condensed matter systems are explored in refs.~\cite{Visser98,GarayEtAl99,VisserEtAl01a,VisserEtAl01b,BarceloEtAl05}.} The sound field behaves similarly to a massless relativistic field in curved backgrounds, and the same arguments which lead to Hawking radiation also predict thermal emission of sound waves in these sonic analogues. In the case of sonic black holes it is known for sure however that the  hydrodynamic description must fail as soon as  the molecular scale is reached. Non-standard dispersion relations have been introduced in order to account for the short-scale modifications induced when approaching the molecular scale.

\begin{subequations}
Unruh \cite{Unruh95} introduced a modified dispersion relation of the form
\begin{equation}
	E = k_0 \tanh^{1/n}{\big[(k/k_0)^n\big]}
\end{equation}
in the comoving frame. Deviations from the massless wave equation become relevant at $k=k_0$. This dispersion relation can be interpreted  in the framework of 2-dimensional sonic or real black holes. Unruh found through a numerical analysis that the evaporation process was not affected by the modified dispersion relation. The result was later confirmed by Brout \emph{et al.} \cite{BroutEtAl95} using a WKB analysis. Corely and Jacobson \cite{CorleyJacobson96} found similar results using the  dispersion relation
\begin{equation}
	E = (k^2 - k^4/k_0^2)^{1/2}.
\end{equation}
See refs.~\cite{Jacobson99,Helfer03} for a complete review of the subject. As a side remark, let us mention that Visser \cite{Visser98} observed that taking into account viscosity in the equations of motion of the fluid amounted to a generalized dispersion relation of the form
\begin{equation}\label{VisserEq}
 	\mathcal E = (k^2 - k^4/k_0^2)^{1/2} - i k^2/k^0.
\end{equation}
\end{subequations}

A similar trans-Planckian problem
problem appears in inflationary cosmology: in most inflationary models the period of inflation lasts so long that the modes responsible
for the large scale structure had length scales much smaller than
the Planck length in the early stages of inflation. Therefore it
is of interest to determine to what extent the properties of the
fluctuation spectrum are sensitive to modifications of the
dispersion relation at the Planck scale. \emph{Ad hoc} modified dispersion relations have been used in the literature, similar to those employed in the corresponding analysis in the black hole case. Martin and Brandenberger \cite{MartinBrandenberger01} found significant deviations from the usual inflationary predictions in certain cases, while Niemeyer and Parentani \cite{Niemeyer00,NiemeyerParentani01} argued that no significant modifications could occur provided modes propagate adiabatically. Subsequent works have further analyzed whether trans-Planckian effects can leave any imprint in the inflationary power spectrum  \cite{NiemeyerEtAl02,MartinBrandenberger03,HassanSloth03}.

\subsection{Fundamental modifications of the dispersion relations}
\index{Dispersion relation!fundamental modifications}
\index{Ultrahigh energy cosmic rays}
\index{GZK limit}
\index{Lorentz symmetry!fundamental breaking}

The dispersion relations employed when analyzing the trans-Planckian problem have been motivated by the condensed matter analogies, and have been usually proposed from the outset, generally for computational convenience. In any case, the underlying idea is that the yet-unknown theory of quantum gravity will possess some fundamental length scale, thereby breaking the exact Lorentz invariance. The modified dispersion relations should in principle be derived as some long-wavelength series expansion of the fundamental quantum gravity theory. 

Indeed several physically-motivated dispersion relations
have been already proposed, inspired by various approaches to quantum
gravity \cite{AlfaroEtAl99,AlfaroEtAl02}, string theory
\cite{KosteleckySamuel89}, non-conmutative field theory
\cite{CarrollEtAl01}, variation of couplings
\cite{KosteleckyEtAl02} and multiverses \cite{Bjorken03}, among others. 
Modified
dispersion relations often break the energy degeneracy for a given
three momentum: the particle energy can become helicity dependent
\cite{AlfaroEtAl99,AlfaroEtAl02} and light can become birefringent
\cite{CarrollEtAl90,KosteleckyMatthew01}.

\label{page:GZK}
Besides the trans-Planckian problem, modified dispersion relations have found application in another context.  There have been controversial observations  of ultrahigh energetic cosmic rays
(with energies higher than $10^{20}\,\mathrm{eV}$) beyond the
Greisen-Zatsepin-Kuzmin (GZK) cutoff \cite{TakedaEtAl98}.
 Modified
dispersion relations have been used in order to provide an
explanation for these observations
\cite{Kifune99,Amelino-CameliaPiran01a,Amelino-CameliaPiran01b}.


A natural form for the quantum-gravity inspired
modified dispersion relations below the Planck scale is
\begin{equation} \label{Eq1}
    E^2 = m^2 + |\vect p|^2 + \sum_{n \ge 3} \eta_n
    \frac{|\vect p|^n}{\Mp^{n-2}},
\end{equation}
where $\Mp$ is Planck's mass and $\eta_n$ are coefficients of
order one if the Planck mass is the relevant scale. Jacobson
\emph{et al.}\ \cite{JacobsonEtAl02,JacobsonEtAl03a,JacobsonEtAl03b,JacobsonEtAl03c,JacobsonEtAl06} have studied the constraints on the possible values
of the parameters $\eta_n$ based on current astrophysical
observational data.
 The
constraints on the modifications of the $|\vect p|^2$ have also been studied
\cite{CarrollEtAl90,KosteleckyMatthew01}.


\subsection{Induced modifications of the dispersion relation}

\index{Dispersion relation!induced modifications}
\index{Lorentz symmetry!effective breaking}

Lorentz invariance might be broken by
quantum gravity at a fundamental level, or, even if Lorentz invariance is a fundamental symmetry of the underlying theory, it may
be broken in an effective way in non-trivial backgrounds only.  
In the previous section we already argued that in backgrounds which posses a preferred reference frame
radiative corrections induce dispersion relations which in general contain terms which
effectively break the local Lorentz invariance, \ie, the Lorentz
invariance in the tangent plane [see \Eqref{ModifiedEffective}].

Let us emphasize that this does not imply any kind of fundamental breaking of the
Lorentz symmetry nor any new physics. Rather by ``effective
breaking of the Lorentz invariance'' we mean that radiative
corrections to the dispersion relation in general contain terms which
depend on vector or tensor fields characterizing the background.

Notice that in a fundamental approach to the Lorentz
symmetry breaking one naturally expects each additional power of
momentum to be suppressed by increasing powers of the
Planck mass [see \Eqref{Eq1}], whereas in the effective breaking approach there may be several relevant characteristic energy scales (such as the temperature, the curvature scale or the particle mass). Hence there is
more freedom in the possible values of the suppression factor. For the same reason one can expect other contributions to the dispersion relation not included in \Eqref{Eq1} such as terms of the form $a_n |\vect p|^n$ with $n<3$.

Another important point to signal is that most quantum gravity-inspired dispersion relations proposed so far have been assumed to be real [see however \Eqref{VisserEq}]. In contrast, we will see that a generic feature of the propagation in non-trivial backgrounds is the emergence of new dissipative phenomenology, which  can be encoded in an imaginary contribution to the modified dispersion relation, as shown in \Eqref{ModifiedEffective}. Therefore, the dispersion relations which are naturally encountered in general have an imaginary dissipative part. 

Since gravitational interactions grow with energy, one may think that gravity-induced modifications of the dispersion relation in non-trivial backgrounds could become very relevant when approaching the Planck scale. In relation with the trans-Planckian problem, gravitational interactions in a black hole geometry could eventually significantly alter the mode propagation when approaching the Planck scale even if the Lorentz symmetry is a fundamental invariance of the theory.  Indeed, ’t Hooft \cite{tHooft96,tHooft06}
has long emphasized that strong gravitational interactions in the nearhorizon
region might alter the semiclassical description of
black hole evaporation. A preliminary realization of this
line of thought has been pursued by Parentani in refs.~\cite{Parentani01a,Parentani01b,Parentani02}.

\index{Gravitational interaction}
Gravitational interactions have two interesting properties that make them interesting for the trans-Planckian problem: they grow with energy, and they affect every particle. However, any interaction in a non-trivial background induces an effective modification to the dispersion relation. The emergence of imaginary parts in the dispersion relations is particularly interesting, since even tiny effects can accumulate over time and significantly alter the propagation of particles over long distances.

\section{Propagation in non-trivial backgrounds: previous work}

Of course it is a well known fact that quantum effects evaluated in
non-trivial backgrounds induce an effective breaking of the
Lorentz invariance of the dispersion relation. Without the aim of being complete, let us briefly review in this section some previous work on this line.

\index{Speed of light!in a magnetic field}

In the case of QED, Adler \cite{Adler71} already recognized
that photons propagating under the presence of a strong magnetic
field would propagate at speeds smaller than $c$, thereby
effectively breaking the local Lorentz symmetry.

\index{Speed of light!in a curved spacetime}
Drummond and Hathrell \cite{DrummondHathrell80} realized that electromagnetic quantum
corrections in curved spacetimes would also alter the
characteristics of propagation of photons. They computed the
modification to the speed of light of low-energy photons
propagating in curved spacetimes, and found the surprising result
that in many physical situations light would travel at speeds
greater than $c$. Shore \cite{Shore02,Shore02b} generalized Drummond
and Hathrell's result in order to include high-energy photons and to
account for dispersion. He found that  the prediction of
superluminal photon velocity was apparently exact to all
frequencies. We address the reader to ref.~\cite{Shore03b} for a review of this
topic and a discussion of the fact that these results do not imply any causality violation.

\index{Speed of light!in a thermal bath}
Similarly, in the presence of a thermal QED heat bath the
effective speed of the photons is lowered \cite{Tarrach83,LatorreEtAl95} and
the fermion dispersion relation is modified
\cite{DonoghueEtAl85}. Finally we can also mention that
Scharnhorst \cite{Scharnhorst90} and Barton \cite{Barton90} worked out the
propagation of light between two Casimir plates and found that the
speed of light was increased.

Additionally, let us notice that working in the context of
brane world scenarios, Burgess \emph{et al.} \cite{BurgessEtAl03} considered the
gravity-mediated modifications of the dispersion relation of
brane-bound fermions and photons. In their case the primary source
of Lorentz violation were some extra-dimensional configurations.  This work can be considered a boundary case, in which the modification to the dispersion relation can be regarded as effective or fundamental depending on the point of view.
%

Essentially two different classes of methods have been employed when studying the propagation in  non-trivial backgrounds. The first group of methods involves an analysis of the dynamics of the perturbed mean field, which can be obtained, among others, from the linear response theory or from the analysis of the 1-particle irreducible (1PI) effective action. The second group involves an analysis of the propagators, in the line of \Eqref{IntrodFeynman}, either directly or through the 2-particle irreducible (2PI) effective action. On the one hand, the first class of methods focuses on the dynamics of classical-like configurations rather than on the dynamics of individual elementary (quasi)particles. On the other hand, as far as we are aware of, the physical significance of this second class of methods is not fully established in non-trivial backgrounds (although reasonable guesses have been inferred from the corresponding vacuum case). We will address these points in chapters 3 and 6.

\section{Overview of the thesis}

As stated in the preface, the main goal of this thesis is to study what can be learned about  particle propagation in non-flat or non-empty backgrounds from the analysis of the interacting 2-point propagators. To achieve this goal we have found necessary to  undertake several preliminary steps and develop some techniques  which we mention in the following. 

First, we discuss the propagation in the flat case. Since we are interested in a field theory description of the propagation of particles in non-trivial backgrounds, we find convenient to analyze how the quasiparticle concept appears from a field theoretic viewpoint in general backgrounds (chapter 3). The analysis is done both by reexamining the spectral representation and by explicitly constructing the quantum states corresponding to the quasiparticle excitations. We discuss the exact meaning of the dispersion relation and analyze whether it can be obtained from the poles of the propagators. In order to appropriately consider field theory in non-vacuum states the closed time path (CTP) approach to field theory is used (appendix \ref{app:CTP}).

While studying the quasiparticle properties a natural system-environment separation shows up: the particle mode is the system, and the rest of the degrees of freedom constitute the environment. We exploit this separation by recalling the theory of open quantum systems (chapter 2) and approximating the dynamics of the mode corresponding to the momentum of the particle by a quantum Brownian motion system. Several relevant particle properties are analyzed through this analogy. 

The results found in chapter 3 are applied to two different physical systems. First, in chapter 4 we consider the propagation in a thermal graviton bath, owing to the special significance of the gravitational interaction for the trans-Planckian problem.  We analyze whether the effective dispersion relation contains Lorentz-breaking terms. Kinematic constrains forbid us to properly discuss the imaginary part of the dispersion relation at one loop. For this reason we move in chapter 5 to a qualitatively similar two-level system which nevertheless exhibits dissipation at one loop. 

Up to chapter 5 the particle properties  are analyzed in a second-quantized framework, through the analysis of the mode corresponding to the momentum of the particle. In chapter 5 we also briefly discuss how  first-quantized and statistical descriptions could be introduced, and their relation with the second-quantized description.

In chapters 6 and 7 the results are generalized for curved backgrounds. In chapter 6, after discussing the particle concept in non-trivial backgrounds, we  extend the results of chapter 3 to curved backgrounds, focusing on the novel physical effects that the curved background introduces. We highlight the similarities and differences arising between the propagation in a medium and the propagation in a curved spacetime. We make a special emphasis in the propagation in cosmological backgrounds, discussing different approximations depending on the ratio of the relevant scales intervening in the problem.

In chapter 7 we analyze the same model as in chapter 5 but in an expanding universe, focusing on the dissipative effects. We first discuss the decay rates in an expanding thermal bath, and then we focus on the new decay channels opened by the expansion of the universe. 

Finally, in the conclusions (chapter 8), besides summarizing and discussing the main results of the thesis, we reconnect with some of the points we have introduced in this chapter.

	\chapter{Linear open quantum systems}

\index{Open quantum system}

A relativistic particle propagating in a background can be studied from the perspective of the theory of open quantum systems \cite{Davies,BreuerPetruccione,GardinerZoller}. The mode corresponding to the momentum of the relativistic particle constitutes the open quantum system, whereas the environment is formed by the rest of the closed quantum system, namely, the other modes of the particle field and any other fields in interaction. The dynamics of the relativistic mode can be fully characterized by the reduced density matrix.

In the next chapter we will pursue this point of view by showing that the dynamics of a given particle mode can, under certain assumptions, be treated as if it were an open quantum system interacting linearly with some environment. It is therefore of interest to begin this thesis by studying the theory of linear open quantum systems. 

\index{Quantum Brownian motion}
\index{QBM models}
\index{Caldeira-Leggett model|see{QBM models}}

The paradigm of linear open quantum system is the quantum Brownian motion (QBM) model, 
whose system of interest is a non-relativistic massive particle interacting linearly with an infinite bath of
harmonic oscillators. This model has had many applications in different contexts, among which one may mention the quantum to classical transition \cite{UnruhZurek89,RomeroPaz97}, the escape from a potential well \cite{CaldeiraLeggett81,CaldeiraLeggett83a,ArteagaEtAl03,CalzettaVerdaguer06}, the Unruh effect \cite{MassarParentaniBrout93} or quantum optics \cite{WallsMilburn,GardinerZoller}. In an influential paper Caldeira and Leggett
\cite{CaldeiraLeggett83b} applied the influence functional model
of Feynman and Vernon \cite{FeynmanVernon63,FeynmanQMPI} to the QBM model and
computed in closed analytic form the propagator for the reduced density
matrix. 

Aside from the main motivation for studying linear open quantum systems expressed above, there are additional reasons that make worth studying a QBM model. First, many of the concepts and techniques that apply to a mode by mode description of field theory can already be discussed within a simpler quantum mechanical system, and the field theory language can be compared to that of open quantum systems.   Second, the system considered is linear and can be solved in a closed analytic form, so that, on the one hand, exact and perturbative solutions can be shown to match, and, on the other hand, properties related to the linear character of the system can be discussed. Third, the QBM model can be generalized to encompass the general class of linear open quantum systems. Finally, the QBM system clearly illustrates the need for a closed time path approach.


It should be mentioned that we will not provide a full account of the theory of linear open quantum systems. Although the basic formalism will be presented in order to make the presentation as self-contained as possible, only those aspects relevant for the rest of the thesis will be discussed in detail. The reader interested in a more detailed presentation can check refs.~\cite{Davies,BreuerPetruccione,GardinerZoller,CaldeiraLeggett83a,CalzettaRouraVerdaguer03,RouraThesis,Weiss}. In particular, we shall not discuss the important aspects of decoherence and time evolution of the reduced density matrix.

\section{A quantum Brownian motion model}

We shall consider an open quantum system composed of a harmonic
oscillator $q(t)$, which will be the subsystem under study,
linearly coupled to a free massless field $\varphi(t,x)$, which will
act as environment or reservoir. The action for the full system
can be decomposed as
\begin{subequations}
\begin{equation}
    S[q,\varphi] = S_\mathrm{sys}[q] + S_\mathrm{env}[\varphi] + S_{\text{int}}[q,\varphi], \label{S}
\end{equation}
where the terms on the right-hand side, which correspond
respectively to the action of the harmonic oscillator, the action
of the scalar field and the interaction term are given by
\begin{align}
    S_\mathrm{sys}[q] & = \int \ud{t} \left[ \fud \dot q^2 - \fud     \omega_0^2 q^2 \right], \\
    S_\mathrm{env}[\varphi] & = \int \ud{t} \ud{x}  \left[ \fud(\partial_t
    \varphi)^2-\fud(\partial_x \varphi)^2 \right], \\
    S_{\text{int}}[q,\varphi] & = g \int \ud{t} \ud{x} \delta(x)  \dot q
    \varphi,  \label{SInt}
\end{align}
\end{subequations}
with  $\omega_0$ being the bare frequency of the
harmonic oscillator  and $g$ being the coupling
constant. The oscillator is taken to have unit mass.

We use a one-dimensional free field as the environment, following the treatment of ref.~\cite{UnruhZurek89}. This is equivalent to the alternative representation
 \cite{CaldeiraLeggett83b} in which the environment is modelled by a
large ensamble of  harmonic oscillators. This
equivalence can be seen  performing a mode decomposition in the
interaction term \eqref{SInt},
\begin{equation}
    S_{\text{int}}[q,\varphi] = \sqrt{L} \int \ud t \frac{\vd p}{2\pi} g \dot q 
    \varphi_p,
\end{equation}
where $\varphi_p(t)$ is proportional to the spatial Fourier transform of the
scalar field,
\begin{equation*}
    \varphi_p(t) = \frac{1}{\sqrt{L}} \int \ud x \expp{- i p x} \varphi(t,x),
\end{equation*}
where $L$ is the length of the real axis (formally infinite). 

\index{Distribution of frequencies}
\index{Distribution of frequencies!ohmic}
The model can be generalized by replacing the delta interaction of
equation \eqref{SInt} by a function $f(x)$. In this case the interaction term is,
\begin{subequations}
\begin{equation}
    S_\text{int}[q,\varphi] = \int \ud t \ud x f(x)g \dot q(t) \varphi(t,x).
\end{equation}
or equivalently in the Fourier space,
\begin{equation}
    S_\text{int}[q,\varphi] = \sqrt{L} \int \ud t  \frac{\ud p}{2\pi} \tilde f(-p) g \dot{q}(t) 
    \varphi_p(t).
\end{equation}
\end{subequations}
We shall see that, when working in Fourier space, one can go from the standard to the generalized model by making the replacement
\[
    \frac{\ud p}{2\pi} \ \to \ \frac{\mathcal I(p) \ud p }{2\pi},
\]
where $\mathcal I(p) := \tilde f(p) \tilde f(-p)$, whenever there is a momentum
integration over the environment field. The real even function $\mathcal I(p)$ will be called the \emph{distribution of frequencies}.\footnote{In the literature the distribution of frequencies is frequently defined as $\omega\mathcal I(p)$.} The product $g^2 \mathcal I(p)$ characterizes the properties of the coupling with the environment. The QBM model this way generalized encompasses the entire class of linearly coupled environments. Following the literature, the standard distribution of frequencies $\mathcal I(p)=1$ will be hereafter referred as the ``ohmic'' environment model. 


\section{The influence functional}
\index{Influence functional}
\index{Influence action}
\index{Density matrix!reduced}

The \emph{reduced density matrix} for an open quantum system is defined
from the density matrix $\hat\rho$ of the whole system by tracing out
the environment degrees of freedom:
\begin{equation}
	\hat\rho_\text{s}(t) = \Tr_\text{env} \hat\rho(t),
\end{equation}
or equivalently, in a coordinate representation,
\begin{equation}
\rho _\text{s}(q_\mathrm f,q_\mathrm f^{\prime },t_\mathrm f)=\int \widetilde{\mathrm d}
\varphi \, \rho (q_\mathrm f,[\varphi],q_\mathrm f^{\prime
},[\varphi],t_\mathrm f)\text{.}
\end{equation}
The functional measure $\widetilde{\mathrm d} \varphi$ goes over all field configurations at a
given time $t_\mathrm  f$. When the system and the environment are
initially uncorrelated, {\em i.e.}, when the initial density
matrix
factorizes ---$\hat{\rho}(t_\mathrm i)=\hat{\rho}_\mathrm{s}(t_\mathrm i)\otimes \hat{\rho}%
_\mathrm{e}(t_\mathrm i)$, where $\hat{\rho}_\text{s}(t_\mathrm i)$
and $\hat{\rho}_\text{e}(t_\mathrm i)$ mean, respectively, the density
matrix operators of the system and the environment at the initial
time---, the evolution for the reduced density matrix, which is in
general nonunitary and even non-Markovian, can be written as
\begin{equation}
    \rho _\text{s}(q_\mathrm f,q_\mathrm f^{\prime },t_\mathrm f)=\int \ud{q_\mathrm i}\ud{q_\mathrm i^{\prime
    }}J(q_\mathrm f,q_\mathrm f^{\prime },t_\mathrm f;q_\mathrm i,q_\mathrm i^{\prime },t_\mathrm i)\rho
    _{r}(q_\mathrm i,q_\mathrm i^{\prime },t_\mathrm i)\text{,}
\end{equation}
where the propagator $J$ is defined in a path integral
representation by
\begin{equation}
    J(q_\mathrm f,q_\mathrm f^{\prime },t_\mathrm f;q_\mathrm i,q_\mathrm i^{\prime
    },t_\mathrm i)=\int\limits_{q(t_\mathrm i)=q_\mathrm i}^{q(t_\mathrm f)=q_\mathrm f}{\cal D}%
    q\int\limits_{q^{\prime }(t_\mathrm i)=q_\mathrm i^{\prime }}^{q^{\prime
    }(t_\mathrm f)=q_\mathrm f^{\prime }}{\cal D}q^{\prime } \expp{i(S[q]-S[q^{\prime
    }]+S_{\mathrm{IF}}[q,q^{\prime }]) }\text{,} 
\end{equation}
with $S_{\mathrm{IF}}[q,q^{\prime }]$ being the \emph{influence action},
which is related to the the influence functional introduced by Feynman and
Vernon \cite{FeynmanVernon63,FeynmanQMPI} through $F[q,q^{\prime
}]=\expp {iS_{\mathrm{IF}}[q,q^{\prime }]}$. In turn, the
influence functional can be expressed in the following way:
\begin{equation} \label{FInfl}
\begin{split}
     F[q,q^{\prime }]  =  \iint & \D \varphi \, \D \varphi'
     \rho_\mathrm{e}([\varphi_\mathrm i],[\varphi'_\mathrm i],t_\mathrm i) \\
     & \times \exp{\left[ i \left(
     S[\varphi] - S[\varphi'] + S_{\text{int}}[q,\varphi] -
     S_{\text{int}}[q',\varphi'] \right) \right]}.
\end{split}
\end{equation}
The path integral has the boundary conditions
$\varphi(x,t_\mathrm i)=\varphi_\mathrm i(x)$, $\varphi'(x,t_\mathrm
i)=\varphi'_\mathrm i(x)$, $\varphi(x,t_\mathrm  f)=\varphi'(x,t_\mathrm
f)=\varphi_\mathrm f(x)$; there is also an implicit sum over initial
and final states, $\varphi_\mathrm i(x)$, $\varphi_\mathrm i'(x)$ and
$\varphi_\mathrm f(x)$.
The influence functional can also be expressed in operator language \cite{RouraThesis}:
\begin{equation}
	F[q,q'] = \Tr_\text{env} \left( \hat \rho\, U^\dag(\tf,\ti;[ q]) U(\tf,\ti;[q'])\right),
\end{equation}
where $U(\tf,\ti,[q])$ is the time evolution operator for the environment with $q(t)$ regarded as an external classical source.

\index{Preparation functions} \label{par:InitialConditions}
Considering a factorized initial state is a rather unphysical hypothesis that leads to suprising results in many circumstances (see for instance sect.~\ref{sect:QBMdynamics} and ref. \cite{HuPazZhang92}). The methods presented in this chapter can be generalized to more natural initial density matrices by the use of the so-called preparation functions \cite{Weiss,GrabertEtAl88}. However the preparation function method does not completely solve all the problems because it is based in a sudden change of the density matrix. A more physical approach involves a continous preparation of the system \cite{AnglinPazZurek96}. In any case, these techniques are increasingly more involved, and we shall be mostly interested in studying the situation in which initial conditions are set in the remote past. In this case the system and environment have had enough time to interact and become entangled, and the precise form of the initial state becomes unimporant.

\index{Gaussian state}
 Besides demanding that the total state is a factorized product state, we shall also require stationary and isotropic Gaussian states for the environment. Gaussian states are analyzed in appendix \ref{app:Gaussian}. Since the environment field is one-dimensional, the requeriment of isotropy simply translates in the equivalence of the positive and negative field modes. The general class of linear systems can be characterized by isotropic environment states; there is no loss of generality in this hypothesis. 

When the initial density matrix of the environment $\rho
_\mathrm{env}([\varphi_\mathrm i],[\varphi_\mathrm i'],t_\mathrm i)$ is
Gaussian, the path integrals can be exactly performed and one
obtains \cite{FeynmanVernon63,CaldeiraLeggett83a,RouraThesis}:
\begin{equation} 
\begin{split}
    S_{\mathrm{IF}}[q,q^{\prime}]
    =&- 2\int_{t_\mathrm i}^{t_\mathrm f} \ud{t} \int_{t_\mathrm i}^{t} \ud{t^{\prime
    }}\Dot \Delta (t) \mathcal D(t,t^{\prime })\dot Q(t^{\prime }) \\ & + \frac{i}{2}\int_{t_\mathrm i}^{t_\mathrm f}\ud{t}%
    \int_{t_\mathrm i}^{t_\mathrm f} \ud{t^{\prime }}\Dot \Delta (t) \mathcal N(t,t^{\prime })\Dot \Delta
    (t^{\prime })\text{,}
\end{split}
\end{equation}
where $\Delta(t) :=  q(t)- q'(t)$ and $Q(t) := [ q(t)+
q'(t)]/2$. 
\index{Dissipation kernel}
\index{Noise kernel}
In the ohmic case the kernels $\mathcal D(t,t')$ and $\mathcal N(t,t')$, which are related to the 
dissipation and noise kernels (defined below), are given by
\cite{CalzettaRouraVerdaguer03}:
\begin{subequations} \
\begin{align}
    \mathcal D(t,t')&=\frac{ig^2}{2} \av{ [\hat \varphi_\mathrm I (t,0), \hat
    \varphi_\mathrm I (t',0)] }    = \frac{ig^2}{2} \int \frac{\vd p}{2\pi}  \av{ [\hat \varphi_{\mathrm I(-p)} (t), \hat
    \varphi_{\mathrm Ip} (t')] }     
     ,\\
    \mathcal N(t,t')&=\frac{g^2}{2} \av { \{\hat \varphi_\mathrm I (t,0), \hat \varphi_\mathrm I (t',0)
    \}} = \frac{g^2}{2} \int \frac{\vd p}{2\pi}  \av{ \{\hat \varphi_{\mathrm I(-p)} (t), \hat
    \varphi_{\mathrm Ip} (t')\} } ,
\end{align}
\end{subequations}
 where $\hat \varphi_\mathrm I (x,t)$ is the field operator in the
interaction picture, $\hat \varphi_{\mathrm Ip} (t)$ is the $p$-mode of the same field operator, and we recall that $\av{\cdot} = \Tr( \hat \rho \ \cdot )$
means quantum mechanical average. In the generic case the kernels can be expressed as
\begin{subequations} \label{mathcals}
\begin{align}
    \mathcal D(t,t')&=   \frac{ig^2}{2} \int \frac{\vd p}{2\pi} \mathcal I(p) \av{ [\hat \varphi_{\mathrm I(-p)} (t), \hat
    \varphi_{\mathrm Ip} (t')] }     
     , \label{mathcalD} \\
    \mathcal N(t,t')&= \frac{g^2}{2} \int \frac{\vd p}{2\pi} \mathcal I(p) \av{ \{\hat \varphi_{\mathrm I(-p)} (t), \hat
    \varphi_{\mathrm Ip} (t')\} } . \label{mathcalN}
\end{align}
\end{subequations}
It is convenient to interpret
these kernels as bidistributions or generalized
functions of two variables.

\index{Dissipation kernel}
\index{Noise kernel}
By integration by parts, the influence action can also be
expressed as
\begin{equation} \label{S_IF2}
\begin{split}
    S_{\mathrm{IF}}[q,q^{\prime}]=
    & \int_{t_\mathrm i}^{t_\mathrm f} \ud{t} \int_{t_\mathrm i}^{t_\mathrm f} \ud{t'} \Delta(t) H(t,t') Q(t')  \\
    & + \frac{i}{2} \int_{t_\mathrm i}^{t_\mathrm f} \ud{t} \int_{t_\mathrm i}^{t_\mathrm f} \ud{t'} \Delta(t) N(t,t')
    \Delta(t'),
\end{split}
\end{equation}
or as
\begin{equation} \label{S_IF3}
\begin{split}
    S_{\mathrm{IF}}[q,q^{\prime}]
    =&- 2\int_{t_\mathrm i}^{t_\mathrm f} \ud{t} \int_{t_\mathrm i}^{t} \ud{t^{\prime
    }}\Delta (t) D(t,t^{\prime }) Q(t^{\prime }) +  \int_{t_i}^{t_f} \ud{t} \delta\omega^2_0 \Delta(t)Q(t) \\ & + \frac{i}{2}\int_{t_\mathrm i}^{t_\mathrm f}\ud{t}%
    \int_{t_\mathrm i}^{t_\mathrm f} \ud{t^{\prime }} \Delta (t) N(t,t^{\prime }) \Delta
    (t^{\prime })\text{,}
\end{split}
\end{equation}
where the different kernels are defined as
\begin{subequations} \label{HDN}
\begin{align} 
	H(t,t') &:=-2 \derp{}{t} \derp{}{t'} [\theta(t-t') \mathcal D(t,t')] \label{kernelH1} \\  &= -2\theta(t-t') D(t-t') + \delta \omega_0^2 \delta(t-t') \label{kernelH},
	\\
	D(t,t') &:=   \derp{}{t} \derp{}{t'}  \mathcal D(t,t')  \label{kernelD},\\
	N(t,t') &:=   \derp{}{t} \derp{}{t'} \mathcal N(t,t') \label{kernelN}.
\end{align}
The kernels $D(t,t')$ and $N(t,t')$ are called respectively
\emph{dissipation} and \emph{noise kernels}. The frequency shift $\delta\omega^2_0$ is a formally divergent quantity given by
\begin{equation}\label{FreqShift}
	\delta \omega_0^2 := 2 \lim_{t\to t'} \derp{\mathcal D(t,t')}{t} .
\end{equation}
\end{subequations}

There is a subtle technical point involving the frequency shift and the dissipation kernel. The kernel $H(t,t')$ involves the product of two distributions [see eqs.~\eqref{kernelH1} and \eqref{kernelH}] and such a product is strictly speaking not well-defined. In practice this means that one has to be very careful when considering contributions coming from the coincidence limit of the dissipation kernel. As we have seen, depending on the way calculations are done the divergent frequency shift $\delta\omega_0^2$ may show up or not. We shall see that, at least with the ohmic environment, the final results are well-defined and 
ultraviolet-finite.\footnote{\emph{Note added in the second printing}. We have recently become aware that QBM models are affected by  logarithmic ultraviolet divergences, other than the associated to the initial state preparation \cite{FlemingEtAl07}. In any case, the results presented here are most likely not affected.} 
This is in contrast with the original presentation of the QBM model in terms of a collection of harmonic oscillators \cite{CaldeiraLeggett83b}, in which the frequency shift $\delta\omega^2_0$ has to be introduced in the original action in order to get finite results. (Except for this point, both treatments of the QBM model are completely equivalent.) In any case it is worth recalling that, as a matter of principle, the physical frequency of the interacting system needs not coincide with $\omega_0$.


Let us end this section by presenting yet another alternative equivalent expression for the influence action:
\begin{equation}
\begin{split} \label{S_IF-Alt}
    S_{\mathrm{IF}}[q,q']= \frac{ig^2}{2} \int_\ti^\tf \ud{t} \int_\ti^\tf \ud{t'}\Big[ & \dot q(t) \Delta_\mathrm F(t,t') \dot q(t')
    + \dot q'(t) \Delta_\mathrm D(t,t') \dot q'(t') \\
    &-  \dot q(t) \Delta_{-}(t,t') \dot q'(t')
    - \dot q'(t) \Delta_{+}(t,t') \dot q(t') \Big].
\end{split}
\end{equation}
In the ohmic case the different kernels are given by
\begin{subequations}
\begin{gather}
    \Delta_\mathrm F(t,t')  = \av{ T \hat \varphi_\mathrm I(t,0) \hat \varphi_\mathrm I(t',0)}, \\
    \Delta_\mathrm D(t,t')  = \av{ \widetilde T \hat \varphi_\mathrm I(t,0) \hat \varphi_\mathrm I(t',0)}, \\
    \Delta_{-}(t,t') = \Delta_{+}(t',t) =  \av{\hat \varphi_\mathrm I(t',0) \hat
    \varphi_\mathrm I(t,0)},
\end{gather}
\end{subequations}
and correspond to the Feynman propagator, the Dyson propagator and
the Whightman functions of the environment, respectively. In the generic case the above expressions should be generalized following the lines of eqs.~\eqref{mathcals}.

\index{Dissipation kernel}
\index{Noise kernel}
\section{Dissipation and noise kernels}\label{sect:DisNoise}

Let us explicitly compute the dissipation and noise kernels following eqs.~\eqref{mathcals} and \eqref{HDN}. Although the results are standard, at least in the ohmic case, the details of the calculation method will be useful for us afterwards. We need the expression of the one-dimensional field
operator in the interaction picture in terms of anihilation and creation operators,\footnote{Recall that, for an interacting theory,
operators in the interaction picture follow the same relations as
the operators of the free theory in the Heisenberg picture.}
\begin{equation}
    \hat \varphi_\mathrm I(t,x) = \sqrt{L} \int \frac{\ud{p}}{2\pi} \frac
    {1}{\sqrt{2|p|}}
    ( \hat a_p
    \expp{-i |p| t + i p x} + \hat a_p^\dag \expp {i |p|t-ipx}
    ),
\end{equation}
where $\hat a_p$ and $\hat a^\dag_p$ satisfy
$[\hat a_p, \hat a^\dag_q] = \delta_{pq} = (2\pi/L) \delta(p-q)$, and the expression for the mode-decomposed field operator in analogous terms,
\begin{equation} \label{ModeDecomp}
    \hat \varphi_{\mathrm Ip}(t) = \frac{1}{\sqrt{L}} \int \ud x \varphi_\mathrm I(t,x) \expp{-ipx} = \frac{1}{\sqrt{2|p|}} (\hat a^\dag_p + \hat a^\dag_ {-p} ).
\end{equation}

Since the system is linear, the dissipation kernel \eqref{kernelD}, which corresponds to the field anticonmutator, is state independent.  Let us check it explicitly. According to eqs.~\eqref{mathcalD} and \eqref{kernelD} the dissipation kernel can be computed as 
\begin{equation}
	D(t,t')= \frac{ig^2}{2} \partial_t \partial_{t'} \int \frac{\vd p}{2\pi} \mathcal I(p) \Tr_\text{env}{\Big(\hat\rho_\text{e} \big[\hat \varphi_{\text Ip}(t), \hat \varphi_{\text I(-p)}(t')\big]\Big)}.
\end{equation}
Introducing the explicit evolution operators one gets
\begin{equation*}
\begin{split}
	D(t,t')&= \frac{ig^2}{2} \partial_t \partial_{t'} \int \frac{\vd p}{2\pi} \mathcal I(p) \\
	&\quad\times\Tr_\text{env} {\Big(\hat\rho_\text{e} \expp{-i\hat H_\text{env} (t_\text i-t)} \hat\varphi_{p} \expp{-i\hat H_\text{env}(t-t')} \hat \varphi_{-p}\expp{-i\hat H_\text{env}(t'-t_\text i)}\Big)}\\
	&- \text{($t \leftrightarrow t'$)}.
\end{split}
\end{equation*}
where $\hat H_\text{env}$ is the Hamiltonian operator for the environment field and the initial time $t_\text i$ is taken in the remote past (the result will be independent of the initial time). Since it is sufficient to trace over the two-mode reduced state, we introduce two resolutions of the identity in the subspace of these two modes, $1_{\pm p} = \sum_{n,n'} |n_pn'_{-p}\rangle \langle n_pn'_{-p} |$, and get
\begin{equation*}
\begin{split}
	D(t,t')&= \frac{ig^2}{2} \partial_t \partial_{t'} \int \frac{\vd p}{2\pi} \mathcal I(p)  \sum_{n,n',m,m'} \expp{ -i (n+n'-m-m')|p|(t-t')}\\&\qquad\times \rho_{m,m'} |\langle m_p m'_{-p}| \hat \varphi_{p} | n_p n'_{-p}\rangle |^2- \text{($t \leftrightarrow t'$)},
\end{split}
\end{equation*}
where $\rho_{m,m'}= \Tr_\text{other modes}\langle m_pm'_{-p} |\hat\rho_e|m_pm'_{-p}\rangle$. Since the environment is stationary the density matrix operator conmutes with the Hamiltonian and is diagonal in the basis of eigenstates of the environment Hamiltonian
$|m_pm'_{p'}\rangle$. Expressing the field operator in terms of the creation and anihilation operators [see \Eqref{ModeDecomp}] the above equation can be reexpressed as
\begin{equation*}
\begin{split}
	D(t,t')&= \frac{ig^2}{2} \partial_t \partial_{t'} \int \frac{\vd p}{2\pi} \frac{\mathcal I(p)}{2|p|}  \\ &\qquad \times \sum_{m,m'}  \left[ \rho_{m,m'}(m+1) \expp{ -i |p|(t-t')}+ \rho_{m,m'} m' \expp{ i |p|(t-t')}\right]
	\\ &\quad - \text{($t \leftrightarrow t'$)}.
\end{split}
\end{equation*}
Considering the 1-mode reduced density matrix $\rho_n =  \Tr_\text{other modes}\langle n_p |\hat \rho_\text{env} | n_p\rangle$, and taking into account the property of isotropy of the environment, the above equation can be simplified to
\begin{equation}
\begin{split}
	D(t,t')&= \frac{ig^2}{2} \partial_t \partial_{t'} \int \frac{\vd p}{2\pi} \frac{\mathcal I(p)}{2|p|}   \sum_n \rho_n \left[ (n+1) \expp{- i |p|(t-t')}+  n \expp{ i |p|(t-t')}\right] 
	\\ &\quad - \text{($t \leftrightarrow t'$)}.
\end{split}
\end{equation}
or, substracting the time-reversed part,
\begin{equation*}
	D(t,t') = \frac{ig^2}{2} \partial_t \partial_{t'} \int \frac{\vd p}{2\pi} \frac{\mathcal I(p)}{2|p|}    \left[ \expp{ -i |p|(t-t')}-  \expp{ i |p|(t-t')}\right]  \sum_n \rho_n.
	\end{equation*}
Since the density matrix is normalized, $\sum_n \rho_n=1$, it can be explicitly seen that the dissipation kernel is indeed state-independent:
\begin{equation*}
	D(t,t')=  \frac{ig^2}{2} \partial_t \partial_{t'} \int \frac{\vd p}{2\pi} \frac{\mathcal I(p)}{2p}    (-2i)\sin {[p(t-t')]}.
\end{equation*}
Introducing the Fourier transform we get the value of the dissipation kernel in the frequency space.
\begin{equation} \label{DisQBM}
    D(\omega) = \frac{i \omega g^2}{2} \mathcal I(\omega).
\end{equation}
\pagebreak
The dissipation kernel is closely related to the kernel $H(\omega)$ [see eqs.~\eqref{kernelH1} and \eqref{kernelH}], which is also state-independent and given by:
\begin{equation}
	H(\omega) = g^2 \int \frac{\vd p}{2\pi} \frac{\omega \mathcal I(\omega)}{\omega-\omega'+i\epsilon} + \delta \omega^2_0.
\end{equation}

\index{Occupation number}
In contrast, the noise kernel \eqref{kernelN} is state dependent. Repeating an analogous calculation, we get the following expression for the noise kernel:
\begin{equation}
\begin{split}
	N(t,t')&= \frac{g^2}{2} \partial_t \partial_{t'} \int \frac{\vd p}{2\pi} \frac{\mathcal I(p)}{2|p|}   \sum_n \rho_n \left[ (n+1) \expp{ -i |p|(t-t')}+  n \expp{ i |p|(t-t')}\right] 
	\\ &\quad + \text{($t \leftrightarrow t'$)}.
\end{split}
\end{equation}
Adding the time-reversed part yields:
\begin{equation}
\begin{split}
	N(t,t')&= \frac{g^2}{2} \partial_t \partial_{t'} \int \frac{\vd p}{2\pi} \frac{\mathcal I(p)}{2|p|}  \left[  \expp{ i |p|(t-t')}+ \expp{ -i |p|(t-t')}\right]  \sum_n \rho_n  (2n+1).
\end{split}
\end{equation}
Defining the environment \emph{occupation number} as
\begin{equation}
	n(|p|) := \Tr{ \left( \hat\rho_\text e \hat a^\dag_p \hat a_p \right)}=\Tr{ \left( \hat\rho_\text e \hat a^\dag_{-p} \hat a_{-p} \right)} = \sum_n \rho_n n ,
\end{equation}
the above equation can be reexpressed as
\begin{equation}
\begin{split}
	N(t,t')&= {g^2} \partial_t \partial_{t'} \int \frac{\vd p}{2\pi} \frac{\mathcal I(p)}{2|p|}  \left[  \expp{ i p(t-t')}+ \expp{ -i p(t-t')}\right] \left[ n(|p|) + \frac{1}{2} \right].
\end{split}
\end{equation}
For a general Gaussian stationary environments, characterized by the occupation numbers $n(|\omega|)$, the noise kernel in Fourier space is given by
\begin{equation}
	N(\omega)=  {g^2|\omega|}\mathcal I(\omega)\left[\frac12 + n(|\omega|)\right].
\end{equation}

\index{Fluctuation-dissipation theorem}
For the particular case of an environment in thermal equilibrium at a temperature $T$ the occupation numbers are given by $n(|\omega|) = 1/(\expp{|\omega|/T}-1)$ and the noise kernel is given by
\begin{equation} \label{NoiseQBM}
    N(\omega) = \frac{g^2|\omega| \mathcal I(\omega)}{2} \coth \left(\frac{|\omega|}{2 T}\right).
\end{equation}
Therefore in a thermal state the noise and dissipation kernels are related through the
\emph{fluctuation-dissipation theorem}, (see also
appendix \ref{app:GenRel}):
\begin{equation}\label{FluctDisTh}
    N(\omega) = - i \sign(\omega) \coth  \left(\frac{|\omega|}{2 
    T}\right)
    D(\omega).
\end{equation}

We shall next quote explicit time expressions in the case of ohmic distribution of frequencies. The dissipation kernel is given by:
\begin{equation}
    D(t,t') = \frac{g^2}{4} \delta' (t-t'), \qquad H(t,t') = - \frac{g^2}{2} \delta' (t-t').
\end{equation}
The noise kernel in the vacuum is
\begin{subequations}
\begin{equation}
    N(t,t') = \frac{g^2}{2\pi} \Pf \frac{-1}{(t-t')^2}.
\end{equation}
For thermal states, a closed analytic expression for the noise kernel in time space can only be given in the high temperature limit $T \gg \omega$:
\begin{equation}
	N(t,t') \approx g^2 T \delta(t-t').
\end{equation}
\end{subequations}

By considering an arbitrary distribution of frequencies $\mathcal I(\omega)$ and an arbitrary Gaussian state for the environment $\hat\rho_\text{e}$ the dissipation and noise kernels may adopt almost any value. In the rest of the chapter we shall try to express all results in terms of the dissipation and noise kernels. Only two restrictions will be imposed: (1) we will consider environments with an infinite continous number of degrees of freedom (\ie, avoiding singular delta-like distributions of frequencies), and (2) we will only consider stationary Gaussian states for the environment (see appendix~\ref{app:Gaussian} for a description of the Gaussian states). Under these assumptions, generic exact results can be obtained by working in the Fourier space. To this end, it will prove useful to reexpress \Eqref{kernelH} in Fourier space:
\begin{subequations}
\begin{equation}
     H(\omega) =
     - 2 \int \frac{\ud {\omega'}}{2\pi}
     \frac{ i D(\omega')}
     {\omega-\omega'+i\epsilon} + \delta\omega^2_0.
\end{equation}
The kernel $H(\omega)$ can be decomposed in its real and imaginary parts as:
\begin{align}
     H_\mathrm R(\omega) &:= \Re H(\omega)= -2\PV \int \frac{\ud {\omega'}}{2\pi} \frac{
     i D(\omega')}{\omega-\omega'}+ \delta\omega^2_0, \label{RealImOm} \\
     H_\mathrm I(\omega) &:= \Im H(\omega) =  i D(\omega).
\end{align}
\end{subequations}
We have used the property $1/(x+i\epsilon) = \PV(1/x) - i \pi
\delta(x)$. Notice that $H(-\omega)=H^*(\omega)=
H_\mathrm R(\omega)-iH_\mathrm I(\omega)$. The real and imaginary parts of the kernel in frequency space correspond, respectively, to the even and odd parts in time space. \index{Kramers-Kronig relation} Notice also the Kramers-Kronig relation betewen the real and imaginary parts of the kernel $H(\omega)$:
\begin{equation}
  H_\mathrm R(\omega) = -2\PV \int \frac{\ud {\omega'}}{2\pi} \frac{
     H_\mathrm I(\omega')}{\omega-\omega'}+ \delta\omega^2_0.
\end{equation}

We end up signaling that the noise kernel in the frequency space is positive-defined, 
\begin{subequations}
\begin{equation}
N(\omega)\geq 0,
\end{equation}
and that the dissipation kernel is imaginary negative for positive frequencies and imaginary positive for negative frequencies,
\begin{equation}
	\Im H(\omega)= iD(\omega) \leq 0 \quad \text{if } \omega > 0, \qquad \Im H (\omega)=iD(\omega) \geq 0, \quad  \text{if }\omega<0.
\end{equation}
\end{subequations}

\section{The closed time path generating functional}

\index{Generating functional!in the QBM models}

As explained in appendix~\ref{app:CTP} [see \Eqref{ZCTPOper}], the CTP generating functional can be expressed as
\begin{equation}
    Z[j_1,j_2] = \Tr \left[  \hat \rho\, \widetilde T \expp {- i \int \ud{t} j_2(t)
    \hat q(t) }T \expp {i \int \ud{t} j_1(t)
    \hat q(t)}   \right],
\end{equation}
where $\hat \rho$ is the initial density matrix for
the whole system and the trace is also taken over the whole
system. In terms of path integrals the generating functional is
written, after integrating out the environment,
\begin{equation}
\begin{split} \label{ZCTP2}
    Z[j_1,j_2] = & \int \ud{q_\mathrm f} \int \ud{q_\mathrm i} \ud{q'_\mathrm i}
    \int \limits_{q(t_\mathrm i)=q_\mathrm i}^{q(t_\mathrm f)=q_\mathrm f} \uD{q}
    \int \limits_{q'(t_\mathrm i)=q'_\mathrm i}^{q'(t_\mathrm f)=q_\mathrm f} \uD{q'}
    \rho_\mathrm r(q_\mathrm i,q'_\mathrm i,t_\mathrm i) \\
    & \times \expp{ i(S[q] - S[q'] + S_{\mathrm{IF}}[q,q'])} \expp{ i\left( \int \ud{t} j_1(t) q(t) - \int \ud{t}
    j_2(t) q'(t) \right)}.
\end{split}
\end{equation}
At least for the moment, we will be interested in the case in which initial
conditions are set in the remote past, so that we shall take the limit
$t_\mathrm i \to -\infty$.

\index{Propagator!retarded}
\index{Propagator!in the QBM models}
Once  the path integrals of
\Eqref{ZCTP2} are performed one can show that, for a Gaussian environment and asymptotic initial boundary
conditions the generating functional can be expressed as\footnote{In the general case there would be a prefactor taking into account the initial conditions for the system. However, since the system has a dissipative dynamics and the initial conditions are given in the remote past, initial conditions for the system turn out to be completely irrelevant.} \cite{CalzettaRouraVerdaguer03}
\begin{equation}
\begin{split} \label{ZCTP}
    Z[j_1,j_2] =  & \
     \expp{  \fud \int \ud{t_1} \ud{t_2} \ud{t_3}
    \ud{t_4}
     j_\Delta(t_1)
    \Gret(t_1,t_2)N(t_2,t_3)j_\Delta(t_4)\Gret(t_4,t_3) } \\
    & \times
    \expp{  -\int \ud{t_1} \ud{t_2} j_\Delta(t_1) \Gret(t_1,t_2)
    j_\Sigma(t_2)},
\end{split}
\end{equation}
where $j_\Sigma(t) := [j_1(t)+j_2(t)]/2$, $j_\Delta(t) :=
j_1(t) - j_2(t)$ and $\Gret(t,t')$ is the retarded propagator of the
kernel
\begin{equation} \label{KernelL}
    L(t,t') = \left( \dert[2]{}{t} + \omega_0^2 \right)
    \delta(t-t') + H(t,t'),
\end{equation}
\ie, is the kernel which verifies
\begin{equation} \label{RetEqDif}
    \int \ud s \Gret(t,s) L(s,t') = - i \delta(t-t'),
    \qquad \Gret(t,t') = 0 \quad \text{if } t<t'.
\end{equation}
Thus in order to have an explicit expression for $Z_{\mathrm{CTP}}$ we only need to know the dissipation kernel (from which the retarded propagator of the kernel $L(t,t')$ can be obtained) and the noise kernel. With expressions in next section is easy to check that the retarded propagator of the kernel $L(t,t')$ also happens to be the retarded propagator of the system:
\begin{equation}
	\GR(t,t') = \av{ [\hat q(t), \hat q(t')]} \theta(t-t').
\end{equation}

\index{Propagator!Feynman}
\index{Propagator!Dyson}
\index{Propagator!Whightman}
\index{Propagator!in the QBM models}

Differentiating the CTP generating functional we  obtain the
different correlation functions. In particular, the
Feynman and Dyson propagators, and the Whightman functions are given by:
\begin{subequations}\label{PropDivers}
\begin{gather}
    G_\mathrm F(t,t') =  \av{T \hat q(t) \hat q(t') } =-  \derff{Z[j_1,j_2]}{j_1(t)}
    {j_1(t')}  \bigg|_{j=j_2=0}, \label{GF} \\
    G_\mathrm D(t,t') =  \av{\widetilde T \hat q(t) \hat q(t') } = -  \derff{Z[j_1,j_2]}{j_2(t)}
    {j_2(t')}  \bigg|_{j=j_2=0}, \\
    G_-(t,t') = G_+(t',t) :=  \av{ \hat q(t') \hat q(t) } =   \derff{Z[j_1,j_2]}{j_1(t)}
    {j_2(t')}  \bigg|_{j=j_2=0}.
\end{gather}
\end{subequations}
The CTP generating functional can also be
expressed as a function of these correlators:
\begin{equation} \label{ZQBMDirectBasis}
\begin{split}
    Z[j_1,j_2] = \exp\bigg[&- \frac{1}{2} \int \ud t \ud {t'}
    \big[  j_1(t) G_\mathrm F(t,t') j_1(t')
    - j_1(t) G_-(t,t') j_2(t') \\
    & - j_2(t) G_+(t,t') j_1(t)
    + j_2(t) G_\mathrm D(t,t') j_2(t') \big]\bigg].
\end{split}
\end{equation}

\section{Correlation functions}
\index{Propagator!in the QBM models}

Let us start by considering the retarded propagator. According to \Eqref{RetEqDif} the computation of the retarded propagator is trivial in Fourier space:
\begin{equation}\label{GretGen}
	\GR(\omega) = \frac{-i}{L(\omega)} = \frac{-i}{-\omega^2 + \omega_0^2  + H(\omega)}.
\end{equation}
The fact that the imaginary part of the kernel $H(\omega)$ is negative (positive) for positive (negative) frequencies, as stated before, implies that the above propagator is indeed retarded. For the ohmic distribution of frequencies one gets
\begin{subequations}
\begin{equation} \label{GretFourier}
    \Gret(t,t')=-i\int \frac{ \ud{\omega} }{2\pi} \frac{1}{L(\omega)} = \int \frac{ \ud{\omega} }{2\pi}
    \frac{i
    \expp{-i\omega(t-t')}}{\omega^2 +i \omega g^2 /2 -
   \omega_0^2},
\end{equation}
or, equivalently, after performing the integral,
\begin{equation}
    \Gret(t,t')= \frac{-i}{\omega_1}\expp{-\gamma(t-t')/2} \sin{
    [\omega_1(t-t')]}
    \: \theta(t-t'), \label{Gret}
\end{equation}
\end{subequations}
where the coefficients $\gamma$ and $\omega_1$ are defined through
\begin{equation} \label{gammaomega}
    \gamma := \frac{g^2}{2}, \qquad \omega_1 := \sqrt{ \omega_0^2 -
    (\gamma/2)^2}.
\end{equation}

Introducing eq.~\eqref{GretGen} into \Eqref{ZCTP}
the CTP generating functional of the
system is obtained as a function of the noise and dissipation kernels. From the generating functional the full set
of propagators of the system can be obtained.

\index{Propagator!Feynman}
\index{Propagator!in the QBM models}

As an example, we proceed now to the computation
of the Feynman propagator. Applying \Eqref{GF},
\begin{equation}
\begin{split}
    G_\mathrm F&(t,t') = \frac{1}{2} \Gret(t,t') + \frac{1}{2}\Gret(t',t)
    - \int \ud{s} \ud{s'}
     \Gret(t,s) N(s,s') \Gret(t',s').
\end{split}
\end{equation}
This last expression is more easily computed in
Fourier space,
\begin{equation*} \label{MagicGF}
\begin{split}
     G_\mathrm F(\omega) = &  \frac{1}{2}  \Gret(\omega)
    +\frac{1}{2}  \Gret(-\omega) - \fud  \Gret(\omega)  {N}(\omega)
    \Gret(-\omega) \\
= &  \frac{1}{2}  \Gret(\omega)
    -\frac{1}{2}  \Gret^*(\omega)  +\fud  \Gret(\omega)  {N}(\omega)
    \Gret^*(\omega) \\
    = & \  i \Im \Gret(\omega) + N(\omega)
    \lvert \Gret(\omega) \rvert^2.
\end{split}
\end{equation*}
With the aid of \Eqref{GretGen}, the Feynman propagator reads
\Eqref{GretGen}:
\begin{equation} \label{FeynmanGenT}
\begin{split}
    G_\mathrm F(\omega) 
    &= \frac{- i \left[ - \omega^2 + \omega_0^2  + H_\mathrm R(\omega) \right]
    + N(\omega)}{\left[ - \omega^2 + \omega_0^2 + H_\mathrm R(\omega)\right]^2 +
    \left[H_\mathrm I(\omega)\right]^2}.
\end{split}
\end{equation}
This last expression is completely general and valid for any state.
If we now introduce the fluctuation-dissipation at zero
temperature [see \Eqref{FluctDisTh}], $[ H_\mathrm I(\omega)]^2=
[N(\omega)]^2$,  we may write
\begin{equation*}
     G_\mathrm F(\omega) = i\frac{ \omega^2 - \omega_0^2  - H_\mathrm R(\omega)
     -i N(\omega)}{\left[ - \omega^2 + \omega_0^2  + H_\mathrm R(\omega)\right]^2 +
    \left[N(\omega)\right]^2}.
\end{equation*}
Factorizing this last equation we find the final expression for
the Feynman propagator of the QBM system at zero temperature:
\begin{subequations}
\begin{equation} \label{FeynmanGen0}
      G_\mathrm F(\omega)  =  \frac{i }{\omega^2-\omega_0^2 -H_\mathrm R(\omega)
     +i N(\omega)}.
\end{equation}
Note that this expression is valid for any distribution of
frequencies in the environment. For the ohmic environment
one get:
\begin{equation}
    G_\mathrm F(\omega) =
    \frac{i }
    {\omega^2 + i(g^2/2) |\omega| -\omega_0^2}.
\end{equation}
\end{subequations}

The other correlators may similarly be computed. Either
we may functionally differentiate the CTP generating functional,
see \Eqref{PropDivers}, or we may apply the relations in appendix \ref{app:GenRel}. In this latter case it is useful to take the retarded propagator \eqref{GretGen} as one of the basic building blocks of the propagators, the other being the Hadamard function, 
\begin{subequations} \label{HadamardQBM}
\begin{align}
	G^{(1)}  (t,t') &=  2\int \ud{s} \ud{s'}
     \Gret(t,s) N(s,s') \Gret(t',s') \\
	G^{(1)} (\omega ) &=2 |\Gret(\omega)|^2 N(\omega).
\end{align}
\end{subequations}
General expressions
valid for any state are:
\begin{subequations} \label{GOtherFourier}
\begin{align}
    G_\mathrm D(\omega)  &= \frac{ i \left[ - \omega^2 + \omega_0^2  + H_\mathrm R(\omega)
    \right]
    + N(\omega)}{\left[ - \omega^2 + \omega_0^2 + H_\mathrm R(\omega)\right]^2 +
    \left[H_\mathrm I(\omega)\right]^2}, \\
    G_-(\omega) &= \frac{ N(\omega) + H_\mathrm I(\omega)}{\left[ - \omega^2 + \omega_0^2 + H_\mathrm R(\omega)\right]^2
    +
    \left[H_\mathrm I(\omega)\right]^2}, \\
    G_+(\omega) &= \frac{ N(\omega) - H_\mathrm I(\omega)}{\left[ - \omega^2 + \omega_0^2 + H_\mathrm R(\omega)\right]^2
    +
    \left[H_\mathrm I(\omega)\right]^2}.
\end{align}
\end{subequations}
The Hadamard functions and Pauli-Jordan propagators are given by:
\index{Propagator!Hadamard}
\index{Propagator!Pauli-Jordan}
\index{Propagator!in the QBM models}
\begin{subequations} \label{GOtherFourier2}
\begin{align}
    G^{(1)}(\omega)  &= \frac{ 2 N(\omega)}{\left[ - \omega^2 + \omega_0^2 + H_\mathrm R(\omega)\right]^2 +
    \left[H_\mathrm I(\omega)\right]^2}, \\
    G(\omega)  &= \frac{ -2H_\text I(\omega)}{\left[ - \omega^2 + \omega_0^2 + H_\mathrm R(\omega)\right]^2 +
    \left[H_\mathrm I(\omega)\right]^2}.
\end{align}
\end{subequations}

Notice that in general only the retarded (and advanced) propagators follow the typical structure $G(\omega)=-i/[-\omega^2+\omega_0^2 + \Sigma(\omega)]$:
\begin{equation} \label{QBMSigma}
	\GR(\omega) = \frac{-i}{-\omega^2+\omega_0^2+\SigmaR(\omega)}, \quad \SigmaR(\omega) = H(\omega).
\end{equation}
The Feynman propagator can only be put in this form  in a vacuum state.

\section{The stochastic approach} \label{sect:stoch}

It is a well known fact that the evolution of the state of a closed quantum system composed entirely by free or linearly coupled harmonic oscillators is completely determined once the solution of the corresponding classical problem is known. This does not mean however that the behavior of the system is classical: it still exhibits genuine quantum properties such as the level discretization or the zero point energy. However, the time-evolution of the quantum system can be expressed in terms of the corresponding classical trajectories. Obviously, these classical trajectories  do not have a physical realization but must be regarded as computational tools.

Similarly, it can be shown that for linearly coupled open quantum systems the evolution of the reduced density matrix can be fully determined by the solution of a corresponding classical stochastic equation. In fact, the correspondence between quantum and stochastic dynamics for linear systems is also well known \cite{GardinerZoller} and was already noticed by Feynman long ago \cite{FeynmanVernon63,FeynmanQMPI}. It was further investigated in refs.~\cite{CalzettaRouraVerdaguer03,RouraThesis}. Again, the fact that the evolution of the system can be determined by stochastic methods does not mean that the system does not exhibit generic quantum properties. The stochastic trajectories do not have a direct physical meaning and should also be regarded as computational tools.

Here we will not give a complete account of the stochastic description for open quantum systems, but instead we shall introduce the stochastic method via a heuristic trick, and show its usefulness with a particular example. The stochastic approach will be also used later in section \ref{sect:QBMdynamics}.

\index{Effective action}
There is a direct relation between the influence functional and
the CTP effective action for an open quantum system:
\cite{CamposVerdaguer96,MartinVerdaguer99b}
\begin{equation} \label{ActionCTPTree}
    \Gamma^{(0)}[q,q']=S[q]-S[q']+S_{\mathrm{IF}}[q,q'],
\end{equation}
where $\Gamma^{(0)}[q,q']$ is the CTP tree-level
effective action, which for a linear open quantum system coincides
with the complete CTP effective action (see appendix \ref{app:CTP}). Thus, it is possible to
obtain equations of motion for the mean value of the Heisenberg
operator $\hat q$, which we will denote simply as $q$, by
demanding
\begin{equation}
    \frac{\delta}{\delta q(t)} \Gamma^{(0)}_{\mathrm{CTP}}[q,q']
    \Big|_{q'=q} = 0.
\end{equation}

\index{Langevin equation}
\index{Effective action!stochastic}
Now here comes the trick
\cite{CamposVerdaguer96,MartinVerdaguer99b}: We define a
\emph{stochastic effective action} as
\begin{equation} \label{StochAction}
    S_{\text{eff}}[q,q';\xi] = S[q] - S[q'] +  \Re S_{\mathrm{IF}}[q,q']+\int \ud t \xi(t)\left[q'(t)- q(t)\right].
\end{equation}
In this expression $\xi(t)$ must be interpreted as a stochastic Gaussian process defined
by the correlators
\begin{subequations} \label{StochasticCorr}
\begin{align}
    \av{ \xi(t) }_\xi &=0, \\
    \av{ \xi(t)\xi(t') }_\xi &=  N(t,t'),
\end{align}
\end{subequations}
where $\av{\cdot}_\xi$ means stochastic average, or equivalently
by the probability density functional
\begin{equation}
    P[\xi] = \exp{\left( - \frac{1}{2} \int \ud t \ud {t'}
    \xi(t) N^{-1}(t,t') \xi(t') \right)}.
\end{equation}
The stochastic average of a functional of $\xi$ is defined as \[\av{A([\xi],t)}_\xi = \int \tilde{\mathrm d} \xi\, P[\xi] A([\xi],t).\]
The effective action may be written as a statistical average,
\begin{equation}
    \av{ \expp{ {i} S_{\text{eff}}[q,q';\xi]}}_\xi =
    \expp{  i \Gamma^{(0)}_{\mathrm{CTP}}[q,q']}.
\end{equation}
Then we may obtain a stochastic equation of motion for the system,
the Langevin equation:
\begin{equation}
    \frac{\delta}{\delta q(t)} S_{\text{eff}}[q,q';\xi]
    \Big|_{q'=q} = 0.
\end{equation}
The introduction of the stochastic approach with this heuristic trick, although useful and appealing, can be misleading, since one could be tempted to think that the validity of the method is restricted to a classical or semiclassical regime, whereas in fact the stochastic method is valid in the fully quantum regime.

According to \Eqref{ZCTP} the corresponding Langevin equation is
\begin{subequations}
\begin{equation} \label{LangevinMPB}
   \ddot q(t) + \int \ud{t'} H(t,t') q(t') +\omega_0^2 q(t) = {\xi(t)},
\end{equation}
or, particularizing for an ohmic frequency distribution,
\begin{equation} \label{LangevinMPBExpl}
   \ddot q(t) + \frac{g^2}{2} \dot q(t) +\omega_0^2 q(t) =  \xi(t),
\end{equation}
\end{subequations}
where $\xi$ is a stochastic process of zero mean and correlation
function
\begin{equation}
    \av{ \xi(t) \xi(t')}_\xi = N(t,t'). 
\end{equation}

The Langevin equation, \Eqref{LangevinMPB}, can be formally
solved:
\begin{subequations}
\begin{equation} \label{LangevinSolution}
	q(t) = q_\mathrm h(t;q_0,v_0) - i \int_{-\infty}^{\infty} \ud {s} \Gret(t,s)
    \xi(s),
\end{equation}
where $q_\mathrm h(t;q_0,v_0)$ is a homogeneous solution of the differential equation \eqref{LangevinMPB} subject to the initial conditions $q_0=q(t_0;q_0,v_0)$ and $v_0=\dot q(t_0;q_0,v_0)$. Particularizing for the ohmic frequency distribution one gets
\begin{equation} \label{SolutionLangevin}
\begin{split}
    q(t)= &\ q_0 \expp{-\gamma (t-t_0)/2} \cos \omega_1 (t-t_0) + \frac{(v_0+ \gamma q_0)}{\omega_1}
    \expp{-\gamma (t-t_0)/2} \sin \omega_1 (t-t_0) \\
    & - i \int_{-\infty}^{\infty} \ud {s} \Gret(t,s)
    \xi(s),
\end{split}
\raisetag{1.2\baselineskip}
\end{equation}
\end{subequations}
where $\gamma$ and $\omega$ are defined in \eqref{gammaomega}, and $\Gret(t,t')$ is given by \Eqref{Gret}.
In the above equation we have assumed that we are in the weak dissipation case,
$\omega_0 > \gamma$.

Generically speaking, the system has a dissipative dynamics, so that it will decay
to its equilibrium state
$q=\dot q=0$ if the stochastic source is turned off. For the ohmic environment the decay time is of order $\gamma^{-1}$. Consequently,
initial conditions will turn out to be unimportant if they are set in
the remote past, $t_0 \to -\infty$. Then we may forget about the
homogeneous part of the solution and concentrate in the last term
of \Eqref{SolutionLangevin}. In this situation we may compute the stochastic correlation
functions for the position $q$
\begin{equation*}
\begin{split}
    \av{ q(t) q(t') }_\xi & = - \left\langle
    \int \ud {s} \Gret(t,s) \xi(s)
    \int \ud {s'} \Gret(t',s')
    \xi(s')\right\rangle_\xi \\
    & = -\int \ud{s} \ud{s'} \Gret(t,s) \av{\xi(s) \xi(s')}_\xi
    \Gret(t',s').
\end{split}
\end{equation*}
Taking into account that $\av{\xi(s) \xi(s')}_\xi = N(s,s')$ [see \Eqref{StochasticCorr}], we can also write
\begin{equation}
    \av{ q(t) q(t') }_\xi = - \int \ud{s} \ud{s'} \Gret(t,s) N(s,s')
    \Gret(t',s'),
\end{equation}
and comparing with \Eqref{HadamardQBM} we conclude that
\begin{equation}
    \av{ q(t) q(t') }_\xi = \fud G^{(1)}(t,t') = \fud \av{ \{ \hat q(t), \hat q(t')\} }.
\end{equation}
Thus, the stochastic two-point correlation corresponds to the symmetrized two-point correlation function (the Hadamard function).

\index{Stochastic gravity}
The equivalence between stochastic correlation functions and some
quantum correlation functions, which we have presented through a
particular example, is indeed completely general. It can be shown on general grounds that stochastic correlation functions correspond to a subset of quantum correlation functions \cite{CalzettaRouraVerdaguer03,RouraThesis}. 
The correspondence between quantum and stochastic correlation functions is at the heart of the theory of stochastic gravity \cite{CalzettaHu94,HuMatacz95,HuSinha95,CamposVerdaguer96,
CalzettaEtAl97,MartinVerdaguer99a,MartinVerdaguer99c,MartinVerdaguer00,HuVerdaguer03,HuVerdaguer04}.

If initial conditions are set at some finite time\footnote{Recall however that factorized initial conditions are assumed, and that this assumption sometimes leads to unphysical results for finite initial times. See the comments in page \pageref{par:InitialConditions}.}  $t_0$, then additionally to the average over the Gaussian stochastic process, an average over the initial conditions must be implemented. Explicitly, in this case the equivalence between quantum and stochastic correlation functions is given by \cite{RouraThesis,CalzettaRouraVerdaguer03}
\begin{equation}
	 \fud \av{ \{ \hat q(t), \hat q(t')\} }  = \Av{ \av{ q(t) q(t') }_\xi }_{q_0,v_0},
\end{equation}
where the average over the initial conditions is defined as
\begin{equation}
	\av{ q(t) q(t')  }_{q_0,v_0} := \int \ud {q_0} \ud {v_0} q(t;q_0,v_0)q(t';q_0,v_0) W_\text s(q_0,v_0;t_0),
\end{equation}
where $ W_\text s(q_0,v_0;\ti)$ is the \emph{reduced Wigner function}, which is an alternative representation of the reduced density matrix $\hat\rho_\text{s}$. 

\index{Wigner function}
In general, given a density matrix $\hat \rho$, the \emph{Wigner function} is defined as \cite{HilleryEtAl84,Wigner32}
\begin{equation} \label{Wigner}
	W(q,p;t):= \frac{1}{2\pi} \int \ud\Delta \expp{ip\Delta} \rho(q-\Delta/2,q+\Delta/2;t).
\end{equation}
The Wigner function has some similarities with a classical distribution function, although it cannot be interpreted as a probability density since the uncertainity principle prevents a simultaneous measure of the position and the momentum. In fact, the Wigner function can adopt negative values. However, the partial distributions $\int \ud q W(q,p,t)$ and $\int \ud p W(p,q,t)$ are true probability densities. Notice the similarity between the definition of the Wigner function and the definition of the inhomogeneous Fourier transformed propagators, \Eqref{MidTime}.

\section{The perturbative approach}\label{sect:QBMPert}
\begin{fmffile}{mpb}

Let us try to recover the results of the previous sections by
using perturbation theory. Since the system
is linear, we expect the perturbative calculation to reproduce the exact results found in the previous sections. For the remaining of this section we restrict
ourselves to the ohmic distribution of frequencies with $\mathcal I (\omega)=1$. That the results can be easily generalized to an arbitrary distribution of frequencies by including an $\mathcal I(p)$ whenever there is a momentum integration.

The reader must be warned that in this section we shall make use of methods and techniques typical of field theory, which will be described in the next chapter.

\subsection{Environment in a vacuum state}

\index{Self-energy!in the vacuum}

\index{Schwinger-Dyson equation}
The perturbative expansion for the dressed propagator of the
system propagator (indicated by a double line) is given by
\begin{equation*}
\parbox{12mm}{
\begin{fmfgraph}(12,5)
    \fmfleft{i1}
    \fmfright{o1}
    \fmf{double}{i1,o1}
\end{fmfgraph}}
\ = \ %
\parbox{12mm}{
\begin{fmfgraph}(12,5)
    \fmfleft{i1}
    \fmfright{o1}
    \fmf{plain}{i1,o1}
\end{fmfgraph}}\ + \ %
\parbox{28mm}{
\begin{fmfgraph}(28,5)
    \fmfleft{i1}
    \fmfright{o1}
    \fmf{plain}{i1,v1}
    \fmf{dashes}{v1,v2}
    \fmf{plain}{v2,o1}
    \fmfdot{v1,v2}
\end{fmfgraph}}
\ + \ %
\parbox{46mm}{
\begin{fmfgraph}(46,5)
    \fmfleft{i1}
    \fmfright{o1}
    \fmf{plain}{i1,v1}
    \fmf{dashes}{v1,v2}
    \fmf{plain}{v2,v3}
    \fmf{dashes}{v3,v4}
    \fmf{plain}{v4,o1}
    \fmfdot{v1,v2,v3,v4}
\end{fmfgraph}}
\ + \ \cdots,
\end{equation*}
where the single line indicates the free system propagator,
\begin{equation}
    G^{(0)}_\mathrm F(\omega)= \frac{i}{m(\omega^2-\omega_0^2)+i\epsilon},
\end{equation}
and the dashed line indicates the free environment propagator,
\begin{equation}
    \Delta_\mathrm F(\omega) = \int \udpi{p}
    \frac{i}{\omega^2-p^2+i\epsilon} = \frac{1}{2|\omega|}.
\end{equation}
The Feynman rules are completed by adding a $i e (\pm i \omega)$
in each vertex linking a system (environment) with an environment
(system) propagator. The above series can be resumed with the aid
of the \emph{self-energy} $\Sigma(\omega)$, which is defined through the
\emph{Schwinger-Dyson equation}
\begin{equation} \label{SD}
    G_\mathrm F(\omega) = G_\mathrm F^{(0)}(\omega) + G_\mathrm F^{(0)}(\omega)
    [-i\Sigma(\omega)] G_\mathrm F(\omega),
\end{equation}
or alternatively,
\begin{equation}
    G_\mathrm F(\omega)= \frac{i}{m(\omega^2-\omega_0^2)-\Sigma(\omega)}.
\end{equation}
Diagrammatically the self-energy is the sum of all one particle
irreducible diagrams with external legs amputated. In our
case there is just one of these diagrams,
\begin{equation*}
    -i \Sigma(\omega) = \ \parbox{20mm}{\begin{fmfgraph}(20,5)
    \fmfleft{i1}
    \fmfright{o1}
    \fmf{plain}{i1,v1}
    \fmf{dashes}{v1,v2}
    \fmf{dashes}{v2,v3}
    \fmf{dashes}{v3,v4}
    \fmf{plain}{v4,o1}
    \fmfdot{v1,v4}
\end{fmfgraph}}
\end{equation*}
(it must be understood amputated) which can be straightforward
computed to give
\begin{equation}
    -i\Sigma{(\omega
    )}= (ie)(i\omega) \Delta_\text{F}(\omega) (ie)
    (-i\omega) = - \frac{g^2 |\omega|}{2},
\end{equation}
and hence the self-energy coincides with that computed in the
previous sections, $\Sigma(\omega) = -{i g^2 |\omega|}/{2}$.

\subsection{Environment in an arbitrary state}

As explained in appendix \ref{app:CTP}, in an arbitrary state  one has to consider the CTP
doubling of the degrees of freedom. There are two kind of
vertices, 1 and 2 vertices, and four kinds of propagators, 11
(Feynman), 22 (Dyson), 12 (negative Whightman) and 21 (positive
Whightman). When doing perturbation theory  all
Feynman diagrams with internal 1 and 2 vertices must be summed, taking into
account that type 2 vertices carry an additional minus sign with
respect to type 1 vertices. Graphically, the perturbative expansion of one of the propagators (\eg\ the Feynman propagator) can be represented as:
\begin{equation*}
\begin{split}
\parbox{12mm}{
\begin{fmfgraph*}(12,5)
    \fmfleft{i1}
    \fmfright{o1}
    \fmf{double}{i1,o1}
	\fmfv{label=\ssmall 1,label.angle=-90}{i1,o1}
\end{fmfgraph*}}
\ &= \ %
\parbox{12mm}{
\begin{fmfgraph*}(12,5)
    \fmfleft{i1}
    \fmfright{o1}
    \fmf{plain}{i1,o1}
	\fmfv{label=\ssmall 1,label.angle=-90}{i1,o1}
\end{fmfgraph*}}\ + \ %
\parbox{28mm}{
\begin{fmfgraph*}(28,5)
    \fmfleft{i1}
    \fmfright{o1}
    \fmf{plain}{i1,v1}
    \fmf{dashes}{v1,v2}
    \fmf{plain}{v2,o1}
	\fmfv{label=\ssmall 1,label.angle=-90}{i1,o1}
	\fmfv{label=\ssmall 1,label.angle=-90}{v1}
	\fmfv{label=\ssmall 1,label.angle=-90}{v2}
    \fmfdot{v1,v2}
\end{fmfgraph*}}
\ + \ \parbox{28mm}{
\begin{fmfgraph*}(28,5)
    \fmfleft{i1}
    \fmfright{o1}
    \fmf{plain}{i1,v1}
    \fmf{dashes}{v1,v2}
    \fmf{plain}{v2,o1}
	\fmfv{label=\ssmall 1,label.angle=-90}{i1,o1}
	\fmfv{label=\ssmall 1,label.angle=-90}{v1}
	\fmfv{label=\ssmall 2,label.angle=-90}{v2}
    \fmfdot{v1,v2}
\end{fmfgraph*}}%
 \\[.3cm]
&\quad + \ \parbox{28mm}{
\begin{fmfgraph*}(28,5)
    \fmfleft{i1}
    \fmfright{o1}
    \fmf{plain}{i1,v1}
    \fmf{dashes}{v1,v2}
    \fmf{plain}{v2,o1}
	\fmfv{label=\ssmall 1,label.angle=-90}{i1,o1}
	\fmfv{label=\ssmall 2,label.angle=-90}{v1}
	\fmfv{label=\ssmall 1,label.angle=-90}{v2}
    \fmfdot{v1,v2}
\end{fmfgraph*}}
\ + \ \parbox{28mm}{
\begin{fmfgraph*}(28,5)
    \fmfleft{i1}
    \fmfright{o1}
    \fmf{plain}{i1,v1}
    \fmf{dashes}{v1,v2}
    \fmf{plain}{v2,o1}
	\fmfv{label=\ssmall 1,label.angle=-90}{i1,o1}
	\fmfv{label=\ssmall 2,label.angle=-90}{v1}
	\fmfv{label=\ssmall 2,label.angle=-90}{v2}
    \fmfdot{v1,v2} 
\end{fmfgraph*}} \ + \ \cdots
\end{split}.
\end{equation*}

\vspace{3mm}

\index{Self-energy!in the QBM models}
\index{Propagator!in the QBM models}
The self-energy is a matrix-valued quantity related with the propagator through equation \eqref{SelfEnergyGeneralApp}. It  can be computed as the following amputated diagram,
\begin{equation*}
    -i \Sigma^{ab}(\omega) = \ \parbox{20mm}{\begin{fmfgraph*}(20,5)
    \fmfleft{i1}
    \fmfright{o1}
    \fmf{plain}{i1,v1}
    \fmf{dashes}{v1,v2}
    \fmf{dashes}{v2,v3}
    \fmf{dashes}{v3,v4}
    \fmf{plain}{v4,o1}
    \fmfdot{v1,v4}
	\fmfv{label=\ssmall $a$,label.angle=-90}{v1}
	\fmfv{label=\ssmall $b$,label.angle=-90}{v4}
\end{fmfgraph*}},
\end{equation*}
or, symbollically, as
\begin{equation}
	-i\Sigma^{ab}(\omega) = c^{ac} c^{bd} (ie) (i\omega) \Delta_{cd} (\omega) (-i\omega) (ie) 
\end{equation}
where $\Delta_{ab}(\omega)$ is the field propagator with the momentum integrated out.

\index{Self-energy!retarded}
\index{Self-energy!Hadamard}
In order to get explicit results, let us particularize to the case of  a thermal state at temperature $\beta^{-1}$:
\begin{equation}
\begin{split}
    \Delta_{ab}(\omega) &= \int \udpi{p} \left[
	\begin{pmatrix}    
		\dfrac{i}{\omega^2-p^2+i\epsilon}  &	2\pi \delta(\omega^2-p^2) \theta(-\omega) \\
		2\pi \delta(\omega^2-p^2) \theta(\omega) & \dfrac{-i}{\omega^2-p^2-i\epsilon} 		
	\end{pmatrix} \right. \\[2mm]
&\qquad\qquad\left. + 2\pi 	\begin{pmatrix}    
		1  &	1\\
		1 & 1	
	\end{pmatrix} n(|\omega|) \delta(p^2-\omega^2)  \right]\\
	&=
	\frac{1}{|\omega|}\begin{pmatrix}
		\fud + n(|\omega|)  & - \theta(-\omega) - n(|\omega|)\\
		- \theta(\omega)-n(|\omega|) & \fud+n(|\omega|)
	   \end{pmatrix},
\end{split}
\end{equation}
where 
\begin{equation*}
	n(E) =  \frac{1}{\expp{\beta E} - 1}.
\end{equation*}
Therefore the $ab$ component of the self-energy is given by
\begin{equation}
\begin{split}
     \Sigma^{ab}{(\omega)} &= - i {g^2 |\omega|} \begin{pmatrix}
		\fud + n(|\omega|)  & - \theta(-\omega) - n(|\omega|)\\
		- \theta(\omega)-n(|\omega|) & -\fud+n(|\omega|)
	   \end{pmatrix}.
\end{split}
\end{equation}
Notice that the different self-energy components verify the relations found in appendix \ref{app:CTP}. We can extract the retarded self-energy.
\begin{equation}
	\SigmaR(\omega) = \Sigma^{11}( \omega) + \Sigma^{12}(\omega) = - i {g^2 |\omega|}\left[\fud - \theta(-\omega)\right] =  - i \frac{g^2 \omega}{2} = H(\omega),
\end{equation}
in agreement with \Eqref{QBMSigma}. Let us also mention that the Hadamard self-energy $\SigmaN(\omega) = \Sigma^{11}( \omega) + \Sigma^{22}(\omega)$ is proportional to the noise kernel:
\begin{equation}
\begin{split}
	\SigmaN(\omega) = - i g^2 |\omega| 2 \left[\fud +n(|\omega|)\right]  = -i g^2 |\omega| \coth{\left(\frac{\beta|\omega|}{2}\right)} =-2i N(\omega).
\end{split}
\end{equation}
The usefulness of the Hadamard self-energy and the relations between the self-energy and the propagator will be explored in more depth in the next chapter. 

\end{fmffile}

\section{Dynamics of perturbations}\label{sect:QBMdynamics}

So far in this chapter we have only considered equilibrium properties, this is to say, we have considered systems having time-independent density matrices. In this section we would like to go a step further, and study the behavior of the reduced system when it is taken out of equilibrium. In fact, as it is shown in refs.~\cite{RouraThesis,CalzettaRouraVerdaguer03}, the methods and techniques presented in this chapter are already suited to treat this more general generic case,  although our presentation was restricted to the equilibrium situation. For reasons that will become clear in the next chapter, we are interested in studying the behavior of the energy of systems exhibiting some particular properties, and in this situation it proves more useful for us to make a separate analysis, which will be based on the stochastic methods presented in  sect.~\ref{sect:stoch}. 

At some time $t_0$ the system is perturbed by displacing the reduced system from its equilibrium situation. We shall assume that the system is such that the perturbation is long-lived, \ie, the characteristic lifetime of the perturbation is much longer than the characteristic oscillation time. Aside from this assumption, the frequency distribution and the state for the environment will be arbitrary. We shall try to compute, first, the energy of the perturbation and, second, its time-evolution. We are interested in studying properties for long times as compared to the characteristic oscillation time.

Regarding the first point, a couple of comments are in order. First, notice that the energy of the perturbation is not defined unambiguously since the perturbation eventually decays, or, in other words, it does not correspond to the eigenstate of any Hamiltonian. However, if the perturbation is sufficiently long-lived we expect the energy to be still a physically meaningful quantity, defined up to some uncertainty given by the decay rate. A sensible measure for the energy would be the expectation value of some Hamiltonian. However, the naive guess that the energy is given by the expectation value of 
\begin{equation*}
	\hat H_\text{naive}(t) = \fud  \dot{\hat q}^2(t) + \fud \omega_0^2 \hat q^2(t)
\end{equation*}
is obviously not necessarily correct, since the frequency $\omega_0$ need not coincide with the the physical frequency of the interacting oscillators, as there might be (finite or infinite) renormalization effects. For this reason, we shall instead compute the expectation value of the operator
\begin{equation}
	\hat H_\text{sys}(t) = \fud  \dot{\hat q}^2(t) + \fud \Omega_1^2 \hat q^2(t)
\end{equation}
with $\Omega_1$ being yet undetermined. We expect to find a physically reasonable criterion for choosing the value of $\Omega_1$.

Since the Hamiltonian operator is a quadratic operator, and since it can be seen as the coincidence limit of a symmetrized correlation function, its quantum expectation value coincides with the stochastic expectation value:
\begin{equation}
	E(t):=\av{\hat H_\text{sys}(t)} = \Av{\av{H_\text{sys}(t)}_\xi}_{q_0,v_0},
\end{equation}
where we recall that the stochastic average goes over the different realizations of the stochastic field and also over the initial Wigner function of the perturbation. The corresponding Langevin equation \eqref{LangevinMPB} is solved by \Eqref{LangevinSolution}.

If the system is in equilibrium with the environment
 the expectation value of the reduced subsystem is given by the inhomogeneous part of the solution of the Langevin equation: \cite{GardinerZoller} 
\begin{equation}
\begin{split}
	E_0 &= \av{\hat H_\text{sys}(t)}_0=  \av{H_\text{sys}}_\xi = \fud \int \frac{\vd\omega}{2\pi} (\omega^2 + \Omega_1^2) |\Gret(\omega)|^2 N(\omega) \\ &= \fud \int \frac{\vd\omega}{2\pi}  \frac{(\omega^2 + \Omega_1^2)N(\omega)}{[\omega^2 - \omega_0^2 - H_\text{R}(\omega)]^2 + H^2_\text{I}(\omega)}
\end{split}
\end{equation}
See ref.~\cite{GardinerZoller} for a further developement of the above equation and its connection with statistical mechanics. As a side note, let us mention that when the system is at equilibrium the density matrix for the reduced subsystem does not correspond to the eigenstate of a Hamiltonian, so that actually there are fluctuations on the energy of the equilibrium system \cite{NagaevButtiker02}. 

When perturbations are included, the energy of the system is given by
\begin{equation}
	E(t)=E_0 +  \frac{1}{2}  \av{\dot q_\text h^2(t;q_0,v_0)}_{q_0,v_0} + \fud \Omega_1^2 \av{ q_\text h^2(t;q_0,v_0)}_{q_0,v_0},
\end{equation}
where we recall that $q_\mathrm h(t;q_0,v_0)$ is a homogeneous solution of the Langevin equation \eqref{LangevinMPB} with initial conditions $q_0$ and $v_0$.

Let us find explicitly the homogeneous solution of the Langevin equation. In order to deal with the non-local term in the Langevin equation, some additional information is needed on the form of the solution. Since we are considering slowly decaying perturbations, we assume that the solution is of the form
\begin{equation} \label{HomAnsatz}
	q_\text h(t;v_0,q_0) = A(t;v_0,q_0) \expp{i\Omega t} +A^*(t;q_0,v_0) \expp{-i\Omega t} ,
\end{equation}
where the phase $\Omega t$ changes much more rapidly than the modulus $A(t;v_0,q_0)$. Under these assumptions the non-local term can be expressed as
\begin{equation*}
	\int \ud s H(t,s) q_\mathrm h(s) = \int  \ud s \frac{\vd \omega}{2\pi} \expp{-i\omega t} 
	\left[ A(s) \expp{i(\Omega-\omega) s} + A^*(s) \expp{-i(\Omega +\omega)s}\right] H(\omega),
\end{equation*}
where, for simplicity, we omitted the dependence on the initial conditions. Let us concentrate on the first term on the right hand side of this equation. Taking into account that $A(s)$ is a slow function of $s$, the time integral only contributes significantly when $\omega \approx \Omega$; otherwise the phase oscillates too rapidly. Therefore  $H(\omega)$ can be approximated by $H(\Omega)$ in the first term. Similarly, in the second term $H(\omega)$ can be approximated by $H(-\Omega)$. Hence,
\begin{equation*}
\begin{split}
	\int \ud s H(t,s) q_\mathrm h(s) &= H(\Omega) \int \ud s \frac{\vd \omega}{2\pi}A(s) \expp{i(\Omega-\omega) s} \\
	&\quad + H(-\Omega) \int \ud s \frac{\vd \omega}{2\pi}A^*(s) \expp{-i(\Omega+\omega) s}\\
	&= H(\Omega) A(t) \expp{i\Omega t} + H(-\Omega) A^*(t) \expp{-i\Omega t}.
\end{split}
\end{equation*}
According to \Eqref{RealImOm}, the real part of the kernel $H(\omega)$ is even while the imaginary part is odd. Therefore the above equation can be reexpressed as
\begin{equation*}
\begin{split}
	\int \ud s H(t,s) q_\mathrm h(s) &= \Re{H(\Omega)} \left[ A(t) \expp{i\Omega t} +  A^*(t) \expp{-i\Omega t} \right] \\ &\quad+ i \Im{H(\Omega)}  \left[ A(t) \expp{i\Omega t}-   A^*(t) \expp{-i\Omega t} \right],
\end{split}
\end{equation*}
or alternatively, neglecting the time-derivatives of $A(t)$ in front of $\Omega A(t)$,
\begin{equation}
\begin{split}
	\int \ud s H(t,s) q_\mathrm h(s) &= \Re{H(\Omega)} q_\mathrm h(t) - \frac{1}{\Omega}  \Im{H(\Omega)}  \dot q_\mathrm h(t).
\end{split}
\end{equation}
Notice that this amounts to a local expansion of the non-local term:
\begin{subequations}
\begin{equation}
	\int \ud s H(t,s) q_\text h(s) = \delta \Omega^2 q_\text h(t) + \Gamma \dot q_\text h(t) 
\end{equation}
with
\begin{equation}
	\delta \Omega^2 := \Re{H(\Omega)}, \qquad \Gamma := - \frac{1}{\Omega}  \Im{H(\Omega)}.
\end{equation}
\end{subequations}

Within this approximation the equation of motion is given by:
\begin{equation}
	\ddot q (t) + \Gamma \dot q(t) + (\omega_0^2 + \delta\Omega^2) q(t) = 0
\end{equation}
and the corresponding solution is [see also \Eqref{SolutionLangevin}]
\begin{equation}
	q_\mathrm h(t;q_0,v_0)= q_0 \expp{-\Gamma (t-t_0)/2} \cos{ \Omega(t-t_0)} + \frac{v_0}{\Omega} \expp{-\Gamma(t-t_0)/2} \sin{\Omega(t-t_0)},
\end{equation}
where we have identified $\Omega$ through the following self-consistent equation
\begin{equation}
	\Omega^2 = \omega_0^2 + \delta\Omega^2 = \omega_0^2 + \Re H(\Omega),
\end{equation}
and where have made the approximation that $\Omega \gg \Gamma$. Notice that the solution is in accordance with the ansatz \eqref{HomAnsatz}, and notice also that the above equation applies to the ohmic case if $\gamma$ is small in comparison with $\omega_0$. In this case $\delta\Omega=0$ and $\Gamma=\gamma$.

With this solution, the expectation value of the energy is given by
\begin{equation}\label{EnergyPertLarge}
\begin{split}
	E(t) = E_0 &+  \fud \expp{-\Gamma(t-t_0)} \bigg[ \av{q_0^2}_{q_0,v_0} \left[ \Omega^2 \sin^2{\Omega(t-t_0)} + \Omega_1^2 \cos^2{\Omega(t-t_0)} \right] \\
	& \quad+\Av{ \frac{v_0^2}{\Omega^2}}_{q_0,v_0}\left[ \Omega^2 \cos^2{\Omega(t-t_0)} + \Omega_1^2 \sin^2{\Omega(t-t_0)} \right]\\
	&\quad + \Av{ \frac{v_0 q_0}{\Omega}}_{q_0,v_0} (\Omega_1^2 -\Omega^2) \cos{\Omega(t-t_0)} \sin{\Omega(t-t_0)}   \bigg].
\end{split}
\end{equation}
In general the above function is rapidly oscillating with a frequency given by $\Omega$. This is not what is physically expected: the energy should be a smooth function, slowly decaying at long times but approximately constant at short timescales. However, recall that the parameter $\Omega_1$ is still undetermined. If we choose $\Omega_1=\Omega$ the time-evolution of the energy is greatly simplified, and we get the expected behavior:\footnote{There is a subtle point concerning \Eqref{QBMEt} and the factorized initial conditions. This equation implies that any perturbation increases the energy of the system. This is true even if we choose as the reduced state for the system one that it is apparently identical to the equilibrium state. The reason for this behavior is the assumption of uncorrelated initial states. In a short time scale the system will become correlated with the environment, thereby inducing a rapid change in the state of the system, after which it will undergo the slow decay described by \eqref{QBMEt}. The fast decoherence process is not described by \eqref{QBMEt} because of our initial adiabatic-like assumption.}
\begin{equation}\label{QBMEt}
	E(t)=E_0 + \Av{ \frac{\Omega^2 q_0^2+ v^2_0}{2\Omega^2}}_{q_0,v_0} \expp{-\Gamma(t-t_0)}.
\end{equation}
Notice that this is similar to what would happen if we tried to compute the energy of an isolated harmonic oscillator using a Hamiltonian with an arbitrary frequency. When the Hamiltonian frequency differs from the physical frequency the expectation value of the fictitious Hamiltonian is not a constant but oscillates with the physical frequency of the oscillator. Only when the frequency coincides with the physical frequency the energy is constant. Similarly, we shall adopt the criterion that the physical frequency of the oscillator is that which makes the energy constant over short timescales.

Let us end up by summarizing the main results of this section. We have seen that for those systems exhibiting long-lived excitations the dynamics of the perturbations is governed by the kernel $H(t,t')$, which is related to the dissipation kernel and which corresponds to the retarded self-energy. At first approximation, the excitations decay exponentially with a decay rate given by the imaginary part of $H(\Omega)$ (which coincides with the value of the Fourier-transformed dissipation kernel). Moreover, the effective frequency of the perturbation is determined by the real part of $H(\Omega)$. 

In the next chapter the Brownian particle will play the role of the quasiparticle mode, $\Omega$ will correspond to the quasiparticle energy, and $\Gamma$ will correspond to the quasiparticle decay rate.

\index{Green function|see{propagator}}
\index{Hadamard function|see{propagator}}
\index{Whightman function|see{propagator}}

\part{Flat spacetime}

	\chapter{Particle-like excitations in generic flat backgrounds}

In this chapter we examine the elementary excitations over generic quantum states, as seen from the viewpoint of quantum field theory. In particular we would like to analyze all the information on the elementary excitations that can be obtained by studying the different two-point correlation functions.

\section{Particles in the vacuum} \label{sect:ParticleVacuum}

\index{Propagator!in the vacuum|(}

We begin by reviewing some aspects of the notion of particle in standard quantum field theory in the Minkowski vacuum. In particular we shall be interested in studying how the basic characteristics of the particle propagation can be obtained from the analytic structure of the propagators. While most results on this section can be found in standard quantum field theory textbooks (see for instance refs.~\cite{BrownQFT,WeinbergQFT,Peskin,ItzyksonZuber,GreinerQFT,Hatfield}), we present some of them in some detail, first, as an introduction to the more generic situations analyzed later on this chapter, and second, in order to clarify some aspects which will prove relevant later on. Except where more detailed references are given, we address the reader to the aforementioned textbooks for the remaining of this section.

\subsection{Fields, operators and states} \label{sect:FieldsStates}
 
Let us consider a real scalar field operator in the Heisemberg picture $\hat\phi(x)$, corresponding to the degrees of freedom of some field theory. If the theory is non-interacting, the field operator can be decomposed in the positive and negative frequency parts (or creation as destruction parts) as
\begin{subequations}
\begin{align}
	\hat\phi(x) &=  \hat\phi^+(x) + \hat\phi^-(x),\\
	\hat\phi^+(x) &= \sqrt{V} \int \udpi[3]{\vect p} \frac{1}{\sqrt{2E_\vect p}} \hat a^\dag_\vect p \expp{iE_\vect pt - i \vect p\cdot \vect x},\\
	\hat\phi^-(x) &= {\sqrt{V}} \int \udpi[3]{\vect p} \frac{1}{\sqrt{2E_\vect p}} \hat a_\vect p \expp{-iE_\vect pt + i \vect p\cdot \vect x}=(\hat\phi^+)^\dag(x),
\end{align}
\end{subequations}
where $V$ is the (formally infinite) space volume, where $E_\vect p^2:= m^2+\vect p^2$ and where  $\hat a^\dag_\vect p$ and $\hat a_\vect p$ are the creation and annihilation operators respectively, which verify the following conmutation relations:
\begin{subequations} \label{ConmutRelat}
\begin{align}
	[\hat{a}_\vect p,\hat{a}_\vect q] &=  [\hat{a}^\dag_\vect p,\hat{a}^\dag_\vect q] = 0,\\
	[\hat{a}_\vect p,\hat{a}^\dag_\vect q] &=  \delta_{\vect q\vect p}= \frac{1}{V} {(2\pi)^3 \delta^{(3)}{(\vect p - \vect q)}}.
\end{align}
\end{subequations}
If the theory is interacting, the same above relations hold in the interaction picture. The normalization is chosen so that it closely resembles the quantum mechanical normalization with a finite number of degrees of freedom.

Equivalently, the Schrödinger picture of the field operator in momentum space,
\begin{equation}
 	\hat\phi_\vect p = \frac{1}{\sqrt{V}} \int \ud[3]{\vect x}  \expp{-i \vect p \cdot \vect x} \hat\phi(\vect x),
 \end{equation}
 is connected to the creation and destruction operators via
  \begin{equation}
 	\hat\phi_\vect p = \frac{1}{\sqrt{2E_\vect p}} \big(\hat a_\vect p + \hat a^\dag_{-\vect p}\big).
 \end{equation}
The factor $V^{-1/2}$ in the definition of $\phi_\vect p$ is chosen so that the propagators verify:
\begin{equation*}
	G(t,t';\vect p) =  \int \ud[3]{\vect x} \expp{-i\vect p \cdot \vect x} \langle 0 | \hat\phi(t,\vect x) \hat\phi(t',\vect 0) |0\rangle = \av{ 0 |\hat\phi_\vect p (t) \hat\phi_{-\vect p}(t') |0}.
\end{equation*}

\index{Fock space}

Let us now consider the Hilbert space of the states of the theory, which has the structure of a Fock space. For non-interacing theories, the Fock space can be easily built with the aid of the creation and destruction operators:
\begin{equation}
	|\vect p_1\cdots\vect p_n\rangle = \hat a^\dag_{\vect p_1} \cdots \hat a^\dag_{\vect p_n} |0\rangle.
\end{equation}
The above equation assumes that all momenta are different; if this is not the case, the right hand side should incorporate a factor $1/m!$ for each repeated momentum, where $m$ is the number of times that it is repeated.  The states this way created  have well-defined momentum and energy:
\begin{subequations}
\begin{align}
	\hat{\vect p} |\vect p_1\cdots\vect p_n\rangle &=(\vect p_1 + \cdots + \vect p_n) |\vect p_1\cdots\vect p_n\rangle,\\
	\hat{H} |\vect p_1\cdots\vect p_n\rangle &=(E_0+E_{\vect p_1} + \cdots + E_{\vect p_n}) |\vect p_1\cdots\vect p_n\rangle,
\end{align}
\end{subequations}
with the momentum and energy operators given by
\begin{subequations}
\begin{align}
	\hat{\vect p} &= \int \udpi[3]{\vect k} \vect k\, \hat a^\dag_\vect k \hat a_\vect k, \\
	\hat{H} &= \int \udpi[3]{\vect k} E_\vect k \left(  \hat a^\dag_\vect k \hat a_\vect k +\frac12 \right).
\end{align}
\end{subequations}
From now on we will assume that the zero point energy $E_0$ has been subtracted so that the vacuum has zero energy.  Any state of the field can be expanded in the Fock space:
\begin{equation}
	|\Psi\rangle = \sum_m \prod_{\vect p_1, \ldots, \vect p_m} f(\vect p_1, \ldots, \vect p_m) |\vect p_1 \cdots \vect p_m \rangle
\end{equation}

\index{Occupation number}
\index{Mode decomposition}
Complementary to the Fock space expansion, the Hilbert space of the field also admits a mode decomposition. Namely, every pure state of the field can be decomposed in the following way:
\begin{equation}
	|\Psi\rangle = \prod_\vect k \sum_{n_\vect k} c(n_\vect k) |n_\vect k\rangle.
\end{equation}
where the occupation numbers $n_\vect k$ of the modes $|n_\vect k\rangle$ verify
\begin{equation}
	\hat{\vect p} |n_\vect k\rangle = n \vect k |n_\vect k\rangle.
\end{equation}
For free theories the vectors $ |n_\vect k\rangle$ are also eigenstates of the Hamiltonian.

\index{In state@\emph{In} state}
\index{Out state@\emph{Out} state}
\index{Mass!physical}
The situation gets more involved when the theory becomes interacting. In this case, the Hilbert space is still spanned by a set of eigenvectors of the momentum operator $\hat{\vect p}$ and the the full Hamiltonian $\hat H$. Among those states one finds the vacuum state and the one-particle state, labeled by the momentum, similarly to the non-interacting case. In contrast, multiparticle states cannot be labeled by the momentum of each particle in the state, because particles are interacting and the momentum of the particles (and even the number of them) changes. However, when particles are well separated the interaction between them is negligible.
Multiparticle states are recovered in the limit in which the distance between the particles is much larger than the typical interaction range.  Labeling by $|\vect p_1\cdots\vect p_n\rangle_\text{in(out)}$  the eigenstate of the full Hamiltonian that corresponds to a multiparticle state in the limit $t\to-\infty$ ($t\to\infty$), one has
\begin{equation}
	\hat{H} |\vect p_1\cdots\vect p_n\rangle_\text{in(out)} =(E_{\vect p_1} + \cdots + E_{\vect p_n}) |\vect p_1\cdots\vect p_n\rangle_\text{in(out)}
\end{equation}
where $E_\vect p^2 = \mph^2 + \vect p^2$, with $\mph$ being the \emph{physical mass} of the particles, which differs in general from the bare mass present in the Lagrangian. Notice that the particles appearing in the \emph{in} or \emph{out} states do not necessarily correspond to the particles appearing in the corresponding free theory: any unstable particle will not appear in the asymptotic states, and we will possibly have to add bound states to the asymptotic states. (For simplicity our notation assumes just one particle species and does not take into account these possibilities.) Notice also that the \emph{in} and \emph{out} states are defined at all times (although they only have special properties in the asymptotic limits). Therefore, either the \emph{in} or \emph{out} Fock spaces built from those states can be chosen as a basis for the Hilbert space of the interacting theory. 

In the remote past and future one can build a free theory that matches the properties of the interacting theory in these regimes,
\begin{equation}
	\hat{H}_0 |\vect p_1\cdots\vect p_n\rangle_\text{0} =(E_{\vect p_1} + \cdots + E_{\vect p_n}) |\vect p_1\cdots\vect p_n\rangle_\text{0}.
\end{equation}
These ``free states'' correspond unitarily to the \emph{in} and \emph{out} states of the interacting theory in the asymptotic limits\footnote{In order to properly define this correspondence one must work with wavepackets, so that there is localization in time and space; otherwise the states of the Fock space are completely delocalized. See ref.~\cite{WeinbergQFT} for more details.} (the correspondence being different in each case). The in and out states are therefore also unitarily related (via the S matrix). See refs.~\cite{Haag,GreinerQFT,WeinbergQFT} for more details.

\subsection{The K\"all\'en-Lehmann spectral representation}

\index{Propagator!Whightman}
\index{Spectral function}
\index{K\"all\'en-Lehmann spectral representation|see{spectral representation}}
\index{Spectral representation}

Let us consider a complete set of orthonormal states $|\alpha\rangle$ spanning the identity
\begin{equation}\label{IdentSpan}
	\hat 1 = \sum_\alpha |\alpha\rangle\langle \alpha|
\end{equation}
We need not specify the exact form of those states.\footnote{See the previous subsection for a discussion of the form of the states. In any case, notice that the summation in eq.~\eqref{IdentSpan} is symbolic, since there is a continuum of states.} We shall only require that they are eigenstates of the four-momentum operator $\hat p = (\hat H,\hat {\vect p})$:
\begin{equation}
	\hat p^\mu |\alpha\rangle = p^\mu_\alpha |\alpha\rangle.
\end{equation}
Taking into account that the time evolution of the Heisemberg operator can be expressed as
\begin{equation*}
	\hat \phi(x) = \expp{-i \hat p \cdot x} \hat \phi(0)  \expp{i \hat p \cdot x}, 
\end{equation*}
and using \eqref{IdentSpan}, it can readly be seen the Whightman function $G_+(x,x')= \langle 0 | \hat \phi(x) \hat\phi(x')|0\rangle$ can be expressed as
\begin{equation}
\begin{split}
	G_+(x,x')&=\int_0^\infty \ud s \int \udpi[4]{p} \theta(p^0) \delta(p^2+s) \expp{ip\cdot(x-x')} \\ &\qquad \times \sum_\alpha (2\pi)^4 \delta^{(4)}(p-p_\alpha) |\langle 0 |\phi(0)|\alpha\rangle|^2.
\end{split}
\end{equation}
The \emph{vacuum spectral function} $\rho(s)$ is defined as
\index{Spectral function} 
\begin{equation}\label{SpectralFunction}
	\rho(-p^2) := 
\frac{1}{2\pi} \sum_\alpha (2\pi)^4 \delta^{(4)}(p-p_\alpha) |\langle 0 |\phi(0)|\alpha\rangle|^2.
\end{equation}
The fact that the vacuum spectral function only depends on the invariant mass is a consequence of the Lorentz invariance of the theory and the vacuum state. Taking into account that the free Whightman function with squared mass parameter $s$ is given by (see appendix~\ref{app:CTP})
\begin{equation}
	G_+\fr(x,x';s) = \int \udpi[4]{p} \theta(p^0) 2\pi \delta(p^2+s) \expp{ip\cdot(x-x')},
\end{equation}
we obtain the \emph{K\"all\'en-Lehmann spectral representation} of the Whightman function:
\begin{equation}
	G_+(x,x') = \int_0^\infty \ud s \rho(s) G_+\fr(x,x';s).
\end{equation}
The vacuum spectral function verifies, among others, the following properties:
\begin{subequations}\label{PropVacSpectral}
\begin{align}
	\rho(s) &\geq 0,\\
	\rho(s) &= 0,\quad s<0, \\
	\int_0^\infty \ud{s} \rho(s) &= 1.
\end{align}
\end{subequations}

We have briefly sketched the derivation of the spectral representation for the case of the unordered propagator, but the same representation holds for any other propagator in the vacuum. In particular, it also holds for the Feynman, the Pauli-Jordan, and the retarded propagators. 
\index{Propagator!Feynman}
\index{Propagator!Pauli-Jordan}
\index{Propagator!retarded}
\begin{subequations}
\begin{align}
	G(x,x') = \langle 0 | [\hat\phi(x), \hat\phi(x')]|0 \rangle &=  \int_0^\infty \ud s \rho(s) G\fr(x,x';s),\\
 	\GR(x,x') = \theta(x^0-x'^0) \langle 0 | [\hat\phi(x), \hat\phi(x')]|0 \rangle &=  \int_0^\infty \ud s \rho(s) \GR\fr(x,x';s),\\
 	\GF(x,x') = \langle 0 |T \hat\phi(x) \hat\phi(x')|0 \rangle &=  \int_0^\infty \ud s \rho(s) \GF\fr(x,x';s).
\end{align}
\end{subequations}

The spectral representation can be also expressed in  momentum space. The corresponding expressions are the following
\begin{subequations}
\begin{align}
	G_+(p) &= 2\pi \rho(-p^2) \theta(p^0), \\
	G(p) &= 2\pi \rho(-p^2) \sign(p^0), \label{PauliJordanSpectral}\\
	\GR(p) &= \int_0^\infty \frac{-i\rho(s)\,\vd s}{-(p^0+i\epsilon)^2+\vect p^2+s},\label{retardedSpectral} \\
	\GF(p) &= \int_0^\infty \frac{-i\rho(s)\,\vd s}{p^2+s-i\epsilon}. \label{FeynmanSpectral}
\end{align}
\end{subequations}
The first  two equations show that in vacuum the Whightman functions and the Pauli-Jordan propagator essentially amount to the spectral function. The latter two equations show that the retarded propagator and the Feynman propagator have well-defined analyticity properties when considered as functions in the complex $p^2$ plane. In fact, they also have analyticity properties in the complex energy plane, as shown in the following two equivalent representations:
\begin{subequations}
\begin{align}
	\GR(p) &= \int \ud {k^0} \left[ \frac{i \rho(k^{02} - \vect p^2 )}{p^0 - {k^0} + i\epsilon} -  \frac{i \rho(k^{02} - \vect p^2 )}{p^0 + {k^0} + i\epsilon} \right], \\
	\GF(p) &= \int \ud {k^0} \left[ \frac{i \rho(k^{02} - \vect p^2 )}{p^0 - {k^0} + i\epsilon} -  \frac{i \rho(k^{02} - \vect p^2 )}{p^0 + {k^0} - i\epsilon} \right] 
\end{align}
\end{subequations}
Taking into account the relation of the spectral function with the Pauli-Jordan propagator, given by eq.~\eqref{PauliJordanSpectral}, one can also write
\begin{equation}\label{SpectralVacuumAlt}
	\GR(p) = \int \frac{\vd k^0}{2\pi}   \frac{i G(k^0,\vect p)}{p^0 - k^0 + i\epsilon}.
\end{equation}

This last equation also follows directly from the definition of the retarded propagator. In fact, using the relations between the different propagators, detailed in appendix \ref{app:GenRel}, provides an alternative way of deriving the spectral representations. This method will be used to derive the spectral representations in a generic background.

\index{Propagator!in the vacuum|)}
\index{Particle!in the vacuum|(}

\subsection{Stable particles}

\index{Mass!physical}

Let us assume that stable particles exist in the vacuum. From all the states of the theory, let us single out the one-particle states corresponding to those stable particles, characterized by the momentum $\vect p$ and the physical mass $\mph$: $\hat{\vect p }|\vect p\rangle = \vect p |\vect p\rangle$, $\hat{H }|\vect p\rangle = E_\vect p|\vect p\rangle$, with $E_\vect p^2 = \vect p^2 + \mph^2$. Notice that only truly stable particles correspond to eigenstates of the Hamiltonian. Considering those states explicitly, the spectral function can be developed as
\begin{equation*}
\begin{split}
	\rho(-p^2) &= \frac{1}{2\pi} \int \udpi[3]{\vect q} \frac{1}{2E_\vect q} (2\pi)^4 \delta^{(3)}(\vect p- \vect q) \delta(E_\vect q - p^0) \abs{\langle 0 |\phi(0)|\vect q\rangle}^2\\ &\quad +
\frac{1}{2\pi} \sum_\text{mult. states} (2\pi)^4 \delta^{(4)}(p-p_\alpha) |\langle 0 |\phi(0)|\alpha\rangle|^2.
\end{split}
\end{equation*}
This equation can be rewritten as 
\begin{equation}
	\rho(-p^2) = Z \delta(p^2 + \mph^2) + \theta(p^2+m_*^2) \sigma(-p^2) ,
\end{equation}
where 
\begin{equation}
	Z=\abs{\langle 0 |\phi(0)|\vect q\rangle}^2  
\end{equation}
is a positive constant (which does not depend on the momentum because of the Lorentz invariance of the theory), $\sigma(-p^2)$ is the contribution from the multiparticle states, and $m_*$ is the minimum rest energy of the multiparticle states. The $Z$ constant is frequently renormalized to one by rescaling the field (which amounts to adding a suitable counterterm to the original action). 

\index{Propagator!analytic structure}

\begin{figure}
\centering
\includegraphics[width=.9\textwidth]{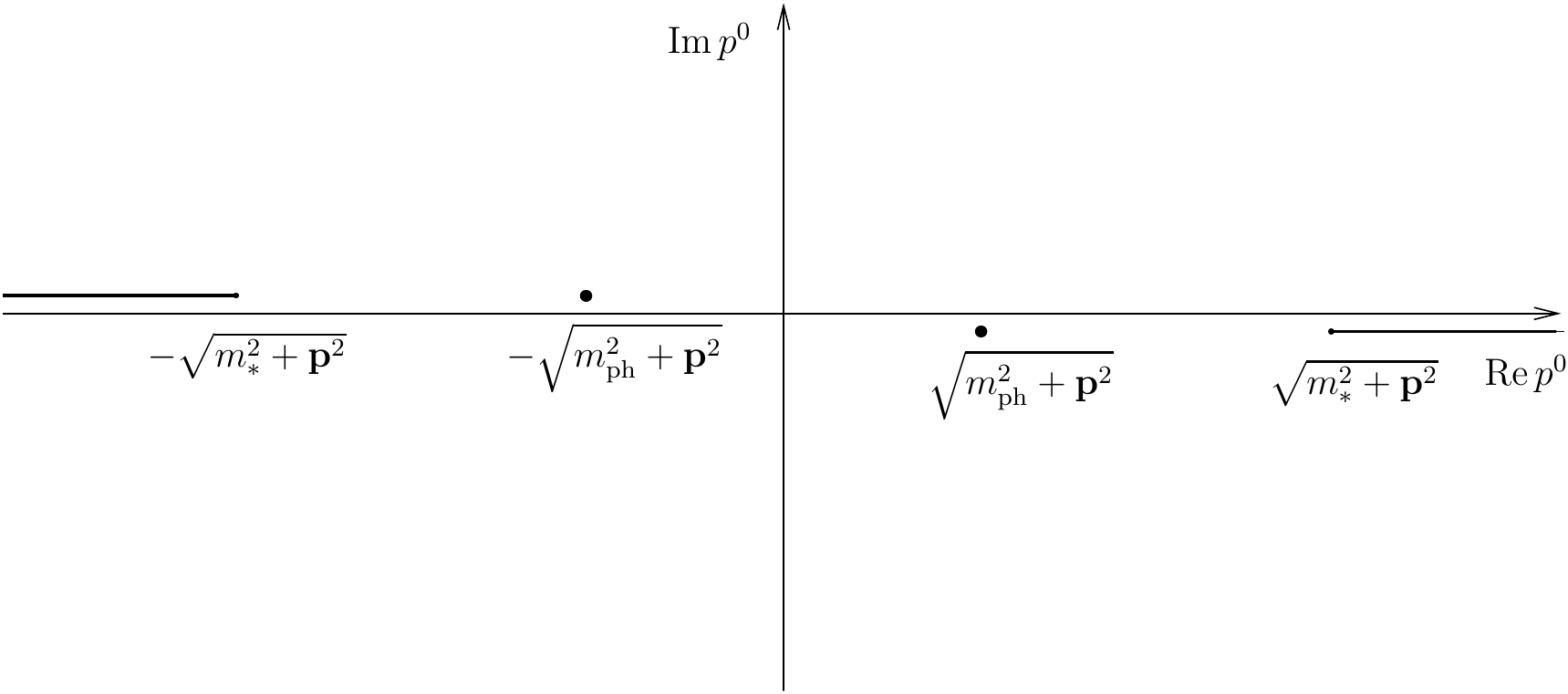}
\caption{Analytic structure of the Feynman propagator in the vacuum, as seen in the complex energy plane. There are two poles corresponding to the stable particle, and two branch cuts, whose branching points indicate the minimum energy for the multiparticle states. Between the poles and the branching points there might be as well other poles corresponding to bound states (not shown in the plot).}
\label{fig:FeynmanVacuum}
\end{figure}

According to eq.~\eqref{FeynmanSpectral}, the Feynman propagator in momentum space has the structure:
\begin{equation}
	\GF(p) = \frac{ -i Z}{p^2 + m^2 - i\epsilon} + \int_{m_*^2}^\infty \vd s \frac{ -i \sigma(s)}{p^2 + s - i\epsilon}.
\end{equation}
Thus, the Feynman propagator is analytical in the complex $p^2$ plane, except for a pole at $p^2 = -\mph^2 + i\epsilon$, and a branch cut starting at $p^2 = -m_*^2 + i\epsilon$ and running parallel to the real axis towards the complex infinity. There are additional poles if bound states can be formed; we shall not take into account this possibility in the following. Although it is customary to analyze the analytic structure of the Feynman propagator in the complex $p^2$ plane, it will prove more useful for us to analyze the analytical structure in the complex energy plane. As shown in figure \ref{fig:FeynmanVacuum}, in the complex energy plane (complex $p^0$ plane), the Feynman propagator is analytical except for the two poles at $p^0 = \pm(E_\vect p - i \epsilon)$ and the two branch cuts, going from $\pm(E^*_\vect p - i \epsilon)$ to the complex infinity parallel to the real axis.

\index{Propagator!analytic structure}

\index{Self-energy!in the vacuum}
In perturbative field theory, the renormalized Feynman propagator can be resummed in the following way
\begin{equation}\label{SelfEnergyVacuum}
	\GF(p) = \frac{-i}{p^2 + m^2 + \Sigma(-p^2)},
\end{equation}
where $m$ is the renormalized mass and $\Sigma(-p^2)$ is the \emph{self-energy}, corresponding to the sum of all one-particle irreducible Feynman diagrams. The locus of the poles and cuts is given by the solution of the equation
\begin{equation*}
	p^2 + m^2 + \Sigma(-p^2) = 0. 
\end{equation*}
The zeros of this equation lie next to the real axis as dictated by the spectral representation \eqref{FeynmanSpectral}. (The self-energy function $\Sigma(-p^2)$ must have the apropriate analytical structure, so the K\"all\'en-Lehmann representation is verified.) The lowest zero of the equation is the pole at $p^2=-\mph^2$. 
\index{On-shell renormalization condition}
With the on-shell renormalization conditions the renormalized mass coincides with the physical mass, $m=\mph^2$, so that $\Sigma(-\mph^2)=0$.

\index{Propagator!analytic structure}

\begin{figure}
\centering
\includegraphics[width=.9\textwidth]{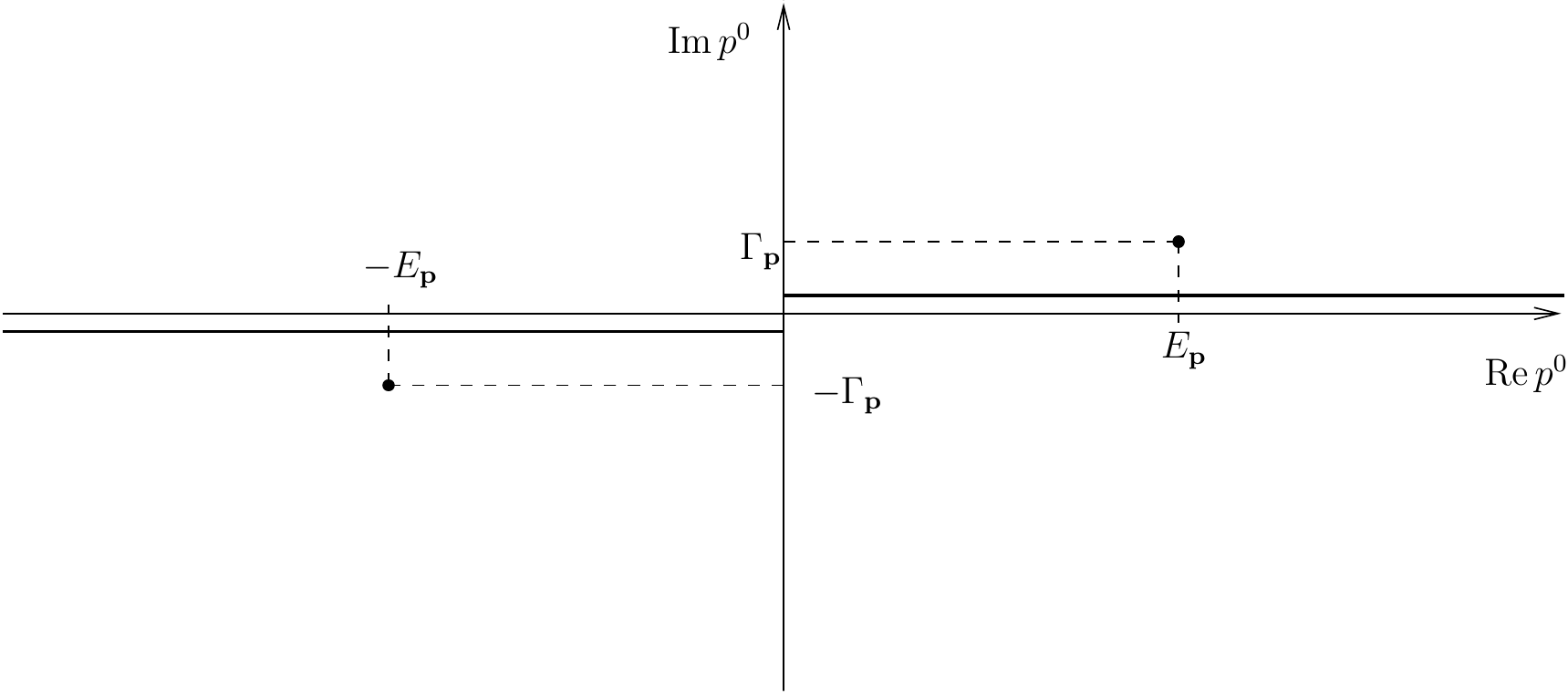}
\caption{Analytic structure of the retarded propagator in the vacuum, as seen in the complex energy plane. See the caption of the preceding figure \ref{fig:FeynmanVacuum}.}
\label{fig:retardedVacuum}
\end{figure}

Although it is usual to consider the Feynman propagator, the analytic structure of other propagators can be studied as well. The retarded propagator will be of special importance for us:
\begin{equation} \label{VacuumRetardedAnalytic}
	\GR(p) = \frac{ -i Z}{p^2+ m^2 - i\epsilon p^0} + \int_{m_*^2}^\infty \vd s \frac{ -i \sigma(s)}{p^2 + s - i\epsilon p^0}.
\end{equation}
As shown in figure \ref{fig:retardedVacuum}, in the complex energy plane (complex $p^0$ plane), the retarded propagator is analytical except for the two poles at $p^0 = \pm E_\vect p - i \epsilon $ and the two branch cuts, going from $\pm E^*_\vect p - i \epsilon$ to the complex infinity parallel to the real axis.

\subsection{Unstable particles}

\index{Particle!unstable}

Recall that unstable particles are not asymptotic states and hence do not correspond to eigenstates of the Hamiltonian nor they belong to the 1-particle sector of the Fock space. Rather, they correspond to combinations of those multiparticle states corresponding to the decay products of the unstable particle. In terms of the K\"all\'en-Lehmann representation of the propagator, this means that unstable states are represented by a branch cut rather than a pole. Hence the spectral function for the multiparticle states is given by 
\begin{equation}
	\rho(-p^2) = \theta(p^2+m_*^2) \rho(-p^2) ,
\end{equation}
where $m_*$ is in this case the minimum energy to create a multiparticle state.

\index{Decay rate}
In any case, the spectral representations of the Feynman and retarded propagators, eqs.\ \eqref{FeynmanSpectral} and \eqref{retardedSpectral}, imply that singularities of the propagator must lie infinitesimally close to the real axis, either in the complex $p^2$ or $p^0$ planes. Apparently, this might seem in contradiction with the usual statement that unstable particles correspond to a pole of the propagator in the complex plane,
\begin{equation}\label{decayGF}
	\GF(p) = \frac{-iZ}{p^2 + m^2 - i m \gamma},
\end{equation}
with $\gamma$ being the decay rate of the unstable particle. In this subsection we will briefly sketch how to reconcile both statements. For more details on this point see refs.~\cite{MatthewsSalam58a,MatthewsSalam58b,BrownQFT,JacobSachs60,Veltman63,Cocolicchio98,GalindoPascual}.

Decomposing the Feynman propagator in the real and imaginary parts, according to the property $1/(x\pm i\epsilon)= P(1/x) \mp i\pi \delta(x)$, we get
\begin{equation}
	\GF(p) = 
		-i\left[ \PV \int \frac{ \rho(s) \vd{s}}{p^2+s} + i\pi \rho(-p^2)\right]
\end{equation}
where $p^2$ is taken real in the above equation. Notice that the propagator $\GF(p)$ is discontinuous across the branch cut, with the discontinuity given by
\begin{equation}
	\Disc \GF(p)  = - 2\pi \rho(-p^2).
\end{equation}
We can continue analytically the Feynman propagator across the branch cut by defining the function
\begin{equation}
	\GF'(p) := 
	\begin{cases}
		\GF(p) & \Im (-p^2), \geq 0, \\
		\GF(p) - 2\pi \rho(-p^2), & \Im (-p^2) < 0,
	\end{cases}
\end{equation}
which is analytical across the cut. Alternatively $\GF'(p)$ and $\GF(p)$ can be regarded as two different Riemann sheets of the same complex function.

The Feynman propagator can be resummed as
\begin{equation}\label{FeynmanUnstable}
	\GF(p) = \frac{-i}{p^2+m^2 + \Sigma(-p^2) } = \frac{-i}{p^2+m^2 + \Re \Sigma(-p^2) + i\Im \Sigma(-p^2) }.
\end{equation}
We now define the mass of the unstable particle as the lowest order solution of
\begin{equation}
	-\mph^2+ m^2 + \Re \Sigma(\mph^2) = 0
\end{equation}
The mass as defined above corresponds approximately to the rest energy of the particle (this assertion will be checked later on). However it should be noted that the energy of an unstable particle fluctuates according to the time-energy uncertainty principle. We would like now to approximate \Eqref{FeynmanUnstable} around $p^2= -\mph^2$. However such an approximation is not meaningful, since the Feynman propagator is discontinuous near the real axis. Instead, we may approximate its analytical continuation $\GF'(p)$ as
\begin{equation}
	\GF'(p) \approx \frac{-iZ}{p^2+\mph^2 + i\Im \Sigma(\mph^2) }.
\end{equation}
Identifying
\begin{equation}
	\gamma =-\frac{1}{\mph} \Im \Sigma(\mph^2)
\end{equation}
as the decay rate of the particles in their rest frame, according to the optical theorem, we recover \Eqref{decayGF}. Thus, the analytic continuation of the Feynman propagator has a pole in the region $\Im \Sigma(p^2) < 0$, whose real part corresponds to the approximate mass of the particle and whose imaginary part corresponds to the decay rate. See refs.~\cite{BrownQFT,GalindoPascual} for additional details.

In summary, in the complex $p^2$ plane the Feynman propagator of an unstable particle has no poles, but has a branch cut next to the real axis, while its analytical continuation across the cut has a pole in the region $\Im{(-p^2)}<0$.  An analogous result applies for the complex $p^0$ plane. In terms of Riemann sheets, one can consider that the Feynman propagator in the complex plane has two Riemman sheets, and that the two complex poles are in the second sheet.

\subsection{Time evolution of the propagator}\label{sect:TimeEvol}

\index{Propagator!time evolution}

So far we have carried out the analysis in the energy-momentum representation. It will prove also illustrative to consider the time-momentum representation of the propagator,
\begin{equation*}
	\GF(t,t';\vect p) = \int\frac{\vd p^0}{2\pi} \expp{-ip^0(t-t')} \GF(p).
\end{equation*}
The aim is to compute the behavior of the propagator for large time lapses. We shall consider both the stable and unstable cases simultaneously. If the particle is unstable, instead of computing the Fourier transform of $\GF(p)$, we will work with its analytic continuation $\GF'(p)$ (the results shall not be affected because $\GF(p)$ and $\GF'(p)$ are identical over the real axis). According to \Eqref{FeynmanUnstable}, the Feynman propagator can be expressed as: 
\begin{equation}
	\GF'(p) \approx \frac{-iZ}{p^2+\mph^2 - i \mph \gamma} + \widetilde G(p),
\end{equation}
where $\gamma=\epsilon$ if the particle is stable. The function $ \widetilde  G(p)$ is analytic function in the vicinity of the pole, but it does develop a singular behavior when approaching the different particle creation thresholds. 

The temporal behavior of the pole can be easily computed:
\begin{equation}
	\GF(t,t';\vect p) = \frac{-iZ}{2\sqrt{\vect p^2+ \mph^2-i\mph\gamma}} \expp{-i\sqrt{\vect p^2 + \mph^2 - i \mph \gamma} |t-t'|} + \widetilde G(t,t';\vect p)
\end{equation}
In order for the particle concept to be meaningful, the condition $\gamma \ll \mph$ must be verified (otherwise one would speak of wide resonances rather than particles). Under these conditions, the above expression can be approximated as
\begin{equation}\label{GFTimeDecay}
	\GF(t,t';\vect p) = \frac{-iZ}{2E_\vect p} \expp{-iE_\vect p|t-t'|}\expp{- \Gamma_\vect p |t-t'|/2} + \widetilde G(t,t';\vect p)
\end{equation}
with
\begin{equation}
	E^2_\vect p = \vect p^2 + \mph^2,
\end{equation}
and where we have defined the decay rate in the laboratory rest frame
\begin{equation}
	\Gamma_\vect p := \frac{ m \gamma}{E_\vect p} = -\frac{1}{E_\vect p} \Im \Sigma(\mph^2).
\end{equation}
\index{Decay rate}

The behavior of remaining piece $\widetilde G(p)$ remains to be determined. In general, it depends on the precise value of the spectral function along the branch cut, which in turn depends on the multiparticle structure of the theory. However, by appealing to the Riemann-Lebesgue theorem, one can show \cite{BrownQFT} that an order of magnitude estimation of $\widetilde G(t,t';\vect p)$ is given by
\begin{equation}
	|\widetilde G(t,t';\vect p)|  \sim \frac{1}{M^{\alpha+1} |t-t'|^\alpha},
\end{equation}
where $M$ is the scale in which the leading multiparticle threshold starts, and where $\alpha$ is some positive coefficient which depends on the particular structure of the theory.

For stable particles the situation is therefore simple. Two time regimes can be distinguished. For short times, of the order of the de Broglie size of the particle $\mph^{-1}$, transient effects dominate, and the time behavior depends on the particular details of the field theory model. For large time lapses ($|t-t'|\gg M^{-1}$), the time behavior is dominated by the pole in the vicinity of the real axis, so that the propagator shows a constant modulus and a rotating phase, whose frequency is given by the energy of the particle.

The situation becomes more involved (and more interesting) for unstable particles. In this case three different time regimes can be distinguished. At short times similar transient effects are found. For relatively large time lapses, the time behavior is dominated by the pole, which in this case is located off the real axis. The modulus of the propagator decreases exponentially with a rate $\Gamma_{\vect p}$. Transient effects are subdominant in this regime, since they decay in a much faster timescale $M^{-1}$. For extremely long times, however, ``transient'' effects dominate again, since any power low decay dominates over an exponential decay for sufficiently long times.

Experimentally the breakdown of the exponential law for very long times is almost never observable, since power-law terms dominate again after many particle lifetimes, when the chances of observing the particle are almost null. 

\subsection{Two-point functions and asymptotic fields}\label{sect:Asympt}

\index{Asymptotic fields}


We have just seen that the 2-point correlation functions correspond to the 2-point correlation function of a corresponding free field plus an additional multiparticle contribution, which vanishes for long times:
\begin{subequations}
\begin{equation}
	G_+(t,t';\vect p) = Z G_+^\text{(1p)}(t,t';\vect p) + \widetilde G_+(t,t';\vect p) \xrightarrow[t-t'\to\infty]{} Z G_+^\text{(1p)}(t,t';\vect p)
\end{equation}
or equivalently
\begin{equation}
	\langle 0 | \hat\phi_\vect{p} U(t,t') \hat\phi_\vect{-p} |0 \rangle  \xrightarrow[t-t'\to\infty]{}  \frac{Z}{2E_\vect p} \, \langle \vect p| U(t,t') | \vect p\rangle
\end{equation}
\end{subequations}
where $|0\rangle$ and $|\vect p\rangle$ are the vacuum and 1-particle state of the interacting theory respectively, and where $U(t,t')$ is the time evolution operator.
We have used the fact that even in interacting theories the existence of a one-particle state with analogous properties to the corresponding free theory is guaranteed by the Lorentz invariance. Therefore, we can make the identification
\begin{equation}\label{PField}
	|\vect p\rangle \cong \frac{\sqrt{2 E_\vect p}}{\sqrt{Z}} \hat\phi_{-\vect p} | 0\rangle .
\end{equation}
The symbol $\cong$ here means equivalence when evaluated in a matrix element in the limit of large time lapses.  Physically, the field operator excites the one-particle state, but also excites the multiparticle state sector. However, when computing a transition element multiparticle excitations are off the mass shell and they decay quickly.

\index{In state@\emph{In} state}
\index{Out state@\emph{Out} state}
\label{Asymptotic field operator}
The heuristic argument given above can be connected to the fact that the particle content of the theory corresponds to that of a free theory in the asymptotic limits.  We define the asymptotic field operator $\phir$,  which by assumption obeys free equations of motion,
\begin{subequations}
\begin{equation} \label{EffeciveFree}
	(\partial_\mu\partial^\mu + \mph^2) \phir =0, \qquad
	- \ddot \phir_\vect p + (\vect p^2+\mph^2)\phir_\vect p=0,
\end{equation}
and which corresponds to the field operator through
\begin{equation}
	\phir(\vect x) \cong   Z^{-1/2} \phi(\vect x), \qquad
	 \phir_\vect p \cong Z^{-1/2} \phi_\vect p.
\end{equation}
\end{subequations}
We also define the corresponding creation and annihilation operators,
\begin{subequations}
\begin{align}
	\hat \ar_\vect p &= \sqrt{2E_\vect p} \left[ \hat\phir_{-\vect p} +   \frac{i}{E_\vect p} \hat\pir_{-\vect p} \right], \\
	\hat \ar^\dag_\vect p &= \sqrt{2E_\vect p} \left[ \hat \phir_{\vect p} -   \frac{i}{E_\vect p} \hat \pir_{\vect p} \right],
\end{align}
\end{subequations}
where $\hat \pir_\vect p$ is the canonical momentum operator and $E_\vect p^2=\vect p^2 + \mph^2$.
The field $\phir$ corresponds to the \emph{in} and \emph{out} fields\footnote{One possible concern is the fact that there is an apparent contradiction between the two canonical conmutation relations $[\hat\phi,\hat\pi]=i$ and $[\hat\phir,\hat\pir]=i$. In reality there is no such contradiction because the two fields are equivalent in the asymptotic limits only when evaluated in matrix elements (weak operator convergence) \cite{GreinerQFT}.  In any case, the reader must be warned  that a fully rigorous analysis is usually not feasible for standard field theories.} (see subsection \ref{sect:FieldsStates}). \emph{In} and \emph{out} Fock spaces can be constructed with the field $\phir$. For further  details refer to \cite{GreinerQFT,Haag,WeinbergQFT}.

With the asymptotic field operator the one-particle state can be represented as
\begin{equation}\label{PFieldAsym}
	|\vect p\rangle \cong  {\sqrt{2 E_\vect p}} \hat\phir_{-\vect p} | 0\rangle_\text{in/out} =\hat\ar^\dag_\vect p | 0\rangle_\text{in/out} 
\end{equation}
If mathematical rigor is not needed, comparing \Eqref{PField} and \Eqref{PFieldAsym} we see that  the distinction between the  field $\phi$ and the the asymptotic field $\phir$ may be blurried provided the constant $Z$ is renormalized to one and assumes that in the asymptotic limit the field obeys effective free equations of motion. In this case one may simply write
\begin{equation}\label{PFieldApprox}
	|\vect p\rangle \cong  {\sqrt{2 E_\vect p}} \hat\phi_{-\vect p} | 0\rangle_\text{in/out} \cong \hat a^\dag_\vect p | 0\rangle_\text{in/out} 
\end{equation}

When particles are unstable the situation is less clear since strictly speaking it does not exist a one-particle sector which also is an eigenstate of the Hamiltonian (the only eigenstates being the multiparticle states corresponding to the decay products of the unstable particles). However, if particles are long-lived one may think of approximate 1-particle states (which will in fact correspond to multiparticle state combinations). We will assume that those are also approximately given by \Eqref{PFieldApprox}.

\subsection{Physical interpretation of the time representation} 

\index{Propagator!time representation}

We have previously seen that in frequency space the poles of the propagator are connected with the energy and lifetime of the particle.
Let us confirm this statement studying the time evolution of the expectation value of the Hamiltonian operator.  

In field theory, strictly speaking only asymptotic properties are completely well-defined. However, as long as one considers sufficiently large time lapses (compared to the relevant inverse energy scales), one can obtain approximate results for finite times. We shall use the asymptotic particle representation in terms of fields, described in the previous subsection, for finite but sufficiently long time lapses. Since anyway the calculations will not be completely rigorous, we shall blurry the distinction between the usual and asymptotic fields, and instead we will suppose that the quantity $Z$ has been renormalized to one.
We shall treat the stable and unstable cases simultaneously.

The relevant system/environment separation will be discussed in sect.~\ref{sect:OQS}, but for the moment let us assume that the reduced Hamiltonian operator for the relevant mode is given by\
\begin{equation}
	\hat H_\vect p = E_\vect p\left( \hat{a}^{\dag}_\vect p \hat{a}_\vect p + \frac{1}{2} \right).
\end{equation}
If in the remote past time $t_{0}$ a particle is introduced into the system, the state for later times will be $|\vect p\rangle \cong a^\dag_{\vect p}|0\rangle$. The evolution of the number of particles in this state is given by
\begin{equation}
	E(t,t_{0};\vect p) := \langle \vect p|\hat H_{\vect p}(t)|\vect p\rangle = E_\vect p 
		\langle 0|\hat a_{\vect p}U(t_{0},t)  \Big(  \hat a^\dag_{\vect p}\hat  a_{\vect p} + \frac{1}{2} \Big) U(t,t_{0}) \hat a^\dag_{\vect p}|0\rangle,
\end{equation}
Introducing a resolution of the identity we obtain:
\begin{equation}
	E(t,t_{0};\vect p) ={E_\vect p}  \sum_{\alpha}		\langle 0|\hat a_{\vect p}U(t_{0},t) \hat a^\dag_{\vect p} |\alpha \rangle \langle \alpha  |\hat  a_{\vect p}U(t,t_{0}) \hat a^\dag_{\vect p}|0\rangle + E_0,
\end{equation}
where $E_0 =( E_\vect p/2) \langle 0 | \hat a^	\dag_\vect p \hat a_\vect p |0 \rangle$ is the vacuum energy. By energy and momentum conservation, only the vacuum survives from the above summation. Therefore:
\begin{equation*}
\begin{split}
	 E(t,t_{0};\vect p) &= E_0+ {4E_\vect p^3} |\langle 0|\hat \phi_{-\vect p} U(t_{0},t) \hat\phi_{\vect p} |0 \rangle|^2
\end{split}
\end{equation*}
from where we obtain
\begin{equation}
	 E(t,t_{0};\vect p) =  E_0 + {4E_\vect p^3} |\GF(t_0,t;\vect p)|^2
\end{equation}
Introducing the explicit value of the propagator, given by \Eqref{GFTimeDecay}, we get the expected result
\begin{equation}
	E(t,t_0;\vect p) = E_0+ E_\vect p \expp{-\Gamma_\vect p(t-t_0)} .
\end{equation}
Particles have energy $E_\vect p$ and decay in  a timescale $\Gamma_\vect p$.

\pagebreak

The above analysis is not completely rigorous in general, but it is even less rigorous for unstable particles. Recall that unstable particles do not correspond to any eigenstate of the Hamiltonian, and thus no asymptotic states can be associated to them. Therefore, strictly speaking, one-particle states are a linear combination of many multiparticle states and one cannot make reference to energy conservation, since energy conservation is associated to asymptotic properties. However, if the lifetime of the particles is long enough one can think of approximate asymptotic states and approximate energy conservation, so that the above calculation would be approximately valid. 

We will reanalyze the same problem with greater care in subsection \ref{sect:Quasitime}.

\index{Particle!in the vacuum|)}
\index{Correlation function|see{propagator}}

\section{Propagators and self-energies in an arbitrary background state} \label{sect:GSigmaGeneral}

For the remaining of this chapter we will be interested in considering field theory over generic states, different than the vacuum, characterized by some density matrix $\hat \rho (t)$. As explained in the introduction, a closed time path approach (CTP) to field theory is naturally needed. The basic aspects of the CTP method are reviewed in appendix \ref{app:CTP}. We will consider only one scalar field, but the results could be trivially extended to any number of scalar fields. We shall focus our attention on the 2-point correlation functions.

\subsection{Two-point correlation functions}

\index{Propagator!in flat backgrounds|(}
\index{Direct basis}
In the vacuum the analysis of the Feynman propagator is usually sufficient. In a generic state this is not usually the case, and an analysis of the different propagators is in order. The Feynman propagator, positive and negative Whightman functions and Dyson propagator,
\index{Propagator!Feynman}
\index{Propagator!Whightman}
\index{Propagator!Dyson}
\begin{subequations} \label{CorrFunct}
\begin{align}
    G_{11}(x,x') &= G_\text{F}(x,x') = \Tr{\big(
\hat\rho\, T \hat\phi(x) \hat\phi(x') \big)
}, \\
    G_{12}(x,x') &= G_{+}(x,x') = \Tr{\big(
\hat\rho\,  \hat\phi(x) \hat\phi(x') \big)
}, \\
    G_{21}(x,x') &=G_{-}(x,x') = \Tr{\big(
\hat\rho\,  \hat\phi(x') \hat\phi(x) \big)
}, \\
    G_{22}(x,x') &=G_\text{D}(x,x') = \Tr{\big(
\hat\rho\,  \widetilde T \hat\phi(x) \hat\phi(x')\big)
 },
\end{align}
\end{subequations}
appear naturally when dealing with perturbation theory in the CTP formalism (see appendix \ref{app:CTP}), and can be conveniently organized in a $2\times 2$ matrix $G_{ab}$, the so-called \emph{direct basis}:
\begin{equation}
	G_{ab}(x,x') = 
		\begin{pmatrix}
			\GF(x,x') & G_-(x,x') \\
			G_+(x,x') & G_\text{D}(x,x')
 		\end{pmatrix}.
\end{equation}

We may also consider the Pauli-Jordan or commutator propagator,
\index{Propagator!Pauli-Jordan}
\begin{subequations}\label{CorrFunct2}
\begin{equation}
 G(x,x') = \Tr{\big(
\hat\rho\,[  \hat\phi(x) ,\hat\phi(x')] \big)
},
\end{equation}
and the Hadamard or anticonmutator function
\index{Propagator!Hadamard}
\begin{equation}
	G^{(1)}(x,x') = \Tr{\big(
\hat\rho\,\{  \hat\phi(x) ,\hat\phi(x') \} \big)}.
\end{equation}
\end{subequations}
For linear systems\footnote{By linear systems we mean systems whose Heisemberg equations of motion are linear. These correspond either to non-interacting systems or to linearly coupled systems.} the Pauli-Jordan propagator is independent of the state and carries information about the system dynamics. Finally, one can also consider the retarded and advanced propagators,
\index{Propagator!retarded}
\index{Propagator!advanced}
\begin{subequations}\label{CorrFunct3}
\begin{align}
    G_\mathrm {R}(x,x') &= \theta(x^0-x'^0) G(x,x') =\theta(x^0-x'^0) \Tr{\big(
\hat\rho\, [\hat \phi(x),\hat\phi(x')]\big)}, \label{RetardedProp}\\
    G_\mathrm {A}(x,x') &= \theta(x'^0-x^0) G(x,x') =\theta(x'^0-x^0) \Tr{\big(
\hat\rho\, [\hat \phi(x),\hat\phi(x')]\big)},
\end{align}
\end{subequations}
which also do not depend on the the state for linear systems. Notice that the Hadamard functions are real, while the Pauli-Jordan, retarded and advanced propagators are purely imaginary in the spacetime representation. Other propagators have both a real and an imaginary part.

\index{Keldysh basis}
\index{Physical basis|see{Keldysh basis}}
As shown in appendix \ref{app:CTP}, by doing a unitary transformation to the propagator matrix it is possible to work in the so-called \emph{physical} or \emph{Keldysh basis} \cite{Das,EijckKobesWeert94},
\begin{equation}\label{Keldysh}
	G'_{a'b'}(x,x') = 
		\begin{pmatrix}
			G^{(1)}(x,x') & G_\text R(x,x') \\
			G_\text A(x,x') & 0
 		\end{pmatrix},
\end{equation}
which is useful for some purposes. Feynman rules have a non-standard, slightly more involved expression in terms of the Keldysh basis \cite{EijckKobesWeert94}, but once we take this fact into account, all matrix relations are valid both in the direct and Keldysh basis.  For linear systems, the non-diagonal components of the Keldysh basis carry information on the system dynamics, whereas the non-vanishing diagonal component has information on the state of the field. We will generally perform calculations in the direct basis, and the physical interpretation will be done either in the direct or the Keldysh basis.

It will be also useful to introduce the correlation functions in momentum space, which are defined as the Fourier transform of the spacetime correlators with respect to the difference variable $\Delta=x-x'$ keeping constant the central point $X=(x+x')/2$:
\begin{equation} \label{MidPoint}
    G_{ab}(p;X) = \int \ud[4]{x} \expp{-i p \cdot \Delta}
    G_{ab}(X+\Delta/2,X-\Delta/2).
\end{equation}
For homogeneous and static backgrounds the Fourier-transformed propagator does not depend on the central point $X$, so that the relation between the spacetime and frequency representations of the propagator correspond to that of the vacuum. Notice that the retarded and advanced propagators, which are purely imaginary in the spacetime representation, develop an imaginary part in the momentum representation.

Obviously not all propagators are independent: the complete set of propagators is determined by a symmetric and an antisymmetric function. From the Feynman propagator the other Green functions can be derived, but, in contrast, the retarded propagator lacks the information about the symmetric part of the correlation function. In appendix \ref{app:GenRel} we sketch the relations between the different propagators, both in the spacetime and frequency representations.

\subsection{Self-energies}

\index{Self-energy!in flat backgrounds|(}
\index{Schwinger-Dyson equation}
As in the vacuum, self-energies can be introduced for interacting systems.
Interaction mixes the two CTP branches, so that  the self-energy has a matrix structure and is implicitly defined through the \emph{Schwinger-Dyson equation}:
\begin{equation} \label{SelfEnergyGeneral}
    G_{ab}(x,x') = G_{ab}^{(0)}(x,x')+
    \int \ud[4]{y} \ud[4]{y'} G_{ac}^{(0)}(x,y) [-i\Sigma^{cd}(y,y')]  G_{db}(y',z),
\end{equation}
where $G^{(0)}_{ab}(x,x')$ are the propagators of the free theory, and $G_{ab}(x,x')$ are the propagators of the full interacting theory. The CTP indices $a,b,c\ldots$ are either 1 or 2, and we use a Einstein summation convention for repeated CTP indices. The $ab$ component of the self-energy can be computed in the direct basis, similarly
to the vacuum case, as the sum of all one-particle irreducible
diagrams with  amputated external legs that begin and end with
type $a$ and type $b$ vertices, respectively (cf. appendix \ref{app:CTP}). 

Two important differences arise with respect to the vacuum case. First, note that eq.~\eqref{SelfEnergyGeneral} is a matrix equation relating the four components of the self-energy with the four components of the propagator. Therefore it does not exist diagonal relation between $G_{11}(x,x')$ and $\Sigma^{11}(y,y')$.  Second, in the vacuum  eq.~\eqref{SelfEnergyGeneral} can be solved in closed analytical form by introducing the Fourier transform, so that the propagator can be expressed in terms of the self-energy. In general this is not possible, as can be shown by expressing \Eqref{SelfEnergyGeneral} in momentum representation:
\begin{equation} \label{SelfEnergyGeneralMomentum}
\begin{split}
    G_{ab}(p;X) &= G_{ab}^{(0)}(p;X)-i
    \int \ud[4]{y} \ud[4]{y'} \ud[4]{\Delta} \int \udpi[4]{q} \udpi[4]{q'} \udpi[4]{q''} \\ &\qquad\times \expp{i q\cdot(X + \Delta/2-y) + i q'\cdot(y-y') + i q''\cdot(y'-X - \Delta/2) - i p \cdot \Delta} \\ &\qquad\times G_{ac}^{(0)}{\left(q\right)} \Sigma^{cd}{\big(q';\tfrac{y+y'}{2}\big)}  G_{db}{\big(q'';\tfrac{X-\Delta/2-y'}2\big)}.
\end{split}
\end{equation}
Therefore, generally there is no simple relation such as \Eqref{SelfEnergyVacuum}.

The self-energy components are not independent but are related through non-perturbative relations. The different relations between the self-energy components are also detailed in appendix \ref{app:GenRel}. A particularly useful combination is the  retarded self-energy,
defined as $\Sigma_\mathrm R(x,x') := \Sigma^{11}(x,x') +
\Sigma^{12}(x,x')$. It  is related to the retarded propagator through
\begin{subequations} \label{SelfEnergyGeneralRetardedAdvanced}
\begin{equation} \label{SelfEnergyGeneralRetarded}
    \GR(x,x') = \GR^{(0)}(x,x')+
    \int \ud[4]{y} \ud[4]{y'} \GR^{(0)}(x,y) [-i\SigmaR(y,y')]  \GR(y',z),
\end{equation}
The above relation, which justifies the name of $\Sigma_\mathrm{R}(x,x')$, can be demonstrated by
expanding the matrix equation \eqref{SelfEnergyGeneral} and using the relations
between the different propagators and self-energies that appear in appendix \ref{app:GenRel}. A similar relation holds for the
advanced propagator $G_\mathrm A(x,x')$ and the
advanced self-energy $\Sigma_\mathrm A(x,x') = \Sigma^{11}(x,x') + \Sigma^{21}(x,x')$:
\begin{equation} \label{SelfEnergyGeneralAdvanced}
    \GA(x,x') = \GA^{(0)}(x,x')+
    \int \ud[4]{y} \ud[4]{y'} \GA^{(0)}(x,y) [-i\SigmaA(y,y')]  \GA(y',z).
\end{equation}
\end{subequations}
Notice that both the retarded and advanced self-energies are real in the spacetime representation (although they are generally complex in the momentum space).

\index{Keldysh basis}

We insist once more that a relation such as that described by eqs.\ \eqref{SelfEnergyGeneralRetardedAdvanced} is exclusive for the retarded and advanced propagators, and does not hold for the other propagators. In fact, it is a consequence of the causality properties of the retarded and advanced propagators, as highlighted in the following alternative derivation, in which we shall make use of the Keldysh basis \eqref{Keldysh}. 
[Bear in mind that relation \eqref{SelfEnergyGeneral} also applies to the Keldysh basis.] In terms of this basis, eq.~\eqref{SelfEnergyGeneral}, applied to the retarded propagator, can be rewritten as 
\begin{equation*}
\begin{split}
	\GR(x,x') = \GR^{(0)}(x,x') 
		&- i \int \ud y \ud {y'} G^{(1)(0)}(x,y) \Sigma'(y,y') \GR(y',x')\\
		&- i \int \ud y \ud {y'} \GR(x,y) \SigmaR(y,y') \GR(y',x')
\end{split}
\end{equation*}
Let us analyze the situation $x^0<x'^0$. Since the retarded propagator is causal, $\GR(x,x')= \GR^{(0)}(x,x')  = 0$. Given that the above equality must hold for any state, the last two terms on the right hand side must vanish independently.  We first consider the second term. If times are ordered as $x^0<y^0>y'^0>x'^0$, the Hadamard function $G^{(1)(0)}(x,y)$ and the retarded propagator $\GR(y',x')$ are non-zero, and therefore $\Sigma'(y,y')$ must vanish when $y>y'$. Repeating the same analysis when times are ordered $x^0<y^0<y'^0>x'^0$, we arrive at the conclusion that $\Sigma'(y,y')$ is always vanishing. Therefore the second term is absent and we recover \eqref{SelfEnergyGeneralRetarded}. Applying the same analysis to the third term we conclude that the retarded self-energy also verifies the causality property
\begin{equation}
	\SigmaR(y,y') = 0 \quad \text{if}\ y^{(0)}<y'^{(0)},
\end{equation}
which also follows from the fact that in fact $\SigmaR(x,x')$ corresponds to an amputated retarded propagator.

\index{Self-energy!Hadamard}
\index{Propagator!Hadamard}
As shown in appendix \ref{app:CTP}, the expression of the self-energy matrix in the Keldysh basis is
\begin{equation}
	\Sigma^{a'b'}(x,x') = \begin{pmatrix}
		0 & \SigmaA (x,x') \\
		\SigmaR(x,x') & \SigmaN(x,x')
	\end{pmatrix},
\end{equation}
where  $\SigmaN(x,x')=\Sigma^{11}(x,x')+\Sigma^{22}(x,x')$ is the Hadamard self-energy. The null component of this matrix corresponds to the fact that $\Sigma'(x,x')$ is always vanishing. The two off-diagonal components of \Eqref{SelfEnergyGeneral} in the Keldysh basis are given by eqs.\ \eqref{SelfEnergyGeneralRetardedAdvanced}, while the non-trivial diagonal component is:
\begin{equation}\label{SelfEnergyHadamard}
\begin{split} 
	\GN(x,x') &= \GN^{(0)}(x,x')- i \int \ud[4]{y}\ud[4]{y'} \\
	&\qquad \times \Big[ \GN^{(0)}(x,y) \SigmaA(y,y') \GA(y',x')\\
	 &\quad\qquad+\GR^{(0)}(x,y) \SigmaR(y,y') \GN(y',x')\\
	&\quad\qquad +i \GR^{(0)}(x,y) \SigmaN(y,y') \GA(y',x')\Big].
\end{split}
\end{equation}
As shown in appendix \ref{app:SelfEnergyHadamard}, this equation can be simplified to
\begin{equation}\label{SelfEnergyHadamardSimp}
	\GN(x,x') =  - i \int \ud[4]{y}\ud[4]{y'} \GR(x,y) \SigmaN(y,y') \GA(y',x').
\end{equation}
In the appendix it is explained that the above form might not be universally correct since it depends on the retarded propagator being the unique inverse of a kernel.

\index{Self-energy|in flat backgrounds|)}

\subsection{(Quasi)homogeneous and (quasi)static states.} \label{sect:QuasiQuasi}

\index{Quasistatic background}
\index{Quasihomogeneous background}

So far, all expressions in this section refer to arbitrary background states $\hat \rho$. In many situations one will encounter states which are homogeneous and static, \ie, states which verify $[\hat\rho,\hat p]=0$.  In other situations one will not have exact homogeneity and/or stationarity but will meet approximate ones, meaning that the typical time or length scales characterizing the breaking of the symmetry will be much larger than the typical scales one is probing, \ie, $[\hat\rho,\hat p] \sim \hat\rho/L$ with $L$ being a large quantity.

For static and homogeneous backgrounds one can solve \Eqref{SelfEnergyGeneralRetarded} for the retarded propagator by going to the momentum representation:
\begin{subequations}
\begin{equation} 	\label{GSigmaDiagonal}
	\GR(p) = \frac{-i}{ p^2 +m^2+\SigmaR(p)-p^0 i\epsilon}.
\end{equation}
Similarly, for the advanced,
\begin{equation}
	\GA(p) = \frac{-i}{ p^2 +m^2+\SigmaA(p)+p^0 i\epsilon}.
\end{equation}
\end{subequations}
Notice that in general the self-energy is a function of the four-momentum $p$, and not only a function of the scalar $p^2$.

\index{Propagator!retarded}
\index{Propagator!advanced}
\index{Propagator!Dyson}
\index{Propagator!Whightman}
\index{Propagator!Hadamard}
\index{Propagator!Pauli-Jordan}
The Pauli-Jordan propagator can be derived from the retarded (see appendix \ref{app:GenRel}) as:
\begin{subequations}
\begin{equation}
	G(p) = 2 \Re \GR(p)= \frac{- 2\Im \Sigma(p)}
	{ [p^2 +m^2+\Re\SigmaR(p)]^2 + [\Im \SigmaR(p)]^2}.
\end{equation}
The Hadamard function admits the following expression:
\begin{equation}
	G^{(1)}(p) = i |\GR(p)|^2 \SigmaN(p) = \frac{ i\SigmaN(p)}
	{ [p^2 +m^2+\Re\SigmaR(p)]^2 + [\Im \SigmaR(p)]^2}
\end{equation}
\end{subequations}
From the above propagators, and the relations in appendix \ref{app:GenRel}, the following expressions can be derived
\begin{subequations}\label{PropsSigma}
\begin{align}
    G_\mathrm F(p)  &= \frac{ -i \left[ p^2 + m^2  + \Re \SigmaR(p)
    \right]
    + i\SigmaN(p)/2}{\left[ p^2 + m^2 + \Re \SigmaR(p)\right]^2 +
    \left[\Im \SigmaR(p)\right]^2}, \\
    G_\mathrm D(p)  &= \frac{ i \left[ p^2 + m^2  + \Re \SigmaR(p)
    \right]
    +i \SigmaN(p)/2}{\left[ p^2 + m^2 + \Re \SigmaR(p)\right]^2 +
    \left[\Im \SigmaR(p)\right]^2}, \\
    G_-(p) &= \frac{ i\SigmaN(p)/2 + \Im \SigmaR(p)}{\left[ p^2 + m^2 + \Re \SigmaR(p)\right]^2
    +
    \left[\Im \SigmaR(p)\right]^2}, \\
    G_+(p) &= \frac{ i\SigmaN(p)/2 - \Im \SigmaR(p)}{\left[ p^2 + m^2 + \Re \SigmaR(p)\right]^2
    +
    \left[\Im \SigmaR(p)\right]^2}.
\end{align}
\end{subequations}
The Whightman functions also admit the following representation:
\begin{subequations}
\begin{align}
    G_-(p) &= \frac{ -i\Sigma^{12}(p)}{\left[ p^2 + m^2 + \Re \SigmaR(p)\right]^2
    +
    \left[\Im \SigmaR(p)\right]^2}, \\
    G_+(p) &= \frac{ -i\Sigma^{21}(p)}{\left[ p^2 + m^2 + \Re \SigmaR(p)\right]^2
    +
    \left[\Im \SigmaR(p)\right]^2}.
\end{align}
\end{subequations}

When the background is only approximately static or homogeneous there are corrections to the above equations. By analyzing \Eqref{SelfEnergyGeneralMomentum} one realizes that the appropriate relation  for the retarded propagator reads
\begin{equation} \label{ApproximateDiagonal}
	\GR(p;X) = \frac{-i}{p^2 +m^2+\SigmaR(p;X)-p^0 i\epsilon} + O(L p),
\end{equation}
where we recall that $L$ is the length scale in which the background varies significantly.
The momentum representation is useful as long as one is analyzing time or length scales much shorter than the scale of inhomogeneity of the background. All equations in this subsection are also valid in the approximately homogeneous case up to order $O(Lp)$.

\index{Thermal background}
\index{Thermal field theory}
The canonical example of static and homogeneous background is the thermal background. In this case the state of the field is
\begin{equation}
	\hat\rho = \frac{\expp{- \beta \hat H}}{\Tr{\big(\expp{- \beta \hat H}\big)}}.
\end{equation}
Thermal field theory \cite{Das,LeBellac,LandsmanWeert87} can thus  be treated as a particular example of field theory over an arbitrary background. This viewpoint corresponds to the so-called real time approach to field theory. We briefly review the subject in appendix \ref{app:CTP}.

\subsection{K\"all\'en-Lehmann spectral representation}

\index{Propagator!Pauli-Jordan}
\index{Spectral function}
\index{Spectral representation}

In a general background the quickest and clearest way to derive the spectral representation is by simply recalling the definition of the propagators. From \Eqref{RetardedProp} we immediately obtain the spectral representation for the retarded propagator,
\begin{equation}
	\GR(p;X) = \int \frac{\vd k^0}{2\pi}   \frac{i G(k^0,\vect p;X)}{p^0 - k^0 + i\epsilon}.
\end{equation}
This equation is identical to its vacuum counterpart, \Eqref{SpectralVacuumAlt}. The advanced propagator follows a similar representation,
\begin{equation}
	\GR(p;X) = \int \frac{\vd k^0}{2\pi}   \frac{i G(k^0,\vect p;X)}{p^0 - k^0 - i\epsilon}.
\end{equation}
Hereafter the Pauli-Jordan function will also be called \emph{spectral function}. The similarities with the vacuum case end here. The retarded and advanced propagators are the only propagators that can be expressed in terms of an integral of the Pauli-Jordan propagator, which can be retarded as the generalization of the vacuum spectral function.  Notice also that in general the spectral representation  can only be expressed as an integral over the energy, and not as an integral over the invariant mass. 

The Pauli-Jordan function verifies the following properties:
\begin{subequations}
\begin{align}
	  G(p;X) &> 0, \quad \text{if}\ p^0>0,\\
	 G(-p;X) &= -G(-p;X), \\
	\int \frac{\vd k^0}{2\pi} k^{0} G(k;X)&=1.
\end{align}
\end{subequations}
The first two properties are a simple consequence of the definition of the propagator. The third property is a sum rule, consequence of the equal time commutation relations,
\begin{equation}
	[\hat\phi(t,\vect x),\dot{\hat\phi}(t,\vect y)] = i \delta^{(3)}(\vect x - \vect y).
\end{equation}

The derivation of the spectral representation  developed in the this section is simple and direct, but hinders much of the physical intuition that can be obtained from the more explicit derivation given in the vacuum case, since we are missing a representation for the Pauli-Jordan function in the line of \Eqref{SpectralFunction}

\index{Spectral representation}
\index{Pauli-Jordan function|see{propagator}}
\index{Propagator!Pauli-Jordan}

Let us see now how for stationary backgrounds an explicit representation for the Pauli-Jordan function can be obtained. The Pauli-Jordan function can be expressed in the basis of eigenstates of the Hamiltonian as 
\begin{equation}
\begin{split}
	G(t,t';\vect p) &= \sum_{\alpha}  \rho_{\alpha} \langle\alpha|[\hat\phi_{-\vect p}(t),\hat\phi_{\vect p}(t')]|\alpha\rangle \\
	 &= \sum_{\alpha}  \rho_{\alpha} \langle\alpha|\hat\phi_{-\vect p} \expp{-i \hat H(t-t')} \hat\phi_{\vect p} \expp{i\hat H(t-t')}|\alpha\rangle - \text{(c.c)},
	 \end{split}
\end{equation}
where $\rho_{\alpha} = \langle \alpha|\hat\rho|\alpha\rangle$. We have used the fact that the state is stationary, so that the density matrix operator commutes with the Hamiltonian and is diagonal in the basis of eigenstates of the Hamiltonian. Let us now introduce the identity operator $1=|\beta\rangle\langle\beta|$:
\begin{equation*}
\begin{split}
	G(t,t';\vect p)
	 &= \sum_{\alpha,\beta}  \rho_{\alpha} \langle\alpha|\hat\phi_{-\vect p} \expp{-i \hat H(t-t')} |\beta\rangle \langle\beta|\hat\phi_{\vect p} \expp{i\hat H(t-t')}|\alpha\rangle - \text{(c.c.)}\\
	 &= \sum_{\alpha,\beta}  \rho_{\alpha} \langle\alpha|\hat\phi_{-\vect p}  |\beta\rangle \langle\beta|\hat\phi_{\vect p}|\alpha\rangle \expp{-i(E_{\beta}-E_{\alpha})(t-t')}- \text{(c.c.)}\\
	 &= \sum_{\alpha,\beta} \rho_{\alpha}\abs{\langle\alpha|\hat\phi_{\vect p}|\beta\rangle}^{2} (-2i) \sin{[(E_{\beta}-E_{\alpha})(t-t')]}
	 \end{split}
\end{equation*}
We have used that $\hat\phi_{\vect p}^\dag = \hat\phi_{-\vect p}$
Going to the frequency space we obtain the desired expression:
\begin{equation} \label{SpectralDetailed}
	G(p) = \sum_{\alpha,\beta} \rho_{\alpha}\abs{\langle\alpha|\hat\phi_{\vect p}|\beta\rangle}^{2} \left[ \delta(p^0 + E_{\alpha} - E_{\beta})-\delta(p^0 - E_{\alpha} + E_{\beta})\right].
\end{equation}
For stationary background states the Pauli-Jordan propagator is proportional to the probability for the field operator of momentum $\vect p$ to induce a transition to a state with higher energy $p^0$, minus the probability to induce a transition to a state of lower energy $p^0$.

\subsection{Perturbation theory, divergences and renormalization}\label{sect:DivRen}

\index{Renormalization}

In most sections of this chapter we discuss general properties of the propagators and self-energies, which do not depend on any perturbative expansion. In practice, however, many times the only way to evaluate the interacting propagators is through a perturbative expansion in the coupling constant. Perturbation theory for field theories in general backgrounds requires the CTP method, reviewed in appendix \ref{app:CTP}. Essentially one has to take into account the CTP doubling of the number of degres of freedom by doubling the number of vertices and multiplying by four the number of propagators. We have already used CTP perturbation theory in sect.~\ref{sect:QBMPert}.

\index{Renormalization}
As long as the background state has finite energy per unit volume, far ultraviolet modes of the field are not be occupied. Therefore, there is an energy scale beyond which the field can be treated as if it were in the vacuum. In the case of thermal field theory this scale is given by the temperature, and  the Bose-Einstein function can be viewed as a natural soft cutoff to the thermal contributions. The counterterms which renormalize the vacuum theory also make the theory ultraviolet-finite in an arbitrary background state. To put it differently, one can first proceed with the usual process of regularization and renormalization with the vacuum theory; then replacing the vacuum state with any physically reasonable state does not generate any additional ultraviolet divergence. 

This does not mean however that the renormalization process is not modified. As we shall see in the following sections, and as it is shown in refs.~\cite{DonoghueHolstein83,DonoghueEtAl85}, finite parts of the counterterms need be adjusted in a different way for each background, if  a physical meaning is to be attributed to the different terms appearing in the action.

\index{Hard thermal loop}
\index{Hard scale}
\index{Soft scale}
The infrared behavior is more subtle. On the one hand, naive perturbation theory may break down under certain circumstances. To be concrete, let us focus on the case of $\lambda\phi^4$ and thermal states (most analysis in the literature are done in the context of thermal field theory). Considering the situation in which the masses are very small as compared to $T$ (the ``hard'' scale) and $g T$ (the ``soft'' scale), by simple dimensional analysis arguments one can convince oneself that the leading contribution to the one loop thermal correction of the propagator is of the order of $g^2 T^2 /p^2$ times the tree level propagator, where $p$ is the external momentum. This leading contribution comes from the situation in which particles with momentum $T$ run inside the loop (``hard thermal loops''). When the external momenta are soft (of the order of $gT$), the one-loop term is of the same order as the tree level term. If the relevant contribution to a given process comes from external soft momenta, in order to do perturbation theory in a meaningful way,  the tree level propagators must be replaced by Braaten and Pisarski's resummed propagator \cite{BraatenPisarski90a,BraatenPisarski90b,Pisarski89}, which incorporates the effect of the hard thermal loops. See refs.~\cite{LeBellac,KraemmerRebhan04} for further details. 

\index{Infrared divergences}
On the other hand, even if resummed propagators are used, infrared divergences may still arise when the masses of the fields are negligible. For thermal bosonic fields the infrared behavior is much worse than in the vacuum, because of the Bose-Einstein factor at low energies goes as $n(E)\approx T/E$. These divergences may show up in the final results of the calculations, or can be hidden in the intermediate stages, leading to finite but incorrect results if they are not properly regularized \cite{Rebhan92} (we will encounter this kind of divergence in the next chapter). The investigation of the infrared divergences at finite temperature is still an open problem \cite{LeBellac,KraemmerRebhan04}.

\index{Propagator!in flat backgrounds|)}

\section{Open quantum system viewpoint} \label{sect:OQS}

\index{Open quantum system}

\subsection{Particle modes seen as reduced systems}

\shortpage

Since any given field theory model is composed by more than one degree of freedom (in fact by an infinite number of them), quantum fields  can be naturally regarded from the viewpoint of open quantum systems. Some  distinguished degrees of freedom of the field constitute the reduced subsystem, while the rest form the environment. The division between the system and the environment is obviously arbitrary, and depends on the properties in which one is interested.  If there are several fields in interaction, it can be interesting to concentrate on one field and trace over the other ones. For instance, in electrodynamics, one can study the so-called Euler-Heisenberg effective action for the photons \cite{ItzyksonZuber}, considering  the electrons as the environment, or the complementary case, in which the electrons are taken as the system of interest and the photons are integrated out \cite{WheelerFeynman49,CaldeiraBarone91,Anastopoulos97,AnastopoulosZoupas97}. Similarly, in stochastic gravity \cite{CalzettaHu94,HuMatacz95,HuSinha95,CamposVerdaguer96,
CalzettaEtAl97,MartinVerdaguer99a,MartinVerdaguer99c,MartinVerdaguer00,HuVerdaguer03,HuVerdaguer04} the system of interest is the gravitational field, and the matter fields are integrated out. In many other circumstances it proves useful to consider as the reduced system those modes of the quantum field which are below some ultraviolet cutoff, with the other ones constituting the environment. This approach  has been used, for instance, studying bubble nucleation \cite{CalzettaRouraVerdaguer01,CalzettaRouraVerdaguer02}, analyzing decoherence in field theory \cite{LombardoMazzitelli96} and in inflationary cosmology \cite{TanakaSakagami97,ZanellaCalzetta06}. The Wilsonian approach to renormalization can also be understood this way \cite{Peskin}.\footnote{In the situations in which there is a clear separation of scales the open quantum system approach essentially reduces to an effective field theory treatment. The characteristic features of open quantum systems, namely decoherence, dissipation an noise, are more manifest in the situations in which there is no clear separation of scales.}

In this thesis we are interested in the properties of individual quasiparticles. Apparently, the division between system and environment that is interesting for us amounts to consider the mode corresponding to the momentum of the particle as the system, while the other modes of the field, as well as the modes of any other fields in interaction, are considered as the environment. 

However, it proves difficult to implement this division for scalar fields, since the mode-decomposed field $\phi_\vect p$ acts on the Hilbert space sector with momentum $\vect p$, but also on the sector with momentum $-\vect p$. To put it differently, naively one would think that the system is the mode $\phi_{\vect p}$, while the environment is formed by the modes $\phi_{\vect q}$ with $\vect q \neq \vect p$. However this naive division is not very convenient, since the mode-decomposed field operator is not real, but is a complex quantity obeying the contraint $\phi_\vect p = \phi^*_{-\vect p}$ and acting on both the sector with momentum $\vect p$ and the sector with momentum $-\vect p$.

Instead of decomposing the field in an exponential Fourier transform, we could perform a sine and cosine Fourier transform:
\begin{equation}
	\phi(\vect x) = \int\limits_{p^3\geq 0} \ud[3]{\vect p} \left[ s_\vect p \sin(\vect p\cdot \vect x) + c_\vect p \cos(\vect p\cdot \vect x) \right]
\end{equation}
Such a transform corresponds to a decomposition in standing waves. Both the coefficients $s_\vect p$ and $c_\vect p$ are real and can be chosen as the reduced system of interest for a given $\vect p$. This is the usual decomposition when studying the dynamics of the modes of the scalar fields in inflation, for instance. However, we are interested in propagating rather than stationary modes. 

\shortpage

If we insist in focusing on the propagating mode with momentum $\vect p$, we may introduce the following canonical transformation:
\begin{subequations}
\begin{align}
	\tilde\phi_\vect p &:= \fud \big(\phi_\vect p + \phi_{-\vect p} \big) + \frac{i}{2 E_\vect p} \big(\pi_\vect p- \pi_{-\vect p} \big), \\
	\tilde\pi_\vect p &:= -\frac{iE_\vect p }{2} \big(\phi_\vect p - \phi_{-\vect p} \big) + \frac{1}{2} \big(\pi_\vect p+ \pi_{-\vect p} \big).
\end{align}
\end{subequations}
The tilde field modes and their canonical conjugate momenta are real, and at least for non-interacting systems, they only act on the Hilbert space sector with momentum $\vect p$. However, using the tilde modes for the description of the field theory processes introduces unwanted additional technical complications ---for instance, the momentum conservation properties become non-manifest, and the relation between the tilde field and its canonical momentum becomes non-trivial.

Instead of focusing on a single mode, given that the field naturally links modes with opposite momentum, we shall choose as the system of interest any two modes with given opposite momenta, and as the environment the remaining modes. Namely, the system degrees of freedom are the two field modes $\phi_\vect p$ and $\phi_{-\vect p}$, and the other modes $\phi_\vect q$, with $\vect q \neq \pm \vect p$, form the environment. The Hilbert space can be decomposed as $\mathcal H = \mathcal H_\text{sys} \otimes \mathcal H_\text{env}$, where in turn $\mathcal H_\text{sys} = \mathcal H_{\vect k} \otimes \mathcal H_{-\vect k}$. Notice that this separation does not correspond to the Fock space decomposition. The entire system is in a state $\hat \rho$; the state of the reduced system is $\hat\rho_\text{s} = \Tr_\text{env} {\hat\rho}$, and the state of the environment is $\hat\rho_\text{e} = \Tr_\text{sys} {\hat\rho}$.  Generally speaking, the state for the entire system is not a factorized product state.
A similar system/environment division is used in refs.~\cite{CampoParentani05a,CampoParentani05b,CampoParentani04} in an inflationary context.

Taking the $\lambda\phi^4$ model as a general example, if no infinities appeared in the calculation, the action could be split in the system action, environment action, interaction action and finite counterterm action as follows:
\begin{subequations}\label{OQSAction}
\begin{align}
	S_\text{sys} &= \int \ud t  \left( \dot \phi_\vect p \dot \phi_{-\vect p} -R^2_\vect p \phi_\vect p \phi_{-\vect p} \right),\\
	S_\text{env} &=  \frac{V{\mathcal Z}_\vect p}{2} \int\limits_{\vect q \neq \pm \vect p} \udpi[3]{\vect q} \int \ud t  \left( \dot \phi_\vect q \dot \phi_{-\vect q} -R^2_\vect p \phi_\vect q \phi_{-\vect q} \right),\\
		S_\text{int} &=\frac{ V^2  \lambda \mathcal Z^2_\vect p}{4!} \int \udpi[3]{\vect q} \udpi[3]{\vect q'} \int \ud t
		\phi_{\vect p} \phi_{\vect q} \phi_{\vect q'} \phi_{-\vect p-\vect q -\vect q'} \\
	S_\text{fc}&=  
	\int \ud t  \left\{ ({\mathcal Z}_\vect p-1)\dot \phi_\vect p \dot \phi_{-\vect p} -\big[{\mathcal Z}_\vect p (\vect p^2+m^2)- R^2_\vect p\big] \phi_\vect p \phi_{-\vect p} \right\},
\end{align}
\end{subequations}
We have allowed for an arbitrary rescaling of the field $\phi_\vect p \to \phi_\vect p/{\mathcal Z}_\vect p$, and we have introduced a finite renormalization of the two-mode frequency, so that the system frequency is $R_\vect p$. For the moment the values of ${\mathcal Z}_\vect p$ and  $R^2_\vect p$ (the latter needs not  be of the form $m^2+\vect p^2$) are left unspecified; we shall look for a physical criterion to fix both quantities. The election of the frequency of the system is not a mere notation issue, since we will be interested in studying the reduced system once the other degrees of freedom are integrated out, and different elections for the mass parameter imply different system Hamiltonian operators,
\begin{equation}\label{OQSH}
	H_\text{sys} = \dot \phi_\vect p \dot \phi_{-\vect p} + R^2_\vect p \phi_\vect p \phi_{-\vect p},
\end{equation}
which in turn  imply different values for the energy of the system. The meaning of ${\mathcal Z}_\vect p$ is less clear at this point, but we shall see that one may need to adjust ${\mathcal Z}_\vect p$ in order for the the notion of particle to be the conventional one. For the moment we just assume that the field is not rescaled, so that $\mathcal Z_\vect p=1$. Let us stress that the freedom in the election of ${\mathcal Z}_\vect p$ and $R_\vect p$ would be present even if there were no infinite contributions to the renormalization process. 

As soon as radiative corrections are considered, additional infinite counterterms to the mass, the field strength and the coupling constant must be introduced:
\begin{equation} \label{InfCount}
\begin{split}
	S_\text{ic} &=  \frac{\mathcal Z_\vect p(\mathcal Z-1 )V}{2} \int\limits_{\vect q \neq \pm \vect p} \udpi[3]{\vect q} \int \ud t  \left( \dot \phi_\vect q \dot \phi_{-\vect q} -R^2_\vect p \phi_\vect q \phi_{-\vect q} \right),\\
		&\quad+	\frac{\lambda_0 \mathcal Z^2 \mathcal Z^2_\vect p - \lambda \mathcal Z^2_\vect p}{4!}V^2\int \udpi[3]{\vect q} \udpi[3]{\vect q'} \int \ud t
		\phi_{\vect p} \phi_{\vect q} \phi_{\vect q'} \phi_{-\vect p-\vect q -\vect q'} \\
	&\quad+
	\int \ud t  \left[ ({\mathcal Z}-1)\mathcal Z_\vect p\dot \phi_\vect p \dot \phi_{-\vect p}-{\mathcal Z}_\vect p\big(\mathcal Z  m_0^2- m^2\big) \phi_\vect p \phi_{-\vect p} \right],
\end{split}
\end{equation}
where $m_0$ is the bare mass, $\lambda_0$ is the bare coupling constant, and $\mathcal Z$ is a field renormalization  parameter, all three being divergent. Notice that  \Eqref{InfCount} is just a slightly unusual way of writing the usual vacuum counterterms for the $\lambda\phi^4$ theory. Divergences are cancelled once for all with the conventional vacuum theory. The renormalized  mass $m$ and the renormalized coupling constant $\lambda$ are fixed with any convenient renormalization scheme in the vacuum, and do not necessarily have any direct physical interpretation.  In contrast, we would like to assign a physical meaning to the mode frequency $R_\vect p$. The properties of all the parameters are summarized in table \ref{tbl:RenPar}. Obviously some parameters are redundant, and could be removed for greater economy; we prefer to introduce all them so it is clear that the divergent structure of the theory coincides with that of the vacuum. Recall that in the following, until further notice, we shall assume that the field renormalization parameter $\mathcal Z_\vect p$ is equal to one, $\mathcal Z_\vect p=1$. 

\index{Renormalization}

\begin{table}
\centering
\begin{tabular}{lccccccc}
	\hline\hline
	Parameter& 	$m_0$ & $\lambda_0$ & $\mathcal Z$ & $m$ & $\lambda$ & $\mathcal Z_\vect p$ & $R_\vect p$ \\ \hline
	Possibly divergent & Y & Y & Y & N & N & N & N \\
	Depends on the state & N & N & N & N & N & Y & Y \\
	Function of the momentum & N & N & N & N & N & Y & Y \\
	Vacuum analysis sufficient & Y & Y & Y & Y & Y & N & N \\
	Directly observable  & N & N & N & N & N & N & Y \\
	\hline\hline
\end{tabular}
\caption{Properties of the different parameters and couplings used in the open quantum system approach to field theory whenever there are divergences. The particular example of $\lambda\phi^4$ is shown, but other theories would be analogous. ``Vacuum analysis sufficient'' refers to the possibility that the value determined in the vacuum is valid for all states. ``Directly observable'' refers to the possibility that the parameter is always the expectation value of a physically relevant quantum operator. See the main text for further clarifications.} \label{tbl:RenPar}
\end{table}

We shall be interested in studying properties of the two-mode with momenta $\pm\vect p$, \ie, properties which only depend on the degree of freedom of the system,. This means that we shall consider only propagators $\hat O_\vect{p}(t)$ acting on the Hilbert space of the system:
\begin{equation*}
	\av{O_\vect{p}(t)} = \Tr{\big[\hat O_\vect{p}(t) \hat\rho\big]} = \Tr_\text{sys}{\big[\hat O_\vect{p}(t) \hat \rho_{\text s}\big]} 
\end{equation*}

At this point we could use all the open quantum system methods presented in the previous chapter, and their non-linear generalizations \cite{RouraThesis}, in order to compute any observable quantity.  However, instead of following that path, we shall take advantage of the standard methods and results of quantum field theory and reinterpret them in the light of  the theory of open quantum systems.

\subsection{The generating functional}

Most relevant information on the reduced quantum system can be extracted from the set of correlation functions, or equivalently from the generating functional for the reduced system, which can be written as
\begin{equation}
\begin{split}
	Z[j_{a,\alpha}]&=\exp\bigg[-\frac{1}{2!}\int \ud t \ud{t'} j_\alpha^{a}(t) j_\beta^{b}(t) G^{\alpha\beta}_{ab}(t,t')\\ &-\frac{1}{4!}\int \ud t \ud{t'}\ud[4]{t''} j_\alpha^{a}(t) j_\beta^{b}(t') j_\gamma^c(t'') j_\delta^d(t''') G^{\mathrm{(C)}\alpha\beta\gamma\delta}_{abcd}(t,t',t'',t''') + \cdots\bigg]
\end{split}
\end{equation}
This expression is somewhat cumbersome and needs some clarification. Indices $a,b,c\ldots$ are CTP indices and take the values 1 and 2 as usual. Indices $\alpha,\beta,\gamma,\ldots$ take the two values $+\vect p$ and $-\vect p$, and make reference to the two field modes $\phi_\vect p$ and $\phi_{-\vect p}$. The propagator $G^{\alpha\beta}_{ab}(t,t')$ is the 2-point propagator connecting CTP indices $a$ and $b$, whose external legs correspond to particles with momenta $\alpha$ and $\beta$. When the state is translation-invariant and isotropic, momentum conservation imposes:
\begin{equation}
\begin{split}
	G^{(+\vect p)(-\vect p)}_{ab}(t,t') &= G^{(-\vect p)(+\vect p)}_{ab}(t,t') =  G_{ab}(t,t';\vect p), \\
	G^{(+\vect p)(+\vect p)}_{ab}(t,t') &= G^{(-\vect p)(-\vect p)}_{ab}(t,t') =  0.
\end{split}
\end{equation}
In turn, $G^{\mathrm{(C)}\alpha\beta\gamma\delta}_{abcd}(t,t',t'',t''';\vect p)$ is the connected part of the 4-point correlation function having  external legs with momenta $\alpha$, $\beta$, $\gamma$ and $\delta$. For translation-invariant states momentum conservation implies that only when momentum is balanced (\ie, two incoming and two outgoing external legs) the correlation function is non-vanishing. 

Obviously the generating functional has terms with arbitrary number of external sources. The open quantum system is non-linear, and a systematic treatment of the generating functional goes beyond the techniques explained in the first chapter. 

However, in many situations one is interested in properties which only depend on the two point correlation functions. In other cases one is doing a perturbative expansion of the generating functional, and connected higher order correlation functions are usually also of higher order in the expansion parameter. Finally there are situations in which one only has access to the two-point correlation functions, and expects (or simply hopes) that connected higher order correlation functions be of less importance. In any of this situations one can be tempted to approximate the generating functional by the following Gaussian expression:
\begin{equation} \label{ZCTPGaussian}
\begin{split}
	Z[j_{a,\alpha}]&\approx\exp\bigg[-\frac{1}{2!}\int \ud t \ud{t'} j_\alpha^{a}(t) j_\beta^{b}(t) G^{\alpha\beta}_{ab}(t,t') \bigg].
\end{split}
\end{equation}
If moreover one is dealing with translation-invariant states, which conserve the momentum, the generating functional can be more simply reexpressed as
\begin{equation} \label{ZDirectBasis}
\begin{split}
	Z[j_{a}]&\approx\exp\bigg[-\frac{1}{2!}\int \ud t \ud{t'} j^{a}(t) j^{b}(t) G_{ab}(t,t';\vect p) \bigg],
\end{split}
\end{equation}
where $j_a$ can either refer to $j_{a(+\vect p)}$ or to $j_{a(-\vect p)}$, implicitly understanding that momentum has to be balanced.

Eq.~\eqref{ZDirectBasis} shows that for translation-invariant states within the Gaussian approximation the reduced two-mode state effectively behaves as a single quantum mechanical degree of freedom. We study the consequences of the Gaussian approximation in the next subsection.

\subsection{Relation with QBM models}
\index{Gaussian approximation}
\index{QBM models}

The Gaussian approximation implies that  the expression of the generating functional of a reduced two-mode in terms of the two-point propagators coincides with that of a quantum Brownian motion model, provided the system is translation-invariant: compare \Eqref{ZQBMDirectBasis} with \Eqref{ZDirectBasis}. Moreover, by comparing eqs.\ \eqref{GretFourier} and  \eqref{GOtherFourier} with eq.\ \eqref{PropsSigma}, we realize that the structure of the two-point propagators is identical in both cases. Notice that this latter fact is independent of the Gaussian approximation.

Therefore, we conclude that there is an equivalent QBM system for every two-mode treated under the Gaussian approximation. In other words,
within the Gaussian approximation, every two-mode of a given quantum field theory can be described as if it were a quantum Brownian particle interacting linearly with some effective environment. We must stress that, similarly as the linear interaction does not coincide with  the real coupling, the effective environment  does not coincide  with the real environment. The precise details of the equivalence are summarized in table\ \ref{tbl:equivalence} for the particular case of the $\lambda\phi^4$ theory.

\begin{table}
\centering
\renewcommand{\arraystretch}{1.3}
\begin{tabular}{lcc}
\hline\hline
& Original system 				& Equivalent linear QBM\\ \hline  
System & two field modes   & harmonic oscillator \\
System d.o.f. & $\phi_\vect p$, $\phi_{-\vect p}$ & 	$q$ \\
Environment & other modes 3-d field  & 1-d field\\
Env. d.o.f & $\phi_\vect q$, $\vect q \neq \pm\vect p$ & $\varphi_p$  	\\
Original freq. & $\sqrt{m^2 + \vect p^2}$ & $\omega_0$ ($=$) \\
Physical freq. & $R_\vect p$ & $\Omega$ ($=$) \\
Coupling & $\frac{\lambda}{4!} \sum_{\vect q\vect q'}
		\phi_{\vect p} \phi_{\vect q} \phi_{\vect q'} \phi_{-\vect p-\vect q -\vect q'}$  & $g_\text{lin}^2 \sum_p \mathcal I(p) \dot q \varphi_p$\\
2-point function & $G_{ab}(t,t';\vect p)$ & $G_{ab}(t,t')$ $(=)$\\
Self-energy & $\Sigma^{ab}(t,t';\vect p)$ & $\Sigma^{ab}(t,t')$ $(=)$\\
\hline\hline
\end{tabular}
\caption{Detail of the equivalent linear QBM system for a $\lambda \phi^4$ quantum field theory. The $\lambda \phi^4$ model has been chosen for concreteness, but the correspondence would be analogous for any other field theory model. The symbol $(=)$ indicates that the original and equivalent quantities are indeed identical despite the name change. } \label{tbl:equivalence}
\end{table}

Let us investigate on the correspondence. 
On the one hand, as we have seen in the previous chapter, the effect of the environment in the QBM system is fully encoded in two kernels, the dissipation kernel $D(t,t')$ ---or $H(t,t')$--- and the noise kernel $N(t,t')$.  The properties of the quantum Brownian particle are fully determined once the frequency  and the noise and dissipation kernels are known. On the other hand, in a quantum field theory the two-point correlation functions are fully characterized by the mass and the self-energy. In turn, the four self-energy components can be determined by the retarded and Hadamard self-energies. By comparing again eqs.\ \eqref{GretFourier} and  \eqref{GOtherFourier} with eq.\ \eqref{PropsSigma} we realize that the precise analogy goes as follows:
\begin{subequations} \label{equivalenceQBM}
\begin{align}
	m^2+\vect p^2\ &\overset{\leftrightarrow}{=}\ \omega^2_{0}\\
	\SigmaR(t,t';\vect p)\ &\overset{\leftrightarrow}{=}\ H(t,t')\\
	\Im \SigmaR(t,t';\vect p)\ &\overset{\leftrightarrow}{=} \ iD(t,t')\\
	\SigmaN(t,t';\vect p)\ &\overset{\leftrightarrow}{=} \ -2iN(t,t')
\end{align}
\end{subequations}
Equivalently, the same conclusions can be reached by simply identifying the self-energies in the original and equivalent systems, as indicated in table \ref{tbl:equivalence}.

This representation provides thus a first rough interpretation for the self-energy components in the Keldysh basis. The retarded self-energy corresponds to the dissipation kernel, indicating that the retarded self-energy determines the dissipative properties of the system and that it is, to some extent, independent of the state of the system. The Hadamard self-energy corresponds to the noise kernel, and thus it is basically related to fluctuations, and more generally to properties which depend on the details of the environment. These statements should be taken with great care, and shall be reanalyzed afterwards.

\index{Dissipation kernel}
\index{Noise kernel}

Although the description in terms of the noise and dissipation kernels is usually sufficient,  if needed, the details of the linear QBM system can be further analyzed. The effective coupling constant can be reproduced from the dissipation kernel (or, as we have just seen, from the imaginary part of the retarded self energy) through
\begin{equation}
	D(\omega) = \frac{ i g^2_\text{lin}}{2} \omega\mathcal I(\omega).
\end{equation}
The  dissipation at energy $\omega$ determines the strength of the coupling of the system with the equivalent environment.
The state for the equivalent environment can be fully reproduced from the noise kernel, which admits the following expression
\begin{equation}
	N(\omega) = g^2_\text{lin} |\omega| \mathcal I(\omega) \left[ \frac{1}{2} + n(|\omega|) \right], 
\end{equation}
where we recall that $n(|\omega|) := \Tr [\hat\rho_\text{e} \hat a_\omega^\dag \hat a_\omega ] $ is the mean occupation number of the $\omega$-mode of the equivalent environment. The knowledge of the occupation numbers fully determine a Gaussian stationary state of the environment. These results follow from eqs.~\eqref{DisQBM} and \eqref{NoiseQBM}.

\index{Occupation number}

In equilibrium, the occupation numbers of the equivalent  environment $n(|\omega|)$ coincide with the occupation number of the sytem $n$ provided that $n(|\omega|)$ is evaluated at the system energy. In turn, the occupation numbers of the original field $n_\vect p$ coincide with these two quantities. Therefore in equilibrium we can also use $n_\vect p = n(|E_\vect p|) = n$ within the equivalence.

Let us end by making several additional comments. First, notice that the analogy depends only on the Gaussian approximation, so that it can be extended to all orders in perturbation theory. Second, notice also the results derived from the QBM interpretation are exact for all those properties which only depend on the two-point correlation functions; for the properties which depend on higher order correlation functions, it will be a correct approximation depending on the validity of the Gaussian approximation, \ie, depending on the relative importance of the connected parts of the correlation functions with respect to the disconnected parts. Third, within the regime of validity of the Gaussian approximation, all the methods developed in the literature for linear quantum systems, including those explained in the second chapter, can be applied to the mode decomposition of quantum field theory. Finally, let us mention that the description in terms of a linear open quantum system allows a unified description of many different quantum field theory systems. The details of the quantum field theory are unimportant once the noise and dissipation kernels are known.

\subsection{Application: interpretation of the imaginary part of the self-energy}

\index{Self-energy!imaginary part}

Let us apply the QBM correspondence  to analyze the physical significance of the imaginary part of the retarded self-energy. Our findings will coincide with the classic result by Weldon \cite{Weldon83}. However, while Weldon's analysis was only valid to first order in perturbation theory, our technique will be valid to all orders. Moreover, our analysis will not be tied to any field theory model. Additionally we will also obtain an interpretation for the Hadamard self-energy.  In this case the QBM analogy is exact since no four point correlation functions are involved. 

To start we need to consider the transition probabilities in the equivalent QBM model. The probability $\Gamma_-$ that an excitation of the Browian particle with positive energy $\omega$ decays into the environment is given by (see \eg\ refs.~\cite{Peskin,LeBellac})
\begin{equation}
	\Gamma_-(\omega) = \frac{1}{2\omega} \int \frac{\vd k}{2\pi2|k|}  2\pi \delta(\omega-|k|) |\mathcal M|^2 [1+n(|k|)],
\end{equation}
where $\mathcal M$ is the amplitude of the transition and $n(|k|)$ is the occupation number of the environment states with energy $|k|$. The factor $1+n(|k|)$ is due to the Bose-Einstein statistics, which enhances the decay probability to those states which are already occupied. To first order in perturbation theory the squared decay amplitude is given by
\begin{equation} \label{matrixM}
	|\mathcal M|^2 = g_\text{lin}^2 \mathcal I(\omega) \omega^2,
\end{equation}
where $g_\text{lin}$ is the linear coupling (which does not coincide in general with the original coupling) and $\mathcal I(\omega)$ is the distribution of frequencies. The factor $\omega^2$ is a consequence of the derivative coupling. The decay probability is
\begin{equation} \label{GammaMinus}
	\Gamma_- = \frac{1}{2} g_\text{lin}^2 \mathcal I(\omega) [1+n(|k|)].
\end{equation}
Likewise, the probability that an excitation of positive energy $\omega$ is created spontaneously from the environment is given by
\begin{equation} \label{GammaPlus}
	\Gamma_+(\omega) = \frac{1}{2\omega} \int \frac{\vd k}{2\pi2|k|}  2\pi \delta(\omega-|k|) |\mathcal M|^2 n(|k|) = \frac{1}{2} g_\text{lin}^2 \mathcal I(\omega) n(|k|).
\end{equation}
In the original system, $\Gamma_-$ can be interpreted as the probability that a (possibly off-shell)  excitation with energy $\omega$ decays into the environment, and $\Gamma_+$ can be interpreted as the probability that an environment spontaneously creates an excitation with energy $\omega$.

Notice that, strictly speaking, the above interpretation in terms of probabilities only has a formal value, since, in general, we are not dealing with stable on-shell particle excitations, as it would be needed for the above expressions to be completely meaningful. The concept of particle in general situations will be analyzed in the following chapters, but we shall not study the theory of collisions as it would be needed  to fully justify the above equations. In any case, the expressions as presented above are only meaningful to leading order in $g_\text{lin}$; going beyond leading order would pose greater conceptual problems.

\index{Decay rate}
We next analyze the self-energy components in the equivalent QBM model. Since the QBM system is linear, the perturbative treatment is exact. Given that $\Im \SigmaR(\omega) = (i/2)[\Sigma^{21}(\omega) - \Sigma^{12}(\omega)]$ (see appendix \ref{app:GenRel}), we start by analyzing $\Sigma^{21}(\omega)$. Applying Feynman rules, described in appendix \ref{app:CTP}, we get
\begin{equation}
	-i\Sigma^{21} (t,t')= - (ig_\text{lin}) \partial_t (i
g_\text{lin}) \partial_{t'} \int \frac{\vd p}{2\pi} \mathcal I(p) \Tr_\text{env}{\big[\hat\rho_\text{e} \hat \varphi_{\text Ip}(t) \hat \varphi_{\text I(-p)}(t')\big]},
\end{equation}
where $\hat \varphi_{\text Ip}$ is $p$-mode of the environment field (see chapter 2) in the interaction picture.  We have exploited the fact that first order perturbation theory yields exact results for the self-energy in linear systems. Notice that the expressions we have to deal with are completely analogous to those we obtained when computing the noise and dissipation kernels in subsect.~\ref{sect:DisNoise} ---indeed, as we have seen the dissipation and noise kernels are closely related to the self-energy components. Following similar steps, the above equation can be developed to
\begin{equation*}
\begin{split}
	\Sigma^{21} (\omega)
&= - ig^2_\text{lin}  \omega^2 \int_{0}^\infty \ud p \frac{\mathcal I(p)}{p}  \sum_n \rho_{p,n} \left[ (n+1) \delta(\omega-p) +  n \delta(\omega+p) \right].
\end{split}
\end{equation*}
Restricting to positive energies, 
\begin{equation}
	\Sigma^{21} (\omega)
= - ig^2_\text{lin}  \omega  \mathcal I(\omega) \sum_n \rho_{\omega,n} (n+1) = - ig^2_\text{lin}  \omega  \mathcal I(\omega) \left[1+n(\omega)\right], \quad \omega > 0,
\end{equation}
we see that $\Sigma^{21}(\omega)$ can be interpreted as being proportional to the decay rate:
\begin{equation}
	\Sigma^{21}(\omega) = -2i\omega \Gamma_-(\omega), \quad \omega > 0.
\end{equation}

Repeating the calculation for  $\Sigma^{21}(\omega)$ we similarly find 
\begin{equation}
	\Sigma^{12} (\omega)
= - ig^2_\text{lin}  \omega  \mathcal I(\omega) \sum_n \rho_{\omega,n} n = -i g_\text{lin}^2 \omega \mathcal I(\omega) n(\omega), \quad \omega > 0,
\end{equation}
having the corresponding interpretation in terms of the creation rate,
\begin{equation}
	\Sigma^{12}(\omega) =- 2i\omega \Gamma_+(\omega), \quad \omega > 0.
\end{equation}

\index{Self-energy!retarded}
The imaginary part of the retarded self-energy is therefore given by
\begin{equation}
	\Im\SigmaR (\omega) = \frac{i}{2} [\Sigma^{21}(\omega) -\Sigma^{12}(\omega)]  = -\frac{1}{2} g^2_\text{lin}  \omega  \mathcal I(\omega) ,
\end{equation}
and can be interpreted as the net decay rate for an excitation of energy $\omega$ ---\ie, decay rate minus creation rate:
\begin{equation} \label{Weldon}
\begin{split}
	\Im\SigmaR(\omega) = -\omega[\Gamma_-(\omega) - \Gamma_+(\omega)].
\end{split}
\end{equation}
We therefore recover Weldon's result.

\index{Self-energy!Hadamard}
We can additionally get an interpretation for the Hadamard self-energy. It is given by
\begin{equation} \label{SigmaNInterp1}
	\SigmaN(\omega)= - \Sigma^{21}(\omega) - \Sigma^{12}(\omega) = i  g^2_\text{lin}  |\omega|  \mathcal I(\omega) [1+2n(|\omega|)],
\end{equation}
and is proportional to the probability of decay plus the probability of creation, 
\begin{equation} \label{SigmaNInterp}
	\SigmaN(\omega)= 2i|\omega|[\Gamma_-(\omega) + \Gamma_+(\omega)].
\end{equation}

In summary, the self-energy components have simple interpretations in terms of the excitation decay and creation rates. The excitations over generic quantum states will be analyzed in the following chapters. Anyway, we can already advance that the interpretation in terms of probabilities is merely formal when the excitations are short-lived. Only when long-lived perturbations can be defined the probability concepts will be really meaningful. 

In any case, the dissipative processes are much richer and frequent in a non-vacuum state than in the vacuum. In the vacuum, the only possible source for dissipation is the decay of a unstable particle into other different particles. In a general background a particle can simply change its momentum because of the interaction with the background, and this also contributes to dissipation, since the relevant degree of freedom is the field mode.

One could be worried by the fact that when computing the transition probabilities we just worked to first order in $g_\text{lin}$. We shall see that when the quasiparticle concept is meaningful the decay probability is very small and therefore the effective coupling with the environment is also very small ---although the original coupling could possibly be very large. Therefore, there would be no significant benefit in going to higher order in $g_\text{lin}$. To put it in a different way, the degree of uncertainty of the notion of decay rate is of the same order of the correction introduced by neglecting higher orders in $g_\text{lin}$.

\section{Quasiparticles in generic quantum states}

\subsection{Quasiparticle excitations}\label{sect:QuasiDef}


\index{Quasiparticle|(}
A quasiparticle can be thought as the generalization of the particle concept to states different than the vacuum. For the purposes of this thesis, we consider that a \emph{quasiparticle}  is a particle-like excitation which travels in some background and which is characterized by the following properties:
\begin{enumerate}
	\item It has some characteristic initial energy $E$. The fluctuations of the energy are much smaller than this characteristic value: $\Delta E \ll |E|$.
	\item It has some characteristic initial momentum $\vect p$. The  fluctuations of the momentum are much smaller than the characteristic energy:  $|\Delta \vect p| \ll |E|$.
	\item It has approximately constant energy and momentum during a long period of time $T=1/\Gamma$, before it starts to decay. Here ``long'' means that the decay rate $\Gamma$ has to be much smaller than the de Broglie frequency of the quasiparticle: $\Gamma \ll |E|$. 
	\item It is elementary, meaning that it cannot be decomposed in the (coherent or incoherent) superposition of two or more entities, having each one separately the same three properties above.
\end{enumerate}
Notice that the third property is somewhat redundant, since by the time-energy uncertainty principle the energy fluctuations are at least given by the decay rate: $\Delta E \gtrsim \Gamma$.
The quasiparticle is essentially characterized by the energy $E$, the momentum $\vect p$ and the decay rate $\Gamma$ ---usually, the dependence of the energy and the decay rate on the momentum is studied. Besides that, the quasiparticle can be characterized by other quantum numbers such as the spin ---although we shall only deal with scalar quasiparticles.


\index{Collective mode}
\index{Quasiparticle!in the strict sense}
Quasiparticle excitations may exist even in strongly correlated systems. The quasiparticles in these systems are usually radically different to the original particles which constitute the medium. For instance, the spin and mass of the quasiparticles might have nothing to do with the original particles: Sometimes a distinction is made between the \emph{collective modes}, which are those excitations which bear no resemblance with the original particles, and which usually involve the system as a whole, and the \emph{quasiparticles in the strict sense}, which are 
those excitations which correspond to the vacuum particles. Sometimes both type of excitations may coexist in the same system. We shall not make such a distinction in the following.

\index{Particle!relation to quasiparticles}
\index{Quasiparticle!relation to particle}
The line separating particles from quasiparticles is arbitrary to some extent. In this thesis we will employ ``quasiparticle'' for the general excitations defined above, and ``particle'' to refer to the particular case in which the occupation numbers of the distinguished mode $n_\vect p$ are vanishing or completely negligible (even if other modes of the same field, or the modes of another field in interaction, can be possibly excited).


If the initial background state has large momentum or energy fluctuations the perturbed state inherits them, and therefore the requirement that the fluctuations of the momentum and energy of the quasiparticles are small cannot be fulfilled. For thermal and, more generally, for Gaussian states, we will see that momentum fluctuations are comparable to the average momentum when the occupation numbers are of order 1. Therefore for bosonic systems the quasiparticle description of Gaussian states requires relatively small occupation numbers ($n_\vect p \ll 1$).

\index{Hydrodynamic description}
When the occupation numbers are of order one or larger, a quasiparticle description might not be very adequate and it might be more useful to move to a hydrodynamic or fluid description \cite{Jeon95,JeonYaffe96,CalzettaEtAl00}. For thermal states the hydrodynamic description applies for the long wavelength modes, while the short wavelength modes can be treated with a quasiparticle description.

In this chapter we shall not attempt to follow the quasiparticle trajectory, but instead will give a description of the dynamics of the mode corresponding to the quasiparticle momentum. A description of the trajectory of the quasiparticle only makes sense in the situation in which there is only one quasiparticle, \ie, when the occupation numbers can be strictly approximated by zero. This situation can be physically realized in a non-trivial way when there are two fields in interaction of very different mass. In this case one can think of going from a second-quantized to a first-quantized description and studying the behavior of the one-particle density matrix $\rho(\vect x,\vect x')$ or the one-particle Wigner function $W(\vect x,\vect p)$. We will briefly explore this point of view in chapter 5.

Quasiparticles are one of the most ubiquitous concepts in condensed matter \cite{Abrikosov}, solid state physics \cite{AshcroftMermin} and thermal field theory \cite{LeBellac,Kapusta}. We shall not by any means attempt to give a complete introduction to the subject in this thesis. Instead, we shall limit ourselves to studying the field-theory description of generic scalar quasiparticles, focusing on the properties that can be extracted from  the two point correlation functions. Moreover, we shall not attempt to follow the trajectory of the decaying piece of the quasiparticle, but will only concentrate on the non-decaying part. Furthermore, no quasiparticle collisions will be studied. 

\subsection{Quasiparticles in free theories} \label{sect:QuasiParticlesFree}

Although in absence of interaction there is no real difference between the particles and the quasiparticles (since the latter have the same properties as the former)  we will see that even in the free case it is not trivial to construct a quantum state which verifies the required quasiparticle properties.

As we have seen, in a non-interacting theory in the vacuum the one-particle state is naturally represented by the action of the creation operator on the vacuum, $|\vect p\rangle  = \hat a^\dag_\vect p|0\rangle$, or equivalently by the action of the field operator: $|\vect p\rangle = \sqrt{2E_\vect p} \hat \phi_{-\vect p} |0\rangle$.

\index{Occupation number}
Over a homogeneous and stationary state $\hat\rho$, the positive-energy quasiparticle state can be represented by the action of the creation operator:
\begin{equation}
\begin{split}
	\hat\rho^\pplus _\vect p &:= \frac{1}{n_\vect p + 1}\, 
\hat a^\dag_\vect p \hat\rho \hat a_\vect p 
	 \\ &= \frac1{n_\vect p + 1} \sum_{n,\{m\}} \rho_{n,\{m\}} (n+1) |(n+1)_\vect p,\{m\} \rangle\langle (n+1)_\vect p,\{m\}|,
\end{split}
\end{equation}
where the second equality is written in a highly schematic notation in order to avoid cumbersome expressions. Here $n_{\vect p}=\Tr{(\hat\rho \hat a^\dag_{\vect p} \hat a_\vect p)}$ represents the mean occupation number of the mode with momentum $\vect p$, $|n_\vect p,\{m\} \rangle$ is a basis of the Hilbert space with the $\vect p$ sector singled out, and $\rho_{n,\{m\}}$ is the diagonal representation of the background state in this basis. Notice that the state is normalized: $\Tr{\hat \rho_\vect p^\pplus } = 1$. The factors $n+1$ on the second equality are a consequence of the Bose-Einstein statistics (the probability to create an additional particle increases with the number of already-existing particles). 

The Bose-Einstein statistics are responsible for the following surprising fact: it is a simple exercise to show that the expectation value of the particle number,
\begin{equation}
	\av{\hat N_\vect p}^\pplus = \Tr{\big(\hat\rho^\pplus _\vect p\hat N_\vect p \big)}  = \Tr{\big(\hat\rho^\pplus _\vect p \hat a^\dag_\vect p \hat a_\vect p\big)}  =  \frac{1}{n_\vect p+1} \av{(\hat N_\vect p+1)^2},
\end{equation}
is actually increased more than one with respect to the unperturbed value:
\begin{equation}\label{surprising}
	\av{\hat N_\vect p}^\pplus = n_\vect p + 1 + \frac{\delta n_\vect p^2}{1+n_\vect p} =: n_\vect p + N_\vect p^\pplus,
\end{equation}
where $N^\pplus_\vect p=1 + \delta n_\vect p^2/(1+n_\vect p)$ is the number of excitations and 
\begin{equation}
	\delta n_\vect p^2 = \av{(\hat N_\vect p-n_\vect p)^2} = \av{\hat N^2_\vect p} - n_\vect p^2 \geq 0
\end{equation}
is the dispersion of the number of particles in the background state. For a particle number eigenstate $\delta n_\vect p^2 =0$ and $N^\pplus_\vect p=1$, and for a Gaussian state (see appendix \ref{app:Gaussian}) $\delta n_\vect p^2 = n_\vect p ( n_\vect p +1)$ and $N^\pplus_\vect p=1 + n_\vect p$. 
Therefore, when the background is in a mixed state, $\hat\rho_\vect p^\pplus$ represents slightly more than one additional quasiparticle. The reason for this is purely statistical, as we have commented: the highly occupied components of the state are more likely to become excited, and therefore they tend to gain statistical weight, thereby increasing the particle number. In other words, the action of the creation operator has two simultaneous effects: on the one hand, adding a quasiparticle to the system; on the other hand, increasing the statistical weight of the highly excited states. The statistical contribution is significant when the occupation numbers are large.

The expectation value of the momentum operator is also affected by this statistical effect:
\begin{subequations}
\begin{equation}
	 \av{\hat {\vect P}}^\pplus =  \Tr{\big(\hat\rho^\pplus _\vect p\, \hat{\vect P} \big)}  = \Tr{\big(\hat\rho^\pplus _\vect p\, {\vect p}\, \hat a^\dag_\vect p \hat a_\vect p\big)}  = \vect p N_\vect p^\pplus.
\end{equation}
Anyway the momentum per excitation is $\vect p$. The energy is similarly affected:
\begin{equation}
	\av{\hat H_0}^\pplus = \Tr{\big(\hat\rho^\pplus  \hat{H}_0 \big)}  =  \Tr{\Big[\hat\rho^\pplus_\vect p \Big( \hat a^\dag_\vect p \hat a_\vect p  + \frac{1}{2} \Big)\Big]}  = E_0 + E_\vect p  N_\vect p^\pplus,
\end{equation}
\end{subequations}
where $E_0=\Tr{(\hat\rho \hat H_0)}$ is the energy of the vacuum and $E_\vect p = \sqrt{\vect p^2+m^2}$ is energy per excitation.

Quasiparticles require small momentum spreads. Without entering into many details, the spread of the momentum in the case of Gaussian states is 
\begin{equation}
 	\left\{\av{\hat {\vect P}^2}^\pplus - [\av{\hat {\vect P}}^\pplus]^2\right\}^{1/2} = |\vect p|(4n_\vect p+5n_\vect p^2)^{1/2},
\end{equation}
which is a small quantity if the occupation numbers are small. Likewise, the spreads in the particle number and the energy have the same corresponding values. Therefore, in order for the state $\hat\rho_\vect p^\pplus$ to adequately represent a quasiparticle occupation numbers must be small as compared to one.

Since the statistical contribution looks awkward, one can imagine a different way of constructing the quasiparticle excitation:
\begin{equation} \label{AlternativeToSurprise}
	\hat\rho^{\text{(alt)}} _\vect p := \sum_{n,\{m\}} \rho_{n,\{m\}} |(n+1)_\vect p,\{m\} \rangle\langle (n+1)_\vect p,\{m\}|.
\end{equation}
This state has essentially identical properties to $\hat\rho_\vect p^\pplus$, except that it contains exactly one additional particle. Therefore it looks like that it corresponds more closely to the quasiparticle concept we discussed before. However, notice that the above state cannot be easily created from the background via the action of the creation and annihilation operators.  We shall argue in the next subsection that the state $\hat\rho^\pplus_\vect p$, and not $\hat\rho^{\text{\tiny (alt)}}_\vect p$, appears naturally when studying quasiparticle creation processes.

\index{Quasiparticle!statistical}
We shall consider that the state $\hat\rho^\pplus_\vect p$ represents exactly one additional real quasiparticle of momentum $\vect p$ and energy $E_\vect p$. The additional contribution to the particle expectation number will be called \emph{statistical quasiparticle} contribution. The statistical quasiparticle contribution can be interpreted as being a consequence of the increased knowledge of the background state which is derived from the creation of a real quasiparticle. 

Another novelty is that negative energy excitations, or holes, can also be defined, represented by the action of the annihilation operator:
\begin{equation}\label{holeState}
\begin{split}
	\hat\rho^\pminus _\vect p &:= \frac{1}{n_\vect p}\, 
\hat a_{-\vect p} \hat\rho \hat a^\dag_{-\vect p} 
	 \\ &= \frac1{n_\vect p} \sum_{n,\{m\}} \rho_{n,\{m\}} n |(n-1)_{-\vect p},\{m\}\rangle\langle (n-1)_{-\vect p},\{m\}|,
\end{split}
\end{equation}
The factors $n$ on the second equality account for the fact that there are more possible particles to annihilate in the more excited sectors of the state. This hole state is also affected by the statistical considerations described above, as it is manifest by showing the expectation value of the number operator:
\begin{equation}
\av{\hat N_\vect p} = \Tr{(\hat\rho^\pminus _\vect p\hat N )}   = n_\vect p - 1 + \frac{\delta n_\vect p^2}{n_\vect p} =: n_\vect p - N_\vect p^\pminus ,
\end{equation} 
where $N^\pminus_\vect p = 1 - {\delta n_\vect p^2}/{n_\vect p}$ is the expected number of negative-energy excitations. Therefore the hole state contains at most one negative-energy excitation. This state has momentum $\vect p$, $\av{\hat{\vect P}}^\pminus  = \vect p N_\vect p$, and has an additional amount of negative energy $E_\vect p$, $\av{\hat{\vect P}}^\pminus  = E_0 - E_\vect p N^\pminus_\vect p$.

For a particle number eigenstate $N_\vect p^\pminus = 1$, but a  is that for a Gaussian state the number of negative energy excitations is actually negative $N_\vect p^\pminus = -n_\vect p$. The reason for this last surprising fact is that the annihilation operator, besides explicitly removing one particle from the state, also enhances the probability of the highly populated sectors of the state, and the last effect is actually dominating. 
We will consider that $\hat\rho_\vect p^\pminus$ represents a single real hole plus some contribution of statistical quasiparticles of opposite momentum.   Again, the statistical quasiparticle contribution can be thought as a consequence of the increased knowledge of the background state coming from the absorption of a quasiparticle (or creation of a hole). In any case, the fact that for Gaussian states the statistical contribution always dominates is an indication that holes cannot be considered true quasiparticles in bosonic systems.
A confirmation of this fact is given by the spread in the momentum, 
\begin{equation}
 	\left\{\av{\hat {\vect P}^2}^\pminus - [\av{\hat {\vect P}}^\pminus]^2 \right\}^{1/2}= |\vect p|(5n_\vect p^2+n_\vect p)^{1/2},
\end{equation}
which is of the same order of the expectation value of the momentum, even if the occupation numbers are small. Even if holes are not true quasiparticles in bosonic systems, it will be important to remember that negative energy excitations are possible when the states are different than the vacuum.

So far we have seen that quasiparticles and holes are respectively created by the creation and destruction operators. The field operator $\hat\phi_{\vect p}$, being a linear combination of creation and destruction operators, creates a coherent superposition of quasiparticles and holes. In effect, the state $\hat\phi_{-\vect p} \hat\rho \hat\phi_{\vect p}$ corresponds to the linear superposition of a quasiparticle and a  hole:
\begin{equation}
	\hat\phi_{-\vect p} \hat\rho \hat\phi_{\vect p} = \frac{1}{2E_{\vect k}} \left(\hat a^\dag_{\vect p} + \hat a_{-\vect p}\right) \hat \rho  \left(\hat a_{\vect p} + \hat a^\dag_{-\vect p}\right).
\end{equation}
For states characterized by low  occupation numbers the quasiparticle contribution dominates over the hole contribution.

\index{Particle detector}

\subsection{Creation of quasiparticles}\label{sect:QuasiCreation}

Let us argue that only the states created with the creation and destruction operators, or equivalently with the field operator, are physically meaningful. 
In a realistic situation quasiparticles are created with the interaction of the field with some external agent. These external agent can be modeled by a ``quasiparticle emitter'' ---which  essentially coincides with the usual particle detector considered in the analysis of the Unruh effect \cite{Unruh76,BirrellDavies}: an external device described by some quantum mechanical degree of freedom $Q$ (which can correspond to a harmonic oscillator or a two-level system). The device starts in an excited state. For simplicity, let us assume that the emitter is linearly coupled with one pair of field modes, $g_Q Q(t) [\phi_\vect p(t)+\phi_\vect{-p}(t)] $.\footnote{Notice that in this simplified model the emitter cannot couple to a single mode of the field because that couping would not be momentum-conserving (or, from another point of view, would not be hermitian).}  The coupling constant $g_Q$ is very small so that the external device, other than emitting quasiparticles, does not significantly perturb the dynamics of the field. The initial state of the field plus detector is assumed to be $\hat\rho \otimes |1_Q\rangle \langle 1_Q|$. The aim is to find the final state for the field $\hat\rho'$ when the detector is measured in its unexcited state $|0\rangle$. 

The time evolution of the entire system under the interaction is given by
\begin{equation}\label{TotalEvolution}
	\hat\rho_\text{total}(t)=
	T \expp{-i\int_{t_0}^t g_Q \hat\phi_\text{I}(s) \hat Q_\text{I}(s) \vd s }
	\hat\rho \otimes |1_Q\rangle \langle 1_Q|
	T \expp{i\int_{t_0}^t g_Q \hat\phi_\text{I}(s) \hat Q_\text{I}(s) \vd s },
\end{equation}
where the subindex $\text I$ indicates interaction picture, and $\phi_\text I := \phi_\vect p+ \phi_{-\vect{p}}$. Since the coupling is small, the above equation can be expanded as
\begin{equation*}
	\hat\rho_\text{total}(t)=
	\left[1-i\int_{t_0}^t g_Q \hat\phi_\text{I}(s) \hat Q_\text{I}(s) \vd s \right]
	\hat\rho \otimes |1_Q\rangle \langle 1_Q|
	\left[1+{i\int_{t_0}^t g_Q \hat \phi_\text{I}(s) \hat Q_\text{I}(s) \vd s }\right],
\end{equation*}
or, developing the expression
\begin{equation}
\begin{split}
	\hat\rho_\text{total}(t)&=
	\hat\rho \otimes |1_Q\rangle \langle 1_Q|
	- i g_Q \int_{t_0}^t \ud s g_Q \big[\hat\phi_\text{I}(s),\hat\rho\big] \otimes \big[ \hat Q_\text{I}(s) ,  |1_Q\rangle \langle 1_Q| \big]\\
	&\quad + g_Q^2 \int_{t_0}^t \ud s \int_{t_0}^t \ud {s'} \hat\phi_\text{I}(s) \hat\rho\, \hat\phi_\text{I}(s') \otimes
	\hat Q_\text{I}(s)  |1_Q\rangle \langle 1_Q| \hat Q_\text{I}(s').
\end{split}
\end{equation}
According to the quantum mechanics postulates \cite{GalindoPascual}, the final state for the system, if the emitter is found unexcited at time $t$, is given by the projection into the ground state of the emitter:
\begin{equation}
	\hat\rho' = \frac{ \Tr_\text{emitter} \left( \hat\rho_\text{total}(t) |0_Q\rangle \langle0_Q| \right)}{ \Tr_\text{total} \left( \hat\rho_\text{total}(t) |0_Q\rangle \langle 0_Q| \right)}.
\end{equation}
Developing the above expression we find
\begin{equation*}
\begin{split}
	\hat\rho' &= N  \int_{t_0}^t \ud s \int_{t_0}^t \ud {s'} \hat\phi_\text{I}(s) \hat\rho\, \hat\phi_\text{I}(s') \otimes 
	\langle 0 | \hat Q_\text{I}(s)  |1_Q\rangle \langle 1_Q| \hat Q_\text{I}(s') | 0 \rangle \\
	&= N'  \int_{t_0}^t \ud s \int_{t_0}^t \ud {s'} \expp{i \Omega(s-s')} \hat\phi_\text{I}(s) \hat\rho\, \hat\phi_\text{I}(s'),
\end{split}
\end{equation*}
where $\Omega$ is the frequency of the harmonic oscillator and $N$ and $N'$ are normalization constants. Expanding in terms of creation and annihilation operators we get:
\begin{equation*}
\begin{split}
	\hat\rho' &= N'' \int_{t_0}^t \ud s \int_{t_0}^t \ud {s'} \expp{i \Omega(s-s')}\Big(  \hat a_\vect p^\dag \hat\rho\, \hat a_\vect p \expp{-iE_\vect p(s-s')}+\hat a^\dag_\vect p \hat\rho\,\hat a_{-\vect p}^\dag \expp{-iE_\vect p(s+s')} 
	\\ &\qquad +  \hat a_{-\vect p} \hat\rho\, \hat a_\vect p \expp{iE_\vect p(s+s')} +  \hat a_{-\vect p} \hat\rho\, \hat a^\dag_{-\vect p} \expp{iE_\vect p(s-s')} 
	+ \hat a_{-\vect p}^\dag \hat\rho\, \hat a_{-\vect p} \expp{-iE_\vect p(s-s')}\\ &\qquad +\hat a^\dag_{-\vect p} \hat\rho\,\hat a_{\vect p}^\dag \expp{-iE_\vect p(s+s')} 
	 +  \hat a_{\vect p} \hat\rho\, \hat a_{-\vect p} \expp{iE_\vect p(s+s')} +  \hat a_{\vect p} \hat\rho\, \hat a^\dag_{\vect p} \expp{iE_\vect p(s-s')}\Big).
\end{split}
\end{equation*}
Let us assume that the frequency of the oscillator is tuned so that $\Omega= E_\vect p$. In this case, for sufficiently large time lapses the dominant value of the integral is given by the stationary value of the integral of the first term, which amounts to considering energy conservation. In that case
\begin{equation} \label{physicalStateQP}
\begin{split}
	\hat\rho' &\approx N''' \big(\hat a_\vect p^\dag \hat\rho\,\hat a_\vect p + \hat a_{-\vect p}^\dag \hat\rho\,\hat a_{-\vect p} \big) = \frac12 \big(\hat\rho_\vect p^\pplus +\hat\rho_{-\vect p}^\pplus\big).
\end{split}
\end{equation}
Therefore, upon desexcitation of the emitter, the system gets excited to a superposition of two quasiparticle states.\footnote{The superposition would be coherent if phases would not have been negected in the above step.} The argument could be repeated with the same measuring device in the ground state, which now is interpreted as a particle detector. When the measuring device gets excited, the state of the field colapses to the hole state $\hat\rho^\pminus_\vect p$. Similar results would be obtained if the coupling between the system and the oscillator would be of different nature (quadratic for instance).

In this section we have attempted to explicitly build the quantum state corresponding to  quasiparticles, and to compute the expectation values of the relevant operators. We have seen that, while particles and holes are generated by the creation and annihilation operators, or equivalently by the field operator, they also create some additional statistical effects when the background state is a mixed state. This has created some unexpected difficulties with the expectation values of the energy and the momentum, because they are inherently subject to the statistical aspects of the problem. An alternative approach is to study specific transition probabilities, which are less dependent on the statistical aspects of the problem. Namely, we can make an analysis of the probability that the field operator creates a state with given momentum and energy, namely a quasiparticle. We have previously seen that the Pauli-Jordan propagator provides this information, and has the added bonus that it is background-independent for free systems. In absence of interaction the Pauli-Jordan function
\begin{equation}
	G(\omega,\vect p) = \frac{1}{2E_\vect p} \big[\delta(\omega-R_{\vect p})-\delta(\omega+R_{\vect p})\big]
\end{equation}
implies that the field operator can excite quasiparticles with energy $E_\vect p$ and holes with energy $-E_\vect p$.
Let us start the analysis of the interacting case by focusing on the spectral structure of the theory.
The explicit construction of interacting quasiparticles will be postponed.

\subsection{Spectral representation of the quasiparticles}

\index{Spectral representation}
\index{Spectral function}

We have seen that for homogeneous states the Pauli-Jordan propagator, which plays the role of the spectral density, is given by \Eqref{SpectralDetailed}, which can be reexpressed as
\begin{equation} 
	G(\omega,\vect p) = \sum_{\alpha,\beta} \rho_{\alpha}\abs{\langle\alpha|\hat\phi_{\vect p}|\beta\rangle}^{2} \left[ \delta\boldsymbol(\omega - (E_\beta-E_\alpha) \boldsymbol)-\delta \boldsymbol(\omega +  (E_\beta-E_\alpha)\boldsymbol)\right].
\end{equation}
The Pauli-Jordan propagator is proportional to the probability that the field operator creates an excitation of momentum $\vect p$ and energy $\omega$  minus the probability that it excites a state of momentum $\vect p$ and energy $-\omega$. Notice that the value of the energy is fixed by the delta function, and that the value of the momentum is fixed by the mode decomposition of the field operator. The mode labels correspond to the physical momentum provided the interaction contains no spatial derivatives. We do not make any distinction between the $\vect p$ and $-\vect p$ modes since we are assuming spatial isotropy.

Let us assume that around some range of energies the Pauli-Jordan propagator has the following approximate structure:
\begin{equation}\label{StableQuasiparticle} 
	G(\omega,\vect p) \approx \frac{Z_{\vect p}}{2R_{\vect p}}\big[\delta(\omega-R_{\vect p})-\delta(\omega+R_{\vect p})], \quad \text{for }|\omega|\sim R_{\vect p}.
\end{equation}
This means that the field operator with momentum $\vect p$ creates an excitation whose energy is exactly $R_{\vect p}$. Since there is no spread in the energy, the excitation must be infinitely lived (must be stable). The corresponding retarded propagator is:
\begin{equation} 
	\GR(\omega,\vect p) \approx \frac{-iZ_{\vect p}}{-\omega^2 +R_{\vect p}^2 - i\omega\epsilon} , \quad \text{for }|\omega|\sim R_{\vect p}.
\end{equation}
The excitation this way created would have exact energy $R_{\vect p}$ and exact momentum $\vect p$. Therefore it corresponds to a stable quasiparticle. Notice that, in contrast to the stable particles in the vacuum, $Z_\vect p$ can depend on the 3-momentum $\vect p$ and $R_\vect p^2$ need not be of the form $\vect p^2+m^2$.

However, stable quasiparticles are an idealization, and do not correspond exactly to any physical situation, since dissipation is a generic feature of non-vacuum states as long as there is interaction. We recall that by ``dissipation'' we do not necessarily mean the quasiparticle decaying into a product of different quasiparticles: most of the times, ``dissipation'' simply means that the quasiparticle changes its momentum. In any case, stable quasiparticles may be a good approximation in many situations in which the effective coupling to the environment is very small (in the sense of the QBM correspondence explained in the previous chapter). However, this does not mean at all that the real coupling must be small, or that the approximation is limited to weakly interacting systems; as a matter of fact, in strongly coupled situations there might be situations in which assuming free quasiparticles might well be a good approximation.

\index{Propagator!analytic structure}
Anyway, since dissipation is a generic feature of non-vacuum field theories, quasiparticles are not stable, and instead decay with some rate $\Gamma_{\vect p}$. The generic form of the retarded propagator in terms of the self-energy is
\begin{equation}
	\GR(\omega,\vect p) = \frac{-i}{-\omega^2 +\omega_{\vect p}^2 + \SigmaR(\omega,\vect p)}.
\end{equation}
Following the vacuum analysis of subsection \ref{sect:TimeEvol}, we known that whenever there is a long-lived quasiparticle the analytic continuation of the retarded propagator can be approximated by:
\begin{equation}
	\GR(\omega,\vect p) = \frac{-i Z_\vect p}{-\omega^2 +R^2_{\vect p} - i \omega \Gamma_{\vect p}} + \widetilde G(\omega,\vect p),
\end{equation}
where $\widetilde G(\omega,\vect p)$ is an analytic function in the vicinity of the pole. This means that for unstable (but long-lived) quasiparticles, the Pauli-Jordan function, instead of being a delta function around the energy of the pole, it is an approximate Lorentzian function, whose width corresponds to the decay rate. The momentum-dependent functions $R_\vect p$ and $\Gamma_\vect p$ are given, in terms of the self-energy, by
\begin{subequations} \label{RGammaBasic}
	\begin{align}
		R_\vect p^2 &= m^2 + \vect p^2  + \Re \SigmaR(R_\vect p,\vect p),\\
		\Gamma_\vect p &= - \frac{1}{R_\vect p} \Im \SigmaR(R_\vect p,\vect p),
	\end{align}
\end{subequations}
under the assumption that $\Gamma_\vect p$ is much smaller than $R_\vect p$. The spectral representation indicates that $R_\vect p$ corresponds to the quasiparticle energy. The physical meaning of $\Gamma_\vect p$ can be extracted from the interpretation of the imaginary part of the self-energy that we developed before: it corresponds to the net decay rate of the quasiparticle.

\index{Propagator!analytic structure}

\begin{figure}
\includegraphics[width=\textwidth]{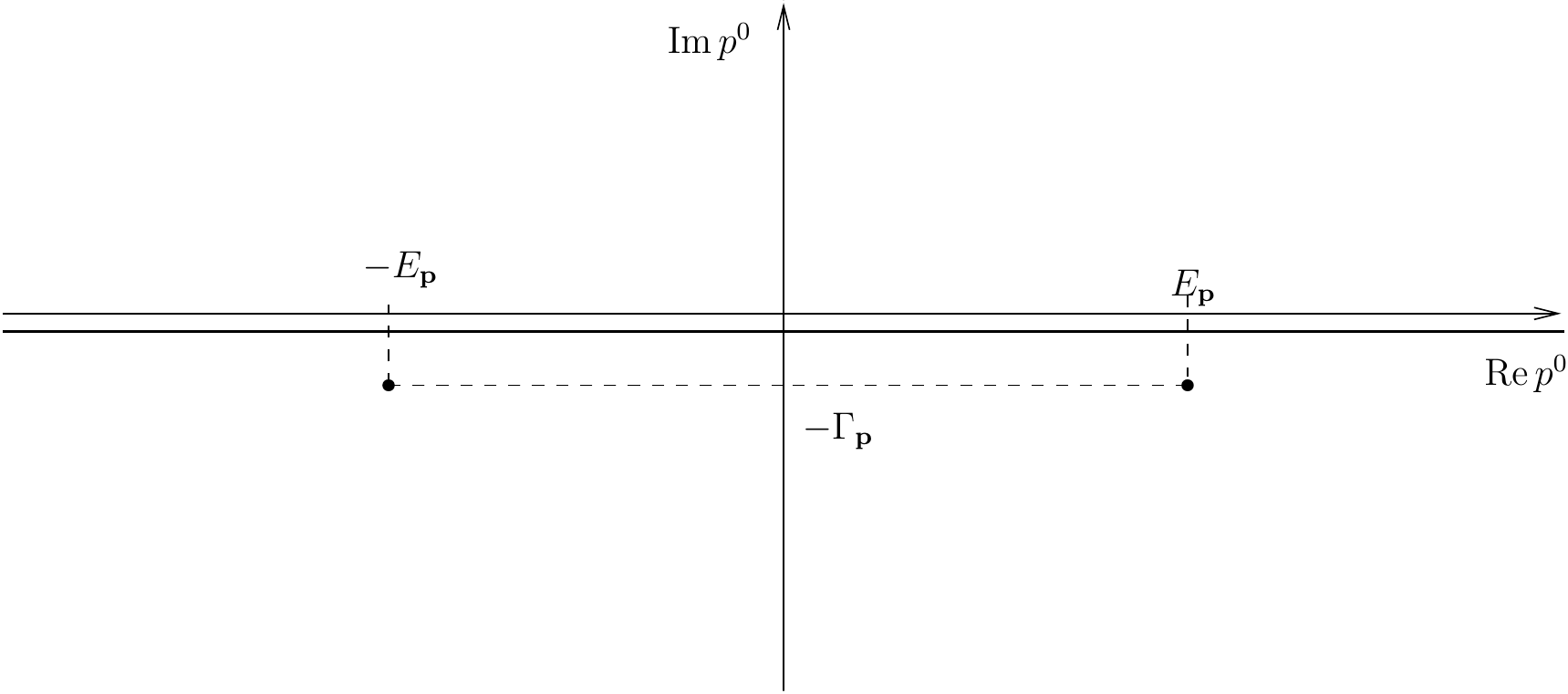}
\caption{Analytic structure of the retarded propagator for quasiparticles in general backgrounds, in the complex energy representation.  The retarded propagator has no poles but only has a cut parallel to the real axis. Its analytic continuation (or second Riemann sheet) has a pair of poles, whose real part corresponds to the energy of the excitation and whose imaginary part corresponds to the decay rate. The cut need not extend from $-\infty$ to $+\infty$: it might be interrupted for some energy sectors.}\label{fig:retarded-alt}
\end{figure}

The structure of the retarded propagator is completely analogous to the vacuum (see sect.~\ref{sect:ParticleVacuum} in this chapter); the analysis will not be repeated here (it is summarized in figure \ref{fig:retarded-alt}). There are however two important differences that it is worth commenting. First, in the vacuum the analysis can be equivalently performed with the Feynman or the retarded propagators, while in generic backgrounds the spectral analysis only applies to the retarded propagator, since in the general case the Feynman propagator has an additional explicit dependence on the background state of the field. Second, in the vacuum one can study the spectral structure either in terms of the energy or in terms of the squared four-momentum. In general,  there is a preferred reference frame and therefore the explicit Lorentz invariance of the results is broken, and, hence, only a spectral analysis based on the energy is meaningful. See refs.~\cite{Weldon99,Weldon02} for a more in depth investigation of the analytic structure of the propagators in thermal field theory.

\index{Propagator!analytic structure}

\index{Propagator!Feynman}
\index{Propagator!Pauli-Jordan}
\index{Propagator!retarded}
\index{Propagator!advanced}
\index{Propagator!Hadamard}
\index{Propagator!retarded}

When the retarded and advanced propagators can be approximated by
\begin{subequations} \label{UnstableQuasi}
\begin{align}
	\GR(\omega,\vect p) &= \frac{-i Z_\vect p}{-\omega^2 +R^2_{\vect p} - i \omega 
	\Gamma_{\vect p}},\\
	\GA(\omega,\vect p) &= \frac{i Z_\vect p}{-\omega^2 +R^2_{\vect p} + i \omega 
	\Gamma_{\vect p}},
\end{align}
the Pauli-Jordan propagator (or spectral function) is given by
\begin{align}
	G^{(1)}(\omega,\vect p) &= \frac{ 2 |\omega| Z_\vect p \Gamma_\vect p   }
	{\left(-\omega^2+R_\vect p^2\right)^2 + \left( \omega \Gamma_\vect p\right)^2}. 
\end{align}
So far we have analyzed the propagators which are independent of the occupation number. Other propagators depend on the occupation number. 
Approximating $n(|\omega|)$ by the value at the pole $n_\vect p=n(R_\vect p)$ we obtain from \Eqref{SigmaNInterp1}
\begin{equation}\label{SigmaNRPole}
	\SigmaN(R_\vect p,T) = -2i \Im \SigmaR(R_\vect p,T)  (1+2n_\vect p).
\end{equation}
Combining the above equation with eqs.~\eqref{PropsSigma} yields
\begin{align}
	G(\omega,\vect p) &= \frac{ 2 \omega Z_\vect p \Gamma_\vect p (1 + 2n_\vect p)  }
	{\left(-\omega^2+R_\vect p^2\right)^2 + \left( \omega \Gamma_\vect p\right)^2}. 
\end{align}
Recall that $n_\vect p$ can be interpreted as the occupation number of the mode with momentum $\vect p$. 
Finally, the Feynman and Dyson propagators and the Whightman functions can be represented by
\begin{align}
	\GF(\omega,\vect p) &= \frac{-i Z_\vect p (-\omega^2+R_\vect p^2)+ |\omega| Z_\vect p  \Gamma_\vect p \left( 2n_\vect p + 1 \right)  }
	{\left(-\omega^2+R_\vect p^2\right)^2 + \left( \omega \Gamma_\vect p\right)^2}, \\
	\GD(\omega,\vect p) &= \frac{i Z_\vect p (-\omega^2+R_\vect p^2)+ |\omega| Z_\vect p  \Gamma_\vect p \left( 2n_\vect p + 1 \right)  }
	{\left(-\omega^2+R_\vect p^2\right)^2 + \left( \omega \Gamma_\vect p\right)^2},\\
	G_+(\omega,\vect p) &= \frac{ 2|\omega| Z_\vect p  \Gamma_\vect p \left[ n_\vect p + \theta(\omega)\right]  }
	{\left(-\omega^2+R_\vect p^2\right)^2 + \left( \omega \Gamma_\vect p\right)^2}, \label{GplusGuay}\\
	G_-(\omega,\vect p) &= \frac{ 2|\omega| Z_\vect p  \Gamma_\vect p \left[ n_\vect p+ \theta(-\omega)\right]   } 
	{\left(-\omega^2+R_\vect p^2\right)^2 + \left( \omega \Gamma_\vect p\right)^2}.
\end{align}

\end{subequations}

\subsection{Quantum states corresponding to quasiparticles}

\index{Quasiparticle!quantum state}

\label{Asymptotic quasiparticles}
When no interaction is present, positive energy quasiparticles are created by the action of the creation operator, and negative energy holes are created by the annihilation operator. If interaction is present and quasiparticles are stable, similarly to the vacuum case  (see section \ref{sect:Asympt}), one can think of defining asymptotic quasiparticle fields and states, from which the two point correlation functions can be reproduced. Namely, quasiparticle states correspond to
\begin{subequations} \label{ThermalAsympt}
\begin{equation}
	\hat\rho^\pplus_\vect p \cong \frac{1}{\bar n_\vect p + 1}\, 
\hat \ar^\dag_\vect p \hat\rho \hat \ar_\vect p ,
\end{equation}
and negative energy holes correspond to
\begin{equation} 
	\hat\rho^\pminus _\vect p \cong \frac{1}{\bar n_\vect p}\, 
\hat \ar_{-\vect p} \hat\rho \hat \ar^\dag_{-\vect p},
\end{equation}
\end{subequations}
where $\bar n_{\vect p}=\Tr{(\hat\rho' \hat \ar^\dag_{\vect p} \hat \ar_\vect p)}$ is the expectation value of the number of asymptotic quasiparticles in the state with momentum $\vect p$, and we recall the symbol $\cong$ means equivalence when evaluated in matrix elements in the asymptotic limit.  The asymptotic creation and destruction operators $\hat\ar$ and $\hat\ar^\dag$ are related to the asymptotic field operator through
\begin{equation}\label{AsymptFieldOp}
	\hat\phir_\vect p = \frac{1}{\sqrt{2 R_\vect p}} \big( \hat \ar _\vect p + \hat \ar^\dag_{-\vect p}\big),
\end{equation}
where $R_\vect p$ is the quasiparticle energy. The asymptotic field, which obeys free equations of motion, $\ddot \phir + (m^2+ R_\vect p^2) \phir = 0$, is connected to the usual field through
\begin{equation}
	\hat\phir_\vect p \cong Z_\vect p^{-1/2} \hat\phi_\vect p,
\end{equation}
where $Z_\vect p$ appears in \Eqref{StableQuasiparticle}.

However, stable quasiparticles do not exist as such, they are only approximations valid in some circumstances. Since quasiparticles are not completely well-defined from a strict point of view, it is not worth pursuing a very formal description. Therefore we will blurry the distinction between the usual and the asymptotic fields, and work with the usual field operator but rescaled a factor  $Z_\vect p^{-1/2}$, \ie, we will set $\mathcal Z_\vect p = Z_\vect p$ [see eqs.~\eqref{OQSAction}]. With this assumption  equations \eqref{ThermalAsympt} can be more simply restated:
\begin{equation}
	\hat\rho^\pplus_\vect p \approx \frac{1}{n_\vect p + 1}\, 
\hat a^\dag_\vect p \hat\rho \hat a_\vect p,  \qquad
	\hat\rho^\pminus _\vect p \approx \frac{1}{n_\vect p}\, 
\hat a_{-\vect p} \hat\rho \hat a^\dag_{-\vect p}. 
\end{equation}
Bear in mind that this representation is only valid when studying (approximately) asymptotic properties.

When we presented the open quantum system analysis of the field modes (see section \ref{sect:OQS}), we also introduced two free parameters, $R_\vect p$ and $\mathcal Z_\vect p$. At this point we have, on the one hand, a criterion to fix their value, and, on the other hand, a physical interpretation for both of them. With respect to the former, $R_\vect p$ must be fixed to the value of the real part of the pole of the propagator, and represents the physical energy of the quasiparticle excitation. With respect to the latter, $Z_\vect p$ measures the probability that the interacting field operator excites the quasiparticle state. Roughly speaking, by rescaling the field a factor $\mathcal Z_\vect p=Z_\vect p$ we  ensure that creation and destruction operators create and destroy quasiparticles with the proper normalization. From now on, we shall assume that such a scaling has been performed so that $\mathcal Z_\vect p=Z_\vect p$. From a practical point of view, this amounts to setting $Z_\vect p=1$ in equations \eqref{UnstableQuasi}.

Similarly to the free case, the states $\hat\rho^\pplus_\vect p$ and $\hat\rho^\pminus_\vect p$, besides representing particles and holes, also contain induced statistical modifications of the background state (see subsect.~\ref{sect:QuasiParticlesFree}). We have already seen that these states have a well-defined momentum, once the statistical effects are properly taken into account ---recall that in absence of derivative couplings the physical momentum operator is not modified. In  the next section we shall see that they have well-defined energy and a small decay rate. Although we have not shown the explicit details, we have also mentioned that the spreads of the energy and the momentum are small provided the occupation numbers are not very large. Additionally, one can check by inspection that both states cannot be decomposed in more elementary states having similar properties. Finally, repeating the analysis in  subsect.~\ref{sect:QuasiParticlesFree} we would find again that upon excitation or desexcitation of a particle emitter, or absorber, the states $\hat\rho^\pplus_\vect p$ and $\hat\rho^\pminus_\vect p$ are naturally generated. Therefore, we conclude $\hat\rho^\pplus_\vect p$ and $\hat\rho^\pminus_\vect p$ are good representations of the quasiparticle excitations.

\subsection{Time evolution of the quasiparticle excitations}\label{sect:Quasitime}

The aim is to find the time evolution of the expectation value of the Hamiltonian of the system. We shall make use of the open quantum system viewpoint that we introduced in the previous section. Let us start with the positive energy excitations, and focus on the state with momentum $\vect p$. The expectation value of the Hamiltonian operator in such a quantum system is given by
\begin{equation}
	E^\pplus(t,t_{0};\vect p) := \av{\hat H_{\vect p}(t)}_\pplus  = \frac{1}{ n_\vect p + 1} \Tr{ \big(\hat a_\vect p^\dag\hat\rho\hat a_\vect p U(t_0,t)\hat H_\vect p U(t,t_{0})   \big) },
\end{equation}
and the Hamiltonian of the 2-mode $\vect p$ is $
	H_\vect p =  \dot \phi_\vect p \dot \phi_{-\vect p} + R_\vect p^2  \phi_\vect p \phi_{-\vect p}
$ [see \Eqref{OQSH}],
with $R_\vect p$ being interpreted as the physical frequency of the mode. 

\index{Wick theorem}
We would like to reanalyze the above expression in the QBM correspondence:
\begin{equation} \label{EplusQBM}
	E^\pplus(t,t_{0};\vect p) \approx \frac{1}{ n + 1} \Tr_\text{sys}{ \big[\hat a^\dag\hat\rho_\text{s}\hat a \,U(t_0,t)\hat H_\text{sys} U(t,t_{0})   \big] }.
\end{equation}
Now $\hat a$ and $\hat a^\dag$ have to be interpreted as the creation and annihilation operators of the equivalent Gaussian QBM system, and $n= \Tr_\text{sys}{(\hat\rho_\text s \hat a^\dag \hat a)}$. The system Hamiltonian is 
\begin{equation*}
	\hat H_{\text{sys}}= \frac{1}{2}\left(\hat p^2 +\Omega\hat q^2\right) = \Omega \Big( \hat a^\dag \hat a + \frac{1}{2} \Big),
\end{equation*}
with $\Omega=R_\vect p$ and $\hat q = (\hat a + \hat a^\dag)/\sqrt{2\Omega}$.  The equality in \Eqref{EplusQBM} is only approximate because we are implicitly  discarding the connected part of the 4-point correlation functions, as it is manifest in the following application of the Wick theorem [see appendix \ref{app:Gaussian} and in particular \Eqref{WickMixed}]:
\begin{equation*}
\begin{split}
	E^\pplus(t,t_{0};\vect p) &\approx \frac{\Omega}{n + 1} \Big\{
	 \Tr_\text{sys}  \big( \hat \rho_\text{s} \hat a \hat a^\dag ) \big[ \Tr_\text{sys}  \big( \hat \rho_\text{s} \hat a^\dag \hat a \big)+\tfrac{1}{2} \big]    \\
	&\quad+ \Tr_\text{sys}  \big[ \hat \rho_\text{s} \, \hat a U(t_0,t) \hat a^\dag U(t,t_0) \big] 
	\Tr_\text{sys}  \big[ \hat \rho_\text{s} \, \hat a^\dag U(t,t_0) \hat a U(t_0,t) \big] \Big\},
\end{split}
\end{equation*}
which can be rewritten as
\begin{equation*}
	E^\pplus(t,t_{0};\vect p) \approx E^{(0)} +\frac{\Omega}{n + 1}
	\left| \Tr_\text{sys}  \big[ \hat \rho_\text{s} \, \hat a U(t_0,t) \hat a^\dag U(t,t_0) \big]\right|^2 ,
\end{equation*}
where $E^{(0)}=\Omega ( n+1/2 )$ is the unperturbed energy, or as
\begin{equation}
	E^\pplus(t,t_{0};\vect p) \approx E^{(0)} +  \frac{\Omega}{n + 1}
	\left|\frac{1}{2\Omega}(\Omega + i\partial_t)(\Omega -i \partial_{t_0}) G_+(t,t_0)\right|^2,
\end{equation}
where we have used the expression of the creation and annihilation operator in terms of the position operator:
\begin{equation}
	\hat a^{(\dag)} (t) = \frac{\big[ \Omega \hat q(t) \pm i {\dot {\hat q}(t)} \big]}{\sqrt{2\Omega}}.
\end{equation}

Undoing the correspondence and using translation invariance, the energy of the perturbation is given by
\begin{subequations}
\begin{equation} \label{EnergyGuay}
	E^\pplus(t,t_{0};\vect p) \approx E^{(0)}_\vect p+ \frac{R_\vect p}{n_\vect p + 1}
	\left|I(t,t_0;\vect p)\right|^2 .
\end{equation}
with
\begin{equation}  \label{IntegralI}
	I(t,t_0;\vect p)=\frac{1}{2R_\vect p}(R_\vect p + i\partial_{t})^2  G_+(t,t_0;\vect p)
\end{equation}
with the unperturbed energy being
\begin{equation*}
	E^{(0)}_\vect p \approx \Omega \Big( n_\vect p+ \frac12 \Big).
\end{equation*}
Using \Eqref{GplusGuay}, the expression $I(t,t_0;\vect p)$ can be written as:
\begin{equation} \label{IntegralI2}
 I(t,t_0;\vect p) :=  \int \frac{\vd \omega}{2\pi} \expp{-i\omega(t-t_0)} \frac{ |\omega| (\omega+ R_\vect p)^2  \Gamma_\vect p \left[ n_\vect p + \theta(\omega)\right]  }{\left(-\omega^2+R_\vect p^2\right)^2 + \left( \omega \Gamma_\vect p\right)^2}.
\end{equation}
\end{subequations}
The above integral is evaluated by integration in the complex plane in appendix \ref{app:Integral}. Neglecting $\Gamma_\vect p$ in front of $R_\vect p$ it is found $ I(t,t_0;\vect p) \approx R_\vect p ( n_\vect p+1) \expp{-\Gamma_\vect p (t-t_0)}$, and therefore the result is
\begin{equation}\label{TimeEvolQuasi1}
	E^\pplus(t,t_{0};\vect p) \approx E^{(0)}_\vect p+ R_\vect p (n_\vect p +1) \expp{-\Gamma_\vect p (t-t_0)}.
\end{equation}
The factor $n_\vect p +1$ is the statistical factor (recall that for a Gaussian state the expected number of excitations is $N_\vect p^\pplus =  n_\vect p +1$). Discounting this statistical factor the energy of the quasiparticle thus evolves as:
\begin{equation}\label{TimeEvolQuasi2}
	E^\pplus_\text{QP}(t,t_{0};\vect p) \approx  R_\vect p  \expp{-\Gamma_\vect p (t-t_0)}.
\end{equation}

For the hole excitations we find a similar result
\begin{equation}
	E^\pminus(t,t_{0};\vect p) \approx E^{(0)}_\vect p+ R_\vect p n_\vect p \expp{-\Gamma_\vect p (t-t_0)}.
\end{equation}
Recall that, because of statistical effects, for Gaussian states the initial expectation value for the energy of the hole excitations is given by $R_\vect p n_\vect p$.

\index{Decay rate}
Eqs.~\eqref{TimeEvolQuasi1} and \eqref{TimeEvolQuasi2}  are the main results of this section. From these equations $R_\vect p$ can be identified as the energy of the quasiparticles, in agreement to the results of the previous section, and $\Gamma_\vect p$ as the decay rate of the quasiparticles, according to the interpretation of the imaginary part of the self-energy as the decay rate ---recall that $R^2_\vect p$ and $R_\vect p\Gamma_\vect p$ correspond to the real and imaginary parts of the self-energy respectively.

It could be argued that the derivation presented in this subsection is somewhat cyclic, because we are assuming from the beginning that $R_\vect p$ is the frequency corresponding to the system Hamiltonian. Following what we did in sect.~\ref{sect:QBMdynamics}, we could assume that the frequency of the system has a different value $R_\vect p'$. We would find a rapid oscillatory behavior for the expectation value of the energy. Only when $R_\vect p'=R_\vect p$ the evolution of the energy is the expected one. See sect.~\ref{sect:QBMdynamics} for more details.

Of course the derivation given here neglects the connected parts of the four order correlation functions. The connected parts of the relevant four order diagrams amount to the contributions coming from quasiparticles that interact with real particles in the background in a way that the initial and final momentum for the quasiparticle is the same.

\subsection{Dispersion relations}

\index{Dispersion relation}
\index{Mass!effective}
\index{Mass!thermal}
The \emph{dispersion relation} is the expression of the energy of the quasiparticle as a function of the momentum, namely
\begin{equation}
	E = R_\vect p,
\end{equation}
where $E$ is the energy of the quasiparticle excitation, and $\vect p$ is the momentum, with the assumption of small spreads. 
 We define the \emph{effective mass} as the value of the energy at zero momentum:
\begin{equation}
	\meff = R_{\vect p= \vect 0}
\end{equation}
 When the states are thermal, the effective mass is called the \emph{thermal mass}. 

Notice that the dispersion relation needs not be of the form $\sqrt{\vect p^2+m^2}$, and therefore the effective mass does not determine by itself the dispersion relation. The background constitutes a preferred reference frame, thus breaking the global Lorentz invariance and, because of the interaction with the environment modes, the quasiparticle energy is affected by the background in a way such that the dispersion relation is no longer Lorentz invariant. We will further pursue these ideas  in chapter 4.

In the particular case of stable particles in the vacuum, the spreads of the momentum and the energy can be made zero by working with plane waves. In general situations this is not possible for two reasons. First, as we saw in section \ref{sect:QuasiDef}, the background state is in general a statistical mixture, and has some uncertainty in the energy and the momentum, uncertainty which propagates to the quasiparticle excitations, and therefore to the dispersion relation. Second, and more importantly, stable quasiparticles do not exist, and hence the energies have an uncertainty which is at least given by the decay rate $\Gamma_\vect p$. 

\index{Dispersion relation!generalized}
For a unstable system with energy $E$ and decay rate $\Gamma$, we define the complex \emph{generalized energy} $\mathcal E$ as follows:
\begin{equation}
	\mathcal E^2 := E^2 - i E \Gamma.
\end{equation}
The \emph{generalized dispersion relation} is the function giving the generalized energy of a quasiparticle in terms of the momentum: 
\begin{equation}
	\mathcal E^2 = R_\vect p^2 - i R_\vect p \Gamma_\vect p.
\end{equation}
Notice that the imaginary part of the generalized dispersion relation places a lower bound on the uncertainty of the real part.

From what we have seen in this chapter, the dispersion relation is given by the real part of the pole of the retarded propagator, or equivalently by the real part of the self-energy,
\begin{equation} 
	E^2 = R^2_\vect p =  m^2 + \vect p^2 + \Re \SigmaR(R_\vect p,\vect p),
\end{equation}
and the generalized dispersion relation is given by the complex location of the pole of the propagator, or equivalently by the  self-energy:
\begin{equation}\label{generalizedDispRel}
	\mathcal E^2 = R_\vect p^2 - i R_\vect p \Gamma_\vect p =  m^2 + \vect p^2 + \SigmaR(R_\vect p,\vect p).
\end{equation}
We will further elaborate on the dispersion relations in sect.~\eqref{sect:GravDiscussion}.

\subsection{Non-stationary, inhomogeneous or anisotropic backgrounds}

\shortpage

So far most expressions in this section have referred to homogeneous, isotropic and stationary backgrounds. In fact, the main results of this section depend, first, on the existence of the diagonal relation \eqref{GSigmaDiagonal} between the self-energy and the propagator in Fourier space (which demands homogeneity and stationarity); second, on the diagonalization of the density matrix in the basis of eigenstates of the Hamiltonian (which is implied by stationarity), and, third, on the equivalence of the $\vect p$ and $-\vect p$ modes (which requires isotropy). Therefore, it appears that the quasiparticle interpretation will be completely spoiled out when the background becomes non-stationary, non-homogeneous or non-isotropic. 

Let us argue that this is not the case, and that a particle interpretation is still feasible in non-homogeneous and non-stationary backgrounds provided that the characteristic scales of variation of the background state are sufficiently large. In the following we shall assume that $L$ is the typical length scale in which the background changes significantly, and $T$ is the typical timescale of evolution of the background. 

As explained in subsection \ref{sect:QuasiQuasi}, when the backgrounds are non-homogeneous or non-stationary there is still an approximate diagonal relation between the retarded propagator and retarded self-energy [see eq.~\eqref{ApproximateDiagonal}], provided $ L \gg 1/E$ and $T \gg 1/E$, where $E$ is the typical energy involved. The retarded propagator will have the usual analytical structure, with the only difference that now the location of the poles will be time and space dependent, and therefore the generalized dispersion relation will also be time and space dependent:
\begin{equation}
	\mathcal E^2 = R_\vect p(t,\vect x) = m^2 + \vect p^2 + \SigmaR\boldsymbol(R_\vect p(t,\vect x),\vect p;t,\vect x\boldsymbol) = 0.
\end{equation}

On the other hand, for timescales much smaller than $T$, the background can be considered stationary. Hence, if one considers energies which correspond to those short timescales, $E \gg 1/T$, the density matrix components corresponding to those energies are diagonal in the basis of eigenstates of the Hamiltonian. Therefore the properties which depend on the diagonalization, such as the spectral representation, are still of application.

Finally, if the origin of the anisotropy is the inhomogeneity of the state, for the relevant energy scales the anisotropy will also be negligible if the condition $L \gg 1/E$ is fulfilled. However, it is  possible that the system is homogeneous but anisotropic, and that the magnitude of the anisotropy is large. In this case some of the expressions we have written down will no longer be quantitatively valid, since we have demanded the equivalence of the forward and backward modes for a given momentum. However in chapter 5 we will show with a particular example that the general picture is not essentially modified.

\section{Alternative approaches}\label{sect:Alternative}

We have seen that quasiparticles can be generated by the action of the field operator, and have studied their properties using two different methods: first, we have analyzed the spectral structure of the excitations generated by the field operator and, second, we have constructed the explicit quasiparticle states, and studied their properties. The quasiparticles can be essentially characterized by their generalized dispersion relations, namely, by the energy and the decay rate as a function of the momentum, and in turn these can be derived from the real and imaginary part of the retarded self-energy.  Let us compare our approaches and results with other techniques found in the literature.

\shortpage

\index{Linear response}
\subsection{Linear response theory}

The standard way of analyzing the quasiparticle properties is with the aid of the linear response theory \cite{Kapusta,LeBellac,FetterWalecka,TheoryOfSolidsIIMIT,Reichl}. In linear response theory the Hamiltonian $H$ of a quantum mechanical system is perturbed with some external perturbation $V(t)$ at time $t_0$, so that the total Hamiltonian is $H + V(t)$ for $t>t_0$. Given some quantum operator $\hat O(t)$, it is a simple exercise to show that to first order in the potential the expectation value of that operator in is given by
\begin{equation}\label{313}
	\av{O(t)} = \av{O(t)}_0 + \delta\av{O(t)} =  \av{O(t)}_0  - i  \int_{t_0}^\infty \Tr{ \big( \hat \rho\, \theta(t-t') \big[\hat O_\text{I}(t),\hat V_\text{I}(s)\big] \big) },
\end{equation}
where I indicates interaction picture with respect to the external perturbation and $\av{\cdot}_0$ indicates the expectation value in absence of the external potential.  

Linear response theory is usually applied to a free or interacting scalar field theory, with the following identifications:
\begin{subequations}
\begin{equation}
	\hat V(t) = \int \ud[3]{\vect x} j(t,\vect x) \hat\phi(\vect x),
	\qquad
	\hat O(t) = \hat\phi(\vect x).
\end{equation}
Then, from \Eqref{313}, one obtains:
\begin{equation}
	 \delta\av{\hat\phi(t,\vect x)} = -i \int \ud[4]{x} \GR(x,x') j(x').
\end{equation}
\end{subequations}
For the case of an impulsive perturbation, $j(x) = j(\vect q) \expp{i\vect q \cdot \vect x} \delta(t)$. If we approximate the retarded propagator as
\begin{equation}
	\GR(\omega,\vect p) \approx \frac{  i Z_\vect p }{(2R_\vect p)(\omega -R_\vect p + i \Gamma_\vect p/2)}, \quad
	\omega \sim R_\vect p,
\end{equation}
the following result for the dynamics of the expectation value of the scalar field is obtained \cite{LeBellac}:
\begin{equation} \label{LinearResponseNaive}
	\delta\av{\hat\phi(t,\vect x)} \approx - 
	\frac{i j(\vect q)}{2 R_\vect p} \theta(t) \expp{ i(\vect q \cdot \vect x - R_\vect p t)} \expp{-\Gamma t/2}.	
\end{equation}
The fluctuation oscillates with a frequency $R_\vect p$ and decays with a rate $\Gamma_\vect p/2$. Those are interpreted as the energy and damping rates of the excitations. Notice that the energy of the oscillation decays with a rate $\Gamma_\vect p$.

The approach based on the linear response theory is simpler than the methods that we have developed in this chapter, highlights the importance of the retarded propagator, and quickly relates the real and imaginary parts of the poles of the propagator to the energy and the decay rate respectively. However, the expectation value of the field operator is not a usual observable in field theory. Moreover, the relation of the classical external perturbation with the particle description is not clear, although it has some analogies with the measuring device we introduced in sect.~\ref{sect:QuasiParticlesFree}. Additionally, the kind of perturbations considered does not correspond to quasiparticles, since the latter, which are of the form $\hat a^\dag \hat\rho \hat a$, have vanishing expectation values for the field operator. Given all that, we believe that in relativistic field theory, a linear response-based approach based on the study of the dynamics of expectation value of the field operator is not  well-suited to study the dynamical properties of quasiparticles in the regime of validity of the quasiparticle description. It might be adequate though to describe the hydrodynamical regime.

In relativistic $N$-body theory, linear response is usually applied to the fermion density, instead of the field itself, and the final result depends on the density correlation functions \cite{FetterWalecka,TheoryOfSolidsIIMIT}. In this case results are physically meaningful because the perturbed quantity corresponds directly to an observable. (See the end of this section for further comments on the non-relativistic problem.) Beyond the thermal field theory context, linear response has many different applications in statistical mechanics \cite{Reichl}, solid state physics \cite{TheoryOfSolidsIIMIT,Reichl}, and even gravitation \cite{AndersonEtAl03}.

\subsection{Effective dynamics of the mean field}


Another possible method to study the quasiparticle properties has been the analysis of the effective dynamics of the mean field. This method has been previously used  in the literature by Weldon \cite{Weldon98} and by Drummond and Hathrell \cite{DrummondHathrell80} in a curved background context. Let us briefly review it and see how it compares to our approach.

\index{Effective action}
The CTP effective action $\Gamma[\bar\phi_1,\bar\phi_2]$ is
defined from the Legendre transform of the CTP generating
functional, in a similar way as the usual in-out effective action
(which can be recovered by setting $\phi_2=0$) and taking into
account the CTP doubling of degrees of freedom. Functionally
differentiating the CTP effective action we get the effective
equations of motion for the expectation value of the field,
$\bar\phi := \Tr (\hat\rho\hat\phi)$:
\begin{equation}\label{CTPEqsMot}
    \left. \derf{\Gamma[\bar\phi_1,\bar\phi_2]}{\bar\phi_{1}(x)}
    \right|_{\bar\phi_1=\bar\phi_2=\bar\phi} = 0.
\end{equation}
These equations of motion are real and causal \citep{Jordan86} because
they correspond to equations of motion of true expectation values. See also appendix \ref{app:CTP}.

The effective action can always be expanded as:
\begin{equation}
\begin{split}
    \Gamma[\bar\phi_1,\bar\phi_2] = \sum_r \frac{1}{r!}\int
    &\  \ud[4]{x_1} \cdots  \vd[4]{x_r}\Gamma^{a_1\cdots a_r} (x_1,\ldots,x_r)
    \bar\phi_{a_1}(x_1)\cdots
    \bar\phi_{a_r}(x_r).
\end{split}
\end{equation}
 The
coefficients $\Gamma^{a_1\cdots a_r} (x_1,\ldots,x_r)$ are called
proper vertices.  A straightforward generalization of the usual
argument (see \eg\ ref.~\cite{Peskin}) shows that this 2-point vertex
corresponds to the inverse propagator,
\begin{equation}
    \Gamma^{ab}(x,y) = i (G^{-1})^{ab}(x,y).
\end{equation}
The Schwinger-Dyson equation, which defines the self-energy $\Sigma^{ab}(x,x')$,
\begin{equation*}
    G_{ab}(x,y) = G_{ab}^{(0)}(x,y) - i \int 
    \ud[4]{z}  \ud[4]{w}  G_{ac}^{(0)}(x,z) \Sigma^{cd}(z,w)
    G_{cb}(w,y),
\end{equation*}
can be manipulated to give
\begin{equation}
    (G^{-1})^{ab}(x,y) = A^{ab}(x,y) +  i\Sigma^{ab}(x,y),
\end{equation}
where $A^{ab}(x,y)$ is the inverse of the free propagator,
\begin{equation} \label{InverseProp}
    A^{ab}(x,y)  =  [(G^{(0)})^{-1}]^{ab}(x,y) = c^{ab}
       i (-\Box^2+m^2) \delta^{(4)}(x-y),
\end{equation}
with $c^{ab}=\diag(1,-1)$. We see that the 2-point vertex can be
expressed as
\begin{equation} \label{2VertexSigma}
    \Gamma^{ab}(x,y) = i A^{ab}(x,y) - \Sigma^{ab}(x,y).
\end{equation}
Hence the 2-point vertex essentially corresponds to the
self-energy. Other proper vertices also have
similar interpretations in terms of 1PI diagrams.

Let us suppose now that the relevant vertex is the 2-particle
vertex $\Gamma^{ab}$.  In this case the effective equations of motion can be expressed as
\begin{equation}
\begin{split}
        \left. \derf{\Gamma}{\bar\phi_{1}}
    \right|_{\bar\phi_1=\bar\phi_2=\bar\phi} &= \int \vd[4]{y} \,[\Gamma^{11}(x,y)+
    \Gamma^{12}(x,y)] \bar\phi(y) = 0   ,
\end{split}
\end{equation}
which, taking into account eqs.~\eqref{InverseProp} and
\eqref{2VertexSigma} can be expanded as
\begin{equation}\label{EqMotion}
       (-\Box_x+m^2)\bar\phi (x)+
     \int \vd[4]{y}  \,
    \Sigma_\mathrm R(x,y) \bar\phi(y)  = 0,
\end{equation}
where $\Sigma_\mathrm R(x,y) = \Sigma^{11}(x,y) +
\Sigma^{12}(x,y)$ is the retarded self-energy. The above equation
of motion is real, because the retarded self-energy is a purely
real quantity when expressed in configuration space (although it
can develop an imaginary part in momentum space).

Introducing the Fourier transform around the mid point, the above equation can be rewritten as
\begin{equation} \label{DispRelNE}
    p^2+m^2+ \Sigma_\mathrm R(p;X) = 0.
\end{equation}
Eq.~\eqref{DispRelNE} amounts to
finding the poles of the retarded propagator. Thus, in flat spacetime the self-energy and effective action methods lead to the same dispersion relation
provided we use a CTP approach in both situations and we neglect
vertices with three external particles or more. However, let us see that the interpretation of the dispersion relation is slightly different. Provided we can approximate,
\begin{equation*}
	p^2+m^2+ \Sigma_\mathrm R(p;X) \approx -\omega^2 + m^2+ \vect p^2 + R_\vect p - i\omega\Gamma_\vect p
\end{equation*}
the solution to $\eqref{EqMotion}$ can be written as
\begin{equation}
	\bar\phi(x) \approx \int \udpi[3]{\vect k} \left [ A(\vect p)
	\expp{-i\vect k \cdot \vect x} 
	+ B(\vect p)
	\expp{i\vect k \cdot \vect x} \right] \expp{-i R_\vect p t} \expp{-\Gamma_\vect p t/2}.
\end{equation}
Therefore the real and imaginary parts of the poles have the role of the frequency of the damping rate of the mean field excitation.

This method has the same essential characteristics as the linear response-based method (see the previous subsection).

\subsection{Bibliographical note}

We end this chapter by quickly commenting other related approaches found in the literature that we are aware of.

In the case of non-relativistic field theory ($N$-body quantum mechanics) one finds analogous results, namely the poles of the retarded propagators determine the energy and decay rate of the excitations \cite{FetterWalecka}. However in that case it is much simpler to reach these conclusions, since the Hamiltonian conmutes with the number operator, and it is clear and unambiguous of what adding a particle exactly means.

Donoghue \emph{et al.} \cite{DonoghueEtAl85} discuss the role of the the real part of the pole of the propagator as related to the energy of the particles, and the fact that it is not necessarily Lorentz-invariant.

Weldon \cite{Weldon99b} studies the energy and momentum of the thermal gluon oscillations. However his method starts with the equations of motion derived from the effective action, and therefore they correspond to hydrodynamic oscillations  rather  than to quasiparticle oscillations.

\label{Hydrodinamic limit}
Relativistic field theory in the hydrodynamic limit has been extensively studied. For instance, let us mention that the evaluation of the hydrodynamic transport coefficients has been carried out by Jeon and Yaffe \cite{Jeon95,JeonYaffe96} and by Calzetta \emph{et al.} \cite{CalzettaHu88}, and that Aarts and Berges study the non-equilibrium time evolution of the spectral function in this regime \cite{AartsBerges01}.

\index{Thermo-field dynamics}
In the context of thermo-field dynamics, Chu and Umezawa \cite{ChuUmezawa93} reformulate thermal field theories in terms of what they call stable quasiparticles. Weldon \cite{Weldon98} attempts a similar reformulation in the real-time approach to thermal field theory. Also in the thermo-field dynamics context, Nakawaki \emph{et al.} \cite{NakawakiEtAl89} study quasiparticle collisions (a subject which is not discussed here).

\index{Kinetic field theory}
\index{Effective action!2PI}
The kinetic approach to field theory by Calzetta and Hu \cite{CalzettaHu88} is the approach most similar to ours. While most other methods rely on the analysis of the mean field, Calzetta and Hu focus on the dynamics of the propagator by studying the 2PI effective action, which leads to equations of motion for the correlation functions. They mainly concentrate on the dynamics of the distribution function (aspect which will  be analyzed in chapter 5).

\addtocontents{toc}{\shortpage[.5]}
	
\chapter{Propagation in a thermal graviton bath}

In this chapter we consider the propagation of a scalar particle
immersed in a thermal bath of gravitons, as an application of the methods and techniques explained in the preceeding two chapters. Our aim is to determine
how the thermal bath affects the propagation of the
scalar particle through radiative corrections. To this end, we
shall compute the thermal corrections to its self-energy.

Since gravitational interactions are
non-renormalizable, the system of the scalar field coupled to
gravity should be conceived as an effective field theory
 \cite{Weinberg79,WeinbergQFT}. Although the full theory is
non-renormalizable, low-energy predictions which do not depend on
the Planck scale behavior of gravity can be extracted
 \cite{Donoghue94a,Donoghue94b,Donoghue95}. On the other hand,
since thermal corrections do not affect the ultraviolet properties
of the theory, thermal field theory fits well with the effective
theory approach.

Let us point out that gravitational interactions can be perturbatively treated in two different ways. First, one can use ordinary perturbation theory. Second, one can consider several different matter fields coupled to gravity, and consider the limit in which the number of fields $N$ is very large \cite{Tomboulis77}. This second perturbative scheme allows to consistently incorporate the backreaction effects of the scalar field on the metric perturbations. The large $N$ expansion of quantum gravity is equivalent \cite{RouraThesis,RouraVerdaguerInPrep})
to stochastic gravity
\cite{CalzettaHu94,HuMatacz95,HuSinha95,CamposVerdaguer96,
CalzettaEtAl97,MartinVerdaguer99a,MartinVerdaguer99c,MartinVerdaguer00,HuVerdaguer03,HuVerdaguer04}. In any case, in this chapter we shall work to lowest order in  ordinary perturbation theory. Hence, we deal with free gravitons; \ie, we
neglect the backreaction of the scalar field on the metric
perturbations. 

\section{The system}

\index{Gravitational interaction}

We consider a minimally coupled real scalar field $\phi$ of mass
$m$ propagating in a spacetime characterized by a metric
$g_{\mu\nu}$. The action for the field is
\begin{subequations}
\begin{equation}
    S_{\phi,g} = - \int \ud[4]x \sqrt{-g} \left( \fud g^{\mu\nu}
    \partial_\mu \phi\, \partial_\nu \phi + \fud m^2 \phi^2
    \right),
\end{equation}
and the action for the metric is
\begin{equation}
    S_g =  \frac{2}{\kappa^2} \int \ud[4]x \sqrt{-g}\, R,
\end{equation}
\end{subequations}
where $R$ is the Ricci scalar, $g$ is the determinant of the
metric, and $\kappa = \sqrt{32 \pi G} = \sqrt{32\pi}\, \Lp$ is the
gravitational coupling constant. Assuming that the metric is a
small perturbation of Minkowski spacetime, $g_{\mu\nu} =
\eta_{\mu\nu} + \kappa h_{\mu\nu}$, the complete action $S =
S_{\phi,g} + S_g$ can be decomposed into the free scalar field,
graviton  and interaction actions as $S = S_\phi + S_h + \Sint$.
These actions, expanded in powers of $\kappa$, are
\index{Graviton!action}
\begin{subequations}
\begin{align}
    S_\phi &=  \int \ud[4]x  \left( - \fud
    \partial_\mu \phi \, \partial^\mu \phi - \fud m^2 \phi^2
    \right), \label{ActPhiClas}
    \\
    \begin{split}
    S_h &= \int \ud[4]x \bigg(- \frac12 \partial^\alpha h^{\mu\nu} \partial_\alpha h_{\mu\nu}
    +  \partial_\nu h^{\mu\nu} \partial^{\alpha}
    h_{\mu\alpha}
     \\ &\qquad\qquad- \partial_\mu h \, \partial_\nu h^{\mu\nu}
    + \frac12 \partial^\mu h\, \partial_\mu h \bigg) + O(\kappa),
    \end{split}\label{GravProp}\\
    \Sint &=  \int \ud[4]x \left( \frac\kappa2 T^{\mu\nu} h_{\mu\nu} +
    \frac{\kappa^2}{4} U^{\mu\nu\alpha\beta} h_{\mu\nu}
    h_{\alpha\beta} \right)
     +  O(\kappa^3),
    \label{IntTerm}
\end{align}
\end{subequations}
where $h=h^{\mu}_{\phantom\mu\mu}$, $T_{\mu\nu}$ is the stress
tensor of the scalar field,
\begin{equation}
    T_{\mu\nu}  =  \partial_\mu \phi \partial_\nu \phi - \fud
    \eta_{\mu\nu} \partial_\alpha \phi \partial^\alpha \phi- \fud \eta_{\mu\nu} m^2
    \phi^2,
\end{equation}
and
\begin{equation}
\begin{split}
    U_{\mu\nu\alpha\beta} = &- 2\eta_{\nu\alpha} \partial_\mu \phi
    \partial_\beta \phi + \eta_{\mu\nu} \partial_\alpha \phi
    \partial_\beta \phi \\
    &+\left(\frac12 \eta_{\mu\alpha} \eta_{\nu\beta} - \frac14
    \eta_{\mu\nu} \eta_{\alpha\beta}\right)(\partial^\sigma\phi
    \partial_\sigma\phi + m^2\phi^2).
\end{split}
\end{equation}
Indices are raised and lowered with the background metric
$\eta_{\mu\nu}$. We have kept only the free terms in the action
for the gravitons $S_h$ because we shall only compute the lowest
order corrections of the self-energy of the $\phi$ field.

\index{Effective field theory}
To compute these corrections we need to introduce counterterms in
order to cancel divergences. Since our system is
non-renormalizable, it has to be understood as an effective field
theory, a low-energy approximation of a more fundamental theory at
the Planck scale
\cite{Donoghue94a,Donoghue94b,Donoghue95,WeinbergQFT}. In order to
compute to a given precision $E^n \kappa^n$, where $E$ is the
energy of the process, one has to introduce all possible
counterterms compatible with the symmetry whose coefficients are
of order $\kappa^n$ at most. In our case the most general action
for the counterterms for the scalar field action up to order
$\kappa^2$ which is compatible with the Poincar\'e symmetry is
\begin{equation}
\begin{split}
    S_\mathrm{count} =  &- \int \ud[4]x  \bigg[  \fud (\mathcal Z m_0^2-m^2)
    \phi^2 +  \fud (\mathcal Z -1) (\partial_\mu\phi \partial^\mu\phi + m^2
     \phi^2)
   \\&\qquad+ \frac14 \kappa^2 C_0 \mathcal Z^2
    (\partial_\mu \partial^\mu \phi)^2 \bigg] + O(\kappa^4),
\end{split}
\end{equation}
where $m_0=m/\sqrt{\mathcal Z}+O(\kappa^2)$ is the bare mass, $\mathcal Z =1+O(\kappa^2)$ is
the field renormalization parameter and $C_0=C/\mathcal Z^2+O(1)$ is a bare
four-derivative coefficient. The finite coefficient $C$ is \emph{a
priori} unknown and constitutes an external input of the theory.
The value of $C$ should be determined by experiments or by
knowledge of the underlying more fundamental theory. On the other
hand, we do not introduce counterterms to the interaction action
because we shall not compute vertex corrections.

\index{Graviton!gauge-fixing term}
Additionally the graviton action \eqref{GravProp} must be
supplemented with a gauge-fixing term. We will work with the
harmonic gauge $\partial_\nu h^{\mu\nu} - (1/2)\partial^\mu h =
0$, whose appropriate gauge-fixing action is
\cite{tHooftVeltman74,Veltman76}
\begin{equation}
    S_\mathrm{gf} = - \int \ud[4] x \left(\partial_\nu h^{\mu\nu} -
    \fud \partial^\mu h \right) \left(\partial^\lambda h_{\mu\lambda} -
    \fud \partial_\mu h \right).
\end{equation}
No Faddeev-Popov ghost fields are needed since we will not
consider graviton self-interactions.

\section{Zero temperature}

The aim in this section is to compute the leading contribution to
the vacuum self-energy of the scalar particle. As shown in the previous chapter, at zero temperature
the self-energy $\Sigma\TO(p^2)$ is related to the Feynman
propagator $G\TO_\mathrm{F}(p)$ through
\begin{equation}\label{Feynman}
    G\TO_\mathrm{F}(p) = \frac{-i}{p^2+m^2+\Sigma\TO(p^2)}.
\end{equation}
We recall that the self-energy can be computed as the sum of all
one-particle irreducible diagrams with amputated external legs. In order to regulate
divergences appearing in the calculation we will use dimensional
regularization \cite{Leibbrandt75,TarrachPascual}.

\begin{figure}
    \centering\hfill
    \includegraphics[width=0.32\textwidth]{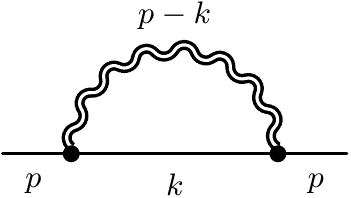}\hfill
    \includegraphics[width=0.32\textwidth]{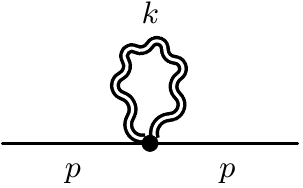}\hfill\mbox{}
    \caption{The two Feynman diagrams needed
    for the calculation of the self-energy: respectively, $\Sigma_{(1)}(p)$ and $\Sigma_{(2)}(p)$.}
    \label{fig:FeyDiag}
\end{figure}

The two diagrams which may contribute to order $\kappa^2$ are
shown in fig.~\ref{fig:FeyDiag}. One must also take into account
the contribution of the counterterms:
\begin{equation}\label{Sigma0}
\begin{split}
    \Sigma\TO(p^2) = &\ (m_0^2 \mathcal Z^2 -m^2) + (\mathcal Z-1) (p^2+m^2) + \kappa^2 \mathcal Z^2 C_0 p^4 \\ &\ + \Sigma\TO_{(1)}(p^2) + \Sigma\TO_{(2)}(p^2) +
    O(\kappa^4).
\end{split}
\end{equation}
We first concentrate on the first diagram $\Sigma\TO_{(1)}(p^2)$.
Applying the Feynman rules described in appendix
\ref{app:FeyRules} we find, in $d$ spacetime dimensions,
\begin{equation}
\begin{split}
    -i\Sigma\TO_{(1)}(p^2)
    =  \mu^\varepsilon\int  &\ \frac{\vd[d]k}{(2\pi)^d}  \tau_{\mu\nu} (p,k) \tau_{\alpha\beta} (k,p)\\
     \times  &\ \frac{-i \mathcal P^{\mu\nu\alpha\beta}}{(p-k)^2-i\epsilon}
    \frac{-i}{k^2+m^2-i\epsilon} ,
\end{split}
\end{equation}
where $\varepsilon=4-d$, $\mathcal P_{\mu\nu\alpha\beta}$ and
$\tau_{\mu\nu}(p,k)$ are given in appendix \ref{app:FeyRules}, and
$\mu$ is an arbitrary mass scale. Developing the products in the
numerator we find
\begin{equation} \label{SigmaDecInt}
\begin{split}
    -i\Sigma\TO_{(1)}(p^2) = &\
     \frac{\kappa^2}{2}
    p^2 \eta^{\mu\nu} I_{\mu\nu}(p) - \kappa^2 m^2 p^\mu
    I_{\mu}(p) \\ &- \kappa^2 m^4 (1+\varepsilon/4)I(p) +
    O(\varepsilon),
\end{split}
\end{equation}
where the momentum integrals $I(p)$, $I_\mu(p)$ and
$I_{\mu\nu}(p)$ are defined and  computed in
appendix~\ref{app:Int}. The result for $\Sigma\TO_{(1)}(p^2)$ is
\begin{equation}
\begin{split}
    \Sigma\TO_{(1)}(p^2) = &\ \frac{\kappa^2}{(4\pi)^2}  \bigg[
    \frac{m^4}{\hat\varepsilon} +  {2m^4}
    +\frac{m^2p^2}{\hat\varepsilon} + 2m^2p^2 \\
    & -\left( \frac{ m^6}{2p^2} +\frac{m^4}{2}\right) \ln\left( 1 + \frac{p^2}{m^2}-i\epsilon\right) \\
    & - \left( m^4 + m^2p^2 \right)
    \ln\left(\frac{p^2+m^2}{\mu}-i\epsilon\right) \bigg] + O(\varepsilon),
\end{split}
\end{equation}
where ${1}/{\hat\varepsilon} = {2}/{\varepsilon} - \gamma +
    \ln 4\pi$.
The second diagram $\Sigma\TO_{(2)}(p^2)$ is a massless tadpole,
and these are identically zero in dimensional regularization
\cite{Leibbrandt75}.

\index{Renormalization}
\index{On-shell renormalization condition}
Once we include the counterterms and take the limit $\varepsilon
\to 0$ the renormalized self-energy is found to be
\begin{equation}\label{SigmaM}
\begin{split}
    \Sigma\TO(p^2) &=
    - \frac{\kappa^2}{(4\pi)^2} \left( \frac{ m^6}{2p^2} + \frac{m^4}{2} \right) \ln\left( 1 +
    \frac{p^2}{m^2}-i\epsilon \right) \\
     &\quad- \frac{\kappa^2}{(4\pi)^2} \left(m^4 + m^2p^2 \right)
    \ln\left(\frac{p^2+m^2}{\mu^2}-i\epsilon\right) \\
    &\quad + C \kappa^2 (p^2+m^2)^2 +
    O(\kappa^4),
\end{split}
\end{equation}
where we have absorbed divergences in the following way:
\begin{subequations}
\begin{align}
    m_0^2 \mathcal Z&= m^2 - C \kappa^2 m^4 + O(\kappa^4)\\
    \mathcal Z &= 1 + 2 C \kappa^2m^2- \frac{\kappa^2m^2}{(4\pi)^2} \left( \frac{1}{\hat\varepsilon} + 2     \right)+ O(\kappa^4),\\
    C_0\mathcal Z^2 &= C + O(\kappa^4).
\end{align}
\end{subequations}
In order to relate the bare and the renormalized parameters we
have imposed the on shell renormalization condition
\begin{equation} \label{RenorCond}
    \Sigma\TO(-m^2) = 0,
\end{equation}
which fixes the position of the pole to be $m^2$. The complete
specification of the on shell renormalization scheme also involves
fixing the residue of the pole, which implies an additional
condition on the derivative of the self-energy
\cite{WeinbergQFT,Hatfield}. However this second condition is
found to be singular due to the divergence of the logarithms when
evaluated on shell. This kind of singularity is generic for
theories with interacting massless particles because of the
absence of well defined asymptotic regions in the presence of
long-range forces. For a discussion of this point in the case of
QED see ref.~\cite{Hatfield}. Anyway in this chapter we are just
interested in the position of the pole, which is always well defined.

\index{On-shell renormalization condition}

\index{Renormalization}
Note that the only divergence arises from the field
renormalization. In particular, the four-derivative coefficient
$C$ does not get renormalized: there are no logarithmic
corrections to the $p^4$ term in the self-energy \eqref{SigmaM},
and $C$ coincides with its classical value $C_0$. Therefore it
would have been consistent not to include the four-derivative
counterterm and simply take $C=0$ from the beginning.

\index{Self-energy!imaginary part}
The self-energy develops a negative imaginary part for $-p^2>m^2$,
\begin{equation} \label{ImSigma}
\begin{split}
    \Im \Sigma\TO (p^2)=\theta(-p^2-m^2)\frac{\kappa^2}{16\pi} \left( \frac{
    m^6}{2p^2}+
    \frac{3m^4}{2} + m^2p^2 \right) \!,
\end{split}
\end{equation}
which, according to the optical theorem \cite{Peskin}, accounts
for the probability of a scalar particle with momentum $p$ to
decay into an on shell scalar and an on shell graviton. Notice
that on shell ($p^2=-m^2$) the imaginary part of the self-energy
vanishes, because there is no phase space available for the
spontaneous emission of a graviton.

\section{Finite temperature}

In this section we compute the thermal contribution to the
self-energy working within the real-time approach to thermal field
theory. Recall that in this approach, which is explained in appendix
\ref{app:CTP}, the number of degrees of freedom is doubled, and
one has to consider four propagators organized in a $2\times2$
matrix $G_{ab}(p)$.  The self-energy also becomes a matrix
$\Sigma^{ab}(p)$.  We will concentrate on the retarded propagator
$G_\mathrm R(p)=G_{11}(p)-G_{12}(p)$.


\subsection{Self-energy: Real part} \label{sect:scalarReal}

Let us proceed to the calculation of the real part of
$\Sigma_\mathrm R(p)$. Since $\Re \Sigma_\mathrm R(p) = \Re
\Sigma^{11}(p)$ [see eq.~\eqref{SigmaR11}], we shall compute the
real part of $\Sigma^{11}(p)$ instead. We have to consider both
diagrams sketched in fig.~\ref{fig:FeyDiag}, where now internal
propagators are taken to be thermal and of type 11:
\begin{equation}\label{Sigma11}
\begin{split}
    \Sigma^{11}(p) = &\ (\mathcal Z^2 m_0^2-m^2) + (\mathcal Z - 1) (p^2+m^2) + \kappa^2 C_0 p^4 \mathcal Z^2 \\ &\ + \Sigma^{11}_{(1)}(p) + \Sigma^{11}_{(2)}(p) +
    O(\kappa^4).
\end{split}
\end{equation}
The first diagram, $\Sigma^{11}_{(1)}(p)$, is given by
\begin{widetext}
\begin{equation}
\begin{split}
    \Sigma_{(1)}^{11}(p) = \frac{i\kappa^2}{2} \mu^\varepsilon\int \udpi[d]{k}
     & \ \left[ \frac{1}{(p-k)^2-i\epsilon} + 2\pi
    i n(|p^0-k^0|) \delta\boldsymbol((p-k)^2\boldsymbol) \right] \\
    \times & \ \left[ \frac{1}{k^2+m^2-i\epsilon} + 2\pi
    i n(|k^0|) \delta(k^2-m^2) \right] \\ \times &\ g_1(p^2,k^2,p\cdot
    k)
\end{split}
\end{equation}
\end{widetext}
where
\begin{equation}
\begin{split}
    g_1(p^2,k^2,p \cdot k) &= -\frac{2}{\kappa^2}
    \tau_{\mu\nu}(p,k) \tau_{\alpha\beta}(k,p)
    \mathcal P^{\mu\nu\alpha\beta}\\
    &= k^2 p^2 - 2 (k \cdot p) m^2 -2(1+\varepsilon/4) m^4,
\end{split}
\end{equation}
the function $n(E)$ is the Bose-Einstein distribution,
\begin{equation}
    n(E) = \frac{1}{1-\expp{E/T}},
\end{equation}
and we recall that $\varepsilon=4-d$ and that $\tau_{\mu\nu}(p,k)$
and $\mathcal P^{\mu\nu\alpha\beta}$ are given in
appendix~\ref{app:FeyRules}. On the other hand, at finite
temperature the tadpole diagram $\Sigma^{11}_{(2)}(p)$ no longer
vanishes since its temperature-dependent part  gives a finite
contribution,
\begin{equation}
    \Sigma_{(2)}^{11}(p) =- \frac{\kappa^2}{2} g_3(p^2) \int \udpi[4]k 2\pi \delta(k^2)
    n(|k^0|),
\end{equation}
where
\begin{equation}
\begin{split}
    g_3(p^2) = -\frac{2i}{\kappa^2} V_{\mu\nu\alpha\beta}(p,p) \mathcal
    P^{\mu\nu\alpha\beta} = 10m^2 + 4p^2,
\end{split}
\end{equation}
with $V_{\mu\nu\alpha\beta}(p,k)$ given in
appendix~\ref{app:FeyRules}.

To order $\kappa^2$ one has
\begin{equation} \label{ReSigmaABC}
\begin{split}
    \Re \Sigma^{11}(p)
    &= \Re \Sigma^{\scriptscriptstyle (T=0)}(p^2)   -
    \frac{\kappa^2}{2} \left[ A(p) + B(p) + C(p^2) \right]+ O(\kappa^4),
\end{split}
\end{equation}
where $\Sigma^{\scriptscriptstyle (T=0)}(p^2)$ is the $T=0$
self-energy [see eqs.~\eqref{Sigma0} and \eqref{SigmaM}] and where
the integrals $A(p)$, $B(p)$ and $C(p^2)$ are defined through
\begin{subequations}
\begin{align}
\begin{split}
    A(p) &= \int \frac{\mathrm d^4 k}{(2\pi)^3} n(|p^0 - k^0|)
    \delta\boldsymbol( (p-k)^2 \boldsymbol) g_1(p^2,k^2,p \cdot k) \PV \frac{1}{k^2+m^2} \label{A},
\end{split}\\
\begin{split}
    B(p) &= \int \frac{\mathrm d^4 k}{(2\pi)^3}  n(|k^0|)
    \delta(k^2+m^2) g_1(p^2,k^2,p \cdot k) \PV \frac{1}{(p-k)^2} ,  \label{B}
\end{split}\\
C(p^2) &= g_3(p^2) \int \frac{\mathrm d^4 k}{(2\pi)^3} n(|k^0|)
\delta(k^2), \label{C}
\end{align}
\end{subequations}
We have set $d=4$ because
thermal contributions are ultraviolet finite. Note that the same
renormalization process which makes the $T=0$ self-energy finite
also renormalizes the $T>0$ self-energy; there is no need to
introduce additional temperature-dependent infinite counterterms. Hereafter
we will concentrate in the on shell results,\footnote{As we will see in section \ref{sect:GravDis}, the on-shell location needs not coincide with the vacuum value. However, since we are doing a first order perturbative calculation, we may evaluate the perturbative terms in the unperturbed on-shell condition.} $p^0 = E_\vect p =
\sqrt{m^2+|\vect p|^2}$, since these are the ones we will need
afterwards for the calculation of the thermal mass and the
modified dispersion relation.

The integrals $A(p)$ and $C(p^2)$ take into account the effect of
the thermal gravitons on the scalar particle. In the on shell case
it is possible to give an explicit expression for these integrals
valid at any temperature (the details of the calculation can be
found in appendix~\ref{app:ABCD}):
\begin{align}
    A(E_\vect p, \vect p)&= -\frac16 m^2 T^2, \label{ResultA}\\
    C(-m^2) &= \frac12 m^2 T^2. \label{ResultC}
\end{align}
Notice that $A(E_\vect p, \vect p)$ does not depend on the
three-momentum $\vect p$. This is a surprise since eq.~\eqref{A}
is not manifestly Lorentz invariant. This simplification, which is
also found in electrodynamics \cite{DonoghueEtAl85}, will have
important consequences  when computing the modified dispersion
relation.

The integral $B(p)$ takes into account the effect of the thermal
scalars in the heat bath. For temperatures well below the mass $m$
there are almost no scalar particles in the bath, and this indeed
shows up in an exponential suppression of $B(E_\vect p, \vect p)$
at low temperatures, $T \ll m$:
\begin{equation} \label{ResultBLowT}
\begin{split}
    B(E_\vect p, \vect p) \approx \sqrt{\frac{ m^5 T^3}{2\pi^3}} \left(
    \frac{m^2+2|\vect p|^2}{3m^2+4|\vect p|^2}\right) \expp{-m/T},
    \quad T \ll m.
\end{split}
\end{equation}
For temperatures of the order of the mass $m$, there are thermal
scalar particles in the bath and they give a significant
contribution to the self-energy. At high temperatures  $T \gg m$,
the leading contribution to the integral is given by
\begin{widetext}
\begin{equation} \label{ResultBHighT}
\begin{split}
    B(E_\vect p, \vect p) &\approx
    \frac{m^2 T^2\sqrt{m^2+ |\vect p|^2}}
{24|\vect p|}
 \ln \left(\frac{ 2\sqrt{m^2+ |\vect
    p|^2}- |\vect p|}{ 2\sqrt{m^2+ |\vect
    p|^2}+ |\vect p|}\right)  + \frac{1}{8} m^2 T^2, \quad T \gg m.
\end{split}
\end{equation}
The details of the calculation for both the low- and
high-temperature limits are given in appendix~\ref{app:ABCD}.

Summarizing, according to eqs.~\eqref{ReSigmaABC} and
\eqref{ResultA}--\eqref{ResultBHighT} the real part of the on
shell self-energy in the low- and high-temperature regimes is
given by:
\begin{equation} \label{ReSigmaT}
\begin{split}
    &\ \Re \Sigma_\mathrm R (E_\vect p, \vect p) \approx\\
    &\quad    \begin{cases}
         -\tfrac{1}{6} \kappa^2 m^2 T^2
        -   \sqrt{\tfrac{m^5 T^3}{{8\pi^3}}}\kappa^2  \expp{-m/T} \left(
        \tfrac{m^2+2|\vect p|^2}{3m^2+4|\vect p|^2}\right) ,
        & T \ll m,\\[3ex]
         \tfrac{1}{48} \kappa^2 m^2 T^2 \left[  - 11
        + \tfrac{ \sqrt{m^2+ |\vect p|^2}}
     {|\vect p|}
         \ln \left(\tfrac{ 2\sqrt{m^2+ |\vect
       p|^2}+ |\vect p|}{ 2\sqrt{m^2+ |\vect
        p|^2}- |\vect p|}\right)\right] , & T \gg m.
    \end{cases}
\end{split}
\end{equation}
Notice that from the high-temperature result we can deduce that
for massless particles the on shell self-energy is exactly zero.
\end{widetext}

\subsection{Self-energy: Imaginary part}\label{sect:scalarIm}

Now we want to compute $\Im \Sigma_\mathrm R(p)$. Similarly to the
previous subsection we could compute $\Sigma^{11}(p)$ and then
make use of eq.~\eqref{SigmaR11}; however, it is somewhat easier
for us to compute $\Sigma^{12}(p)$ and $\Sigma^{21}(p)$ and make
use of relation \eqref{cut} instead. The self-energy
$\Sigma^{12}(p)$ is given by
\begin{equation}
\begin{split}
    \Sigma^{12}(p) &= i \frac{\kappa^2}{2} \int \frac{\mathrm d^4 k}{(2\pi)^2} n(k^0) \sign(k^0)  n(p^0-k^0) \sign(p^0-k^0) \\
    &\quad \times \delta\boldsymbol( (p-k)^2 \boldsymbol) \delta(k^2+m^2) g_1(p^2,k^2,p \cdot
    k),
\end{split}
\end{equation}
where we used the property $\theta(-p^0) + n(|p^0|) =
\sign(p^0)n(p^0)$. In a similar way we find for $\Sigma^{21}$(p):
\begin{equation}
\begin{split}
    \Sigma^{21}(p) &= i \frac{\kappa^2}{2} \int \frac{\mathrm d^4 k}{(2\pi)^2}[1+ n(k^0)] \sign(k^0) [1+n(p^0-k^0)] \sign(p^0-k^0) \\
    &\quad \times \delta\boldsymbol( (p-k)^2 \boldsymbol) \delta(k^2+m^2) g_1(p^2,k^2,p \cdot
    k),
\end{split}
\end{equation}
where now we used $\theta(p^0) + n(|p^0|) = \sign(p^0)[1+n(p^0)]$.
Thus from eq.~\eqref{cut} and the two previous equations we get
\begin{equation} \label{ImSigmaR}
    \Im \Sigma_\mathrm R(p) = -\frac{\kappa^2}{4} D(p),
\end{equation}
where $D(p)$ is defined by
\begin{equation} \label{D}
\begin{split}
    D(p) &=  \int \frac{\mathrm d^4 k}{(2\pi)^2}F(p^0,k^0) g_1(p^2,k^2,p \cdot
    k)
   \delta\boldsymbol( (p-k)^2 \boldsymbol) \delta(k^2+m^2)
\end{split}
\end{equation}
with
\begin{equation}
    \begin{split}
    F(p^0,k^0) &=  \sign(k^0)\sign(p^0-k^0) \left[ 1 + n(k^0)
    + n(p^0-k^0)  \right],
    \end{split}
\end{equation}
which can be developed to give
\begin{equation}
    F(p^0,k^0)= \fud
    \sinh\left(\frac{p^0}{2T}\right)\csch\left(\frac{|p^0-k^0|}{2T}\right)\csch\left(\frac{|k^0|}{2T}\right).
\end{equation}
After manipulating the integral $D(p)$ (the details can be found
in appendix~\ref{app:ABCD}) the imaginary part of the self-energy
can be expressed as the following phase-space integral:
\index{Self-energy!imaginary part}
\begin{equation} \label{IntImSigma}
    \Im \Sigma_\mathrm R(p) = \frac{\kappa^2m^2(m^2+2p^2)}{32\pi |\vect{p}|} \left| \int_{Q_1}^{Q_2}  \ud Q
    F(p^0,Q)\right|.
\end{equation}

Let us now evaluate the integral in \Eqref{IntImSigma} for an
arbitrary temperature. For simplicity, we restrict ourselves to
the case $p^0 > |\vect p|$, but including both $(p^0)^2 > m^2 +
|\vect p|^2$ and $(p^0)^2 < m^2 + |\vect p|^2$. In this situation
the integral can be performed analytically to give
\begin{widetext}
\begin{equation} \label{ImSigmaT}
\begin{split}
    \Im \Sigma_\mathrm R(p) &= \frac{\kappa^2T m^2(m^2+2p^2)}{32\pi |\vect p|} \\ &\quad \times 
    \ln \left[\frac{
    \sinh \left(\frac{(p^0+|\vect p|)^2+m^2}{4T(p^0+|\vect p|)}\right)
    \sinh \left(\frac{(p^0)^2-|\vect p|^2-m^2}{4T(p^0-|\vect p|)}\right)}
    {\sinh \left(\frac{(p^0-|\vect p|)^2+m^2}{4T(p^0-|\vect p|)}\right)
    \sinh \left(\frac{(p^0)^2-|\vect p|^2-m^2}{4T(p^0+|\vect p|)}\right)}
    \right].
\end{split}
\end{equation}
\end{widetext}
Making use of the property
\[
    \ln (\sinh x) \approx |x| - \ln 2 - i \pi \theta(-x) - \expp{-2|x|} ,  \quad
    |x| \gg 1,
\]
we easily obtain the low-temperature approximation
\begin{equation}
\begin{split}
     \Im \Sigma_\mathrm R(p) = &\ \theta(-p^2-m^2) \frac{\kappa^2}{16\pi}
      \left( \frac{m^6}{2p^2} + \frac{3m^4}{2} + m^2 p^2
    \right)+O(\expp{-m/T}),
\end{split}
\end{equation}
which is in agreement with the zero-temperature result of
eq.~\eqref{ImSigma}. 


At one loop a real particle can neither emit nor absorb a
real graviton, because these processes are kinematically
forbidden. Hence, at this order, the imaginary part of the
self-energy must vanish on shell. However in the on-shell limit
$p^0 \to E_\vect p=\sqrt{|\vect p|^2+m^2}$, we obtain, from
eq.~\eqref{ImSigmaT},
\begin{equation} \label{ImDOnShell}
\begin{split}
    \Im \Sigma_\mathrm R(p) \xrightarrow[p^0 \to E_\vect p]{} -&\ \frac{\kappa^2 m^4T}{32\pi |\vect p|
    }
    \ln \left(\frac{{\sqrt{m^2 + |\vect p|^2}+|\vect p| }}
      {{\sqrt{m^2 + |\vect p|^2}-|\vect p|}}\right),
\end{split}
\end{equation}
which is nonzero if $T\neq 0$.

This non zero result for the on shell self-energy is an artifact
of not having introduced an infrared regularization; see
refs.~\cite{Weldon99,Rebhan92} and subsects.~\ref{sect:DivRen} and \ref{sect:NeedHTL}. To illustrate this point one can
take the on-shell limit directly in eq.~\eqref{IntImSigma}: in
this limit $Q_1,Q_2 \to 0$, so that the phase space of the
integral vanishes while the thermal function $F(p^0,0)$ diverges.
Hence the non vanishing result we obtained is a direct consequence
of the infrared divergence of the Bose-Einstein distribution. When
regulating the infrared behavior ---for instance by giving a tiny
mass to the graviton--- no imaginary part is found  at $p^2=-m^2$.
Additionally, had we worked in an arbitrary gauge we could have
verified that the result of eq.~\eqref{ImDOnShell} is not even
gauge invariant \cite{Weldon99,Rebhan92}. 

\begin{figure}
    \hfill
    \includegraphics[width=0.25\textwidth]{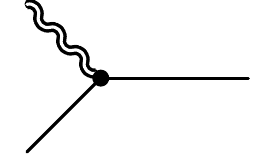}
    \hfill
    \includegraphics[width=0.25\textwidth]{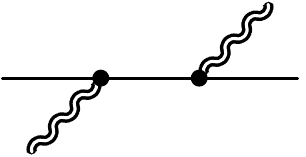}
    \hfill\mbox{}
    \caption{Diagrams such as the one depicted on the left would
    contribute at one loop to the thermalization rate at finite temperature, and hence to the
    imaginary part of the self-energy. However all those one-vertex
    diagrams are identically zero because of kinematical reasons.
    The first non-vanishing contribution would be due to
    Compton-like diagrams such as the one depicted on the right,
    and hence it would be a two loop contribution.}
    \label{fig:cut}

\end{figure}

\index{Self-energy!imaginary part}
Whereas the real part of the self-energy gives the change in
energy of the particle and hence the dispersion relation, the
imaginary part accounts for the dissipative effects and gives the
thermalization rate. At one loop
the imaginary part of the self-energy is zero. We do
not see the thermalization effect since all processes which could
contribute to it, such as the one shown in fig.~\ref{fig:cut} (on
the left), are kinematically forbidden. In order to account for
thermalization effects we should compute the imaginary part of the
self-energy with a perturbative calculation in terms of the hard thermal loop resummed graviton propagator (see following subsection), which incorporates two-loop effects. At two loops, the particle can
exchange momentum with the thermal bath through processes such as
the one to the right of fig.~\ref{fig:cut}, which would correspond
to the Compton scattering in electrodynamics. Notice here a
significant difference with respect to the vacuum case: while at
zero temperature the self-energy is purely real to all orders of
perturbation theory (because the particle is stable), at finite
temperature it acquires an imaginary contribution at two loops.

\subsection{Need for hard thermal loop resummation}\label{sect:NeedHTL}
\index{Hard thermal loop}

For massless gauge theories Braaten and Pisarski \cite{Pisarski89,BraatenPisarski90a,BraatenPisarski90b} showed that the one-loop correction for the propagator of a gauge field is of the same order of magnitude as the tree-level propagator whenever the external momenta are small (see subsect.~\ref{sect:DivRen}). Therefore, the leading contribution to the one-loop correction, the hard thermal loop contribution, must be resummed to the tree level propagator in order to get meaningful results when gauge particles with small momenta give a significant contribution to any given diagram. If resummation is not performed it is possible, for instance, that the leading contribution to a two-loop diagram is of the same perturbative order than the one-loop term. Resummation is possibly also needed in the case of gravity, as it is highlighted in the following heuristic argument. The leading contribution to the graviton self-energy is of the order of $\kappa^2 T^4$ \cite{BrandtFrenkel93,BrandtFrenkel98,BrandtFrenkel99}. Therefore, the one-loop correction to the propagator of the gravitons is of the order of $\kappa^2 T^4 / p^4$, where $p$ is the momentum, or $\kappa^2 T^4 / p^2$ times the tree level contribution. Thus, when the momentum of the gravitons is of the order of $\kappa T^2$ the one loop contribution is  of the same order as the tree level contribution. Although $\kappa T^2\sim T^2/\Mp$ is in general a very small energy scale, if gravitons with momenta of the order of $T^2/\Mp$ made an important contribution to the integrals resummation would be needed. Resummation could also be needed for scalar particles if their masses are below that energy scale.


In the calculation of the correction to the real part of the scalar self-energy, a careful examination of the computations shows that the leading contribution to the self-energy integrals (see appendix \ref{app:ABCD}) is given in the regime in which the graviton momentum is of the order of $T$. Since $T \gg T^2/\Mp$ no graviton resummation is needed. If the mass of the scalar particles is very small, the leading contribution to the integrals comes from scalars whose energy is of the order of $T$, so no resummation is needed either. Therefore, there is no additional contribution to order $\kappa^2$ to the graviton self-energy coming from the resummed propagator, or, in other words, there is no contribution to order $\kappa^2$ coming from two-loop diagrams. 

In contrast, we have seen that the imaginary part of the self-energy gives a non-zero result if not properly regularized, result which is due to the infrared divergence of the integrand. We have discarded that contribution because of the kinematical constraints in the one-loop diagrams. In any case, it appears that the infrared energy scale $T^2/\Mp$ might be relevant for the calculation of the decay rate, and therefore a resummed graviton propagator should be used for the perturbative evaluation of the decay rates. It is therefore possible that there is a $\kappa^2$ contribution to the decay rate coming from two-loop diagrams. 

\section{Dispersion relation} \label{sect:GravDis} \label{sect:GravDiscussion}

\index{Dispersion relation}
At zero temperature, the position of the pole of the propagator
gives the energy of the state and hence defines the dispersion
relation. The position of the pole is
found to be
\begin{equation}
\begin{split}
    E^2  &= m^2 +|\vect p|^2 + \Sigma\TO(-m^2) = m^2 + |\vect p|^2,
\end{split}
\end{equation}
where the second equality is a consequence of the renormalization
condition \eqref{RenorCond}. The dispersion relation is clearly
Lorentz invariant, as expected.

\index{Thermal mass}
Similarly, at finite temperature, we argued in the previous chapter that  the location of the poles of the
retarded propagator determines the inertial properties of the
particle. The location of the poles is given by
\begin{equation} \label{DispRel}
     E^2 =R_\vect p^2  = m^2 + |\vect p|^2 + \Re \SigmaR(p^0,\vect p).
\end{equation}
The thermal mass is obtained by setting $\vect p=\vect 0$:
\begin{equation}
    \mth^2 = m^2 + \Re \SigmaR(\mth,\vect0).
\end{equation}
In a Lorentz invariant situation one would simply have $E^2 =
\mth^2+ |\vect p|^2$, but in general there can be additional
dependence on the three-momentum $\vect{p}$ on the right-hand
side,
\begin{equation} \label{RelDispTh}
    E^2 = R_\vect p^2 = \mth^2+|\vect p|^2 + \mathcal F(\kappa,T,\mth,\vect p).
\end{equation}
The Lorentz-breaking additional term in the dispersion relation
leads to modifications of the group velocity of the particles
$\vect v = \mathrm dp^0/\mathrm d\vect p$:
\begin{equation}\label{GroupVelocity}
    \vect v = \frac{\vect p}{p^0} + \frac{1}{2p^0}\derp{\mathcal F}{\vect
    p}.
\end{equation}

\index{Lorentz symmetry!effective breaking}

Let us find the explicit form of the thermal mass and
Lorentz-breaking terms both in the low and high temperature
regimes. Equation \eqref{DispRel} can be solved perturbatively:
\begin{equation} \label{DispRelGen}
     E^2 = m^2 + |\vect p|^2 +  \Re \Sigma_\mathrm R(E_\vect p,\vect
    p) + O(\kappa^4),
\end{equation}
where we recall that $E_\vect p = \sqrt{ m^2 + |\vect p|^2}$. At
low temperatures the modified dispersion relation, according to
eqs.~\eqref{ReSigmaT} and \eqref{DispRelGen}, is approximately
given by
\index{Dispersion relation!modified}
\begin{equation}
\begin{split} \label{RelDispT}
    E^2 &\approx \mth^2 +|\vect p|^2
        -   \kappa^2 \sqrt{\dfrac{ m^5 T^3
        }{{2\pi^3}}} \expp{-m/T}
        \left( \dfrac{|\vect p|^2}{3m^2+4|\vect p|^2} \right),
\end{split}
\end{equation}
where the leading contribution to the thermal mass is
\begin{equation} \label{PhysMassT}
    \mth^2 \approx m^2 - \frac{1}{6}  \kappa^2 m^2 T^2.
\end{equation}
In the high three-momentum limit the term which modifies the
dispersion relation becomes a constant and can be reabsorbed in
the thermal mass. The group velocity is modified according to
\Eqref{GroupVelocity}:
\begin{equation}
    \vect v \approx \frac{\vect p}{p^0} \left( 1 - \frac{  {\kappa }^2 m^{9/2} {T  }^{3/2} \expp{-m/T} }
    {{\sqrt{2 \pi^3}}\,{\left( 3m^2 + 4|\vect p|^2  \right)}^2}
       \right).
\end{equation}
Notice that at low temperatures the Lorentz breaking term carries
a Boltzmann factor $\expp{-m/T}$. This is due to the mentioned
fact that the non-trivial momentum dependence comes from the
thermal scalar particles whose abundance is exponentially
suppressed at low temperatures. Analogously to what happens in
electrodynamics \cite{DonoghueEtAl85}, the effect of the graviton
bath only shows up in the thermal mass.

At high temperature, $T\gg m$, the modified dispersion relation is
found to be
\begin{equation} \label{RelDispHighT}
\begin{split}
     E^2 &\approx \mth^2 + |\vect p|^2  +  \frac{\kappa^2 m^2 T^2}{48}
         \left[ \frac{ E_\vect p}{|\vect p|} \ln \left(\frac{ 2E_\vect p+ |\vect p|}{ 2E_\vect p- |\vect p|}\right) - 1 \right],
\end{split}
\end{equation}
\index{Thermal mass}
and the thermal mass is
\begin{equation}
    \mth^2 \approx m^2 - \frac{11}{48} \kappa^2 m^2 T^2.
\end{equation}
At high three-momentum the modification of the dispersion relation
can be also reabsorbed in the thermal mass. The group velocity is
given by
\begin{equation} \label{SpeedHighT}
\begin{split}
    \vect v \approx \frac{\vect p}{p^0} \bigg[ 1 &+ \frac{1}{96} \frac{
    \kappa^2 T^2 m^4}{(4m^2+3|\vect p|^2)|\vect p|^2} -
    \frac{1}{96} \frac{\kappa^2 T^2  m^4}{|\vect p|^3 E_\vect p} \ln \left( \frac{ 2
    E_\vect p + |\vect p|}{2 E_\vect p - |\vect p|} \right)
    \bigg].
\end{split}
\end{equation}
At high temperature the terms which break the Lorentz symmetry are
no longer exponentially suppressed. Notice also that eq.~(\ref{RelDispHighT}) shows that there is no modification to the
dispersion relation for massless scalars.

The Lorentz breaking term is negative in the low-temperature case,
which implies that the speed of propagation is lowered with
respect to the standard relativistic case. In contrast, the
Lorentz breaking corrections have a positive sign at high
temperatures, so that the speed of propagation is increased.
However, it is always lower than the speed of light, as can be
seen by expanding \Eqref{SpeedHighT} in the ultrarelativistic
limit:
\begin{equation}
    \vect v \approx \frac{\vect p}{|\vect p|} \left(1 - \frac{m^2\left[ 48 + {\kappa }^2 T^2 \left( \ln 3 -1 \right)
       \right] }{96 {|\vect p|}^2} + \cdots \right)  .
\end{equation}

We should emphasize that the above violation of the Lorentz
symmetry is only an effective violation, in the following sense: if one applies a Lorentz
transformation simultaneously to the particle and the bath, all
results are Lorentz invariant. This can be seen can by introducing
the unit vector $l^\mu$ which gives the four-velocity of the
thermal bath. Then the energy and the three-momentum of the
particle with respect to the bath are
\begin{equation}
    p^0 = -l_\mu p^\mu, \qquad |\vect p| = \sqrt{(l_\mu
    p^\mu)^2+p_\mu p^\mu}\, .
\end{equation}
In this article for obvious reasons of simplicity we worked in the
bath rest frame where $l^\mu=(1,0,0,0)$.

For low values of the momentum the effective dispersion relation
can be expanded in powers of $\vect p$. For instance in the
low-temperature case, for momenta satisfying $|\vect p|<m$, the
effective dispersion relation \eqref{RelDispT} can be expanded as
\begin{equation}
\begin{split} \label{RelDispTExp}
    E^2 \approx \mth^2 +|\vect p|^2
        &-   \dfrac{\kappa^2 m^{5/2} T^{3/2}\expp{-m/T}
        }{\sqrt{2\pi^3}}  \frac{|{\vect p}|^2}{3m^2} \\ &+ \dfrac{\kappa^2 m^{5/2} T^{3/2}\expp{-m/T}
        }{\sqrt{2\pi^3}}
        \frac{2|{\vect p}|^4}{27m^4} + \cdots .
\end{split}
\end{equation}
It is worth mentioning the difference between this expansion and
that of eq.~\eqref{Eq1}. In a fundamental approach to Lorentz
symmetry breaking one expects each additional power of momentum to
be suppressed by increasing powers of the Planck mass, whereas in
the effective breaking approach we have pursued there are several
energy scales. Hence there is more freedom in the possible values
of the suppression factor. For instance, in the low-temperature
regime ---\ie, in \Eqref{RelDispTExp}--- the $|\vect p|^4$ term is
suppressed by $(T/m)^{3/2}e^{-m/T}$. The
fact that in the latter approach there are more energy scales also
explains why modifications to the $|\vect p|^2$ term are present
in eq.~\eqref{RelDispTExp} while they are not included in
eq.~\eqref{Eq1}. Notice also that the corrections that we found
only contain even powers of the momentum, and thus they are in
agreement with the results of ref.~\cite{Lehnert03}.


In the present universe the effects we discussed are completely
negligible when applied to electrons or protons, both because they
are proportional to the Planck length square and because they are
exponentially suppressed. To obtain relevant effects, one should
consider Planckian temperatures. However, in this case
perturbation theory around a flat-spacetime approximation probably
fails. Therefore the results obtained here should be considered an
indication that effective violations of Lorentz invariance indeed
occur, and that gravitational interactions cannot be neglected
when exploring highly energetic regions of dispersion relations.

Let us end with a short summary of the main points of this chapter.
We have illustrated with a particular example how local Lorentz
symmetry is effectively violated because of the interactions with
a non-trivial ensemble of metric fluctuations, even if Lorentz
symmetry holds at a fundamental level. We have also shown how this
effective violation can be addressed at low energies within
standard physics. The main quantitative result of the chapter are
eqs.~\eqref{RelDispT} and \eqref{RelDispHighT}, which explicitly
show the modifications of the dispersion relation. As in the
electromagnetic case, this effect is exponentially suppressed at
low temperatures. Moreover, the modifications of the dispersion
relation are suppressed when the three-momentum of the massive
particle (defined in the heat bath rest frame) is much larger than
the temperature and the mass. This last result is somewhat
unexpected since the gravitational coupling grows with the energy,
unlike the electromagnetic case. Therefore, in spite of the
derivative coupling, no violation of Lorentz invariance is found in
the high-momentum limit, at least at one loop. Finally we have
also shown that gravitational interactions do not generate a
four-derivative term in the scalar field action.

	\chapter{Dissipative effects in the propagation: a three-field model}
\shortpage

\index{3-field model}
In the previous chapter we studied the propagation of a scalar particle in a thermal bath of gravitons. We computed the one-loop correction to the position of the pole of the retarded propagator. Due to kinematic constraints in the relevant one-loop diagrams this correction was purely real. This prevented us to  properly consider the dissipative effects, associated with the imaginary part. However, there is no reason to expect that the imaginary part will vanish when studying two-loop diagrams or using resummed perturbation theory.

In this chapter we analyze a simpler physical system 
which is free of the kinematic constraints that made the imaginary part of the pole vanish, 
but nevertheless shows the same kind of dissipative behavior we could expect with the 
gravitational interaction at higher loops. The system contains two massive fields 
which interact with a massless radiation field. The same system will be analyzed in a cosmological background in chapter 7.

We first analyze an equilibrium situation: the propagation in a thermal background.
We then analyze the same system in a different background, 
consisting of a thermal bath of gravitons plus a one-particle excitation. 
It is an interesting example of a background which is non-thermal and non-isotropic. In this way we can check the usefulness of the methods developed in chapters 2 and 3 for backgrounds other than a thermal bath. Most results of section 3 were derived for isotropic backgrounds; we will check wether they also hold for non-isotropic systems. In particular, we will verify whether the 
interpretation of the imaginary part of the poles of the retarded propagator as rates 
of approach to thermal equilibrium still holds.

In this chapter, besides applying the results of chapter 3 to a particular field theory model, we also consider further theoretical developments. First, using heuristic methods, we comment on the time-evolution of an  ensemble of quasiparticles when it is slightly taken out of equilibrium. Second, we generalize the results to general non-equilibrium distribution functions. In this way we obtain a relativistic Boltzmann equation that describes the evolution of the quasiparticles. Finally, we extend these  results to the study of the dynamics of a single excitation, as opposed to the dynamics of the ensemble of particles.

\subsubsection*{A remark on the notation}\label{NotationRemark}

For greater clarity of the expressions we will somewhat modify the notation in this chapter. Mode labels, instead of appearing as a subindex, will frequently appear as an argument of the function, to avoid cluttering the subindices. For instance, the decay rates are written in other chapters as $\Gamma_\vect p$, but here they will be written as $\Gamma(\vect p)$. 

In order to differentiate the occupation number of the radiation field from the occupation number of the scalar fields, the latter will be denoted with a $f$ instead of a $n$. Therefore, what in other chapters is $n_\vect p$ here will be written as $f(\vect p)$.

\section{A three-field model}

\index{3-field model}

We consider a quantum field model consisting of two massive scalar fields $\phiM$ and $\phim$, of masses $M$ and $m$, respectively, and a massless scalar field $\chi$, which interact via a trilinear interaction, with  a coupling constant $\tilde g$. The action for the model is:
\begin{equation}
\begin{split}
    S &= -\fud \int \ud[4]x \big( \partial_\mu \phiM \partial^\mu \phiM + M^2 \phiM^2 + \partial_\mu \phim \partial^\mu \phim + m^2 \phim^2 \\ &\qquad + \partial_\mu \chi \partial^\mu \chi - 2 \tilde g \phiM \phim \chi
        \big),
\end{split}
\end{equation}
where we consider that $M>m$. Notice that in this model 
the coupling constant $\tilde g$ has dimensions of energy. 
We prefer to work with the dimensionless coupling constant $g=\tilde g/m$.

The model allows for two equivalent alternative interpretations: the excitations of the two 
fields $\phiM$ and $\phim$ may correspond to two different particles, or either they may 
correspond to two different internal degrees of freedom of the same particle. In other words, 
the doublet $(\phim,\phiM)$ may represent a particle with two internal states, one of them 
being more energetic than the other. 

We will be interested in considering the limit in which the masses of the two fields are 
very large, while their mass difference  $\dm = M-m$ is of the order of magnitude of
the energies $E$ involved in the problem: $M,m \gg \dm$ and $\dm \sim E$. In this limit the 
system can be considered a relativistic two-level atom of mass $m$, 
represented by $(\phiM,\phim)$, interacting with a radiation field.
This model was already used in ref.~\cite{Parentani95} to study the consequences of the 
recoils of a uniformly accelerated two-level atom subject to the Unruh effect \cite{Unruh76}.

  From now on we will alternate the field-theoretic and atomic-like description and notations, often naming the particle represented by the doublet $(\phim,\phiM)$ as the ``atom'', and indicating quantities referring to the excited state $\phiM$ with a star and quantities related to the fundamental state $\phim$ without the star.

\section{Self-energy and decay rates in the vacuum}

\begin{figure}
	\centering
	\includegraphics[width=0.40\textwidth]{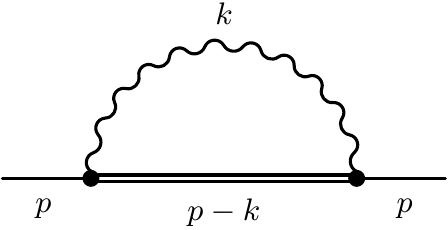} \hspace{3ex} \includegraphics[width=0.40\textwidth]{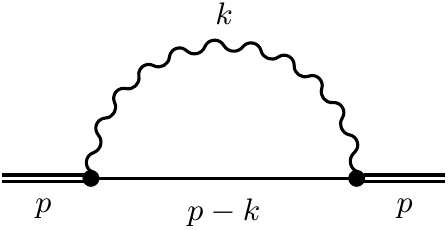}
	\caption{Feynman diagrams leading to the one-loop self-energy of the fields $\phim$ and $\phiM$ (respectively the fundamental and excited atomic states). Single straight lines represent the fundamental state, double lines the excited state, and curly lines the massless particle.}\label{fig:SigmaVac}
\end{figure}

Recall that at zero temperature the self-energy $\Sigma\TO(p^2)$ of the field $\phim$ is defined
through
\begin{equation}\label{FeynmanProp}
    G_\mathrm{F}\TO(p) = \frac{-i}{p^2+m^2+\Sigma\TO(p^2)},
\end{equation}
where $G_\mathrm{F}\TO(p)$ is the zero temperature Feynman propagator of $\phim$. Recall also that the self-energy can be computed as the sum of all one-particle irreducible diagrams with amputated external legs. At one loop, the self energy corresponds to the diagrams in  fig.~\ref{fig:SigmaVac}, and for the case of the fundamental atomic state it is found to be:
\begin{equation}
\begin{split}
   \Sigma\TO(p^2) &=  \frac{g^2 m^2}{(4\pi)^2} \bigg[
     \frac{M^2}{p^2} \ln \left(1+\frac{p^2}{M^2} -i\epsilon \right) \\
	 &\qquad -  \frac{M^2}{m^2} \ln \left(1-\frac{m^2}{M^2} \right) + \ln
     \left(\frac{p^2+M^2}{M^2-m^2}-i\epsilon\right)\bigg].
\end{split}
\end{equation}
The calculation is standard, and the details closely follow those of the previous chapter.  For the excited state $\phiM$ the self-energy is found just exchanging $M$ by $m$: ${\Sigma\TO}^*(p^2)  = \Sigma\TO(p^2)[M \leftrightarrow m]$.

\index{On-shell renormalization condition}
As in the previous chapter, we have used an on-shell renormalization scheme in order to absorb the divergences, so in this case the real part of the self-energy is zero at the mass-shell: $\Re \Sigma\TO(-m^2)=0$. This ensures, according to the K\"allen-Lehman representation of the propagator, that the masses $m$ and $M$ correspond to the true masses of the two states of the atom.\footnote{Strictly speaking, this is only true for the stable states such as $\phim$, which are the only ones that have well-defined asymptotic modes. }

The decay rate, defined as the probability of decay per unit time, is proportional to the imaginary part of the self-energy at the mass-shell, according to the optical theorem \citep{Peskin,WeinbergQFT}. For the excited stated the decay rate $\gamma^*_-$ is proportional to the mass gap:
\begin{subequations}
\begin{equation}
     \gamma_-^* = -\frac{1}{M} \Im {\Sigma\TO}^*(-M^2) \approx  \frac{g^2}{8\pi} \dm,
\end{equation}
while the low-mass state is of course stable:
\begin{equation}\label{OffShell-Decay}
    \gamma_- = -\frac{1}{m} \Im \Sigma\TO(-m^2) = 0.
\end{equation}
\end{subequations}
Notice that the decay rate of the heavy state, as expected, is governed by the mass gap, and does not depend on the mass of the atom.

A similar analysis can be carried out for the self-energy $\Pi$ of
radiation field $\chi$. Since there is no gauge invariance
protecting the mass of the ``scalar photon'', the renormalization
process will generate a mass term for it. We will 
put to zero the renormalized mass, so that the real part of the self-energy vanishes. 
The imaginary part of the self-energy automatically vanishes on-shell,
since a real massless particle cannot decay. However, the imaginary part of the self-energy is non-zero off-shell, because the virtual excitations of the radiation field decay with a rate given by
\begin{equation} \label{PiVacOff}
    \Gamma^\text{(rad)}(q^0,\vect q) = -\frac{1}{q^0} \Im \Pi\TO(q^2)
\end{equation}
in the laboratory rest frame.

\index{Decay rate}

\section{Self-energy and decay rates in a thermal background}


In this section we consider the situation in which the radiation field is not in its vacuum state, but in a thermally excited state corresponding to a temperature of the order of the mass gap $\dm$. This means that the massive scalar fields remain unexcited in the vacuum.

As explained in chapters 3 and 4, at finite temperature the real and imaginary parts of the poles of the retarded propagator, given by the on-shell values of the retarded self-energy, admit interpretations as the dispersion relations and decay rates, respectively. Although we will also compute the real part of the self-energy, in this chapter we will mainly concentrate on the analysis of the imaginary part.

As an aside, let us mention that since the scalar photons involved have momenta of the order of the mass gap, $p \sim \dm \sim T \gg gT$, no Braaten and Pisarski resummation is needed (see subsect.~\ref{sect:DivRen}).

\subsection{Thermal decay rates and distribution functions}

\index{Decay rate}
\longpage

\index{Detailed balance condition}
Recall that the imaginary part of the retarded self-energy can be expanded in general as 
\begin{equation}\label{ImSigmaRGamma}
    \Im\SigmaR(E_\vect p, \vect p) = - R_\vect p[\Gamma_-(\vect p) - \Gamma_+(\vect p)],
\end{equation}
where $\Gamma_-(\vect p)$ and $\Gamma_+(\vect p)$ are respectively the  decay and creation rates, which depend on the temperature $T$, and where $R_\vect p$ is the energy of the excitation. The decay rate $\Gamma_-(\vect p)$ is the probability per unit time for an incoming particle with momentum $\vect p$ to decay to any other state into the thermal bath (this includes simply going to a different momentum state). Similarly, the creation rate $\Gamma_+(\vect p)$ is the probability per unit time that the state of momentum $\vect p$ is spontaneously created from the thermal bath. The decay and creation rates verify the detailed balance condition $\Gamma_+(\vect p) = \expp{- E_\vect p\vect/T} \Gamma_-(\vect p)$. Unlike in the vacuum case, they are usually computed in the laboratory reference frame.

\index{Distribution function}

\longpage
For an ensemble of particles obeying the Bose-Einstein statistics 
 and described by the distribution function $f(\vect p,t)$,
the total amount of decay is proportional
 to the actual value of the distribution. 
 Similarly, the rate of increase will be proportional to the creation rate, and, 
 because of the Bose-Einstein statistics, will be favored for those states already populated. 
These heuristic arguments suggest that the time evolution of 
its distribution function $f(\vect p,t)$, slightly departed 
from thermal equilibrium, is given by \citep{Weldon83}
\begin{equation}\label{DecayThermalEq}
    \derp{f(\vect p,t)}{t} = - \Gamma_- (\vect p) f(\vect p,t) + \Gamma_+ (\vect p) [1+f(\vect p,t)].
\end{equation}
Notice that this equation is only valid for distribution functions $f(\vect p,t)$  not far from thermal equilibrium; if the deviation from thermal equilibrium is larger, the decay and creation rates have to be computed with the out of equilibrium distributions, as we will see in the next section. Solving the above differential equation we get
\begin{equation}
	f(\vect p, t) = \frac{1}{\expp{E_\vect p/T} -1} - \Delta f_0 (\vect p) \expp{-\Gamma(\vect p)t},
\end{equation}
where $\Delta f_0(\vect p)$ is the initial departure from equilibrium, and 
\begin{equation} \label{GammaExp}
	\Gamma(\vect p)=\Gamma_-(\vect p) - \Gamma_+(\vect p)
\end{equation}
is the rate of approach to the equilibrium. Comparing eq.~\eqref{ImSigmaRGamma} with eq.~\eqref{GammaExp} we recover the result that the imaginary part of the self-energy is proportional to the rate of approach to the equilibrium.

\index{Decay rate}
In the case considered here the temperature $T$ is much lower than the atom mass $m$, so that there are no thermal atoms in the bath. Since an atom cannot be spontaneously created from a thermal radiation field, $\Gamma_+(\vect p)=\Gamma_+^*(\vect p)=0$, and the thermalization rate for the atomic states, $\Gamma(\vect p)$ and $\Gamma^*(\vect p)$, coincides with the net decay rate $\Gamma_-(\vect p)$ and $\Gamma_-^*(\vect p)$.  In terms of the decay amplitude $\mathcal M$, to leading order in the coupling constant the decay rates can be computed as (see   \eg~\cite{Peskin,LeBellac})
\begin{subequations} \label{DirectGamma}
\begin{align}
\begin{split}
	\Gamma^*_-(\vect p) &= \frac{1}{2 E^*_\vect p} \int \udpi[3]{{\vect q}} \frac{1+n(|\vect q|)}{2 |\vect q|} \int \udpi[3]{\vect p'}	
     \frac{1}{2 E_{\vect p'}}
    |\mathcal M|^2 \\
	&\qquad \times (2\pi)^4 \delta^{(3)}(\vect p + \vect q -\vect p') \delta(E^*_\vect p - |\vect q| - E_{\vect p'}) ,
\end{split}\\
\begin{split}
    \Gamma_-(\vect p) &= \frac{1}{2 E_\vect p} \int \udpi[3]{{\vect q}} \frac{n(|\vect q|)}{2 |\vect q|} \int \udpi[3]{\vect p'}	
     \frac{1}{2 E^*_{\vect p'}}
    |\mathcal M|^2 \\
	&\qquad \times (2\pi)^4 \delta^{(3)}(\vect p + \vect q -\vect p') \delta(E_\vect p + |\vect q| - E^*_{\vect p'}) ,
\end{split}
\end{align}
\end{subequations}
where $E_\vect p = \sqrt{m^2 + \vect p^2}$ and $E_\vect p^{*} = \sqrt{M^2 + \vect p^2}$, and where $n(E)$ is the Bose-Einstein distribution function, 
\begin{equation}
    n(E)=\frac{1}{\expp{E/T} -1}.
\end{equation}
These two processes correspond to the Feynman diagrams shown in fig.~\ref{fig:decayT}. The statistical factors $n(|\vect q|)$ and $1+n(|\vect q|)$ take into account the presence of radiation in the bath. The squared decay amplitudes simply correspond to $|\mathcal M|^2 = 
\tilde g^2 =m^2 g^2$.

\index{Decay rate}
\subsection{Explicit values of the self-energy and decay rates}

The real and imaginary parts of the retarded self-energy can be
obtained from the various self-energy components. 
The one-loop $\phim$ self-energy 
is (see fig.~\ref{fig:SigmaVac}), 
\begin{equation}
    -i\Sigma^{ab}(p)= - {g^2 m^2} \int \udpi[4]{k} G^{(0)*}_{ab} (p-k) \Delta_{ab}^{(0)}(k), 
\end{equation}
where $G^{(0)*}_{ab} (p)$ is the free vacuum propagator for the massive field $\phiM$,  $\Delta_{ab}^{(0)}(p)$ is the free thermal propagator for the radiation field, and no implicit summation is assumed in this equation. Since the details of the calculation are very similar to those in the previous chapter, here we will just present the results.

Concerning the real part, the one-loop contribution is the following:
\begin{equation}\label{RealThreeFields}
    \Re\SigmaR(E_\vect p, \vect p) \approx -\frac{g^2 }{6} \frac{T^2m}{ \dm}, \qquad
    \Re\SigmaR^*(E^*_\vect p, \vect p) \approx \frac{g^2 }{6} \frac{T^2m}{\dm},
\end{equation}
The fact that both contribution have opposite sign is simply due to the fact that both contributions are related by the exchange $m \leftrightarrow M$. Notice that the result is momentum-independent, similarly to what happens in electrodynamics \citep{DonoghueEtAl85} and gravitation (see the previous chapter). Hence the real part of the self-energy can be absorbed in the thermal mass
\begin{equation}
    \mth^2 \approx m^2-\frac{g^2}{6} \frac{T^2m}{\dm}, \qquad
    \Mth^2 \approx M^2+ \frac{g^2}{6} \frac{T^2m}{\dm}.
\end{equation}
The real dispersion relations $E^2 = \mth^2 + \vect p^2$ and $E^{*2} = \Mth^2 + \vect p^2$ are thus Lorentz-invariant. 

As in the case of gravity and electrodynamics, the finite temperature contribution to the dispersion relation of the atomic states can be absorbed in a thermal contribution to the mass, although this is not what one would expect \emph{a priori}, since the thermal bath introduces a preferred reference frame and there is no immediate reason why the effective dispersion relation should preserve the frame independence. Indeed we will see that in non-thermal situations there are effectively contributions to the dispersion relation that cannot be absorbed in the thermal mass. This points towards the direction that the absorption of the modified dispersion relation in the mass is a property linked to the particular features of thermal field theory.

\begin{figure}
	\centering
	\mbox{} \hfill \includegraphics[width=0.3\textwidth]{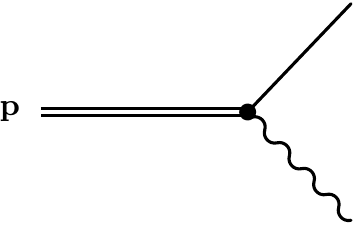} \hfill \includegraphics[width=0.3\textwidth]{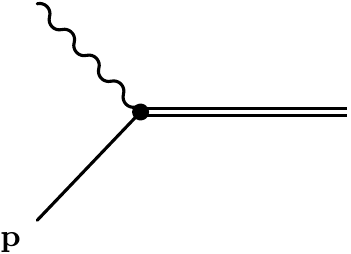} \hfill \mbox{}
	\caption{Feynman diagrams giving the decay rate of the excited and fundamental atom states with momentum $\vect p$ respectively. On the left, the excited state decays to the fundamental state by emitting a massless particle into the thermal bath. On the right, the fundamental state absorbs a massless particle from the bath and gets excited.} \label{fig:decayT}
\end{figure}

Concerning the imaginary part, and for the sake of simplicity, we quote the results in the limit $\dm \ll m$. For the excited state we have
\begin{equation}\label{3fieldsDecayGeneral2}
	\Im\SigmaR^*(E^*_\vect p,\vect p) = -\frac{g^2 m^2 T}{16 \pi |\vect p|  } \ln \left(\frac{\cosh \Big[\frac{(-2 \dm+m) E^*_\vect p}{2 m T}\Big]-\cosh
   \Big[\frac{mE^*_\vect p+{2 \dm |\vect p|}}{2m T}\Big]}{\cosh \Big[\frac{(-2
   \dm+m) E^*_\vect p}{2 m T}\Big]-\cosh \Big[\frac{mE^*_\vect p -2 \dm |\vect p|}{2 m
   T}\Big]}\right)
\end{equation}
and for the low mass state
\begin{equation}\label{3fieldsDecayGeneral}
	\Im\SigmaR(E_\vect,\vect p) =- \frac{g^2 m^2 T}{16 \pi |\vect p|  } \ln \left(\frac{\cosh \Big[\frac{(2 \dm+m) E_\vect p}{2 m T}\Big]-\cosh
   \Big[\frac{mE_\vect p-{2 \dm |\vect p|}}{2m T}\Big]}{\cosh \Big[\frac{(2
   \dm+m) E_\vect p}{2 m T}\Big]-\cosh \Big[\frac{mE_\vect p +2 \dm |\vect p|}{2 m
   T}\Big]}\right)
\end{equation}
The corresponding decay rates, which correspond to the Feynman diagrams in fig.~\ref{fig:decayT}, go as follows:
\begin{subequations}
\begin{align}
\begin{split}
    \Gamma_-^*(\vect p) &= -  \frac{1}{E^*_\vect p} \Im \SigmaR^*(E^*_\vect p,\vect p) \\ &=  \frac{g^2 m^2 T}{16 \pi |\vect p| E^*_\vect p } \ln \left(\frac{\cosh \Big[\frac{(-2 \dm+m) E^*_\vect p}{2 m T}\Big]-\cosh
   \Big[\frac{mE^*_\vect p+{2 \dm |\vect p|}}{2m T}\Big]}{\cosh \Big[\frac{(-2
   \dm+m) E^*_\vect p}{2 m T}\Big]-\cosh \Big[\frac{mE_\vect p -2 \dm |\vect p|}{2 m
   T}\Big]}\right)
\end{split}	 \\
\begin{split} \label{DecayRateT}
    \Gamma_-(\vect p) &= -  \frac{1}{E_\vect p} \Im \SigmaR(E_\vect p,\vect p)\\ &=  \frac{m^2 T}{16 \pi |\vect p| E_\vect p } \ln \left(\frac{\cosh \Big[\frac{(2 \dm+m) E_\vect p}{2 m T}\Big]-\cosh
   \Big[\frac{mE_\vect p-{2 \dm |\vect p|}}{2m T}\Big]}{\cosh \Big[\frac{(2
   \dm+m) E_\vect p}{2 m T}\Big]-\cosh \Big[\frac{mE_\vect p +2 \dm |\vect p|}{2 m
   T}\Big]}\right)
,
\end{split}
\end{align}
\end{subequations}
When the atom is at rest, or the kinetic energy can be neglected in front of the rest energy, the above equations reduce to:
\begin{subequations} \label{RestGamma}
\begin{align}
    \Gamma_-^*=\Gamma_-^*(\vect 0)  &=  \frac{g^2}{8\pi} \dm \left[ 1 + n(\Delta m) \right]\\
    \Gamma_-=\Gamma_-(\vect 0) &=  \frac{g^2}{8\pi} \dm \, n(\Delta m) 
\end{align}
\end{subequations}
The result obtained from the self energy coincides with the one computed directly 
from the transition rates of eqs.~\ref{DirectGamma}. 
One can also check that we recover the vacuum results in the zero temperature limit. 

\begin{figure}
	\centering
	\includegraphics[width=.7\textwidth]{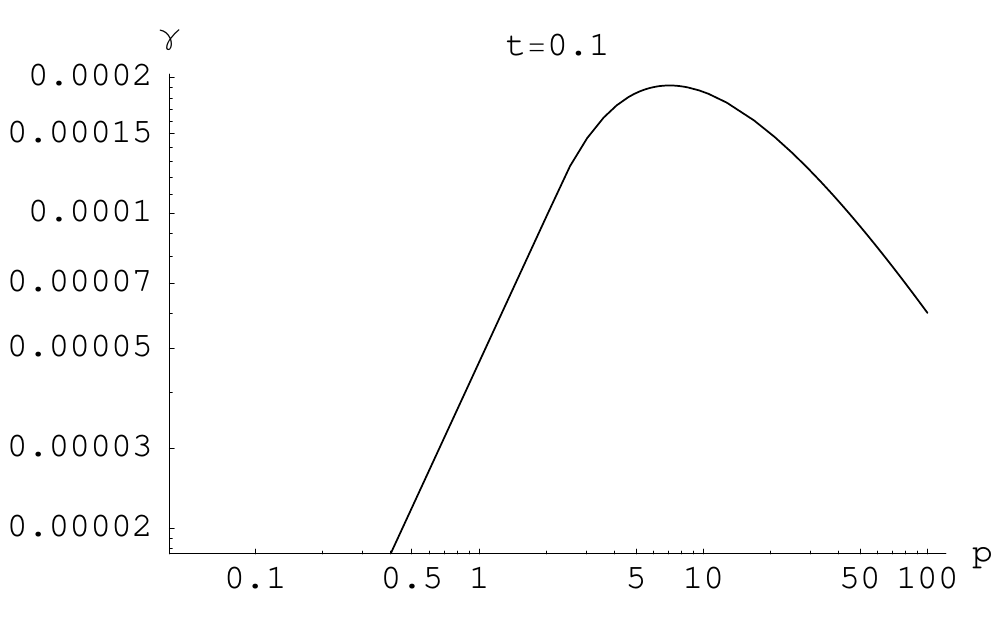}
	\caption{Decay rate as a function of the momentum in the particle, for a temperature $T=\dm/10$. We have chosen  to quote the decay rate in the particle rest frame, $\gamma_-(\vect p) = (E_\vect p/m) \Gamma_-(\vect p)$ to discount for the time dilation effects. Momentum is in units of $m$, temperature in units of $\dm$ and the decay rate in units of $g^2 \dm$. Notice that for low momenta the decay rate is negligible, but when the boosted temperature is of the order of $\dm$ we recover the GZK limit.}\label{fig:GZK}
\end{figure}

\index{Detailed balance condition}

It is important to point out that eqs.~\eqref{RestGamma} admit a double interpretation. On the one hand, from the point of view of field theory, they represent the decay rates of the particle excitations in a thermal bath. We would like to emphasize that in this context ``decay'' simply means going to a different state, and not necessarily to a lower mass state. Notice also that the interaction preserves the total number of particle excitations, so that $\phiM$ decays into $\phim$ and viceversa. On the other hand, from the point of view of the internal atomic states, $\Gamma_-$ describes the probability of excitation of the fundamental state in a thermal bath, and $\Gamma_-^*$ describes the probability of (stimulated) decay of the excited state into the fundamental. For this reason, $\Gamma_-$ and $\Gamma_-^*$ also 
verify the ``atomic'' detailed balance condition $\Gamma^*_- = \expp{-\dm/T} \Gamma_-$,
wherein only the mass gap enters.

\index{GZK limit}
\label{page:GZK2}

The decay rate of the fundamental state in the ultrarelativistic limit should be studied from the general expression \eqref{DecayRateT}.
 In figure \ref{fig:GZK} we plot the decay rate as a function of the momentum in the case of  relatively low temperatures, recovering the GZK limit (see page \pageref{page:GZK}): decay rates increase sharply when the boosted temperature is of the order of the mass gap.

Finally, let us simply mention that at one loop there is no thermal contribution to the self-energy of the massless field $\Pi$ because of the hypothesis that the massive fields $\phim$ and $\phiM$, which are interacting with $\chi$, remain thermally unexcited.

We end this section by summarizing the results obtained so far. By means of the analysis of the poles of the retarded propagators of the two fields $\phim$ and $\phiM$ we have recovered the basic properties of the atomic levels. At zero temperature, the fundamental state of the atom is stable, while the excited state has a finite lifetime, inversely proportional to the mass gap  between the two states $\dm$. At finite temperature the fundamental state is no longer stable, but has a characteristic excitation time $\Gamma_-^{-1}$, proportional to the density of radiation. 
The decay rate of the excited state is enhanced because of the Bose-Einstein statistics.
These results are of course perfectly known, 
but it is interesting
to see how the atomic description emerges from the field-theoretic language.
For instance,
from the point of view of field theory, $\Gamma_-$ and $\Gamma_-^*$ are interpreted as the decay rates of the particle excitations of the fields $\phim$ and $\phiM$, respectively, whereas from the point of view of the internal atomic states, $\Gamma_-$ and $\Gamma_-^*$ correspond to the internal transition rates as explained above. Since the atom only has two levels  $\Gamma_-$ and $\Gamma_-^*$ obey a detailed balance condition. Notice that the field creation rates $\Gamma_+$ and $\Gamma_+^*$ vanish since no atoms can be spontaneously created from a thermal bath.

\section{Decay rates and distribution functions in general backgrounds}

\index{Decay rate}
\index{Distribution function}

Following the same kind of heuristic arguments that lead 
to \Eqref{DecayThermalEq}, one can see that the rate of approach to equilibrium of a large ensemble 
of atoms is described by
\begin{subequations} \label{DecayOutT}
\begin{align}
    \derp{f(\vect p,t)}{t} &=     - \Gamma_- (\vect p,[f^*],t) f(\vect p,t) + \Gamma_+ (\vect p,[f^*],t) [1+f(\vect p,t)], \\
	\derp{f^*(\vect p,t)}{t} &= - \Gamma^*_- (\vect p,[f],t) f^*(\vect p,t) + \Gamma^*_+ (\vect p,[f],t) [1+f^*(\vect p,t)]
\end{align}
\end{subequations}
The functions $f(\vect p,t)$ and $f^*(\vect p,t)$ are the distribution functions of the atoms, in its fundamental and excited atomic states, respectively. The decay rate $ \Gamma_- (\vect p,[f^*],t)$ is the probability per unit time for an incoming light state of momentum $\vect p$ to decay into any other state on top of a background characterized by a distribution function of heavy states $f^*(\vect p,t)$. Similarly, the creation rate $\Gamma_+ (\vect p,[f^*],t)$ is the probability per unit time for a particle with momentum $\vect p$ to be spontaneously created from the distribution of heavy particles characterized by $f^*(\vect p,t)$. Analogous definitions apply for $\Gamma^*_- (\vect p,[f],t)$ and $\Gamma^*_+ (\vect p,[f],t)$. In terms of decay amplitudes, these are expressed as
\begin{subequations} \label{DirectGammaOutEq}
\begin{align}
\begin{split}
	\Gamma^*_-(\vect p,[f]) &= \frac{1}{2 E^*_\vect p} \int \udpi[3]{{\vect q}} \frac{1+n(|\vect q|)}{2 |\vect q|} \udpi[3]{\vect k}	
     \frac{[1+f(\vect k)]}{2 E_{\vect k}}
    |\mathcal M|^2 \\
	&\qquad \times (2\pi)^4 \delta^{(3)}(\vect p - \vect q -\vect k) \delta(E^*_\vect p - |\vect q| - E_{\vect k}),
\end{split} \\
\begin{split}
    \Gamma_-(\vect p,[f^*]) &= \frac{1}{2 E_\vect p} \int \udpi[3]{{\vect q}} \frac{n(|\vect q|)}{2 |\vect q|} \udpi[3]{\vect k}	
     \frac{[1+f^*(\vect k)]}{2 E^*_{\vect k}}|\mathcal M|^2  \\
	 &\qquad \times
    (2\pi)^4 \delta^{(3)}(\vect p + \vect q -\vect k) \delta(E_\vect p + |\vect q| - E^*_{\vect k}) ,
\end{split} \\
\begin{split}
	\Gamma^*_+(\vect p,[f]) &= \frac{1}{2 E^*_\vect p} \int \udpi[3]{{\vect q}} \frac{1+n(|\vect q|)}{2 |\vect q|} \udpi[3]{\vect k}	
     \frac{f(\vect k)}{2 E_{\vect k}}
    |\mathcal M|^2 \\
	 &\qquad \times (2\pi)^4 \delta^{(3)}(\vect p - \vect q -\vect k) \delta(E^*_\vect p - |\vect q| - E_{\vect k}),
\end{split} \\
\begin{split}
    \Gamma_+(\vect p,[f^*]) &= \frac{1}{2 E_\vect p} \int \udpi[3]{{\vect q}} \frac{n(|\vect q|)}{2 |\vect q|} \udpi[3]{\vect k}	
     \frac{f^*(\vect k)}{2 E^*_{\vect k}}
    |\mathcal M|^2 \\
	 &\qquad \times (2\pi)^4 \delta^{(3)}(\vect p + \vect q -\vect k) \delta(E_\vect p + |\vect q| - E^*_{\vect k}).
\end{split}
\end{align}
\end{subequations}
For simplicity we have considered the radiation field to remain in thermal equilibrium, but expressions could be easily generalized in order to allow for an arbitrary distribution function of the radiation, instead of the thermal Bose-Einstein distribution function $n(|\vect q|)$. 

The distribution functions are normalized to the total number of atoms $N$ multiplied by a factor which depends on the (formally infinite) volume:
\begin{equation}
	\int \udpi[3]{\vect p} [ f(\vect p,t) + f^*(\vect p,t) ] = N \frac{(2\pi)^3}{V}.
\end{equation}
By using eqs.~\eqref{DecayOutT} and \eqref{DirectGammaOutEq} we verify that the total number of particles is a conserved quantity:
\begin{equation}
	\dert{}{t} \int \udpi[3]{\vect p} [ f(\vect p,t) + f^*(\vect p,t) ] =  \frac{(2\pi)^3}{V} \dert{N}{t}=0.
\end{equation}

Several comments are in order. First, notice that these equations are not derived from first principles, but are based on a reasonable physical considerations. Second, they assume that scattering theory applies, so that approximately asymptotic regions are implicitly assumed. This means that two characteristic timescales are considered, the interaction timescale (which is of the order of the inverse mass gap) and the evolution timescale, and that it is assumed that the evolution timescale is much larger than the interaction timescale. In other words, we can assume that time can be divided into cells which are much larger than the interaction timescale but much smaller than the evolution timescale \cite{CalzettaHu88}; inside each one of these cells it is assumed that asymptotic regions can be defined and that standard scattering theory applies.    Third, notice  the presence of the statistical factor $ 1+f(\vect p,t)$ in eq.~\eqref{DecayOutT}, which implies that the we are considering a distribution function containing many particles, $N\gg1$.  Finally, notice also that no spatial dependence on the distribution function is considered. This means that we are considering distribution functions which are initially homogeneous and which remain so during the time evolution.

\index{Boltzmann equation}
Eqs.~\eqref{DecayOutT} can be generalized to allow for non-homogeneous distributions in the following way:
\begin{subequations} \label{Boltzmann}
\begin{align}
    \left[ \derp{}{t} + \vect v \cdot \derp{}{\vect x} \right] f(\vect p,x)&=     - \Gamma_- (\vect p,[f^*],x) f(\vect p,x) + \Gamma_+ (\vect p,[f^*]x) [1+f(\vect p,x)], \\
	\left[ \derp{}{t} + \vect v \cdot \derp{}{\vect x} \right] f^*(\vect p,x)&= - \Gamma^*_- (\vect p,[f],x) f^*(\vect p,x) + \Gamma^*_+ (\vect p,[f],x) [1+f^*(\vect p,x)],
\end{align}
\end{subequations}
where $\vect v$ is the group velocity of the particles, see \Eqref{GroupVelocity}. If interaction with the bath can be neglected for this term (for instance when performing a perturbative calculation) $\vect v= \vect p/E_\vect p$. This is an effective Boltzmann equation for the system of interest and coincides essentially with that derived by Calzetta and Hu \cite{CalzettaHu88}. For this equation to be valid the inhomogeneity length scale  must be much larger than the characteristic interaction scale. To put it differently, these equations require that the entire spacetime can be divided into cells which are sufficiently large so that asymptotic collision theory holds, but sufficiently small so that the distribution function is stationary and homogeneous inside therm.

\index{Wigner function}

If instead of a distribution function of a large number of particles we are dealing with a single particle, we may be tempted to rewrite the Boltzmann equation as
\begin{subequations} \label{BoltzmannOne}
\begin{align}
    \left[ \derp{}{t} + \vect v \cdot \derp{}{\vect x} \right] f(\vect p,x)&=     - \Gamma_- (\vect p,[f^*],t) f(\vect p,x) + \Gamma_+ (\vect p,[f^*],t) , \\
	\left[ \derp{}{t} + \vect v \cdot \derp{}{\vect x} \right] f^*(\vect p,x)&= - \Gamma^*_- (\vect p,[f],t) f^*(\vect p,x) + \Gamma^*_+ (\vect p,[f],t),
\end{align}
\end{subequations}
where now the function $f^*(\vect p,x)$ can be regarded as being proportional to the reduced Wigner function corresponding the state of the particle,
\begin{equation*}
	f(\vect p,x) = N  W_\text{s}(\vect x,\vect p;t)
\end{equation*}
(with $N$ being a normalization constant), and similarly for $f^*(\vect p,x)$.  These equations are appealing because they connect the second-quantized description, which is based on the field modes, with a first-quantized description, which is based on the particles themselves. Several remarks are to be done however. First, recall that the above equations are not derived from first principles. Second, notice that they only make sense for highly delocalized particle states. Third, they do not take into account the possible coherence of the state. For all these reasons we believe that the passage from the second to the first-quantized descriptions in the case of a single particle should be undertaken with greater care.

Anyway, in chapter 3 we saw that even for inhomogeneous and non-stationary states the decay and creation rates can be obtained from the self-energy, provided the scales of inhomogeneity and non-stationarity are much larger that the microscopic scales, as it is assumed in eqs.~\eqref{DecayOutT} and \eqref{Boltzmann}. For a distribution function of particles characterized by the function $f(\vect p,x)$ we have
\shortpage
\begin{subequations}
\begin{align}
    \Im\SigmaR(E_\vect p, \vect p;[f,f^*];x) &= - R_\vect p(x)  [\Gamma_-(\vect p,[f^*],x) - \Gamma_+(\vect p,[f^*],x)], \\
	\Im\SigmaR^*(E_\vect p, \vect p;[f,f^*];x) &= - R^*_\vect p \vect(x)  [\Gamma^*_-(\vect p,[f],x) - \Gamma^*_+(\vect p,[f],x)].
\end{align}
\end{subequations}
The notation is somewhat cumbersome because we have added the explicit dependence on the distribution function, but this is nothing more that the usual relation \eqref{Weldon} between the self-energy and the decay rates. As a side note, recall that in chapter 3 we saw that the occupation numbers $f(\vect p,x)$  fully determine the reduced state of the field modes to Gaussian order.

\index{Kinetic field theory}
\index{Effective action!2PI}
Summing up, we have studied the equations governing the time-evolution of the distribution function of the two atomic states, $f$ and $f^*$, as a function of the decay and creation rates, both in a standard near-equilibrium situation [eq.~\eqref{DecayThermalEq}] and in an out-of-equilibrium situation [eqs.~\eqref{DecayOutT}]. The extension to the inhomogeneous case has led us to the Boltzmann equation \eqref{Boltzmann}. The basic physical assumption is the two-scale separation: the time and length scales characteristic of the time evolution of the distribution function are assumed to be much larger than the microscopic interaction scales. The same scale separation was used in chapter 3 when studying the quasiparticle excitations in general backgrounds. Calzetta and Hu \cite{CalzettaHu87} derive an equivalent Boltzmann equation within kinetic field theory, using a two-particle irreducible approach to the effective action and an analogous scale separation ---recall that our approach is based on reasonable physical assumptions.

We have also tried to extend the Boltzmann equation to deal with one-particle situations. In this case the Boltzmann equation can be interpreted as the effective Focker-Planck equation for the evolution of the one-particle (first-quantized) Wigner function. In this context, the calculation of the self-energy in a one-particle background gives us the relevant decay rates which could be used to analyze the first stages of a particle evolution. Anyway we have argued that the one-particle Boltzmann equation does not preserve the information on the phases of the particles (does not preserve the particle coherence), and we believe that the approach presented in chapter 3 to analyze the quasiparticle dynamics (and which has been used in the rest of this chapter) is best suited in this context. In any case it would be worth having an approach which allows a systematic derivation of the one-particle Fokker-Planck equation from the field-theoretical second-quantized description. As an aside, notice that such first-quantized description would be only appropriate when considering situations in which the particle number is conserved, such in this case.

Let us now look for a particular application, analyzing the self-energy and decay rates in the case of having an initial particle in the background state.

\shortpage
\section{Self-energy and decay rates in a one-particle background}

\index{Decay rate}
In this section we consider departure from thermal equilibrium situation: we compute the 
Green functions in a background state characterized by the presence of an atom in 
its fundamental state, on top of a thermal radiation bath, \ie, we consider $f(\vect p)\propto \delta^{(3)}(\vect p- \vect p') $ and $f^*(\vect p)=0$.
Since studying the propagators in this background involves creating 
an additional test particle,
in total there will be two massive particles interacting with
the thermal radiation. 

In order to obtain the retarded propagator in this background, we will first compute  the retarded self-energy  $\Sigma_\mathrm R(x,x')=\Sigma^{11}(x,x')+\Sigma^{12}(x,x')$ with the usual CTP perturbation theory, and at the end we will use \Eqref{SelfEnergyGeneral} to connect the retarded self-energy to the retarded propagator.
To this end we also need the free propagators in the one-particle background state, see appendix \ref{app:FreeOne}. 
The additional particle give rises to an additional on-shell term in the expression of the free propagators.

Notice that the state is homogeneous and translation-invariant. Considering a non-dynamical background state is an approximation, since the background evolves. However, recall that we are considering physical systems in which the microscopic interaction timescale is much shorter than the evolution timescale. The calculation of the self-energy or the decay rates corresponds to the microscopic timescale, and therefore is a good approximation to assume the stationarity of the background.

Since the background is homogeneous and stationary for short scales,
the Fourier transform of the retarded propagator 
has the form given by eq.~\eqref{ApproximateDiagonal}:
\begin{subequations} 
\begin{equation}
    G_\mathrm{R}(p;{\vect p'}) = \frac{-i}{p^2+m^2+\SigmaR(p;{\vect p'})}.
\end{equation}
By $G_\mathrm{R}(p;{\vect p'})$ we denote the retarded propagator in the state characterized by 
the extra on shell particle with momentum ${\vect p'}$. Analogous relations 
apply for the propagator and self-energy of 
of the excited state, $G^*$ and $\Sigma^*$, and those of 
the radiation field, $\Delta$ and $\Pi$:
\begin{equation}
    G^*_\mathrm{R}(p;{\vect p'}) = \frac{-i}{p^2+M^2+\SigmaR^*(p;{\vect p'})},\qquad
	\Delta^*_\mathrm{R}(p;{\vect p'}) = \frac{-i}{p^2+\PiR^*(q;{\vect p'})}
\end{equation}
\end{subequations}
Notice that the background, despite being homogeneous and stationary, is non-isotropic. One question we want investigate is whether in this background the imaginary part of the 
self-energy still admits an interpretation as a decay rate or a rate of approach to the 
equilibrium, since the expressions in chapter 3 were restricted to the isotropic case.

In the sequel, 
we will compute the modification of the self-energies of the radiation field $\chi$ and the 
massive field $\phiM$. 
At one loop, the low-mass state $\phim$ will not receive modifications.

\subsection{Radiation field}

\begin{figure}
	\centering
	\includegraphics[width=0.70\textwidth]{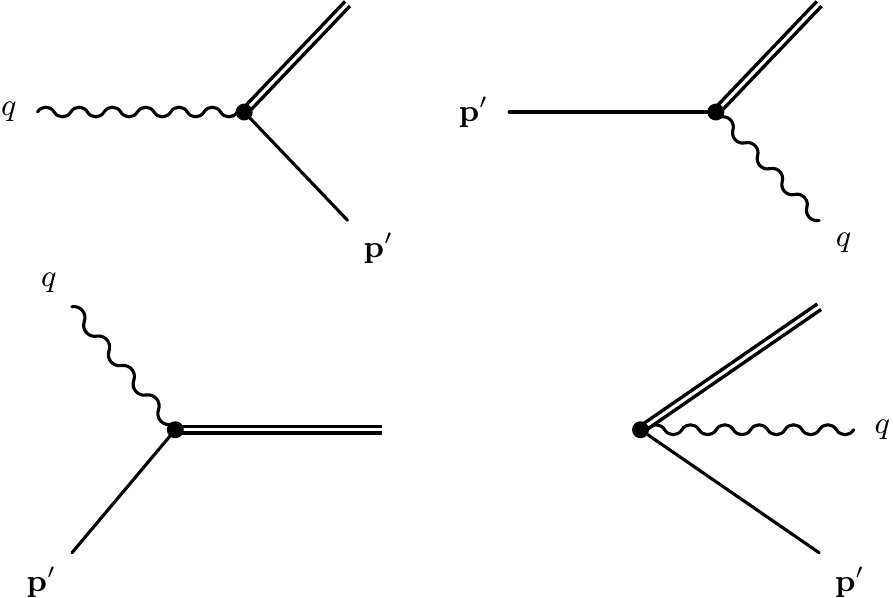}
	\caption{Diagrams contributing to the off-shell imaginary part of the self-energy of the radiation field in a background characterized by the presence of an atom in its fundamental state. The two diagrams on the left show the contribution in form of  decay rate, and the diagrams on the right the contribution as creation rates. The only diagram contributing on shell, $q=(|\vect q|,\vect q)$, is the one at the bottom on the left.}\label{fig:chi_decay}
\end{figure}

For the radiation field we focus on the imaginary part of the retarded self-energy. Using relation \eqref{cut},
\begin{equation}
    \Im \PiR(q;{\vect p'}) = \frac{i}{2} \left[ \Pi^{21}(q;{\vect p'}) - \Pi^{12}(q;{\vect p'}) \right],
\end{equation}
the self-energy component $\Pi^{12}$ is given by:
\begin{equation}
    \Pi^{12}(q;{\vect p'}) = -ig^2 m^2 \int \udpi[4]{k} G^{(0)}_{12}(k;{\vect p'}) G^{(0)*}_{12}(q-k),
\end{equation}
where $G^{(0)*}_{12}(p)$ corresponds to the free vacuum propagator of $\phiM$ and $G^{(0)}_{12}(k;{\vect p'})$ corresponds to the free propagator of $\phim$ in the background state $|{\vect p'}\rangle$ (see appendix \ref{app:FreeOne}). Restricting for the moment to the on-shell value, and discarding the purely vacuum contribution (because the vacuum contribution corresponds to the decay rate of a massless particle in the vacuum, and we know that this process cannot happen on-shell), we get:
\begin{equation}
\begin{split}
    \Pi^{12}(|\vect q|,\vect q;{\vect p'})  &= ig^2 m^2 \int \udpi[4]{k} \frac{2\pi}{2E^*_\vect{q-k}}
    \delta (q^0 - k^0 + E_\vect{q-k}^*)  \\
    &\quad \times \frac{(2\pi)^4}{2E_\vect kV}
    \left[ \delta^{(3)}(\vect{k-p'}) \delta(k^0-E_\vect k) + \delta^{(3)}(\vect k+{\vect p'}) \delta(k^0+E_\vect k) \right].
\end{split}
\end{equation}
Similarly for $\Pi^{21}$:
\begin{equation}
\begin{split}
    \Pi^{21}(|\vect q|,\vect q;{\vect p'})  &= ig^2 m^2 \int \udpi[4]{k} \frac{2\pi}{2E^*_\vect{q-k}}
    \delta (q^0 - k^0 - E_\vect{q-k}^*)  \\
    &\quad \times \frac{(2\pi)^4}{2E_\vect kV}
    \left[ \delta^{(3)}(\vect{k-p'}) \delta(k^0-E_\vect k) + \delta^{(3)}(\vect k+{\vect p'}) \delta(k^0+E_\vect k) \right].
\end{split}
\end{equation}
Adding the two previous results 
and developing the integrals with the delta functions one finds:

\begin{equation}\label{intImPi}
\begin{split}
    \Im \PiR (|\vect q|,\vect q;{\vect p'})  &= -\frac{g^2 m^2}{4E_{\vect p'}V} \left[ \frac{2\pi}{2 E^*_\vect {q-p'} } \delta(|\vect q| - E_{\vect p'} - E^*_\vect{q-p'})  \right.
    \\ &\qquad
    - \frac{2\pi}{2 E^*_\vect {q-p'} } \delta(|\vect q| - E_{\vect p'} + E^*_\vect{q-p'})    \\
    &\qquad +\frac{2\pi}{2 E^*_\vect {q+p} } \delta(|\vect q| + E_{\vect p'} - E^*_\vect{q-p'}) 
    \\ &\qquad -\left.\frac{2\pi}{2 E^*_\vect {q+p} } \delta(|\vect q| + E_{\vect p'} + E^*_\vect{q+p}) \right].
\end{split}
\end{equation}
From the four deltas, only one can contribute as:

\begin{equation}
    \Im \PiR (|\vect q|,\vect q;{\vect p'})  = -\frac{g^2 m^2}{8V E^*_\vect {q+p} E_{\vect p'}} 2\pi \delta(|\vect q| + E_{\vect p'} - E^*_\vect{q+p}).
\end{equation}
One can assign a decay rate to this self-energy,
\begin{equation}
    \Gamma^\text{(rad)}_-(|\vect q|,\vect q;{\vect p'})= - \frac{1}{|\vect q|}\Im \PiR (|\vect q|,\vect q;{\vect p'})  = \frac{g^2 m^2}{8V |\vect q| E^*_\vect {q+p} E_{\vect p'}} 2\pi \delta(|\vect q| + E_{\vect p'} - E^*_\vect{q+p}).
\end{equation}
	This decay rate in fact corresponds to the absorption rate of the radiation field by the unexcited 
	atom in the background. The appearance of the spacetime volume $V$  
	follows from the hypothesis that the atom is delocalized.
	In a more physical situation it would be replaced by the characteristic volume 
	of the wavefunction of the atom, or by the inverse 
	density of particles per unit volume. 
	In the case in which the atom is originally at rest, we get:
\begin{equation}
    \Gamma_-^\text{(rad)}(|\vect q|,\vect q;{\vect p'}) \approx  \frac{g^2 }{8V |\vect q|} 2\pi \delta(|\vect q| - \dm).
\end{equation}
We can check that this self-energy truly corresponds to a decay rate computing directly the relevant Feynman diagram (shown in fig.~\ref{fig:chi_decay}) by using expressions analogous to those of eqs.~\eqref{DirectGammaOutEq}.

In eq.~\eqref{intImPi}, only one of the four deltas contributed on-shell. 
We can go somewhat further by reexpressing this equation off the photon mass shell:
\begin{equation}\label{intImPiOffShell}
\begin{split}
     \Delta\!\Im \PiR (q;{\vect p'}) &=  -\frac{g^2 m^2}{4E_{\vect p'}V} \left[ \frac{2\pi}{2 E^*_\vect {q-p'} } \delta(q^0 - E_{\vect p'} - E^*_\vect{q-p'}) \right. \\ 
    &\qquad- \frac{2\pi}{2 E^*_\vect {q-p'} } \delta(q^0 - E_{\vect p'} + E^*_\vect{q-p'})     \\
    &\qquad +\frac{2\pi}{2 E^*_\vect {q+p} } \delta(q^0 + E_{\vect p'} - E^*_\vect{q+p}) \\
    &\qquad - \left.\frac{2\pi}{2 E^*_\vect {q+p} } \delta(q^0 + E_{\vect p'} + E^*_\vect{q+p}) \right],
\end{split}
\end{equation}
where $\Delta\!\Im\PiR(q;{\vect p'})$ is the correction to the equilibrium off-shell self-energy.
 The first and third terms on the right hand side of this equation give the modification of the 
 decay rate due to the background particle with momentum ${\vect p'}$, while the second and 
 fourth term give the modification of the creation rate, taking into account the Bose-Einstein 
 statistics of the particles, as it can be seen in fig.~\ref{fig:chi_decay}. So the imaginary 
 part of the retarded self-energy is proportional to the decay rate minus the creation rate, 
\begin{equation}
	\Im \PiR (q;{\vect p'}) = -\frac{1}{q^0}\left[ \Gamma_-^\text{(rad)}(q;{\vect p'}) - 
	\gamma_+(q;{\vect p'}) \right], 
\end{equation}
thereby recovering the same relation of eq.~\eqref{ImSigmaRGamma}
which applied to massive fields. Compare also the above equation with the corresponding expression in the vacuum,
eq.~\eqref{PiVacOff}.

\subsection{Massive field $\phiM$}

In this case we start by the computation of the real part of the self-energy. Using the property $\Re \SigmaR^*(p;\vect p') = \Re \Sigma^{11*}(p;\vect p')$ [see eqs.~\eqref{KeldyshCompDef}], and evaluating non-equilibrium contribution to the second Feynman diagram in fig.~\ref{fig:SigmaVac}, we get
\begin{equation}
\begin{split}
	&\ \Delta\!\Re \SigmaR^*(p;\vect p') = - m^2 g^2 \int \udpi[4]{k} \frac{1}{-(p^0-k^0)^2+(\vect p + \vect k)^2 } \\
	&\qquad \times \frac{(2\pi)^4}{2E_{\vect p'} V} \left[ \delta^{(3)}(\vect k - \vect p') \delta(k^0-E_{\vect p'}) + \delta^{(3)}(\vect k + \vect p') \delta(k^0-E_{\vect p'} )\right],
\end{split}
\end{equation}
which gives
\begin{equation}
\begin{split}
	\Delta\!\Re \SigmaR^*(p;\vect p') &= \frac{m^2 g^2}{2E_{\vect p'}	} \bigg( \frac{1}{(\vect p - \vect p')^2 - (p^0 - E_{\vect k'} )^2}  \\ &\qquad + \frac{1}{(\vect p + \vect p')^2 - (p^0 + E_{\vect k'} )^2}\bigg).
\end{split}
\end{equation}
Evaluated in the mass shell, and in the approximation of slowly moving heavy atoms, one has:
\begin{equation}
	\Delta\!\Re \SigmaR^*(E^*_\vect p,\vect p;\vect p') = \frac{g^2}{2}  \frac{m/V}{(\vect p - \vect p')^2 - (\dm)^2} . 
\end{equation}
Notice that the above expression cannot be absorbed in the mass term.

\begin{figure}
	\centering
	\includegraphics[width=0.75\textwidth]{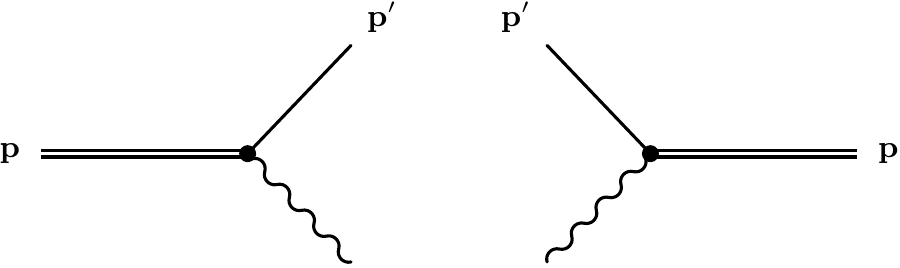}
	\caption{Diagrams contributing to the off-shell imaginary part of the self-energy of an excited atom, with momentum $\vect p$, in a background characterized by the presence of another atom in its fundamental state, with momentum ${\vect p'}$. The diagram on the left corresponds to the decay of the excited atom, and the diagram on the right to the creation.}\label{fig:Mdecay}
\end{figure}

For the case of the imaginary part of the self energy, we can repeat a similar calculation to the previous subsection. Now we have to take into account that the light field is in a one-particle state, but also that the massless field is in a thermal state. Following similar same steps as in the previous case we get:
\begin{equation}\label{intIm_SigmaM}
\begin{split}
     \Delta\!\Im \SigmaR^* (p;{\vect p'}) &=  -\frac{g^2 m^2}{4E_{\vect p'}V} 
	  \bigg[ \frac{2\pi}{2 |\vect{p-p'}| } \delta(p^0 - E_{\vect p'} - |\vect{p-p'}|)[1+n(|\vect{p-p'}|)] \\
    &\qquad- \frac{2\pi}{2 |\vect{p-p'}| } \delta(p^0 - E_{\vect p'} + |\vect{p-p'}|) [1+n(|\vect{p-p'}|)]     \\
    &\qquad +\frac{2\pi}{2 |\vect{p+p'}| } \delta(p^0 + E_{\vect p'} - |\vect{p+p'}|) [1+n(|\vect{p+p'}|)] \\
    &\qquad- \frac{2\pi}{2 |\vect{p+p'}| } \delta(p^0 + E_{\vect p'} + |\vect{p+p'}|)[1+n(|\vect{p+p'}|)] \\
	&\qquad+ \frac{2\pi}{2 |\vect{p-p'}| } \delta(p^0 - E_{\vect p'} + |\vect{p-p'}|)n(|\vect{p-p'}|) \\
    &\qquad- \frac{2\pi}{2 |\vect{p-p'}| } \delta(p^0 - E_{\vect p'} - |\vect{p-p'}|)n(|\vect{p-p'}|)     \\
    &\qquad +\frac{2\pi}{2 |\vect{p+p'}| } \delta(p^0 + E_{\vect p'} + |\vect{p+p'}|)n(|\vect{p+p'}|) \\
    &\qquad- \frac{2\pi}{2 |\vect{p+p'}| } \delta(p^0 + E_{\vect p'} - |\vect{p+p'}|)n(|\vect{p+p'}|) 
	 \bigg]. \raisetag{2em}
\end{split}
\end{equation}

This eight terms can be similarly interpreted as the modification to the decay and creation rates of the massive state due to the presence of the extra atom and the thermal massless particle. The terms  with a global negative sign correspond to decay and the ones with a global positive sign to creation. So again we have
\begin{equation}
	\Delta\!\Im \SigmaR^* (p;\vect p') = -\frac{1}{p^0}\left[ \Gamma^*_-(p;{\vect p'}) - \Gamma^*_+(p;{\vect p'}) \right].
\end{equation}
On-shell only the first and sixth terms contribute as
\begin{equation}
\begin{split}
     \Delta\!\Im \SigmaR^* (E_{\vect p},\vect p;{\vect p'}) &=  -\frac{g^2 m^2}{4E_{\vect p'}V} 
	  \bigg[ \frac{2\pi}{2 |\vect {p-p'}| } \delta(E_{\vect p} - E_{\vect p'} - |\vect{p-p'}|)
	  [1+n(|\vect{p-p'}|)] \\
    &\qquad- \frac{2\pi}{2 |\vect {p-p'}| } \delta(E_{\vect p} - E_{\vect p'} - |\vect{p-p'}|)n(|\vect{p-p'}|)  \bigg].
\end{split}
\end{equation}
The two Feynman diagrams corresponding to these two terms are shown in fig.~\ref{fig:Mdecay}, the first being a decay term and the second a creation term. The two terms can be added to give a net rate of approach to equilibrium:
\begin{equation}
    \Delta\Gamma^*(\vect p;{\vect p'})  = \frac{g^2 m^2}{8V |\vect p-{\vect p'}| E^*_\vect{p} E_{\vect p'}} 2\pi \delta(E^*_\vect p - E_{\vect p'} - |\vect{p-p'}|).
\end{equation}
Notice that the thermal contributions cancel exactly. In the case in which the atom is at rest,
\begin{equation}
	\Delta\Gamma^* = \frac{g^2}{8V \dm} 2\pi \delta(\dm - |\vect p - \vect p'|).
\end{equation}

To sum up, the results show that the usual interpretation of the imaginary part of the self-energy as 
a difference between destruction and creation rates  holds 
both for the radiation field and the massive field.
In the case of a massless particle the self-energy acquires an imaginary part, which reflects the fact that radiation can be absorbed by the atom in the background. In the case of an exited atom, its effective decay rate increases because of the presence of the additional atom. The contribution to the effective rate can be split as the difference of two terms: a  decay rate minus a creation rate, the second being 
the reason why
the atom in the fundamental state can become exited. It is interesting to note that the increase of the effective decay rate is independent of the temperature, although each term separately depends on it. This is a consequence of the fact that the system effectively behaves as if it were linear 
(in linear systems the retarded propagator is independent on the background state).

\part{Curved spacetime}

\addtocontents{toc}{\shortpage[.5]}
	\chapter{Particle-like excitations in curved backgrounds}

Much attention has been given in the literature to the concept of particle in curved spacetimes (see \cite{BirrellDavies,WaldQFT,Fulling} and references therein). Most analysis have however dealt with non-interacting particles, \ie, with particles which are free, except for the classical gravitational interaction with the curved background. Although in the  next section we will  review free particles as an introduction,  in this chapter we will mainly focus on interacting particles, generalizing the results of chapter 3 to the case of curved backgrounds.

\section{Free particles in curved spacetimes}\label{sect:FreeCurved}

\index{Particle!in curved spacetimes|(}

Different related free particle concepts exist in curved spacetimes. In this section we briefly review some of them.  We shall not provide a complete account: we will only focus on some aspects relevant for this thesis. Except where more detailed references are provided, we address the reader to refs.~\cite{BirrellDavies,WaldQFT,Fulling,JacobsonQFT} for the remaining of this section. 

Let us start by recalling some basic equations. Non-interacting field operators obey the equations of motion
\begin{equation}\label{FreeFieldEq}
	\big[ \Box_x + m^2 + \xi R(x)] \hat\phi(x) = 0,
\end{equation}
where $\xi$ is the conformal parameter and where 
\begin{equation*}
	\Box_x \hat\phi(x)= g^{\mu\nu} \nabla_\mu\nabla_\nu \hat\phi(x) = \frac{1}{\sqrt{-g}} \partial_\mu \left[ \sqrt{-g} g^{\mu\nu} \partial_\nu \hat\phi(x)\right].
\end{equation*}
For globally hyperbolic spacetimes, the field operator can be decomposed in modes,
\begin{equation}\label{ModeDecompCurved}
	\hat \phi(x)  = \sum_\alpha \big[ \hat a_\alpha   u_\alpha (x) + \hat a_\alpha ^\dag u_\alpha ^*(x) \big],
\end{equation}
where $u_\alpha(x)$ are a orthonormal complete set of modes satisfying \Eqref{FreeFieldEq}, and where the creation and  annihilation operators satisfy the usual conmutation relations \eqref{ConmutRelat}. The orthonormality is defined with respect to the scalar product
\begin{equation*}
	(u_\alpha,u_\beta) = -i\int_\Sigma u_\alpha(x) \overleftrightarrow{\partial_\mu} u_\beta(x) \sqrt{-g_\Sigma(x)} \vd{\Sigma^\mu} = \delta_{\alpha\beta}.
\end{equation*}

\subsection{Particles associated with asymptotic observers}

\index{Particle!associated to asymptotic observers}

When the spacetime is such that at least one asymptotically flat Minkowski region exists, one can speak of particles associated to asymptotic observers. The field operator can be decomposed in terms of the  modes $u_\vect k$,
\begin{equation}\label{ModeDecompCurvedAsymp}
	\hat \phi(x)  = \sum_\vect k \big[ \hat a_\vect k u_\vect k(x) + a_\vect k^\dag u_\vect k^*(x) \big],
\end{equation}
which  reduce to the standard Minkowski modes when evaluated in the asymptotic region. Without mathematical precision,
\begin{equation*}
	u_\vect k(x) \approx \frac{1}{\sqrt{2 E_\vect k}} \expp{-i E_\vect k  t+ i\vect k \cdot \vect x}, \quad \text{when $x$ is in the asymptotic region}.
\end{equation*}
The modes are eigenstates of the coordinate vectors $(\partial_t,\partial_i)$, which in turn are Killing vectors in the asymptotic region:
\begin{equation*}
	 \quad \Lie_{\partial_t} u_\vect k = -i E_\vect k u_\vect k, 
	\quad \Lie_{\partial_i} u_\vect k = i k_i u_\vect k, \quad{\text{in the asymptotic region}}.
\end{equation*}
With respect to this mode decomposition one can construct a vacuum state, $\hat a_\vect p |0\rangle = 0$, a one-particle state $|\vect p\rangle=\hat a_\vect p^\dag |0\rangle$, and, in general, the entire Fock space. The states thus constructed are defined in the entire spacetime, but  have a  particle meaning  in the asymptotic region only.

The vacuum $|0\rangle$, associated to the modes $u_\vect k$, is invariant under the whole Poincar\'e group in the asymptotic region:  not only the vacuum is invariant under time and space translations, it is also invariant under rotations and boosts, although this is not manifest from the above construction. An easy way to prove this fact is to realize that the vacuum Green functions $G^{(0)}_{ab}(x,x')$, which characterize the vacuum state, depend only on the relativistic invariants $(x-x')^2$ and $\theta(x^0 - x'^0)$. Notice that there are many different decompositions  \eqref{ModeDecompCurvedAsymp}  which are equivalent under the action of the Lorentz group.

\index{Bogolubov transformation}
Whenever two different asymptotic regions exist, a second mode decomposition of the field operator can be introduced, with another associated set of  modes $\bar u_{\bar{\vect p}}(x)$, chosen so that it reduces to the standard Minkowski set in the second region. Both sets of modes are related through the Bogolubov transformation
\begin{equation}
\bar u_\beta = \sum_{\alpha}(\alpha_{\beta\alpha} u_\alpha + \beta_{\beta\alpha} u^*_\alpha).
\end{equation}
As it is well known, when $\beta_{\beta\alpha}\neq 0$ the two vacua $|0\rangle$ and $|\bar 0\rangle$ do not  coincide, meaning that what is perceived as the vacuum in one asymptotic region is seen as an excited particle state in the other sector, leading to the well-known phenomenon of particle creation in curved spacetimes. When $\beta_{\beta\alpha}= 0$ the vacua are equivalent, and the two Fock spaces are related by the action of the Lorentz group.

\index{Particle!global concept}
\subsection{Global particle concepts}

The notion of particles associated to asymptotic observers, although clear both from the mathematical and the physical points of view, is somewhat restrictive because most spacetimes do not have asymptotic regions and, despite that, one would like to define particles.

One possibility is that the spacetime has no asymptotic regions, but that it  has some global symmetries under a certain class of transformations. By choosing a set of modes which are eigenstates of the Killing vectors associated to those symmetries we are led to global particle concepts. For instance, if $\xi$ is a timelike  Killing vector and the modes satisfy
\begin{equation}
	\Lie_\xi u_\alpha = -i \omega_\alpha u_\alpha,\quad \text{for some $\omega_\alpha > 0$, }
\end{equation}
we may think of a global particles states, defined by $|\alpha\rangle = \hat a_\alpha^\dag |0\rangle$, where $\hat a^\dag_\alpha$ is the creation operator associated to the mode $u_\alpha$ and $|0\rangle$ is the vacuum state verifying $\hat a_\alpha |0\rangle = 0$. Notice however that, if no further symmetries are present, these global particles are incomplete, in  the sense that only the energy is well-defined for them, and not the momentum. It is usually convenient to choose a coordinate system adapted to the relevant Killing vectors. 

\index{Unruh effect}
Even in Minkowski spacetime different global particle definitions may exist. For instance, instead of working with Cartesian coordinates and eigenstates of the space translation Killing vectors, one may work with spherical coordinates and eigenstates of the angular and radial momenta. The spherical modes are  labeled by the radial momentum and the total and azimuthal angular momenta. In this case one would speak of ``spherical particles''. The vacua associated to the Cartesian and spherical modes is anyway equivalent.
Also in Minkowski, a system of coordinates $(\eta,\xi,y,z)$ associated to an ensamble of uniformly accelerating observers can be introduced, called the Rindler coordinates, 
in terms of which the metric is $\vd s^2 = \expp{2a\xi} ( -\vd \eta^2 + \vd \xi^2 ) + \vd y^2 + \vd z^2$. 
They cover a quadrant of the Minkowski spacetime (the Rindler wedge). Positive-frequency modes with respect to the $\eta$ coordinate can be defined: $\Lie_{\partial_\eta} u_\alpha = -i \omega_\alpha u_\alpha$ ($\partial_\eta$ is a timelike Killing vector). A Rindler vacuum is associated to the Rindler modes, and global Rindler particles can be obtained by the action of the creation operator. The vacuum associated to the usual Minkowski modes corresponds to a thermal bath of Rindler particles (Unruh effect). 

\index{Particle!conformal}
\index{Propagator!conformal fields}

There may also be a spacetime which is not Minkowski, but which is conformally equivalent to Minkowski: $g_{\mu\nu} = \Omega^2(\eta,x^i) \eta_{\mu\nu}$. In this case there exist four conformal Killing vectors $(\partial_\eta, \partial_i)$ associated to time and space translations. For conformally coupled particles (\ie, particles with no mass and $\xi = 1/6$), the field operator  can be expanded in terms of
\begin{equation}
	\hat \phi(x)  = \sum_\vect k [ \hat a_\vect k u_\vect k(x) + a_\vect k^\dag u_\vect k^*(x) ], \quad u_\vect k(\eta,x^i) =  \frac{1}{\Omega(\eta,x^i) \sqrt{2|\vect k|}} \expp{-i|\vect k| \eta + i k_i x^i}.
\end{equation}
A conformal Fock space can be associated to them: $a_\vect k|0\rangle = 0$, $|\vect k\rangle = a_\vect k^\dag|0\rangle$. The conformal modes verify $\Lie_{\partial_\eta} u_\vect k = - |\vect k| u_\vect k$ and  $\Lie_{\partial_i} u_\vect k =  k^i u_\vect k$, and are thus defined by their conformal energy $|\vect k|$ and their conformal momentum $\vect k$. The propagators associated to conformal particles can be computed from the Minkowski propagators: $G_{ab}(x,x') = \Omega^{-1}(x) G^\text{\tiny (Mink)}_{ab}(x,x') \Omega^{-1}(x')$. Of particular interest is the cosmological case in which $\Omega(\eta,x^i) = a(\eta)$. In this case the vectors $\partial_i$ are true Killing vectors, and $\partial_\eta$ remains a conformal Killing vector. 

When a spacetime has many degrees of symmetry, such is the case of the maximally symmetric de Sitter or anti-de Sitter spacetimes, one usually chooses modes adapted to a particular subset of the symmetries of the spacetime. As in Minkowski spacetime, several different possibilities exist. First, one can  construct a vacuum state which is invariant under the full symmetry group, or alternatively a vacuum which is only invariant under a subgroup of it. Second, global particles created as excitations of this vacuum ca have different characterizations depending on the symmetries of the modes. A large amount of literature has been devoted to the analysis of the vacuum states and Fock space structure of the de Sitter spacetime; we will not enter into  further details.

\subsection{Quasilocal particles}

\index{Particle!quasilocal}

Global particle concepts are mathematically appealing, but  the relation of the global particles with the familiar Minkowski particles  is not always transparent. There must be some sort of approximation in curved spaces for which the concept of particle resembles that of Minkowski since, after all, the expansion of our universe does not interfere with the usual particle experiments. 

\index{Equivalence principle}

As we know from the equivalence principle, an inertial observer cannot tell whether the spacetime is flat or curved as long as he or she only performs ``quasilocal'' experiments, \ie, experiments whose typical time and length scales are much smaller than the typical curvature radius $L$ at the location of the observer.

\index{Riemann normal coordinates}

\index{Tetrad}
\index{Vierbein|see{tetrad}}

One can associate to an inertial observer located at a spacetime point $x_0$ a orthogonal tetrad of vectors of unit norm $e_{a}^{\ \mu}$. The timelike vector $e_{0}^{\ \mu}$ is tangent to the worldline of the observer, and the spacelike vectors $(e_{1}^{\ \mu},e_{2}^{\ \mu},e_{2}^{\ \mu})$ determine the three spatial directions of the observer. In the reference frame of the tetrad the metric at the point $x_0$ can be written as
\begin{equation}
	g_{ab}(x_0) = e_{a}^{\ \mu} g_{\mu\nu}(x_0) e_{b}^{\ \nu} = \eta_{ab},
\end{equation}
where $g_{ab}$ is the metric in the orthonormal reference frame and where the second equality is a consequence of the orthonormality of the tetrad. Without entering into technical details, by considering the geodesics tangent to the vectors of the tetrad, a local Cartesian coordinate system can be defined: the Riemann normal coordinates. The metric expressed in Riemann normal coordinates is
\begin{equation}
	g_{ab}(y) = \eta_{ab} +  \frac13 R_{acbd} y^c y^d + O(y^4/L^4),
\end{equation}
where the coordinates $y^a$ have the origin at a point $x_0$, the Riemann tensor has been evaluated at this point and the roman Latin subindices $a,b,c\ldots$ have been used to stress that the Riemann normal coordinates are associated to the tetrad. The Riemann tensor is of order $| R_{abcd}| \sim L^{-2}$, where we recall that $L$ is the typical curvature radius. If the experiments are done in a region which is much smaller than the typical curvature radius, the second term in the above equation can be neglected and $g_{ab} = \eta_{ab} + O(y^2/L^2)$.

In a region surrounding the point $x_0$, the field operator satisfies \cite{Massar93}
\begin{equation}
	\left[ -?\eta^mn? \partial_m \partial_n + m^2  +\xi R - \frac13 ?R_a^n? y^a \partial_n + \frac13 ?R^m_a^n_b? y^a y^b \partial_m \partial_n \right] \hat\phi(y) +O(\tfrac{y^4}{L^4}) = 0
\end{equation}
and can be expanded in modes:
\begin{subequations}
\begin{align}
	\hat\phi(y) &= \sum_\vect p \left[ \hat a_\vect p(x_0) u_\vect p(y)
	+\hat a^\dag_\vect p(x_0) u^*_\vect p(y) \right], \\
\begin{split}
	u_\vect p (y) &= \frac{\expp{ip^a p_a} }{\sqrt{2p^0}} \bigg[1 + \frac{\xi R}{m^2} + \frac{i}{3m^2} ?R_a^n? y^a p_n - \frac{i}{3m^2} ?R^m_a^n_b? y^a y^b p_m p_n \\ &\qquad \frac{4i}{3m^2} ?R^m_a^n^s? y^a p_m p_n p_s + \frac{8}{3m^8} ?R^mnsr? p_m p_n p_s p_r \bigg] \\ &\quad+ O(y^4/L^4).
\end{split}
\end{align}
\end{subequations}
where in the above equation $p^0 = E_\vect p  = \sqrt{m^2 + \vect p^2}$, with $\vect p = (p^1,p^2,p^3)$. A quasilocal vacuum $|0,x_0\rangle$ can be defined by demanding $\hat a_\vect p(x_0) |0,x_0\rangle = 0$, and quasilocal particle excitations can be defined by $|\vect p,x_0\rangle = \hat a_\vect p^\dag(x_0) |0,x_0\rangle$. As long as one considers a small patch surrounding the point $x_0$, the quasilocal particles have the same properties as the usual Minkowski particles. Most of the times it will be enough to retain the leading term, which corresponds to the Minkowski expression.

The quasilocal particles concept only makes sense for particles whose wavelength is much smaller than the typical curvature radius, or, in other words, for modes verifying $E_\vect p^2 \gg |R_{abcd}|$. Furthermore, one can only study their properties  while they remain within the local spacetime patch. Whenever the time or distance traveled by the particles become comparable to the local curvature radius the quasilocal approximation loses its validity. 
The degree of uncertainty of the quasilocal particle concept depends on the expansion order: if  the metric in the Riemann coordinates is computed to order $n$, the degree of relative uncertainty of the quantities derived within the quasilocal formalism is of the order of  $(E_\vect p L)^{-(n+1)}$ at most. Ultimately, one could think of computing the expansion in Riemann coordinates to infinite order; in this case there could still be some uncertainty, but it would be non-analytical in $(E_\vect p L)^{-1}$.

Correlation functions $G_{ab}(x,x')$ can also be considered in the quasilocal approximation. A quasilocal momentum representation can be introduced  in the neighborhood of the points with coordinates $x$ and $x'$, by considering  Riemann normal coordinates with the origin at some point in that neighborhood. Bunch and Parker \cite{BunchParker79} introduce that system of coordinates at  $x'$. However, introducing the fixed point at the extreme of the interval breaks the properties of the propagators which we have used in this thesis, and which are detailed in appendix \ref{app:GenRel}. To connect with the work in the previous chapters, and following ref.~\cite{CalzettaEtAl88}, it will prove more convenient for us to introduce the system of normal coordinates around an arbitrary point $x_0$.  Namely, we shall consider
\begin{subequations}
\begin{equation}\label{LocalMomentum}
\begin{split}
	\bar G^{(0)(x_0)}_{ab}(p, Y)
	= \int \ud[4]{y} \expp{-ip_a y^a} \bar G^{(0)}_{ab}(Y+y/2,Y-y/2)
\end{split}
\end{equation}
where $y^a$ are normal coordinates around the point $x_0$ and $Y-y/2$ and $Y-y'/2$  are, respectively, the coordinates of the points $x$ and $x'$  in the system of Riemann normal coordinates.
A particular case of interest will be when the point $x_0$ is chosen so that it is at half the geodetic distance between $x$ and $x'$:
\begin{equation}\label{LocalMomentum2}
\begin{split}
	\bar G^{(0)}_{ab}(p, X) &= \int \ud[4]{y} \expp{-ip_a y^a} \bar G^{(0)}_{ab}(x,x') = \int \ud[4]{y} \expp{-ip_a y^a} \bar G^{(0)}_{ab}(y/2,-y/2)
\end{split}
\end{equation}
\end{subequations} 
where $\bar G^{(0)}_{ab}(p, X)$ is a shorthand expression for  $\bar G^{(0)(X)}_{ab}(p, 0) $.
Notice that in order to avoid cumbersome notations  we are using the same symbol for the propagator in arbitrary coordinates, the propagator in Riemann normal coordinates and the Fourier transformed propagators. The bar over the propagator indicates the following rescaling:
\begin{equation} \label{FourierCurved}
	\bar G^{(0)}_{ab}(x,x')  :=[-g(x)]^{1/4}   { G^{(0)}_{ab}(x,x') } [-g(x')]^{1/4}.
\end{equation}
The  usefulness  of this rescaling will be discussed in the following sections. Unbarred propagators have the usual definitions, \eg, $ G_{11}(x,x') =   G_\text F(x,x') = \Tr{ [\hat \rho T \hat \phi(x) \hat \phi(x')]}$. 

To first order in the curvature we find:
\begin{equation}
\begin{split}
	\bar G^{(0)}_\text{R} (p, X) &=  \frac{-i}{p^2 + m^2 - i p^0 \epsilon} 
	+ \frac{-iR(\frac{1}{8} - \xi)}{(p^2 + m^2)^2 - i p^0 \epsilon} 
	\\ \quad &+ \frac{-iR_{ab} p^a p^b}{(p^2 + m^2)^3 - i p^0 \epsilon} + O(p^{-4} L^{-4}).
\end{split}
\end{equation}
This expression is valid for generic states, provided there is no quantum interaction. For quasilocal vacuum states, we can give explicit expressions for the other propagators as well. For instance, for the Feynman propagator,
\begin{equation}
	\bar G^{(0)}_\text{F} (p, X) =  \frac{-i}{p^2 + m^2 - i \epsilon} 
	+ \frac{-iR(\frac{1}{8} - \xi)}{(p^2 + m^2)^2 - i \epsilon} 
	+ \frac{-iR_{ab} p^a p^b}{(p^2 + m^2)^3 - i  \epsilon} + O(p^{-4} L^{-4}).
\end{equation}
The above equation can also be thought as defining the quasilocal vacuum.


Notice two different conditions must be fulfilled for the momentum representation \eqref{LocalMomentum} to be valid. First, the energy of the particles must be sufficiently large: $E_\vect p\gg L^{-1}$.  Second, the geodetic distance between $x$ and $x'$ must be much smaller than the local curvature radius, or what is the same,  $|y^a y_a| \ll L^2$.\footnote{This requirement applies both for timelike and spacelike separated points, but for null separations it is obviously meaningless. In this case one has to require that the affine distance along a null geodesic is much smaller than the local curvature radius.}


\subsection{Adiabatic particles}\label{sect:adiabaticParticles}

\index{Particle!adiabatic}
Quasiparticle local concepts are physically meaningful, but the requirement that the time and distances considered is small as compared with the size of the quasilocal patch can be an obstacle when following the dynamics of a particle over large periods of time. Let us see how this condition can be lifted in the cosmological case. The basic idea is that we shall consider spacetimes which are slowly evolving as compared to the characteristic particle energy.

\index{Hubble parameter}
Let us consider a Friedman-Lemaître-Robertson-Walker (FLRW) spacetime with flat spatial sections:
\begin{equation}\label{FLRW}
	\vd s^2 = -\vd t^2 + a(t)(\vd x^2 + \vd y^2 + \vd z^2).
\end{equation}
The field operator can be expanded in conformal modes $u_\vect k(x)$, with $\vect k$ being the conformal momentum. The equation for the modes is separable, $u_\vect k(t,x^i) = \expp{i \vect k \cdot \vect x} \chi_\vect k(t)$, where the function $\chi_\vect k(t)$ verifies
\begin{equation}
	\left[ \frac{1}{a^3} \dert{}{t}{} \left(a^3 \dert{}{t}{} \right) + m^2 + \frac{\vect k^2}{a^2} + \xi \left(  4 \frac{\ddot a}{a} + 2\frac{\dot a^2}{a^2} \right) \right] 
	\chi_\vect k(t) = 0,
\end{equation}
with the dot meaning time derivative. The rate of evolution of the scale factor is given by the Hubble parameter $H(t) = \dot a(t)/a(t)$. We assume that during the spacetime evolution $H(t)$ has the characteristic value $H$. We introduce a dimensionless time parameter $u = Ht$, in terms of which the above equation can be expressed as
\begin{equation}
	\left[ \frac{H^2}{a^3} \dert{}{u}{} \left(a^3 \dert{}{u}{} \right) + m^2 + \frac{\vect k^2}{a^2} + \xi \left(  4 \frac{H^2}{a} \dert[2]{a}{u} + 2\frac{H^2}{a^2} \dert{a}{u} \right) \right] 
	\chi_\vect k(t) = 0.
\end{equation}
and we look for solutions to the above equation of the Wentzer-Kramers-Brillouin (WKB) form,
\begin{equation*}
	\chi_\vect k = \frac{1}{\sqrt{2 \partial_u B(u)}} \exp {\left( \frac{-iB(u)}{H}  \right)},
\end{equation*}
evaluating $B(u)$ in a perturbative expansion: $B =B_0 + HB_1+ H^2 B_2 + \cdots$. The leading order term yields
\begin{equation*}
		\left( \dert{B_0(u)}{u} \right)^2 = m^2 + \frac{\vect k^2}{a^2(u)} =: E^2_\vect k(u),
\end{equation*}
and the first order term vanishes. The solution for $B(u)$ is
\begin{equation*}
	B(u) = \int^u_{u_0} \ud {u'} E_\vect k(u') + O(H^2/E_\vect k^2) = H \int^t_{t_0} \ud {t'}  E_\vect k(t') + O(H^2/E_\vect k^2),
\end{equation*}
with the time $t_0$ being arbitrary (it only adds a phase). Therefore, to leading adiabatic order, the mode solutions are
\begin{equation}
	u_\vect k(t,x^i) = \frac{1}{\sqrt{2E_\vect k}(t)} \expp{ -i\int^t_{t_0} \ud {t'}  E_\vect k(t') + i \vect k \cdot \vect x}.
\end{equation}
This equation is valid for all times, with the only requirement that the evolution is sufficiently slow. As we see, to first adiabatic order there is no dependence on the conformal coupling parameter. We shall not go beyond the leading adiabatic order.

Once the adiabatic modes are obtained, the usual Fock space construction can take place. A vacuum state can be associated to the adiabatic modes by demanding $\hat a_\vect k|0\rangle = 0$, and adiabatic particles can be built by setting $|\vect k\rangle = \hat a^\dag_\vect k |0\rangle$. Notice that adiabatic particles are labeled by their conformal momentum. 

\index{Propagator!adiabatic|(}
Propagators can be computed as usual from the modes. The retarded propagator, to first adiabatic order, reads:
\begin{equation}\label{FreeRetardedAdiabatic}
\begin{split}
	\bar G^{(0)}_\text{R}(t_1,t_2;\vect k) = \frac{-i}{  \sqrt{ E_\vect k(t_1) E_\vect k(t_2)}} \sin\left({\int^{t_1}_{t_2} \ud{t'} E_\vect k(t')} \right) \theta(t_1-t_2).
\end{split}
\end{equation}
 Recall however that the retarded propagator has the interesting property that it is independent of the state. Other vacuum propagators can be computed in a similar way.

We present in the following an alternative 
derivation of equation \eqref{FreeRetardedAdiabatic}  
based on the retarded equation of motion, because it will be useful for us later on. The equation of motion of
the 
retarded propagator $\bar\GR^{(0)}(t_1,t_2;\vect p)$ of the rescaled fields is
\begin{equation}
	\left[ \frac{1}{a^3(t)} \derp{}{t}{} \left(a^3(t) \derp{}{t}{} \right) + m^2 + \frac{\vect k^2}{a^2(t)} \right]
	\left[ \frac{\bar\GR^{(0)}(t,t';\vect k)}{a^{3/2}(t) a^{3/2}(t')} \right] = \frac{-i}{a^3(t)} \delta(t-t').
\end{equation}
For simplicity of the expressions we have assumed that the conformal coupling parameter is zero; it could be added without difficulty, but it plays no role to first order in the adiabatic expansion.
Introducing the dimensionless time coordinate $u = Ht$, the equation reads:
\begin{equation*}
	\left[ \frac{H^2}{a^3(u)} \derp{}{u}{} \left(a^3(u) \derp{}{u}{} \right) + m^2  + \frac{\vect k^2}{a^2(u)} \right]
	\left[ \frac{\bar\GR^{(0)}(u,u';\vect k)}{a^{3/2}(u) a^{3/2}(u')} \right] = -\frac{iH}{a^3(u)} \delta(u-u'),
\end{equation*}
We choose the following WKB-like ansatz for the solution:
\begin{equation} \label{Ansatz}
	\bar\GR^{(0)}(u,u';\vect p)  = \frac{-i}{\sqrt{-\partial_u A(u,u'; \vect k) \partial_{u'} A(u,u'; \vect k)}} \sin{\left( \frac{A(u,u';\vect k)}{h} \right) } \theta(u-u'),
\end{equation}
for some function $A(u,u'; \vect k)$ to be determined.
The singular part of the equation
is automatically verified 
when $\partial_u A(u,u'; \vect k)= - \partial_{u'} A(u,u';\vect k)$ in the limit $u\to u'$. As for the regular part we make a series expansion in $H$: $A=A_0+H A_1 + H^2 A_2 \cdots$. The term in $H^0$ gives:
\begin{equation*}
	\left(\derp{A_0(u,u';\vect k)}{u}\right)^2 = m^2 + 
	\frac{\vect k^2}{a(u)^2}. 
\end{equation*}
The solution with the appropriate initial conditions is
\begin{equation*}
	A_0(u,u';p) = \int_{u'}^u \ud{u''} \sqrt{ m^2 + \frac{\vect k^2}{a^2(u'')}}  = H \int_{t'}^t \ud{t''} \sqrt{ m^2 + \frac{\vect k^2}{a^2(t'')}} .
\end{equation*}
The next term $A_1$ vanishes, so that the first correction to the above expression in the series expansion of $A$ is of order $H^2/E_\vect k^2$.  We therefore recover eq.~\eqref{FreeRetardedAdiabatic}.

Let us now compare the adiabatic and quasilocal vacua and particles through the analysis of the retarded Green functions. 
In the cosmological background  the Fourier transformed propagator is given by
\begin{equation}
	\bar\GR^\text{\tiny(QL)}(\omega,\vect p;T)  = \frac{-i}{-\omega^2 + \vect p^2 + m^2+i\epsilon} + O(E_\vect p^{-1} L^{-1}),
\end{equation}
We recall that the Fourier transform is defined with respect to the Riemann normal coordinates. In this case we have chosen a system of Riemann normal coordinates adapted to the comoving observer. 
In the above expression $\omega$ and $\vect p$ can respectively be interpreted as the physical energy and momentum. The time representation of the above equation is given by
\begin{equation*}
	\bar\GR^\text{\tiny(QL)} (t,t';\vect p)  = \frac{1}{\sqrt{m^2+\vect p^2}} \sin{\left[\sqrt{m^2 + \vect p^2}(t-t')\right]} \theta (t-t'),
\end{equation*} 
or, recalling that the physical momentum $\vect p$ is related to the conformal momentum $\vect k$ via $\vect p= \vect k/a(T)$, 
\begin{equation}\label{QLShortTime}
	\bar\GR^\text{\tiny(QL)} (t,t';\vect p)  = \frac{1}{\sqrt{m^2+\vect k^2/a^2(\frac{t+t'}{2})}} \sin{\left[\sqrt{m^2 + \frac{\vect k^2}{a^2\big(\frac{t+t'}{2}\big)}}(t-t')\right]} \theta (t-t').
\end{equation} 
This expression should be compared with the corresponding adiabatic expression, \Eqref{FreeRetardedAdiabatic}. It explicitly shows that the adiabatic and the quasilocal expansions coincide provided that the time difference $(t-t')$ is much smaller than the typical expansion time $H^{-1}$. Therefore, we confirm that the quasilocal and adiabatic expansions have a similar range of validity, the only difference being that the adiabatic expansion allows for arbitrarily large time lapses.

Let us end this subsection by mentioning that the conformal vacuum is an adiabatic vacuum of infinite order for the conformal particles \cite{BirrellDavies}.
\index{Propagator!adiabatic|)}

\subsection{Particle detectors}\label{sect:ParticleDetectors}
\index{Particle detector}

Given the elusive nature of the notion of particle in curved spacetimes, one alternative approach is to move to an operational definition of the particle concept. To this end, Unruh \cite{Unruh76} introduced a very simple model of particle detector, consisting of an idealized point particle coupled to the field. In fact, we already introduced a particle detector model in subsect.~\ref{sect:QuasiCreation} to study the quantum state corresponding to the quasiparticle excitation. In curved spacetimes the quasiparticle detector is usually employed with a slightly different and simpler role: to elucidate whether a given quantum state in the curved background contains particles, as seen from the perspective of the detector.

As in chapter 3, the detector consists of a harmonic oscillator $Q$ linearly coupled to the field (a two-level system can be used instead). In chapter 3 we employed a fixed harmonic oscillator coupled to a pair of modes of the detector. In curved spacetime one usually considers a moving detector (so that the particle concept for arbitrary observers can be analyzed) coupled to the spacetime representation of the field (so that the introduction of the momentum representation is avoided). Thus, we shall replace the coupling $g_Q Q(t) [\phi_\vect p(t)+ \phi_{-\vect p}(t)]$ used in subsect.~\ref{sect:QuasiCreation} by a coupling $g_Q Q(\tau) \phi\boldsymbol(x(\tau)\boldsymbol)$, where $\tau$ is the detector proper time.

\index{Response function}
The initial state of the field plus detector is assumed to be $\hat\rho \otimes |0_Q\rangle \langle 0_Q|$. As in subsect. \ref{sect:QuasiCreation} [see \Eqref{TotalEvolution}], if the state of the field is initially $\hat\rho$, the density matrix for the entire system at some later time $t$ is
\begin{equation}
	\hat\rho_\text{total}(t)=
	T \expp{-i\int_{\tau_0}^\tau g_Q \hat\phi_\text{I}\boldsymbol(x(\tau')\boldsymbol) \hat Q_\text{I}(\tau') \vd {\tau '} }
	\hat\rho \otimes |0_Q\rangle \langle 0_Q|
	\widetilde T \expp{i\int_{t_0}^t g_Q \hat\phi_\text{I}\boldsymbol(x(\tau')\boldsymbol) \hat Q_\text{I}(\tau') \vd {\tau '} },
\end{equation}
where the subindex $\text I$ indicates interaction picture. An expansion in the coupling constant leads to second order that
\begin{equation*}
\begin{split}
	\hat\rho_\text{total}(t)&=
	\hat\rho \otimes |0_Q\rangle \langle 0_Q|
	- i \int_{\tau_0}^\tau \ud {\tau'} g_Q \big[\hat\phi_\text{I}\boldsymbol(x({\tau'})\boldsymbol),\hat\rho\big] \otimes \big[ \hat Q_\text{I}({\tau'}) ,  |0_Q\rangle \langle 0_Q| \big]\\
	&\quad + g_Q^2\int_{\tau_0}^\tau \ud {\tau'} \int_{\tau_0}^\tau \ud {{\tau''}} \hat\phi_\text{I}\boldsymbol(x(\tau')\boldsymbol) \hat\rho\, \hat\phi_\text{I}\boldsymbol(x(\tau'')\boldsymbol) \otimes
	\hat Q_\text{I}({\tau'})  |0_Q\rangle \langle 0_Q| \hat Q_\text{I}({\tau''}) .
\end{split}
\end{equation*}
The probability that the detector gets excited is given by the expectation value of the projector $|0_Q\rangle \langle 0_Q|$ (the transition probability to higher states is negligible):
\begin{equation}
\begin{split}
	P_+(\tau) &= \Tr { \left(\hat\rho_\text{total}|1_Q\rangle \langle 1_Q| \right) } \\ &= 
	g_Q^2\int_{\tau_0}^\tau \ud {\tau'} \int_{\tau_0}^\tau \ud {{\tau''}}  \expp{ i\Omega(\tau'-\tau'')} \Tr {\big[ \hat\rho\, \hat\phi_\text{I}\boldsymbol(x(\tau'')\boldsymbol) \hat\phi_\text{I}\boldsymbol(x(\tau')\boldsymbol)  \big]},
\end{split}
\end{equation}
where $\Omega$ is the frequency of the detector. Therefore the probability of excitation is given by $P_+(\tau) = g_Q^2 \mathcal F(\Omega,[x^\mu],\tau,\tau_0)$, where 
\begin{equation}	
	\mathcal F(\Omega) = \int_{\tau_0}^\tau \ud {\tau'} \int_{\tau_0}^\tau \ud {{\tau''}}  \expp{ -i\Omega(\tau'-\tau'')} G_+  \boldsymbol(x(\tau'), x(\tau'')\boldsymbol) 
\end{equation}
is the detector response function.

For a given state $\hat\rho$ in an arbitrary spacetime, one can therefore say that a given observer $x^\mu(\tau)$ observes particles if the detector response function is different from zero. The Unruh particle detector is usually employed to analyze the particle concept for accelerating observers (Unruh effect), where one verifies that the response of the detector for accelerating observers in the vacuum coincides that of inertial observers in a thermal bath. 

Let us next indicate some relevant properties of the response function for inertial observers. The response function can be developed as 
\begin{equation*}	
	\mathcal F(\Omega) = \int_{\tau_0}^\tau \ud {T} \int_{-2 \min{(T-\tau_0,\tau-T)}}^{2 \min{(T-\tau_0,\tau-T)}} \ud \Delta \expp{ -i\Omega \Delta } G_+  \boldsymbol(x(T+\Delta/2), x(T-\Delta/2)\boldsymbol) .
\end{equation*}
There always exists a system of normal coordinates around the point $x(T)$ in which the coordinates of the inertial trajectory $x^\mu(\tau')$ can be represented by $y^a(\Delta/2) = (\Delta/2,0,0,0)$. Using these normal coordinates the above equation can be rewritten as
\begin{equation*}	
	\mathcal F(\Omega) = \int_{\tau_0}^\tau \ud {T} \int_{-2 \min{(T-\tau_0,\tau-T)}}^{2 \min{(T-\tau_0,\tau-T)}} \ud \Delta \expp{ -i\Omega \Delta } G_+(\Delta/2,\vect 0;-\Delta/2,\vect 0) .
\end{equation*}
Introducing the Fourier transform with respect to the normal coordinates we obtain
\begin{equation}	\label{ResponseFunction}
\begin{split}
	\mathcal F(\Omega) \approx \int_{\tau_0}^\tau \ud{T} \int \udpi[3]{{\vect p}}  G_+\boldsymbol(-\Omega,\vect p;x(T)\boldsymbol).
\end{split}
\end{equation}
It is easy to verify that for quasilocal vacuum states, whose Whightman functions are $G_+(p) = 2\pi \delta(p^2 + m^2) \theta(p^0) + O(L^{-2} \Omega^{-2})$ the response functions are $\mathcal F(\Omega) =  O(L^{-2} \Omega^{-2})$. For thermal states, 
$G_+(p) = 2\pi \delta(p^2 + m^2) [\theta(p^0) + n(|p^0|)] + O(L^{-2} \Omega^{-2})$ and the response function is given by $\mathcal F(\Omega) = [{\Omega n(\Omega)}/({2\pi})]\theta(\Omega-m) + O(L^{-2} \Omega^{-2})$.

Therefore, working with quasilocal or adiabatic vacua of order 1, the probability of detecting a particle decreases as $L^{-2} \Omega^{-2}$. If we worked with a quasilocal or adiabatic vacuum of order $n$, the probability of detecting a particle would decrease as $(L \Omega)^{n+1}$. Even with adiabatic vacuums of infinite order particles can be detected with a certain probability which is non-analytical in $(L \Omega)^{-1}$. We will encounter an example in the next chapter.


Let us finally compare the coupling $g_Q Q(\tau) \phi\boldsymbol(x(\tau)\boldsymbol)$, which we used in this section, with the coupling $g_Q Q(t)[ \phi_\vect p(t)+\phi_{-\vect p}(t)]$, used in subsection \ref{sect:QuasiCreation}. For inertial trajectories in the normal coordinates adapted to them the coupling $g_Q Q(\tau) \phi\boldsymbol(x(\tau)\boldsymbol)$ reduces to
$g_Q Q(t) \phi(t,\vect 0)$, which can be reexpressed as
\begin{equation}
	g_Q Q(t) \phi(t,\vect 0)	= g_Q Q(t)\sum_{\vect p} \phi_\vect p(t).
\end{equation}
Therefore, with the coupling used in this section, the detector couples to all field modes, whereas with the coupling used in subsect.~\eqref{sect:QuasiCreation} the field couples to the modes with momentum $\pm\vect p$. In any case, notice that for sufficiently large times the detector only couples with particles with energy $\Omega$. For large enough energies both couplings lead to the same results.
\index{Particle!in curved spacetimes|)}

\section{Interacting fields in curved spacetimes}

In the following we consider the general situation in which there is an interacting scalar field theory in a globally hyperbolic spacetime, characterized by some metric $g_{\mu\nu}$ in a generic state $\hat\rho$. For concreteness, we consider that the scalar field theory consists of one self-interacting field $\phi$, but all expressions could be trivially extended to any number of interacting scalar fields. 

Most work studying interacting fields over general curved backgrounds was developed during the late 70s and early 80s, and was focused on the study of the renormalizability of the theories (see \cite{BirrellDavies} and references therein). See  refs.~\cite{FriedmanEtAl92,HollandsWald01,HollandsWald02,HollandsWald03,HollandsWald05} for more recent works focusing on general properties of interacting fields in backgrounds. In the last years most work has been focused on de Sitter and anti-de Sitter spacetimes; see \eg\ \cite{BrosEtAl94,BrosMoschella96,EinhornLarsen03,BrosEtAl07} for some de Sitter references. We will not attempt to make any review of the subject, nor make a complete presentation of interacting quantum fields in curved backgrounds. We will simply highlight some aspects relevant for us.


\subsection{Propagators and self-energies}

In curved spacetimes, propagators are defined as in flat spacetime: see section \ref{sect:GSigmaGeneral}, and in particular eqs.~\eqref{CorrFunct}, \eqref{CorrFunct2} and \eqref{CorrFunct3}. They can similarly be organized in the $2\times2$ matrices $G_{ab}(x,x')$ (direct basis) or $G'_{a'b'}(x,x')$ (Keldysh basis). Propagators are biscalars and thus independent of the choice of the coordinate system ---in particular notice that the temporal orderings appearing in the definitions of some propagators are coordinate-independent. The general relations between the different propagators, derived from their definitions and detailed in appendix \ref{app:GenRel}, also hold in curved spacetimes. Notice also that in curved spacetime, even if the field is in an asymptotic vacuum state, asymptotic vacua do not generally coincide, and thus the CTP formalism is needed in every situation. Therefore considering just the Feynman propagator is usually not sufficient.

\index{Schwinger-Dyson equation}
Self-energies $\Sigma_{ab}(x,x')$ are similarly introduced through the Schwinger-Dyson equation:
\begin{equation} \label{SelfEnergyGeneralCurved}
\begin{split}
    G_{ab}(x,x') = G_{ab}^{(0)}(x,x') &+
    \int {\vd[4]{z}}\sqrt{-g(z)}\, \vd[4]{z'}\sqrt{-g(z')} \\ &\quad \times G_{ac}^{(0)}(x,z) [-i\Sigma^{cd}(z,z')]  G_{db}(z',x'),
\end{split}
\end{equation}
where $G^{(0)}_{ab}(x,x')$ are the propagators of the corresponding free theory. The self-energy components $\Sigma_{ab}(x,x')$ are also biscalars, and hence independent of the coordinate system. It will be useful to work with rescaled fields, propagators and self-energies as follows:
\begin{subequations}\label{rescaling}
	\begin{align}
		\bar\phi(x) &:= [-g(x)]^{1/4}, \\
		\bar G_{ab}(x,x') &:= [-g(x)]^{1/4}  G_{ab}(x,x') [-g(x')]^{1/4}, \\
		\bar \Sigma_{ab}(x,x') &:= [-g(x)]^{1/4}  \Sigma_{ab}(x,x') [-g(x')]^{1/4}.
	\end{align}
\end{subequations}
With these definitions $\bar\phi(x)$, $\bar G_{ab}(x,x')$ and  $\bar \Sigma_{ab}(x,x')$ are (bi)scalar densities of weight 1/2. In terms of the bar quantities, the relation between propagators and self-energies becomes identical to flat spacetime:
\begin{equation} \label{SelfEnergyGeneralCurved2}
\begin{split}
    \bar G_{ab}(x,x') = \bar G_{ab}^{(0)}(x,x') &+
    \int {\vd[4]{z}}\, \vd[4]{z'} \ \bar G_{ac}^{(0)}(x,z) [-i\bar\Sigma^{cd}(z,z')]  \bar G_{db}(z',x').
\end{split}
\end{equation}
As in flat spacetime, the retarded propagator obeys a direct relation with the retarded self-energy:
\begin{equation} \label{SelfEnergyGeneralRetardedCurved}
    \bar\GR(x,x') = \bar\GR^{(0)}(x,x')+
    \int \ud[4]{z} \ud[4]{z'} \bar\GR^{(0)}(x,z) [-i\bar\SigmaR(z,z')]  \bar\GR(z',x'),
\end{equation}

In general, all properties relating the propagators and self-energies explained in section \ref{sect:GSigmaGeneral} and  appendix \ref{app:GenRel} still apply in curved backgrounds; we shall not repeat them here. Those properties which are expressed in the Fourier space are also valid for curved spacetime when using a Fourier transformation in a system of normal coordinates centered around some point $x_0$ in the neighborhood of $x$ and $x'$:\footnote{To be precise, they are valid in the region of the spacetime which is adequately covered by the Riemann normal coordinates. Notice that they are valid to all order in the curvature and not only in the leading order expansion that we shall introduce in subsection \ref{sect:smallCurv}.}
\begin{equation}\label{localFourier}
	\bar G^{(x_0)}_{ab}(p;X) = \int \udpi[4]{y} \expp{-ip_a y^a}  \bar G_{ab}(Y+y/2,Y-y/2),
\end{equation}
where $Y+y/2$ and $Y-y/2$ are the points $x_1$ and $x_2$ in normal coordinates. When  the normal coordinates are chosen with origin at $X$, the above expression can be simplified to
\begin{equation}
	\bar G_{ab}(p;X) = \int \udpi[4]{y} \expp{-ip_a y^a}  \bar G_{ab}(y/2,-y/2),
\end{equation}
where we recall that $\bar G_{ab}(p;X)$ is a shorthand for $\bar G^{(X)}_{ab}(p;0)$. 

\subsection{Perturbation theory, divergences and  effective field theories in curved spacetimes}
\index{Renormalization}
\index{Effective field theory}
\index{Infrared divergences}

In principle, the perturbative evaluation of the propagators in curved spacetime can be performed in the same way as in Minkowski, taking into account the CTP doubling of the number of degrees of freedom. However, several difficulties arise in practice. First, in curved spacetime even the free propagators are not usually expressible in closed analytic form; in general, evaluation of loop corrections in curved spacetimes is technically challenging. Second, the interpretation of the propagators and self-energies in curved spacetimes does not necessarily coincide with that of flat spacetime. Finally, naive perturbation theory may be spoiled  both because of ultraviolet and infrared divergences.
We will see how to deal with the first and second points in the following sections. Let us now address the last point: the divergent structure of the theory in curved spacetime.

For a start one would be tempted to say that the ultraviolet divergences of an interacting field theory in a curved spacetime are the same as the divergences of the same field theory in flat spacetime, since for sufficiently high energy scales the spacetime looks flat. However this is not true, because there are subleading divergences which may depend on the local curvature of the manifold. In the case of the $\lambda\phi^4$ theory ---which is the most well studied case (see \cite{BirrellDavies} and references therein)---, besides the usual counterterms for the mass, the field strength and the coupling constant, counterterms for the conformal coupling parameter $\xi$ must be included to make the theory renormalizable in the old sense.

In general, in the spirit of effective field theories, one has to consider the most general Lagrangian which is compatible with the symmetries of the problem. For the case of $\lambda\phi^4$  the following Lagrangian should be considered:
\begin{equation}
\begin{split}
	S_\phi &= - \int \vd[4]{x} \sqrt{-g} \left[ \frac12 g^{\mu\nu} \partial_\mu \phi \partial_\nu\phi + \frac{1}{2} (m^2 + \xi R) \phi^2 + \frac{\lambda}{4!} \phi^4 \right]	
	 \\
	&\quad - \int \vd[4]{x} \sqrt{-g} \bigg[ \frac{8\pi C}{\Mp^2} (g^{\mu\nu} \partial_\mu \partial_\nu\phi)^2 + \frac{\alpha_1 R g^{\mu\nu} + \alpha_2 R^{\mu\nu}}{2\Mp^2} \partial_\mu \phi \partial_\nu \phi \\ &\qquad +  \frac{\beta_1 R}{4!\Mp^2} \phi^4 + \frac{\beta_2}{4!\Mp^2} (\partial_\mu \phi \partial^\mu\phi) \phi^2 \bigg] + O(1/\Mp^{4}).
\end{split}
\end{equation}
The coefficients $m$, $\xi$ and $\lambda$ correspond to relevant or marginal couplings, and the coefficients $C$, $\alpha_1$, $\alpha_2$, $\beta_1$ and  $\beta_2$ correspond to irrelevant couplings, and are thus suppressed by a higher mass scale (which in this case naturally corresponds to the Planck mass). All the coefficients are considered to be finite (renormalized), and therefore the action has to be supplemented by the corresponding infinite counterterms (including also a counterterm for the field strength). The effective field theory approach to gravity has been extensively studied by Donoghue and collaborators \cite{Bjerrum-BohrEtAl03, Donoghue94a,Donoghue94b,Donoghue95} and we have already used it in chapter 4. 

Notice that ultraviolet properties set a difference between fields in curved spacetimes and fields in non-trivial Minkowski backgrounds, because the ultraviolet properties of the latter correspond exactly to those of the Minkowski vacuum. Intuitively, the reason for that is the following: curvature effects decay as a power law as the energy increases (or the length decreases), while the occupation numbers decay exponentially with the energy for most physical states (such as thermal states).

Besides the divergences associated to the renormalization of the couplings of the theory, in curved spacetime another kind of ultraviolet divergences arise, associated to the stress-energy tensor and present even for free fields.  In flat spacetime the energy density of the vacuum is a divergent quantity; however this divergence can be trivially substracted by adopting a normal-ordering procedure. By contrast, in curved spacetime the subleading divergences mentioned above make the substraction much more involved. One needs to consider a generalized gravitational action including terms up to second order in the curvature (which can be also understood in the spirit of effective field theories). The understanding of the divergent structure of the  stress-energy tensor is a major cornerstone for the theory of semiclassical gravity, and much work has been devoted in the literature to the study of those divergences and the methods to regularize and renormalized them. They   constitute now a standard subject \cite{BirrellDavies,Fulling}, and we will not consider them any further.

\index{Infrared divergences}

We have so far considered the ultraviolet divergences, but infrared divergences may also arise. Calculations which are formally correct might be spoiled out by the divergent infrared behavior of the theory, even if final results are not explicitly infrared divergent.  As opposed to ultraviolet divergences, which share a common structure for all theories, infrared divergences depend on the large scale structure of the spacetime, and therefore not many general things can be said about them. There are additional reasons indicating that they may be a difficult subject to deal with. On the one hand, we already saw that, even in flat spacetime with a non-vacuum background state infrared divergences are rather subtle and in fact modify the usual perturbation theory.  On the other hand, in flat spacetime the resolution of the infrared divergences often depends on the particle interpretation of the far infrared modes of massless particles, and a particle interpretation does not generally exist for long wavelength modes in curved spacetime. Infrared divergences usually appear when dealing with masless (or effectively massless) fields. Heuristically, one can make sure that the infrared divergences do not play an important role by checking that the relevant contribution to all intermediate expressions is not governed by the far infrared modes. If this is the case, one can be reasonably confident of the infrared stability of the results; otherwise one has to deal with the infrared divergences in a case by case basis.

Notice that three different energy scales appear in the action: the curvature scale $L^{-1}$, the mass scale $m$, and the Planck scale $\Mp$. Furthermore, when one considers the dynamics of particles, a fourth energy scale appears: the particle energy $E$. So far we have exploited the fact that the Planck scale is much higher than the other scales.  In the following, wee shall mostly consider situations in which the energy of the particle is much lower than the Planck mass, but still larger than the curvature scale: $\Mp \gg E \gg L^{-1}$. Let us analyze the consequences of such scale division. 

\subsection{Small curvature approximation}\label{sect:smallCurv}

\longpage

Whenever the energy of the particles is much larger than the inverse curvature radius of the spacetime, the Schwinger-Dyson equation \eqref{SelfEnergyGeneralCurved} can be evaluated in a local momentum representation. Let us show it explicitly. First, reexpress eq.~\eqref{SelfEnergyGeneralCurved} in a normal system of coordinates $y^a$ around any point in the vicinity of $x$ and $x'$ (for concreteness, we  choose the point $X$ at half the geodetic distance of $x$ and $x'$):
\begin{equation} \label{SelfEnergyGeneralRiemann}
\begin{split}
    \bar G_{ab}(y/2,-y/2) &= \bar G_{ab}^{(0)}(y/2,y') \\ &\quad+
    \int \ud[4]{y''} \ud[4]{y'''} \bar G_{ac}^{(0)}(y/2,y'') [-i\bar\Sigma^{cd}(y'',y''')]  \bar G_{db}(y''',-y/2). \raisetag{2.5\baselineskip}
\end{split}
\end{equation}
Next, introduce the Fourier transform with respect to the $y$ coordinates in the above equation:
\begin{equation} \label{SelfEnergyGeneralMomentumRiemann}
\begin{split}
    \bar \GR^{(X)}(p;0) &= \bar \GR^{(0)(X)}(p;0)-i
    \int \ud[4]{y} \ud[4]{y'} \ud[4]{\Delta} \int \udpi[4]{q} \udpi[4]{q'} \udpi[4]{q''} \\ &\qquad\times \expp{i q\cdot(\Delta/2-y/2) - i q''\cdot(y/2+  \Delta/2) - i p \cdot \Delta} \\ &\qquad\times \bar \GR^{(0)(X)}{\big(q;\tfrac{\Delta/2+y/2}{2}\big)} \SigmaR^{(X)}{\big(q';0\big)}  \bar \GR^{(X)}{\big(q'';\tfrac{-\Delta/2+y/2}2\big)}.
\end{split}
\end{equation}
The different functions appearing in the above equation, which so far is universally valid, can be approximated around the point $X$ in the following way:
\begin{equation*}
	f^{(X)}(p;x) = f^{(X)}(p;0) + y^\mu \derp{f^{(X)}(p;x)}{{y^\mu}}\bigg|_{x=0} + O(x^2),
\end{equation*}
where $f$ stands either for the propagator or the self-energy and where $x$ is the distance from $X$ to the point where the function is evaluated. In the Fourier transform, the leading contribution to the integrals are given by coordinate separations $x^a$ which are of the order of $l_\text{int}$, where $l_\text{int}$ is the typical time or length scale in which interaction takes place. On the other hand, the scale of variation of the propagators and self-energies is $\partial f^{(X)}/\partial x \sim f^{(X)}/L$, where $L$ is the radius of curvature of the spacetime at the point $X$ (or the typical scale of homogeneity of the state, if this latter quantity is smaller). Whenever $l_\text{int} \ll L$, to first order in the inverse curvature radius the following result is obtained:
\begin{equation}\label{GSigmaCurved}
	\bar\GR(p;X) = \frac{-i}{p^2 + m^2 + \bar\SigmaR(p;X)} + O(l_\text{int}/L,p^{-2}L^{-2}) ,
\end{equation}
where we recall that $\GR(p;X) = \GR^{(X)}(p;0)$ and $\SigmaR(p;X) = \SigmaR^{(X)}(p;0)$. We will do a more careful analysis of the conditions of validity of the corresponding equation in the cosmological case. 

\longpage


\index{Spectral representation}

Most expressions derived in the flat case are also applicable to the first order quasilocal expression of  propagator. In particular, eqs.~\eqref{CorrFunct}, \eqref{CorrFunct2} and \eqref{CorrFunct3} also apply in curved spacetimes. A spectral representation can also be derived:
\begin{equation}
	\GR(p;X) = \int \frac{\vd k^0}{2\pi}   \frac{i G(k^0,\vect p;X)}{p^0 - k^0 + i\epsilon}.
\end{equation}
In the above equation the $k^0$ integral also runs over small energy values, where the quasilocal expansion is not justified. However, as we will argue in the next section, within the range of validity of the quasilocal expansion the spectral function will be significantly different from zero only for values of the energy much larger than the inverse curvature radius. The Pauli-Jordan propagator also admits an interpretation in terms of the excitation probability of the field operator:
\begin{equation} \label{SpectralDetailedCurved}
	G(p) \approx \sum_{\alpha,\beta} \rho_{\alpha}\abs{\langle\alpha|\hat\phi_{\vect p}|\beta\rangle}^{2} \left[ \delta(p^0 + E_{\alpha} - E_{\beta})-\delta(p^0 - E_{\alpha} + E_{\beta})\right].
\end{equation}
The above equation is approximate because it depends on the simultaneous diagonalization of the density matrix and Hamiltonian operators, diagonalization which is only approximate for non-stationary spacetimes or non-stationary states.

Within the first order quasilocal approximation for the propagator the QBM analogy that we presented in sect.~\ref{sect:OQS} is also of application for curved spacetimes.

\section{Interacting quasiparticles in curved spacetimes} \label{sect:InteractingQuasi}

\index{Quasiparticle!in curved spacetime|(}

In this section we  consider the generalization of the results of sect.~\ref{sect:FreeCurved} in two different ways. First, we extend them for the case  of interacting field theories. Second, we consider quasiparticle excitations over a generic background state $\hat \rho$. We do not give detailed derivations of the results which are analogous to the flat case and have been already discussed in chapter 3.

As in flat spacetime, we would like to define  quasiparticles as long-lived elementary excitations carrying some momentum and some energy, propagating in general situations. From the different approximations to the particle concept that we introduced in  the case of free fields, the ones that are more closely adapted to this idea are the quasilocal and adiabatic particle concepts, and, to lesser extent, the particle detector approach. In this section we will mostly concentrate on the quasilocal  approach, but at the end we will comment on the other approximations. The analysis of adiabatic particles in cosmology will be postponed to the next section.


\subsection{Interacting quasilocal quasiparticles}

\index{Quasiparticle!quasilocal}

If the region of interest is small enough it is sufficient to remain at leading order in the quasilocal expansion. It will be useful to recall the expressions for the free retarded propagator, which reduces to the Minkowski expression:
\begin{equation}
	\bar G^{(0)}_\text{R} (\omega,\vect p; x) =  \frac{-i}{-\omega^2 + \vect p^2 + m^2 - i \omega \epsilon} +
	 O(p^{-2} L^{-2}).
\end{equation}
In subsect.~\ref{sect:QuasiParticlesFree} we focused on the retarded propagator, which does not depend on the state for free fields. In this section we will make use of other propagators, such as the Hadamard propagator:
\begin{equation}
	\bar G^{(1)(0)} (\omega,\vect p; x) =  2\pi \delta(-\omega^2 + \vect p^2 + m^2) [1 + 2n_\vect p(x)]+
	 O(p^{-2} L^{-2}).
\end{equation}
where $n_\vect p(x)$ is the occupation number of the state with momentum $\vect p$ around the point $x$.

Let us now consider interacting quasiparticles. As in chapter 3, two different approaches can be used: a frequency-based approach, and a real-time approach. We start with the frequency analysis.
Assume that  quasilocal spectral function, evaluated around some point $x$ in some quantum state $\hat\rho$, is strongly peaked around some value $R_\vect p$:
\begin{equation}
	\bar G(\omega,\vect p;x) \approx \frac{2 Z_\vect p(x) \omega \Gamma_\vect p(x)}
	{[-\omega^2 + R_\vect p^2(x)]^2 + [\omega \Gamma_\vect p(x)]^2}, \quad \omega \sim R_\vect p.
\end{equation}
We assume that $Z_\vect p$, $\Gamma_\vect p$ and $R_\vect p$ change very slowly with the spacetime point $x$. The retarded propagator derived from the above spectral representation is
\begin{equation}\label{RetardedQL}
	\bar\GR(\omega,\vect p;x) \approx \frac{-i Z_\vect p(x)}
	{-\omega^2+R_\vect p^2(x)-i\omega\Gamma_\vect p(x)},\quad 	\omega \sim R_\vect p.
\end{equation}
Comparing with \Eqref{GSigmaCurved}, we obtain
\begin{subequations}\label{RGammaQL}
\begin{align}
	R_\vect p^2 (x) &= \Re \SigmaR\boldsymbol(R_\vect p(x),\vect p;x\boldsymbol), \\
	\Gamma_\vect p (x) &= -\frac{1}{R_\vect p(x)}\Im \SigmaR
\boldsymbol(R_\vect p(x),\vect p;x\boldsymbol).
\end{align}
\end{subequations}
As in flat spacetimes, $R_\vect p(x)$ can be interpreted as the quasiparticle energy and $\Gamma_\vect p(x)$ as the decay rate of the quasiparticles. The interpretation of $R_\vect p(x)$ is based on the spectral representation of the propagator, and  the interpretation of $\Gamma_\vect p(x)$ is based on the interpretation of the imaginary part of the self-energy explained in chapter 3.
\index{Decay rate}

Several different approximations  are implied by the above equations. Firstly, they correspond to the leading order quasilocal expansion and therefore there are corrections of the order of $R_\vect p^{-2} L^{-2}$, where we recall that $L$ is the radius of curvature of the spacetime. Secondly, it is implied that the interaction takes place in a region $l_\text{int}$ much smaller than the quasiflat region; there are corrections of the order of $l_\text{int}/L$. Finally, the form of the spectral function is merely an approximation around the quasiparticle energy $R_\vect p$, and, as in flat spacetime, in general there are additional contributions at higher energies corresponding to the multiparticle sector. 

One can wonder whether any further physical effects can be encoded in $R_\vect p$ and $\Gamma_\vect p$, beyond those found in flat spacetime. One can imagine contributions to the self-energy which depend on the local curvature of the spacetime. Those contributions need not be necessarily suppressed by a power $(R_\vect p L)^{-1}$, since there might by another sufficiently low mass scale in the theory $\dm$, such that $(\dm L)^{-1}$ is of order one. Indeed we will find this phenomenon in the field theory system we shall introduce in the next chapter. However, the interaction length scales associated to those terms will be of the order of the curvature radius of the spacetime, $t_\text{int}\sim L$, and therefore the quasilocal expansion \eqref{RetardedQL} would not be valid in those circumstances. We will see how to handle those contributions in the cosmological case.

\index{Quasiparticle!quantum state}
Let us now move to the complementary analysis in terms of the evolution of the quasiparticle state. As in flat spacetimes, the quantum state corresponding to quasiparticles is given by
\begin{equation}
	\hat\rho_\vect p^\pplus(t) \approx \frac{1}{n_\vect p(x) + 1} U(t,t_0) \hat a_\vect p^\dag(x) \hat\rho(t_0) \hat a_\vect p(x) U(t_0,t), \quad \text{(S.P.)}
\end{equation}
 and where the relation between the creation and annihilation operators and the field operator depends on the spacetime point: 
\begin{equation}
\hat\phi_\vect p = \frac{1}{\sqrt{2R_\vect p(x)}}\big[\hat a^\dag_\vect p(x) + \hat a_\vect p(x)\big]. \quad \text{(S.P.)}
\end{equation}
The label (S.P.) indicates Schrödinger picture. We assume that the field operator has been rescaled a factor $Z_\vect p^{1/2}(x)$. Recall that the  representation is only valid when evaluated inside an expectation value for sufficiently large time lapses. We shall assume that the spacetime can be divided in sufficiently large cells of size $l_\text{obs}$, which are much smaller than the local curvature radius and at the same time much larger than the inverse energy of the particles. Asymptotic expressions can be thought to hold inside these boxes. 

The time evolution of the quasiparticle energy can be computed as in flat spacetime (see subsect.~\ref{sect:Quasitime}). The energy of a quasiparticle following a geodesic in the vicinity of a point $x$ is
\begin{equation}
	E^\pplus(\tau;\vect p;x) = 
	E_\vect p^{(0)}+ [n_\vect p(x) +1] \expp{-\Gamma_\vect p(x)\tau}.
\end{equation}
The time $\tau$ must be chosen in a way that it is much longer than the typical interaction timescale, but that at the same time  much shorter than the typical size of the quasilocal patch. 

In summary, the properties of the particles in a quasilocal approximation closely correspond to those in flat spacetime provided the interaction takes place in a quasilocal patch much smaller than the local curvature radius, and provided also that we remain inside this patch.  The different relations implied can be summarized as follows: $R_\vect p^{-1}, l_\text{int} \ll l_\text{obs} \ll L$.

At the leading order in the curvature expansion no more physical effects are found beyond those present in a flat background, provided interaction times are small enough. One could think of incorporating curvature effects by going beyond  the linear order in the curvature expansion. This can be certainly done, but notice that in order for the quasilocal expansion to be meaningful both the interaction and the de Broglie length scales must be much smaller than the local curvature radius, and this therefore places a strong constraint to the validity of the quasilocal approximation to any order. 

\subsection{Particle detectors and other approaches}
\index{Particle detector}

\shortpage

In absence of interaction we introduced five different approaches to the particle concept, namely: asymptotic particles, global particles, quasilocal particles, adiabatic particles and particles as measured by detectors. When considering interaction, in this section we have focused on the quasilocal approach, and in the next section we will consider the adiabatic approach in cosmology, which is a natural extension of the quasilocal formalism allowing for arbitrarily large observation times. Let us now briefly comment on the extension of the other three approaches for interacting quasiparticles.

Asymptotic approaches to the particle concept in interacting situations are considered in ref.~\cite{BirrellDavies}. The  S-matrix approach of standard field theory in flat spacetime can be adapted with relatively minor modifications to the situation of having a curved spacetime with asymptotic regions. It is a suitable approach for studying, \eg, particle creation due to interaction. However, as we have already commented, most spacetimes do not have asymptotically flat regions.

Global approximations to particles interacting de Sitter have been pursued, for instance, in ref.~\cite{BrosEtAl07}. As we already mentioned, global particles do not necessarily resemble the usual Minkowskian particles.

Within the detector-based approach, recall that a given state is said to contain particles depending on the probability for a particle detector to get excited. In absence of interaction we argued that when the resonant frequency of the detector was much higher than the inverse radius of curvature of the spacetime, the response of the detector corresponded to the particle content of the state in the quasilocal approximation. In other words, whenever the quasilocal particle concept is meaningful, the quasilocal and particle detector approaches are  equivalent. 

Adding interaction does not significantly alter this image. In fact, the analysis of sect.~\ref{sect:ParticleDetectors}, and in particular the derivation of \Eqref{ResponseFunction}, is not limited to free fields. By developing \Eqref{ResponseFunction} one can realize that the response function of a particle detector in the presence of a quasilocal quasiparticle consists of two parts: a part depending on the background state, and a part depending on the quasiparticle excitation on top of that background. This latter part will only be present when the detector is tuned to the quasiparticle energy: $\Omega \approx R_\vect p$. This calculation closely follows the computation of the average value of the quasiparticle energy, which is summarized above and is detailed in subsect.~\ref{sect:Quasitime}; it will not be developed here. Notice that we have also implicitly used particle detectors (in the role of ``particle emitters'') when introducing the quasiparticle state.

\shortpage
To sum up, the detector and quasilocal approaches to the particle concept are equivalent provided the quasiparticle energy is large enough as compared to the curvature radius. However, particle detectors could be  interesting when extending the 
quasiparticle concept to  energies smaller than the inverse local curvature radius. 

Another completely different method is that of Drummond and Hathrell \cite{DrummondHathrell80} (see also the subsequent analysis by Shore \cite{Shore02a,Shore02b,Shore03,Shore03b}), who studied the quantum modifications to the trajectory of the light rays when propagating in a curved background. As we explained in sect.~\ref{sect:Alternative}, theirs was a mean field approach based on the analysis of the effective action, and therefore one cannot properly speak of particles or quasiparticles. The results of  sect.~\ref{sect:Alternative} can be easily adapted to curved spacetime. The equation of motion for the mean fields is\footnote{In chapter 3 we denoted the mean fields with a bar. Unfortunately in this chapter we have used the bar to indicate the rescaled fields. To avoid confusions, here the mean field will be simply denoted by $\phi$.} [see \Eqref{EqMotion}]
\begin{equation}\label{EqMotionCurved}
       (-\Box_x+m^2)\phi (x)+
     \int \vd[4]{y}  \, \sqrt{-g(y)}
    \Sigma_\mathrm R(x,y) \phi(y)  = 0.
\end{equation}
By introducing the Fourier transform in a system of normal coordinates we recover the correct dispersion relations
\begin{equation} \label{DispRelNECurved}
    p^2+m^2+ \Sigma_\mathrm R(p;x) = 0,
\end{equation}
although its interpretation is slightly different, as explained in \ref{sect:Alternative}. The advantage of Drummond and Hathrell's method is that the light ray trajectories can be followed beyond the quasilocal patch, so that the modifications to the null geodesics can be computed for long distances.


\section{Interacting quasiparticles in cosmology}

\index{Quasiparticle!adiabatic|(}

The results in last section show that quasiparticles in curved spacetimes, when treated in the leading order quasilocal expansion, exhibit essentially the same properties as in a flat background provided both the characteristic interaction and observation regions are much smaller than the locally flat region.
Let us see how in cosmology both restrictions can be lifted by considering particles in the adiabatic approximation.


In cosmological backgrounds the field operator can be expanded in conformal modes, $\phi(t,\vect x) = \sum_\vect k\phi_\vect k(t) \expp{i \vect k \cdot \vect x}$. In practice this means that each conformal mode can be analyzed separately and that we can exploit conservation of the conformal momentum.

Eq.~\eqref{SelfEnergyGeneralCurved}can be particularized to the cosmological case:
\begin{equation}
	\bar G_{ab}(t,t';\vect k) = \bar G_{ab}^{(0)}(t,t';\vect k) -i 
	\int \ud s \ud {s'}  \bar G_{ac}^{(0)}(t,s;\vect k) 
	\bar\Sigma^{cd}(s,s';\vect k) \bar G_{db}(s',t';\vect k),
\end{equation}
where the rescaled fields are $\bar \phi_\vect k(t) = a^{3/2}(t) \phi_\vect k(t)$. Two particular relations will be of interest: the one corresponding to the retarded propagator,
\begin{subequations}
\begin{equation}\label{Dyson}
	\bar \GR(t,t';\vect k) = \GR^{(0)}(t,t';\vect k) -i 
	\int \ud s \ud {s'}  \bar \GR^{(0)}(t,s;\vect k) 
	\bar\SigmaR(s,s';\vect k) \bar \GR(s',t';\vect k),
\end{equation}
and the one corresponding to the Hadamard function [see \Eqref{SelfEnergyHadamard}],
\begin{equation} \label{SelfEnergyHadamard2}
	\bar \GN(t,t';\vect k) = -i 
	\int \ud s \ud {s'}  \bar \GR(t,s;\vect k) 
	\bar\SigmaN(s,s';\vect k) \bar \GA(s',t';\vect k).
\end{equation}
\end{subequations}

\subsection{Short observation times}

In expanding universes the propagators, though space-translation invariant, are not
time-translation invariant. 
Nevertheless, we can always 
express the propagator 
in a
Fourier transform 
with respect to the difference variable $\Delta=t-t'$, 
while keeping  $T=(t+t')/2$ constant:
\begin{equation} \label{mixed}
	\bar\GR(\omega,T;\vect k):=\int \ud \Delta \expp{i\omega\Delta} 
	\bar\GR(T+\Delta/2,T-\Delta/2;\vect k) \, .
\end{equation}
By reexpressing \eqref{Dyson} in terms of the Fourier-transformed quantities we obtain the following cumbersome expression:
\begin{equation} \label{DysonFourier}
\begin{split}
	&\ \bar \GR(\omega,T;\vect k) = \bar\GR^{(0)}(\omega,T;\vect k) \\ &\quad - i \int \ud \Delta \ud s \ud{s'} \int \frac{ \vd{\omega_1} \vd{\omega_2} \vd{\omega_3}  }{(2\pi)^3} \expp{-i\omega_1(T-\Delta/2-s)-i\omega_2(s-s') - i\omega_3(s'-T-\Delta/2)}
	\\ &\qquad \times
	\bar\GR^{(0)}(\omega_1,\tfrac{T}{2} + \tfrac{\Delta}{4} + s;\vect k) \bar\SigmaR(\omega_2,\tfrac{s+s'}{2};\vect k) \bar\GR(\omega_3,\tfrac{s'}{2}+\tfrac{T}{2}-\tfrac{\Delta}{4};\vect k).
\end{split}
\end{equation}

We shall undertake several steps  to simplify this last equation, following what we did in the case of the quasilocal approximation, but taking further care in the approximations.
First, notice that, because of the use of the
retarded propagators,
the different times 
involved in the above relation are ordered as:
\begin{equation*}
	t=T + \frac{\Delta}{2} \geq s \geq s' \geq T - \frac{\Delta}{2}=t'.
\end{equation*}
Notice also that
the free retarded propagators and the self-energy can 
all be expanded around the time $T$ as
\begin{equation*}
	f(T+ \delta t) =f(T) + \delta t \derp{f(T)}{T} + O(\delta t^2).
\end{equation*}
The propagators and self-energies appearing in the second term in the right hand side 
of eq.~\eqref{DysonFourier} are evaluated at times which differ from $T$ by quantities 
$\delta t$ which are of the order of  the typical observation time $t_\text{obs}$ (for the case of the propagators) and of the order of $t_\text{int}$ (in the case of the self-energy). Therefore, if we are able to establish that
\begin{equation} \label{shortim}
	 \frac{t_\text{obs}}{\bar\GR}\derp{\bar\GR}{T}, 
	 \frac{t_\text{obs}}{\bar\GR^{(0)}}\derp{\bar\GR^{(0)}}{T}, \frac{t_\text{int}}{\bar\SigmaR}\derp{\bar\SigmaR}{T} 
	 \ll 1,
\end{equation}
then we can approximate the last line of eq.~\eqref{DysonFourier} by evaluating the three functions 
at time $T$. 
Under this assumption  \eqref{DysonFourier} can be greatly simplified  to
\begin{equation} \label{shortime}
	\bar\GR(\omega,T;\vect k) = \frac{ -i}
	{[-i \bar G^{(0)}(\omega,T;\vect k)]^{-1} + 
	\bar\SigmaR(\omega,T;\vect k)  }\, ,
\end{equation}
where 
the
free propagator is approximated by
\begin{equation}
	 [-i \bar G^{(0)}(\omega,T;\vect k)]^{-1} \approx - \omega^2 + \frac{\vect k^2 }{a^2(T)} +  m^2 .
\end{equation}
Thus, in order to check the validity of eq.~(\ref{shortime})
it is sufficient 
to check 
the inequalities
\eqref{shortim}. 
By differentiating the above equation, the variation of the free propagator can be approximated by:
\begin{equation}
	\frac{t_\text{int}}{\bar\GR^{(0)}} \derp{\bar\GR^{(0)}}{T}
	 \sim \frac{\vect k^2 t_\text{int}}{a^2(T) E_\vect k } H t_\text{int}  \ll 1,
\end{equation}
where we have used the time-energy uncertainty relation, $\omega - E_\vect{p} \sim 1/t_\text{obs}$. 
If we consider observation times $t_\text{obs}$ much smaller than the typical expansion timescale $H^{-1}$, 
and momenta $\vect k$ which are at most of the order of $(t_\text{obs}^2 H)^{-1}$ this relation is indeed verified. The self-energy evolves according to
\begin{equation}
	\frac{t_\text{int}}{\SigmaR} \derp{\bar\SigmaR}{T} 
	\sim t_\text{int} H \ll 1,
\end{equation}
which is also verified if $H^{-1} \gg t_\text{int}$. 
Finally, since the interacting propagator is obtained 
from the free propagator and the self-energy, 
it also verifies the inequality if the other two do.

It is important to stress that eq.~(\ref{shortim})
is stronger than the adiabatic approximation, since 
the latter only requires the masses to be much larger than the expansion rate. It actually corresponds to the quasilocal approximation developed in the previous section.
In the cosmological context we will refer to eq.~(\ref{shortim})
as the short-time approximation. 

\index{Expansion rate|see{Hubble parameter}}

Under this short-time assumption, one can find a time representation of 
the propagator by Fourier-transforming eq.~\eqref{shortime}. 
For long time differences, compared with the inverse mass, we can further approximate the 
self-energy by its value at the pole.
Assuming a small decay rate, one finds 
\begin{equation}\label{shortimetime}
	\bar G_\text{R}(t,t';\vect k) = \frac{-i}{ R_\vect k(T)} \sin\left[R_\vect k(T)(t-t')\right] \expp{-\Gamma_\vect k(T)(t-t')/2} \theta(t-t'),
\end{equation}
with
\begin{subequations}
\begin{equation} \label{R}
	R^2_\vect k(t) :=  m^2 + \frac{\vect k^2}{a^2(t)} + \Re\bar\SigmaR\boldsymbol(R_\vect k(t),T;\vect k\boldsymbol),
\end{equation}
and 
\begin{equation} \label{gamma}
	\Gamma_\vect k(t) := -\frac{1}{R_\vect k(t)} \Im\bar\SigmaR\boldsymbol(R_\vect k(t),t;\vect k\boldsymbol).
\end{equation}
\end{subequations}
Therefore one recovers the quasilocal expressions [see \Eqref{RGammaQL}].

Expressions for the other propagators are essentially identical to those we found in the quasilocal case, provided we take into account that the occupation number $n_\vect k$ is an adiabatic invariant and hence  time-independent. The construction and analysis of the quasiparticle states will be also equivalent to that in the previous section.

In summary, the short-time approximation, which is equivalent to the quasilocal approximation developed in the previous section, is valid provided the following scale separation is verified:
$
	R_\vect p^{-1}, t_\text{int} \ll t_\text{obs} \ll H^{-1}.
$
For ultrarelativistic particles there is an additional condition on the momenta: $|\vect p| \ll (t_\text{int}^2 H)^{-1}$.

\subsection{Long observation times}

\index{Propagator!adiabatic|(}
We wish to extend the above results for the case of arbitrarily large observation times, provided we remain in the adiabatic regime and the interaction time is small. This is to say, we want to extend the results of the previous subsection to the case $R_\vect p^{-1}, t_\text{int} \ll t_\text{obs} , H^{-1}$.

Let us consider the equation of motion of the interacting propagator.
By acting with the differential operator 
\[
	\frac{1}{a^3(t)} \derp{}{t}{} \left(a^3(t) \derp{}{t}{} \right) + m^2 + \xi R(t) + \frac{\vect k^2}{a^2(t)}
\]
on equation \eqref{Dyson}, we get the equation of motion for the retarded propagator:
\begin{equation} \label{EqRet}
\begin{split}
	&\left[ \frac{1}{a^3(t)} \derp{}{t}{} \left(a^3(t) \derp{}{t}{} \right) + m^2 + \xi R(t) + \frac{\vect k^2}{a^2(t)} \right]
	\left[ \frac{\bar\GR(t,t';\vect k)}{a^{3/2}(t) a^{3/2}(t')} \right]  \\
	&\qquad\qquad + \frac{1}{{a^{3/2}(t) a^{3/2}(t')} } \int \ud{s} \bar\SigmaR(t,s;\vect k)  \bar\GR(s,t';\vect k) = \frac{-i}{a^3(t)} \delta(t-t')
\end{split}
\end{equation}
This equation is exact and does not rely
on the adiabatic or short-time approximations. 
Our first task in this subsection is to find the 
WKB-like solution to this equation. 
Since we are interested in a first order adiabatic solution, 
we start by discarding all terms of the equation of motion which are of higher 
order in $H/R_\vect p$. We thus have 
\begin{equation} \label{eqc}
\begin{split}
	&\left[ \derp[2]{}{t} + m^2 + \frac{\vect k^2}{a^2(t)} \right]
	 \bar\GR(t,t';\vect k)  + \int \ud{s} \bar\SigmaR(t,s;\vect k)  \bar\GR(s,t';\vect k) = -i \delta(t-t').
\end{split}
\end{equation}
The equation of motion for the retarded propagator in the rescaled time $u = Ht$ is:
\begin{equation*}
\begin{split}
	\Big[ H^2 \derp[2]{}{u} &+ m^2 + \frac{\vect k^2}{a^2(u)} \Big]
	 \bar\GR(u,u';\vect k)  \\ &+ \frac{1}{H} \int \ud{v} \bar\SigmaR(u,v;\vect k)  \bar\GR(v,u';\vect k) = -i H \delta(u-u').
\end{split}
\end{equation*}
Expanding the non-local term as 
\begin{equation} \label{expansion}
\begin{split}
	N(u,u') &:=\frac{1}{H} \int \ud{v} \bar\SigmaR(u,v;\vect k)  \bar\GR(v,u';\vect k)\\ &=: \left[ \delta R^2_\vect k(u) + H \Gamma_\vect k(u) \derp{}{u} \right] \bar\GR(u,u';\vect k) + O(H^2),
\end{split}
\end{equation}
the equation of motion  can be approximated by
\begin{equation} \label{operatorRetardedAdiabatic}
	\left[ H^2 \derp[2]{}{u} + H \Gamma_\vect k(u) \derp{}{u} + R_\vect k^2(u)\right]
	 \bar\GR(u,u';\vect k)  = -i H \delta(u-u'),
\end{equation}
where $R_\vect k^2 = E_\vect k^2 + \delta R^2_\vect k$. The leading order adiabatic solution is (assuming that $\Gamma_\vect k$ is much smaller than $R_\vect k$)
\begin{equation*}
\begin{split}
	\bar G_\text{R}(u,u';\vect k) &= \frac{-i}{  \sqrt{ R_\vect k(u) R_\vect k(u')}} \sin\left({\frac{1}{H}\int^{u}_{u'} \ud{v} {R_\vect k(v)}} \right) \\ &\qquad \times \expp{-{\frac{1}{2H}\int^{u}_{u'} \ud{v} \Gamma_\vect k(v)} }  \theta(u-u'),
\end{split}
\end{equation*}
or, going back to the original time representation,
\begin{equation}\label{InteractingWKB}
\begin{split}
	\bar G_\text{R}(t,t';\vect k) &= \frac{-i}{  \sqrt{ R_\vect k(t) R_\vect k(t')}} \sin\left({\int^{t}_{t'} \ud{s} {R_\vect k(s)}} \right) \expp{-{{}\int^{t}_{t'} \ud{t} \Gamma_\vect k(t)/2} }  \theta(t-t').
\end{split}
\end{equation}
Compare eq.~\eqref{InteractingWKB} with eq.~\eqref{FreeRetardedAdiabatic}: the interacting propagator can be expressed in terms of the adiabatic evolution of the quasiparticle energy and decay rates.
\index{Decay rate}

Let us now investigate how $\delta R_\vect k$ and $\Gamma_\vect k$ can be expressed in terms of the self-energy. To this end, we reconsider the non-local term:
\begin{equation*} 
\begin{split}
	N(u,u')&= \frac1H \int \ud{v}\frac{-i \bar\SigmaR(u,v;\vect k)}{  \sqrt{ R_\vect k(v) R_\vect k(u')}} \sin\left({\frac{1}{H}\int^{u}_{u'} \ud{v'} {R_\vect k(v')}} \right) \\ &\qquad \times \expp{-{\int^{v}_{u'} \ud{v'} \Gamma_\vect k(v')/(2H)} }  \theta(v-u').
\end{split}
\end{equation*}
Splitting the integration range $[u',v]$ in the segment $[u,v]$ minus the segment $[u,u']$, and using that $\sin(A-B)= \sin A \cos B - \sin B \cos A$ we obtain
\begin{equation*} 
\begin{split}
	N(u,u')&= \frac1H \int \ud{v}\frac{-i \bar\SigmaR(u,v;\vect k)}{  \sqrt{ R_\vect k(v) R_\vect k(u')}}\expp{-{\int^{v}_{u'} \ud{v'} \Gamma_\vect k(v')/(2H)} }  \theta(v-u')\\
	 &\qquad \bigg[ \sin\left({\frac{1}{H}\int^{u}_{u'} \ud{v'} {R_\vect k(v')}} \right)  \cos\left({\frac{1}{H}\int^{u}_{v} \ud{v} {R_\vect k(v')}} \right) \\ &\qquad- \cos \left({\frac{1}{H}\int^{u}_{u'} \ud{v'} {R_\vect k(v')}} \right) \sin\left({\frac{1}{H}\int^{u}_{v} \ud{v} {R_\vect k(v')}}  \right) \bigg].
\end{split}
\end{equation*}
Since the typical interaction time $t_\text{int}$ is much smaller than the typical expansion time, the self-energy is different from zero only when $u-v \lesssim t_\text{int} \ll H^{-1}$, and in the above equation we may replace $v$ by $u$, everywhere except in the argument of the oscillating functions. Doing that we find:
\begin{equation*} 
\begin{split}
	N(u,u')&\approx \frac1H \int \ud{v}\frac{-i \bar\SigmaR(u,v;\vect k)}{  \sqrt{ R_\vect k(u) R_\vect k(u')}}\expp{-{\int^{u}_{u'} \ud{v'} \Gamma_\vect k(v')/(2H)} }  \theta(u-u')\\
	 &\qquad \bigg[ \sin\left({\frac{1}{H}\int^{u}_{u'} \ud{v'} {R_\vect k(v')}} \right)  \cos\left({\frac{1}{H}\int^{u}_{v} \ud{v'} {R_\vect k(v')}} \right) \\ &\qquad- \cos \left({\frac{1}{H}\int^{u}_{u'} \ud{v'} {R_\vect k(v')}} \right) \sin\left({\frac{1}{H}\int^{u}_{v} \ud{v'} {R_\vect k(v')}}  \right) \bigg],
\end{split}
\end{equation*}
or alternatively,
\begin{equation}  \label{NonLocalGeneral}
\begin{split} 
	N(u,u')&\approx \left[ \frac{1}{H}\int \ud v   \bar\SigmaR(u,v;\vect k) \cos\left({\frac{1}{H}\int^{u}_{v} \ud{v'} {R_\vect k(v')}} \right) \right]\bar\GR(u,u';\vect k) \\
	&\quad-\left[ \frac{H}{R_\vect k}\int \ud v   \bar\SigmaR(u,v;\vect k) \sin\left({\frac{1}{H}\int^{u}_{v} \ud{v'} {R_\vect k(v')}} \right) \right] \derp{\bar\GR(u,u';\vect k)}{u}.
\end{split}
\end{equation}
Comparing \Eqref{expansion} with \eqref{NonLocalGeneral}, and going back to the original time $t$, we identify
\begin{subequations}
\begin{align}
	\delta R^2_\vect k(t) &= \int \ud s   \bar\SigmaR(t,s;\vect k) \cos\left({\int^{t}_{s} \ud{s'} {R_\vect k(s')}} \right), \\
	\Gamma_\vect k(t) &= -\frac{1}{R_\vect k}\int \ud {s}   \bar\SigmaR(t,s;\vect k) \sin\left({\int^{t}_{s} \ud{s'} {R_\vect k(s')}} \right) .
\end{align}
\end{subequations}
Since the typical interaction time $t-s$ is much smaller than the local curvature radius, the energies and decay rates are approximately constant during the interaction process and  the above equations can be reexpressed as 
\begin{subequations} \label{quasilocaldRGamma}
\begin{align}
	\delta R^2_\vect k(t) &= \int \ud \Delta  \bar\SigmaR(t+\Delta/2,t-\Delta/2;\vect k) \cos\left[  R_\vect k(t) \Delta \right] ,\\
	\Gamma_\vect k(t) &= -\frac{1}{R_\vect k} \int \ud \Delta
	\bar\SigmaR(t+\Delta/2,t-\Delta/2;\vect k) \sin\left[  R_\vect k(t) \Delta \right],
\end{align}
\end{subequations}
or, what is the same, we recover eqs.~\eqref{R} and \eqref{gamma}.

The equations for the time evolution of other propagators can be similarly derived. For instance, in appendix \ref{app:adiabaticHadamard} we show that 
the time evolution of the Hadamard function is given by
\begin{equation}\label{adiabaticHadamard}
	G^{(1)}(t,t';\vect k) = \frac{ 1+ 2n_\vect k }{\sqrt{R_\vect p(t)R_\vect p(t')}} \cos\left({\int^{t}_{t'} \ud{s} {R_\vect k(s)}} \right) \expp{-\left|\int^{t}_{t'} \ud{s} \Gamma_\vect k(s)/2\right| }.
\end{equation}
where $n_\vect k$ is the occupation number of the state with conformal momentum $\vect k$. For comparison, the corresponding free propagator is:
\begin{equation}\label{HadamardAdiabatic}
\begin{split}
	(\bar G^{(1)})^{(0)}(t,t';\vect k) = \frac{1+2n_\vect k}{  \sqrt{ E_\vect k(t) E_\vect k(t')}} \cos\left({\int^{t}_{t'} \ud{s} E_\vect k(s)}\right). 
\end{split}
\end{equation}

In general, the above equations require small interaction times:
$
	R_\vect p^{-1},t_\text{int} \ll H^{-1}, t_\text{obs}.
$
However for non-relativistic particles the requirement that the interaction time is small as compared to the typical expansion time can be somewhat relaxed, since the precise condition  is
\begin{equation}
	\frac{t_\text{int}}{R_\vect p}\derp{R_\vect p}{t} \sim \frac{\vect k^2/a^2}{R_\vect p} t_\text{int} H \ll 1.
\end{equation}

Notice that, beyond the adiabatic and short interaction time approximations, eqs.~\eqref{InteractingWKB} and \eqref{quasilocaldRGamma} also assume that the self-energy can be adequately approximated by a local expansion. We expect this approximation to be valid for sufficiently large times, similarly as in in flat spacetime, where we saw that the near-pole approximation was valid for long times, leading to the well known exponential decay. It is possible  however that the non-local contributions to the self-energy might be dominant for extremely long times, similarly as it happens in flat spacetimes, where the exponential decay law ultimately breaks. However, the flat spacetime analogy suggests that when \Eqref{InteractingWKB} might loose its validity the quasiparticle has completely decayed.


\index{Quasiparticle!quantum state}

Let us now consider the quantum states corresponding to quasiparticle excitations in an adiabatic context. The field operator can be decomposed in the creation and annihilation operators through (see subsect.~\ref{sect:Asympt})
\begin{equation}
	\bar\phi_\vect p \approx \frac{1}{\sqrt{2 R_\vect p(t)}} \big[ \hat a_\vect p(t) + \hat a^\dag_{-\vect p}(t)\big]. \quad \text{(S.P.)}
\end{equation}
Several remarks are in order. First, recall that the symbol (S.P.) indicates that all operators appearing are in the Schrödinger picture, despite the time dependence of the creation and annihilation operators. Second, notice that strictly speaking this representation is of application for the asymptotic fields, not for the interacting field itself (see subsect.~\ref{sect:QuasiCreation} for further details). However, as we did in the flat spacetime case, we will blurry the distinction between the interacting and asymptotic fields, taking into account that the above relation is only an approximate equality when evaluated inside  a matrix element for large time lapses. Finally, notice that, as in the previous subsection, we have already assumed that the field has been rescaled a factor $Z_\vect p^{-1/2}$. With all this in mind, the quasiparticle state can be constructed: 
\begin{equation}
	\hat\rho^\pplus_\vect p(t) \approx \frac{1}{n_\vect p + 1}\, 
U(t,t_0)\hat a^\dag_\vect p(t_0) \hat\rho(t_0) \hat a_\vect p(t_0) U(t_0,t). \quad \text{(S.P.)}
\end{equation}
The background state is assumed to evolve in cosmological timescales.

The time evolution of the expectation value of the Hamiltonian is given by
\begin{equation}
	E^\pplus(t,t_{0};\vect p) =  \frac{1}{ n_\vect p + 1} \Tr{ \big[\hat a_\vect p^\dag(t_0)\hat\rho(t_0)\hat a_\vect p(t_0) U(t_0,t)\hat H_\vect p U(t,t_{0})   \big] }.
\end{equation}
where the Hamiltonian of the 2-mode $\pm \vect p$ is given by $
	H_\vect p =  \dot \phi_\vect p \dot \phi_{-\vect p} + R_\vect p^2  \phi_\vect p \phi_{-\vect p}
$. Following similar steps as in subsect.~\ref{sect:Quasitime}, and in particular implicitly using the QBM analogy, but taking into account  the additional time dependence, we find
\begin{equation}\label{EnergyGuayAdiabatic}
	E^\pplus(t,t_{0};\vect p) \approx E^{(0)}(t) +  \frac{R_\vect p(t)}{n_\vect p + 1}
	\left|\frac{[R_\vect p(t) + i\partial_t][R_\vect p(t_0) -i \partial_{t_0}]}{2\sqrt{R_\vect p(t)R_\vect p(t_0)}} G_+(t,t_0)\right|^2,
\end{equation}
where $E^{(0)}(t)$ is the energy of the background state. This is the expression analogous to \Eqref{EnergyGuay} for cosmological backgrounds.
In the adiabatic approach we cannot use the Fourier transform methods;  however, by taking into account that $G_+(t,t') = [\GR(t,t') + G^{(1)}(t,t')]/2$ for $t>t'$, introducing eqs.~\eqref{InteractingWKB} and \eqref{HadamardAdiabatic} in \eqref{EnergyGuayAdiabatic} we find 
\begin{equation} 
	E^\pplus(t,t_{0};\vect p) \approx E^{(0)}_\vect p(t)+ R_\vect p(t)(1+n_\vect p)\expp{-\int_{t_0}^t \ud s \Gamma(s)}.
\end{equation}
From this equation we identify $R_\vect p(t)$ as the energy of the quasiparticles, which is slowly evolving with the background, and $\Gamma_\vect p(t)$ as their decay rate. Remember that the factor $(1+n_\vect p)$ is caused by the fact that the quasiparticle state actually contains more than one particle excitation.

\index{Dispersion relation}

Notice that even when the observation times are long the particle properties can be adequately condensed in a dispersion relation
\begin{equation}
\begin{split}
	\mathcal E^2(t) &= m^2 + \frac{\vect k^2}{a^2(t)} + \SigmaR\boldsymbol(R_\vect k(t),t;\vect k\boldsymbol) =  R^2_\vect p(t) - i R_\vect k(t) \Gamma_\vect k(t).
\end{split}
\end{equation}

\subsection{Long interaction and observation times}

So far we have restricted the analysis to interaction times much smaller than the typical curvature radius. However, there are interesting situations in which the typical interaction timescale is of the order of the curvature radius, such as the case of the quantum effects induced by the curvature. Therefore, it is of interest to consider the extension of the above results to the case of interaction times of the order of the expansion timescale $H^{-1}$, requiring that the observation times are much longer than the interaction time. We will therefore consider the situation  $R_\vect p^{-1} \ll t_\text{int},  H^{-1} \ll t_\text{obs}$, making also  the approximation that the total amount of decay during the interaction time is negligible (otherwise the particle would have completely decayed at the observation point). 

Since the particle will be observed at large times it makes sense to approximate the non-local term by a local expansion:
\begin{equation} \label{expansionL}
\begin{split}
	N(t,t') &:=\frac{1}{H} \int \ud{v} \bar\SigmaR(u,v;\vect k)  \bar\GR(v,u';\vect k)\\ &=: \left[ (\delta R\ret_\vect p)^2(u) + H \Gamma\ret_\vect k(u) \derp{}{u} \right] \bar\GR(u,u';\vect k) + O(H^2).
\end{split}
\end{equation}
The ``ret'' indication has been added for later convenience. Notice that the reescaled times verify $u-u' \gg 1$. The solution to the equation of motion is therefore the same as before
\begin{equation}\label{InteractingWKBL}
\begin{split}
	\bar G_\text{R}(t,t';\vect k) &= \frac{-i}{  \sqrt{ R\ret_\vect k(t) R\ret_\vect k(t')}} \sin\left({\int^{t}_{t'} \ud{s} {R\ret_\vect k(s)}} \right) \\ &\quad \times \expp{-{{}\int^{t}_{t'} \ud{t} \Gamma\ret_\vect k(t)/2} }  \theta(t-t'),
\end{split}
\end{equation}
\shortpage
where $[R\ret_\vect k(t)]^2 = E^2_\vect k(t) + [\delta R\ret_\vect k(t)]^2$. We have so far recovered a result analogous to that in the previous section. 
Let us now determine  how $\delta R\ret_\vect k$ and $\Gamma\ret_\vect k$ can be expressed in terms of the self-energy, by reconsidering the non-local term. Splitting similarly the integration range $[u',v]$ in the segment $[u,v]$ minus the segment $[u,u']$ yields
\begin{equation*} 
\begin{split}
	N(u,u')&= \frac1H \int \ud{v}\frac{-i \bar\SigmaR(u,v;\vect k)}{  \sqrt{ R\ret_\vect k(v) R\ret_\vect k(u')}}\expp{-{\int^{v}_{u'} \ud{v'} \Gamma_\vect k(v')/(2H)} }  \theta(v-u')\\
	 &\qquad \bigg[ \sin\left({\frac{1}{H}\int^{u}_{u'} \ud{v'} {R\ret_\vect k(v')}} \right)  \cos\left({\frac{1}{H}\int^{u}_{v} \ud{v} {R\ret_\vect k(v')}} \right) \\ &\qquad- \cos \left({\frac{1}{H}\int^{u}_{u'} \ud{v'} {R\ret_\vect k(v')}} \right) \sin\left({\frac{1}{H}\int^{u}_{v} \ud{v} {R\ret_\vect k(v')}}  \right) \bigg].
\end{split}
\end{equation*}
Now we may no longer replace $v$ by $u$ in the energy prefactors since the energy may have significantly changed during the interaction time. We may still make the replacement in the argument of the exponential function since we have made the hypothesis that the total amount of decay during the interaction time is negligible. We thus find: 
\begin{equation}  \label{NonLocalGeneralL}
\begin{split} 
	N(u,u')&\approx (\delta R\ret_\vect k)^2(u) \bar\GR(u,u';\vect k) +H \Gamma\ret_\vect k(u)  \derp{\bar\GR(u,u';\vect k)}{u},
\end{split}
\end{equation}
where in non-rescaled time $t$,
\begin{subequations}\label{GammaRSuperGen}
\begin{align}
	(\delta R\ret_\vect k)^2(t) &= \int \ud s   \frac{R_\vect k\ret(t)\bar\SigmaR(t,s;\vect k) }{\sqrt{R\ret_\vect k(t)R\ret_\vect k(s)}} \cos\left({\int^{t}_{s} \ud{s'} {R\ret_\vect k(s')}} \right), \\
	\Gamma\ret_\vect k(t) &= -\int \ud {s} \frac{\bar\SigmaR(t,s;\vect k)}{\sqrt{R\ret_\vect k(t)R\ret_\vect k(s)}}   \sin\left({\int^{t}_{s} \ud{s'} {R\ret_\vect k(s')}} \right).
\end{align}
\end{subequations}
Several remarks should be done with respect to the above expressions.
First, notice that the self-energy is evaluated in a kind of frequency representation, but with the  frequency varying along the interaction range, because the on-shell position changes signifficantly during the interaction time. Second, due to the retarded character of the equations, the frequency representation is not evaluated around time $t$, but rather it is evaluated in a time interval ending at time $t$.  Finally, notice also the presence of the square root prefactor, which accounts for the fact that the propagator prefactor significantly changes during the interaction time.

\index{Decay rate}
Let us concentrate on the decay rate in the following. Recalling that the above quantities will only be meaningful when observed for large periods of time, let us average the decay rate for timescales much smaller than the observation time. In other words, we choose a timescale $t_\text{av}$ much larger than $H^{-1}$ but still much smaller than $t_\text{obs}$, and average the decay rate over that scale:\begin{equation}
\begin{split}
	\Gamma_\vect k(t) &=\frac{1}{t_\text{av}}\int_{t-t_\text{av}/2}^{t+t_\text{av}/2} \ud{t'} \Gamma\ret_\vect k(t') \\
	  &=-\frac{1}{t_\text{av}}\int_{t-t_\text{av}/2}^{t+t_\text{av}/2}\ud{t'}\int \ud {s} \frac{\bar\SigmaR(t',s;\vect k)}{\sqrt{R\ret_\vect k(t')R\ret_\vect k(s)}}   \sin\left({\int^{t'}_{s} \ud{s'} {R\ret_\vect k(s')}} \right)
\end{split}
\end{equation}
Taking into account that $t_\text{av}$ is much larger than the interaction time, and introducing the semisum and semidifference coordinates we get:
\begin{equation*}
\begin{split}
	\Gamma_\vect k(t) 
	  &\approx-\frac{1}{t_\text{av}}\int_{t-t_\text{av}/2}^{t+t_\text{av}/2}  \ud{T}\int \ud {\Delta} \frac{\bar\SigmaR(T+\Delta/2,T-\Delta/2;\vect k)}{\sqrt{R\ret_\vect k(T+\Delta/2)R\ret_\vect k(T-\Delta/2)}}  \\ &\qquad\times \sin\left({\int^{T+\Delta/2}_{T-\Delta/2} \ud{s'} {R\ret_\vect k(s')}} \right),
	  \end{split}
\end{equation*}
or, equivalently
\begin{equation}
\begin{split}
	\Gamma_\vect k(t) 
	  &\approx-\Im \int \ud {\Delta}  \frac{\bar\SigmaR(t+\Delta/2,t-\Delta/2;\vect k)}{\sqrt{R_\vect k(t+\Delta/2)R_\vect k(t-\Delta/2)}} \expp{i\int^{t+\Delta/2}_{t-\Delta/2} \ud{s'} {R_\vect k(s')}},
	  \end{split}
\end{equation}
where in the above equation we removed the ``ret'' indication for the easy of notation.

The above equation corresponds to the averaged decay rate of a particle as observed in the cosmological reference frame. The square root of the product of energies can be interpreted as the on-shell condition evolving during the interaction time. 
It is therefore natural to conjecture that the decay rate in the particle rest frame is given by:
\begin{equation}
\begin{split}
	\gamma_\vect k(t) 
	  &=-\frac{1}{m}\Im \int \ud {\Delta} \bar\SigmaR(t+\Delta/2,t-\Delta/2;\vect k)  \expp{i\int^{t+\Delta/2}_{t-\Delta/2} \ud{s'} {R_\vect k(s')}}.
	  \end{split}
\end{equation}
\index{Improved frequency representation}

In terms of the following ``improved'' frequency representation for the self-energy,
\begin{equation}\label{frequencyImproved}
	\SigmaImp([R_\vect k],t;\vect k) := \int \ud \Delta \bar\SigmaR(t+\Delta/2,t-\Delta/2;\vect k) \expp{i\int_{t-\Delta/2}^{t+\Delta/2} \ud s R_\vect k(s)}
\end{equation}
the decay rate in the particle rest frame can be expressed as
\begin{equation}
	\gamma_\vect k = -\frac1m \Im\SigmaImp([R_\vect k],t;\vect k). \end{equation}
Notice that the improved frequency representation has been defined only for on-shell values of the frequency; in this sense it cannot be considered a proper integral transform.  

To end, let us notice that for comoving particles the improved frequency representation yields the same results as the standard frequency representation.

\index{Quasiparticle!adiabatic|)}
\index{Quasiparticle!in curved spacetime|)}
\index{Propagator!adiabatic|)}
	\chapter[Dissipative effects in the cosmological propagation]{Dissipative effects in the cosmological propagation: a three-field model}

\index{3-field model}

\longpage

In this chapter we study the quantum propagation of 
particles in a cosmological background, applying the results of the previous chapter. We are particularly interested in 
understanding 
the dissipative phenomena  
related to the time
dependence of the 
metric. To this end, we analyze a massive particle interacting with a massless radiation field in an expanding universe. As we discussed in the introduction, this issue 
may play an important role in justifying the non-trivial dispersion relations which have been used when addressing the trans-Planckian question in cosmology. To extract the dissipative effects, we will compute the imaginary part of the self-energy, first when the massless field is at finite temperature, and then when the massless field is in the vacuum.

As explained in chapters 4 and 5, in 
theories such as QED or perturbative quantum gravity,  dissipative effects 
appear 
only
at two loops in flat spacetime, 
because the one-loop 
diagrams which  
could have
led to dissipation 
vanish on the mass shell. 
Although in curved spacetime this restriction is lifted in some  situations (since energy needs not be conserved in an expanding universe),
we prefer to analyze a field theory system whose dissipative properties in flat spacetime can be understood at one loop, 
in order to be able to compare the curved and the flat spacetime results. For this reason we use the same three-field model which we analyzed in chapter 5.

We 
will adopt the theoretical framework described in the previous chapter. In particular, the masses of the fields will be assumed to be much larger than the expansion rate of the universe. This is a key assumption, because it allows to introduce the adiabatic (WKB) approximation, which not only makes the problem solvable, but also allows for a well-defined particle concept, as explained in the previous chapter. In order to have interesting dynamics, the mass gap between the two massive states will be taken to be much lower than the masses of the fields.

\section{The model}

We 
consider 
spatially isotropic and homogeneous
Friedmann-Lemaître-Robertson-Walker (FLRW) models with flat spatial sections. 
The metric can be expressed as
\begin{subequations}
\begin{equation}
	\vd s^2 = - \vd t^2 + a^2(t) \vd {\vect x}^2,
\end{equation}
in physical time $t$, or as 
\begin{equation}
	\vd s^2 =  a^2(\eta) (-\vd \eta^2 + \vd {\vect x}^2),
\end{equation}
\end{subequations}
in conformal time $\eta$, 
where $a \vd \eta = \vd t$.
The scale factor $a$ is 
arbitrary for the moment.

We introduce the same model as in chapter 5, namely
two massive fields $\phim$, 
and $\phiM$, interacting with a massless field, $\chi$, via a trilinear coupling. 
The total action 
is
\begin{subequations}
\begin{align}
	S &= S_m + S_M + S_\chi + S_\text{int}, \\
	S_m &= \frac{1}{2} \int \ud t \ud[3] {\vect x} a^3(t)  \left( (\partial_t \phim)^2 - \frac1{a^2(t)} (\partial_\vect x \phim)^2 - m^2 \phim^2 \right),\\
	S_M &= \frac{1}{2} \int \ud t \ud[3] {\vect x} a^3(t)  \left( (\partial_t \phiM)^2 - \frac1{a^2(t)} (\partial_\vect x \phiM)^2 -  M^2 \phiM^2 \right),\\
	S_\chi &= \frac{1}{2} \int \ud t \ud[3]{\vect x} a^3(t) 
	\left(  (\partial_t \chi)^2 - \frac1{a^2(t)} 
	(\partial_\vect x \chi)^2 - \xi R(t) 
	\chi^2 \right) ,\\
	S_\text{int} &= g M \int \ud t \ud[3]{\vect x} a^3(t)  \phim \phiM \chi,
\end{align}
\end{subequations}
where $R(t)$ is the Ricci scalar. We 
assume that the massless field is conformally coupled to gravity, so that $\xi = 1/6$.

We consider the two massive fields with large masses but with a small mass difference $\dm := M-m \ll M $. 
As it is shown in chapter 5, the model can be 
interpreted as a 
field-theory description of a 
relativistic
two-level atom 
(of rest mass $m$ and energy gap $\dm$)
interacting with a 
scalar radiation field $\chi$. 
This model was
used in ref.~\cite{Parentani95} to study the recoil
effects of an accelerated 
two-level atom subject to the Unruh effect.

The radiation field $\chi$ is assumed to be at some conformal temperature $\theta$. The corresponding physical temperature is chosen to be 
much smaller than the masses of the fields.
We also choose the 
Hubble rate $H(t) := \dot a(t)/a(t)$ to be 
much smaller than the masses, $M,m \gg H$.
These restrictions ensure that
the number of massive particles is strictly conserved.
The non-trivial dynamics concerns the transitions between the two massive 
fields accompanied by emission/absorption of massless quanta.

As we have seen in the previous chapter,
it is 
useful to 
work with rescaled massive fields 
defined by, $\bar\phi (t,\vect x):= [-g(t,\vect x)]^{1/4}\phi(t,\vect x) = 
a^{3/2}(t) \phi(t,\vect x)$. In terms of these new fields 
the quadratic actions become
\begin{subequations}
\begin{equation}
	S_m = \frac12 \int \ud t \ud[3] {\vect x}  \left( a^3(t) 
	\{\partial_t [ a^{-3/2}(t) \bar\phim]\}^2 - \frac1{a^2(t)} 
	(\partial_\vect x \bar\phim)^2 -  
	m^2
	\bar\phim^2 \right),
\end{equation}
and similarly for the massive field $\bar\phiM$, whereas the interaction term
becomes	
\begin{equation}
	S_\text{int} = g M \int \ud t \ud[3]{\vect x}   \bar\phim \, \bar\phiM \, 
	\chi. \label{BarInt}
\end{equation}
\end{subequations}

\index{Propagator!adiabatic approximation}
\index{Propagator!conformal fields}
Recall that in a curved spacetime it is not a trivial task
to compute free vacuum 
propagators. 
For massless conformally coupled fields there is a natural vacuum state,
the conformal vacuum. Propagators in this vacuum, when expressed in conformal time, essentially correspond to the flat spacetime propagators. As a first example, the retarded propagator for the radiation field is given by
\begin{equation}
	\Delta_\mathrm{R}^{(0)}(\eta_1,\eta_2;\vect k)  =  \frac{-i}{ a(\eta_1) a(\eta_2)  k} 
	\sin{[(\eta_1-\eta_2) k ]}\, 
\theta(\eta_1-\eta_2),
\end{equation}
where $k=|\vect k|$. In physical time the corresponding expression is simply:
\begin{equation}
	\Delta_\mathrm{R}^{(0)}(t_1,t_2;\vect k) =    
	\frac{-i}{ a(t_1) a(t_2)  k} \sin{\left( \int_{t_2}^{t_2} 
	 \ud t \frac{k  }{a(t)} %
	  \right)}\, \theta(t_1-t_2), %
 \end{equation}
All other propagators can be obtained in a completely analogous way. 
As a second example,
the positive Wightman function in a 
thermal bath is
\begin{equation}
	\Delta_+^{(0)}(\eta_1,\eta_2;\vect k) =    \frac{1}{ a(\eta_1) a(\eta_2)  2k} \left( \expp{-i(\eta_1-\eta_2) k }[1+ n_\theta(k)] + \expp{i(\eta_1-\eta_2) k} n_\theta(k) \right)
\end{equation}
where $n_\theta(\varepsilon)$ is the Bose-Einstein function corresponding to 
a (constant) 
conformal temperature $\theta$:
\begin{equation}
	n_\theta(\varepsilon) := \frac{1}{\expp{\varepsilon/\theta} - 1}.
\end{equation}

For the
massive 
fields, 
rather than attempting to find the exact 
propagator, we will exploit 
the fact that their de Broglie wavelengths 
are much smaller than 
the Hubble 
length $H^{-1}$.  
In this regime, the adiabatic 
approximation, described in the previous chapter, is valid.
Since translation invariance along the spatial axes is preserved, 
we will study propagators of modes labeled by 
their comoving (and conserved) momentum $\vect k$. The free retarded propagator is given by
\Eqref{FreeRetardedAdiabatic}, and the positive Whightman function in the vacuum is
\begin{equation}\label{FreeWhightmanAdiabatic7}
\begin{split}
	\bar G^{(0)}_+(t_1,t_2;\vect k) = \frac{1}{  2\sqrt{ E_\vect k(t_1) E_\vect k(t_2)}} \expp{-i\int^{t_1}_{t_2} \ud{t'} E_\vect k(t')}  .
\end{split}
\end{equation}

\section{Interacting propagators. The self-energy.}

The aim of this section is to compute the
interacting retarded Green function 
\begin{equation}
\bar G_\text{R}(t,t';\vect k)=\theta(t-t')
\langle{[\hat{\bar\phi}_{m\vect k}(t),\hat{\bar\phi}_{m\vect k}(t')]}\rangle 
\end{equation}
within the adiabatic approximation.
We will use perturbation theory in the CTP framework. 
We will not 
detail the aspects 
related to 
the CTP method, which we have already described in chapters 3, 4 and 5 and appendix \ref{app:CTP}.
Instead, we 
concentrate on the novel aspects
induced by 
the universe expansion.

\begin{figure}
	\centering
	\includegraphics[width=0.50\textwidth]{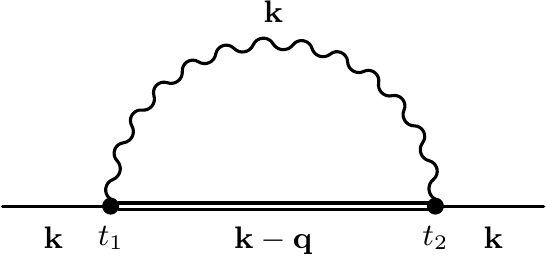}
	\caption{Feynman diagram leading to the one-loop self-energy of the less massive particle. Straight lines represent the fundamental sate, double lines the excited state and curly lines the massless particle.} \label{fig:SigmaCurved}
\end{figure}

In the previous chapter we saw that the retarded propagator is related to the retarded self-energy through eq.~\eqref{SelfEnergyGeneralRetardedCurved}. To one loop, the self-energy can be computed by using CTP perturbation theory as (see figure \ref{fig:SigmaCurved})
\begin{equation}
	- i \bar\Sigma^{ab}(t_1,t_2;\vect k) ={(ig M)^2} c^{aa} c^{bb} \int \udpi[3]{\vect q} \bar G^{*(0)}_{ab}(t_1,t_2;\vect k-\vect q)  \Delta^{(0)}_{ab}(t_1,t_2;\vect q) ,
\end{equation}
where we recall that $a,b,c\ldots$ indices refer to the different CTP branches, $c^{ab}= \text{diag}(1,-1)$; there is no implicit summation in the above equation. The retarded self-energy corresponds to $\SigmaR(t_1,t_2;\vect k) = \Sigma^{11}(t_1,t_2;\vect k) - \Sigma^{12}(t_1,t_2;\vect k)$. 

\longpage

The self-energy can be evaluated either in the standard or improved frequency representations (see the previous chapter):
\begin{subequations}
\begin{align}
	\bar\Sigma^{ab}\boldsymbol(E_\vect k(T),T;\vect k\boldsymbol) &= \int \ud \Delta \bar\Sigma^{ab}(T+\tfrac\Delta2,T-\tfrac\Delta2;\vect k) \expp{iE_\vect k(T) \Delta}, \\
	\widetilde \Sigma^{ab}([E_\vect k],T;\vect k) &= \int \ud \Delta \bar\Sigma^{ab}(T+\tfrac\Delta2,T-\tfrac\Delta2;\vect k) \expp{i\int_{T-\Delta/2}^{T+\Delta/2} \ud s E_\vect k(s)}.
\end{align}
\end{subequations}
When the interaction timescale $t_\text{int}$  is much smaller than the expansion timescale $H^{-1}$, it is equivalent to use the original or improved frequency representations.

\subsection{Imaginary part of the self-energy}

\begin{figure}
	\centering
	\hfill\includegraphics[width=0.3\textwidth]{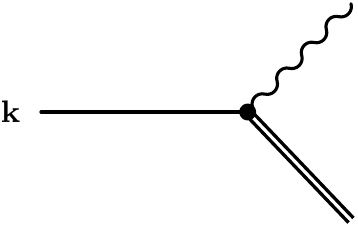}
	\hfill\includegraphics[width=0.3\textwidth]{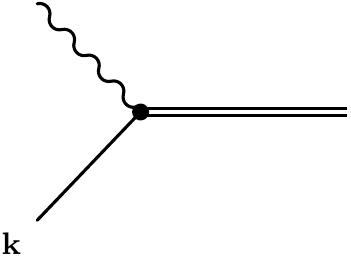}
	\hfill\mbox{}
	\caption{The two Feynman diagrams contributing to the decay rate of the field $\phim$. The diagram on the right corresponds to the usual absorption of a massless particle from the thermal bath, and the diagram on the left represents the emission of a massless particle. The diagram on the left violates energy conservation, and it is hence forbidden in flat spacetime. However this needs not be the case in an expanding universe.}\label{fig:ImSigmaCosCurved}
\end{figure}

The imaginary part
\footnote{Notice that here we are abusing the notation because in the time representation the retarded self-energy is purely real. ``Imaginary part'' here refers to the frequency representation we shall introduce next. In the time representation, it corresponds to the odd part of $\bar\SigmaR$ under the exchange of $t_1$ and $t_2$.}  
of the self-energy is given by 
\begin{equation}
	\Im\bar\SigmaR(t_1,t_2) = \frac{1}{2i} \left[ \bar\Sigma^{12}(t_1,t_2) -\bar\Sigma^{21}(t_1,t_2) \right] \, .
\end{equation}
By applying the CTP Feynman rules in the physical time 
representation to the diagram in figure \ref{fig:SigmaCurved}, and using the free propagators 
given in the previous section, 
to order $g^2$ we get:
\begin{equation} \
\begin{split}
	\bar\Sigma^{21} (t_1,t_2;\vect k) &= 
	\frac{i g^2  M^2 
	}{ a(t_1)a(t_2)} \int \udpi[3] {\vect q} \frac{\expp{ -i \int_{t_2}^{t_1} \ud {t'} E_\vect{p-k}^*(t')}}{ 2\sqrt{ E^*_\vect{p-k} (t_1)E^*_\vect{p-k}(t_2)}} \frac{1}{2|\vect q|}  \\
	 &\quad\times \left[ \expp{-i\int_{t_2}^{t_1} \ud t' |\vect q|/a(t')} [ 1 + n_\theta(|\vect q|) ] + \expp{i\int_{t_2}^{t_1} \ud t' |\vect q|/a(t')} n_\theta(|\vect q|) \right].
\end{split}
\end{equation}
with $E^{*2}_{\vect k-\vect q}(t)=(\vect k-\vect q)^2/a^2(t) +  M^2$. 
The two terms in the last line of \eqref{ImSigmaCos} can be interpreted in terms of two Feynman diagrams, represented in figure \ref{fig:ImSigmaCosCurved},
corresponding respectively to the emission and absorption of a photon (the first 
diagram would  
vanish on shell in flat spacetime). In the (standard or improved) frequency representations,
\begin{equation}
\begin{split}
	\widetilde\Sigma^{21} ([E_\vect k],T;\vect k) \approx- 2i \Im \SigmaImp ([E_\vect k],T;\vect k).
\end{split}
\end{equation}
The equality is valid because the Fourier transform of $\bar\Sigma^{12}$ gives exponentially damped contributions (in $ a m/\theta$ and $m/H$) in the high mass limit we are using. Therefore
\begin{equation} \label{ImSigmaCos}
\begin{split}
	\Im \SigmaImp ([E_\vect k],T;\vect k) &= 
	\frac{-g^2  M^2 
	}{ 2a(t_1)a(t_2)} \int \ud \Delta \int \udpi[3] {\vect q} \frac{\expp{ i \int_{t_2}^{t_1} \ud {t'}[E_\vect{p}(t')- E_\vect{p-k}^*(t')]}}{ 2\sqrt{ E^*_\vect{p-k} (t_1)E^*_\vect{p-k}(t_2)}} \frac{1}{2|\vect q|}  \\
	 &\quad\times \left[ \expp{-i\int_{t_2}^{t_1} \ud t' |\vect q|/a(t')} [ 1 + n_\theta(|\vect q|) ] + \expp{i\int_{t_2}^{t_1} \ud t' |\vect q|/a(t')} n_\theta(|\vect q|) \right].
\end{split}
\end{equation}
where $t_1 = T+\Delta/2$ and $t_2 = T-\Delta/2$.

\longpage

\index{Comoving limit}
The above equation is simplified in the comoving limit, in which $E_\vect {p-k}^*(t) ) M$ and $E_\vect p(t)=m$. In this limit,
introducing spherical coordinates and integrating the trivial angular coordinate, we get: 
\begin{equation} \label{parentEquationDecay}
\begin{split}
	\Im\bar\SigmaR(m,T;\vect 0) &= -\int \ud \Delta \expp{i(m-M)\Delta} 
	\frac{g^2  M}{16\pi^2 a(t_1)a(t_2)}  
	\int_0^\infty k 
	\ud k  \\ 
	 &\qquad\times
	 \left[ \expp{-i\int_{t_2}^{t_1} \ud {t'} k/a(t')} [ 1 + n_\theta(k) ] + \expp{i\int_{t_2}^{t_1} \ud t' k/a(t')} n_\theta(k) \right].
\end{split}
\end{equation}
Notice that in the comoving limit the original and improved frequency representations coincide. 

To compute
self-energy we have to choose a 
specific model 
for the evolution of the scale factor $a(t)$. 
This is what we do in the next sections.

\index{Decay rate}
Before continuing, let us briefly recall the results of the previous chapter, concerning the interpretation of the imaginary part of the self-energy in cosmological contexts. When the interaction times are small, as compared to the expansion rate $H^{-1}$, or  the particles are comoving, the imaginary part of the self-energy can be interpreted as the net particle decay rate in the comoving frame:
\begin{subequations} \label{DecayRates7}
\begin{equation}
	\Gamma_\vect p(t) = - \frac{1}{E_\vect p(t)} \Im \bar\SigmaR \boldsymbol(E_\vect k(t),t;\vect k\boldsymbol).
\end{equation}
When the interaction times are of the order of the expansion rate, and the particles are in motion, the improved frequency representation should be used. In this case the improved frequency representation corresponds to the decay rate in the particle rest frame:
\begin{equation}
	\gamma_\vect p(t) = - \frac{1}{m} \Im \SigmaImp ([E_\vect k],t;\vect k) .
\end{equation}
For large interaction times the decay rate in the comoving frame can be obtained from the following modified version of the self-energy:
\begin{equation}\label{DecayRateNew}
	\Gamma_\vect p(t) = \int \ud \Delta \frac{\Im\bar\SigmaR(t+\tfrac\Delta2,t-\tfrac\Delta2;\vect k)}{\sqrt{E_\vect p(t+\Delta/2)E_\vect p(t-\Delta/2)}} \expp{i\int_{t-\Delta/2}^{t+\Delta/2} \ud s E_\vect k(s)}.
\end{equation}
\end{subequations}

\section{Temperature effects}

As a first step, we approximate the evolution of the scale factor by a linear expansion:
\begin{equation}\label{firstord}
	a(t) =  a(T)[1+H(T)(t-T) + O\boldsymbol(H^2(t-T)^2\boldsymbol)].
\end{equation}
We keep all terms linear in $H$ and drop
terms which contain 
a higher power in $H$. 
For this approximation to be valid, 
timescales much shorter than the Hubble time should be considered. 
At finite temperature, the relevant timescales appearing in 
the Feynman diagrams are of the order of the inverse physical temperature. 
Therefore,
the above expression for the scale factor is appropriate when 
considering physical temperatures which are much larger 
than the expansion rate 
(but still much smaller than the 
fields masses). 
For the remaining of this section the scale hierarchy $M,m \gg \dm, \theta/a \gg H$ will be assumed.

Let us explicitly compute the comoving limit as an example. The imaginary part of the self-energy is then given by:
\begin{equation}
\begin{split}
	\Im \bar\SigmaR (m,T;\vect 0) &= -\frac{g^2 M}{16\pi^2 a^2(T)} \int \ud \Delta \expp{i(m-M)\Delta}  \int_0^\infty \ud k k \, n_\theta(k)
	  \expp{ik \Delta /a(T) }  \, ,
\end{split}
\end{equation}
which yields
\begin{equation}
	\Im \bar\SigmaR (\omega,T;\vect 0) = -  \frac{g^2}{8\pi} M (M-m) 
	\, n_{\theta/a(T)} (M-\omega).
\end{equation}
It is important to remark that the linear dependence on the 
Hubble parameter $H$ cancels out, so that the result only depends on 
the scale factor evaluated at time $T$ at this level of approximation. This directly follows from the odd character of $\Im\SigmaR(t_1,t_2;\vect k)$ under the exchange of $t_1$ and $t_2$.
The first corrections to the above expression are thus quadratic in $H$, in  units of $M$, $\dm$ or $\theta/a(T)$.
Therefore, in the static limit one recovers the flat space-time results of chapter 5 [see eqs.~\eqref{RestGamma}] with a time-dependent temperature:
\begin{equation}
\begin{split}
	\Gamma_\vect 0 = -\frac{1}{m} \Im \bar\SigmaR (m,T;\vect 0) &= \frac{g^2}{8\pi} \dm \, 
	n_{\theta/a(T)} (\dm)\, .
\end{split}
\end{equation}

It is not a matter of chance that we recover the flat spacetime results. The approximation \eqref{firstord} for the scale factor corresponds to the quasistatic or short time approximations we introduced in the previous chapter. We 
have adopted the theoretical framework described in the previous chapter. In particular, the masses of the fields were assumed to be much larger than the expansion rate of the universe. This is a key assumption, because it allows to introduce the adiabatic (WKB) approximation, which not only makes the problem solvable, but also allows for a well-defined particle concept. In order to have interesting dynamics, the mass gap between the two massive states was taken to be much lower than the masses of the fields. For instance, under \Eqref{firstord} the free propagator becomes
\begin{equation}
	\begin{split}
	\bar G^{(0)}_\text{R}(t_1,t_2;\vect k) &= \frac{-i}{  \sqrt{ E_\vect k(t_1) E_\vect k(t_2)}} \sin\left({\int^{t_1}_{t_2} \ud{t'} E_\vect k(t')} \right) \theta(t_1-t_2)\\&= \frac{-i}{ E_\vect k(T)} \sin\left[{E_\vect k(T)(t_1-t_2)} \right] \theta(t_1-t_2)+O(H^2/M^2),
\end{split}
\end{equation}
which is identical to the short time expansion for the propagator [\Eqref{QLShortTime}].
This means that the free propagators appearing inside the Feynman diagrams can be replaced by their quasilocal approximations.
Therefore, the  self-energy in an expanding universe can be obtained from its flat-spacetime counterpart by replacing
\begin{equation} \label{Substit}
	\vect p \to \vect k /a(T),
	 \quad \text{temperature}\to \theta/a(T)\,.
\end{equation}
This is true as long as the physical temperature $\theta/a(T)$ is larger than the expansion rate $H$, as explained above.
It is important to notice that we are not restricted to any short time approximation in the final results: if desired, the interacting propagator can be computed for arbitrarily long time lapses following \Eqref{InteractingWKB}.

\index{GZK limit}

\begin{figure}[t]
	\centering
	\includegraphics[width=\textwidth]{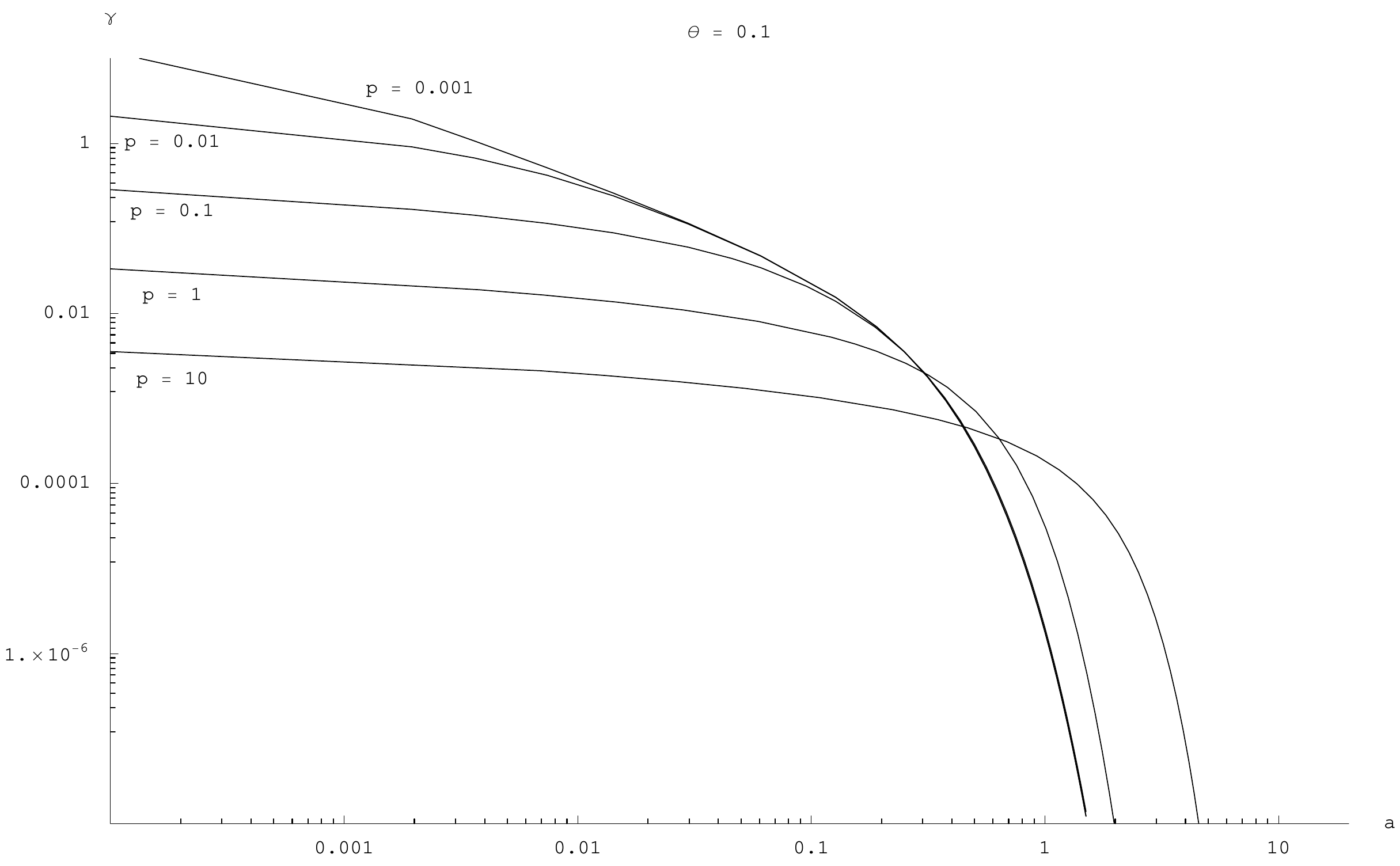}
	\caption{Decay rate  as a function of the scale factor in an expanding universe at conformal temperature $\theta = 0.1 \dm$. In order to discount time dilation effects we plot the decay rate in the reference frame of the particle. The decay rate is in units of $g^2\dm$, the temperature in units of $\dm$ and the conformal momentum in units of $m$. The mass gap is much smaller than the mass.}\label{fig:DecayT}
\end{figure}

We already computed the imaginary part of the self-energy for arbitrary values of the momentum in chapter 5. Making the replacement \eqref{Substit} in \Eqref{3fieldsDecayGeneral} we find the value of the self-energy as a function of the conformal momentum, the energy and the scale factor.  The dependence of the decay rate on the scale factor is plotted in figure \ref{fig:DecayT}. This figure shows that, as we go backwards in time, slowly moving particles become unstable once they reach a physical temperature $\theta/a$  of the order of the mass gap $\dm$. Ultrarelativistic particles become unstable for smaller values of the temperature, once the boosted temperature $(E/p) \theta/a$ is of the order of the mass gap. Notice that for ultrarelativistic particles, as we go backwards in time, both the temperature and the energy of the particles increase. These two effects combine to enhance the decay rate. Loosely speaking, this phenomenon could be interpreted as the GZK limit being reached in a dynamical way (see pags.~\pageref{page:GZK} and \pageref{page:GZK2}).

From the flat space-time calculation of chapter 5 [see \Eqref{RealThreeFields}] we also deduce that the on-shell real part of the retarded self-energy is given by:
\begin{equation}
	\Re \bar\SigmaR(E_\vect k,T;\vect k) = - \frac{g^2}{6} 
	 \frac{\theta^2}{a^2(T)}\frac{m}{\dm} +O(H^2/M^2) . 
\end{equation}

To sum up, the results of this section confirm the well-known result that whenever the temperature is much larger than the expansion rate the particles feel the local temperature at each moment of the expansion.

\section{Vacuum effects in the self-energy}

When the expansion rate of the universe is of the order of the temperature or larger vacuum effects become relevant. In this case we expect a contribution from the first diagram on the left of figure \ref{fig:ImSigmaCosCurved}, which can be interpreted as the particle emitting a massless particle and becoming excited. This second contribution is a consequence of the
absence of energy conservation at those energy scales comparable to the expansion rate.

 Let us evaluate this contribution explicitly for two different models of the universe evolution.

\index{de Sitter}

\subsection{Vacuum effects in de Sitter}

Let us consider flat coordinates in the de Sitter spacetime, which corresponds to a flat cosmological model with 
\begin{equation}
	a(t) = a(T) \expp{H(t-T)}.
\end{equation}
For the sake of simplicity, we will show the details of the calculation in the comoving case, and simply quote the result for the case of arbitrary geodesic motion. 

From \Eqref{ImSigmaCos}, the vacuum contribution to the imaginary part of the self-energy in de Sitter is given by:
\begin{equation}
\begin{split}
	\Im\bar\SigmaR(m,T;\vect 0) &= -\int \ud \Delta \int_0^\infty \ud k \frac{g^2 m k}{16 \pi^2  a^2(T)} \\ &\qquad\times \exp{\left[-i\int_{T-\Delta/2}^{T+\Delta/2}\ud t\left( \dm + \frac{k}{a(T)} \expp{H(T-t)} \right)\right] }.
\end{split}
\end{equation}
Integrating the argument of the exponential gives
\begin{equation*}
	\Im\bar\SigmaR(m,T;\vect 0) = -\int \ud \Delta \int_0^\infty \ud k \frac{g^2 m k}{16 \pi^2  a^2(T)} \expp{-i \dm \Delta - \frac{ik}{a(T)H} \sinh\left( \frac{H\Delta}{2} \right)  - \epsilon k }.
\end{equation*}
We have added a small $-\epsilon k$ term to ensure convergence of the $k$ integral: 
\begin{equation*}
	\Im\bar\SigmaR(m,T;\vect 0) = \int \ud \Delta 
	\frac{g^2 m H^2}{16\pi^2} \frac{\expp{(-i \dm + H) \Delta}}{(-1+\expp{H\Delta}-i \epsilon)^2}.
\end{equation*}
The $\Delta$ integration can be performed in the complex plane taking into account the pole at $\Delta=i\epsilon$. The result is the following:
\begin{equation}
	\Im\bar\SigmaR(m,T;\vect 0) = - \frac{g^2}{8\pi}  \frac{m \dm}{\expp{2\pi\dm /H}-1} = - \frac{g^2}{8\pi}  m \dm n_{H/2\pi}(\dm).
\end{equation}
This results coincides with the imaginary part of the self-energy in a Minkowski thermal bath at physical temperature $H/(2\pi)$ [see eqs.~\eqref{RestGamma}]. A decay rate can be associated to this imaginary part of the self-energy:
\begin{equation}
	\Gamma_\vect 0(T) = - \frac{1}{m} \Im\bar\SigmaR(m,T;\vect 0) 	 = \frac{g^2}{8\pi}   \dm n_{H/2\pi}(\dm).
\end{equation}
Recall that in the comoving case there is no difference between the standard and improved self-energies.
\index{Decay rate}

\index{de Sitter effective temperature}
The result is not at all unexpected, since it is well known \cite{BirrellDavies} that a comoving particle detector in de Sitter conformal vacuum perceives a bath of thermal radiation at a temperature  $H/(2\pi)$. The fact that the decay rate has an exact thermal character can be traced to the existence of thermalization theorems under the presence of event horizons \cite{Takagi86}. Our results are also consistent with those of Bros \emph{et al.}~\cite{BrosEtAl07}, who similarly find inestabilities for massive particles propagating in de Sitter. Bros \emph{et al.} perform a calculation which is in some aspects comparable to ours, but using a global rather than an adiabatic approach to the particle concept.

This result can be extended for moving particles with arbitrary momentum, following the same steps above. Notice that in this case it is important to use the improved frequency representation because the interaction timescale is of order $H^{-1}$.  However the resulting expressions are too complicated to be evaluated in full generality, so one must restore to some series expansion. We have performed a non-relativistic expansion in $\vect k^2/m^2$. Interestingly, we have found that the additional momentum-dependent terms cancel and that the result is 
\begin{equation}
	\Im\SigmaImp([E_\vect k],T;\vect k) = \frac{g^2}{8\pi}  m \dm\, n_{H/2\pi}(\dm)
\end{equation}
at least to order $\vect k^{10}/m^{10}$. It is therefore reasonable to assume that the above result is valid to all orders in perturbation theory. The result is not unexpected, since the decay rate in the particle rest frame 
\begin{equation}\label{decayRateDeSitter}
	\gamma = -\frac{1}{m} \Im\SigmaImp([E_\vect k],T;\vect k)  = \frac{g^2}{8\pi}  \dm\, n_{H/2\pi}(\dm)
\end{equation} 
is time- and momentum- independent as required by the de Sitter invariance of the problem (bear in mind that both the adiabatic and conformal vacua are de Sitter invariant). Since the invariance is not manifest from our expressions, this provides a non-trivial check of the calculation. We have also computed the decay rate in the comoving frame, according to \Eqref{DecayRateNew}. The result is the following:
\begin{equation}
	\Gamma_\vect k(t) = \frac{g^2}{8\pi}  \dm\, n_{H/2\pi}(\dm) \left( 1 - \frac{\vect k^2}{2 a^2(t) m^2} \right) + O(\vect k^4/m^4).
\end{equation}


\subsection{Power-law inflation}
\index{Power-law inflation}
\index{Decay rate}

Let us now consider the case of power-law inflation, where the scale factor evolves according to
\begin{equation}
	a(t) = a(T) \left( \frac{t}{T} \right)^\alpha.
\end{equation}
The Hubble rate is given by $H(T) = \alpha/T$. Since expressions rapidly become  cumbersome, we shall only sketch the calculation in the simplest situations, and will present the result for the other cases.

As before, let us first consider comoving particles. According to eq.~\eqref{ImSigmaCos}, the imaginary part of the self-energy is given by
\begin{equation}
\begin{split}
	\Im\bar\SigmaR(m,T;\vect 0) &= -\int \ud \Delta \int_0^\infty \ud k \frac{g^2 m k}{16 \pi^2  a^2(T)} \left( 1- \frac{\Delta^2}{4T^2} \right)^{-\alpha} \\ &\qquad\times \exp{\left\{-i\int_{T-\Delta/2}^{T+\Delta/2} \ud t \left[ \dm + \frac{k}{a(T)} \left( \frac{T}{t} \right)^\alpha \right]\right\} }.
\end{split}
\end{equation}
Integrating the argument of the exponential, and  performing the integral over $k$ yields (it is necessary to add a  $-\epsilon k$ term to ensure convergence)
\begin{equation}
\begin{split}
	\Im\bar\SigmaR(m,T;\vect 0) &= \frac{2 g^2 m (-1+\alpha)^2}{\pi^2} \int \ud \Delta 
	\expp{-i \dm \Delta} \\
	&\qquad \times\frac{ 
	\left( 1- \frac{\Delta^2}{4T^2} \right)^{-\alpha}
	(2T + \Delta)^{2\alpha}}
	{\{ T^\alpha(2 T + \Delta) - (2T-\Delta)^{1-\alpha}
	[T(2T+\Delta)]^\alpha\}^2}.\raisetag{4em}	
\end{split}
\end{equation}

In order to proceed further a particular value of $\alpha$ has to be chosen. Closed analytic expressions can be obtained for positive integers values of $\alpha$. For the case of $\alpha=4$ we obtain:
\begin{equation*}
\begin{split}
	\Im\bar\SigmaR(m,T;\vect 0) &= \frac{9g^2 m }{16\pi^2} \int \ud \Delta 
	\frac{ \expp{-i \dm \Delta} 
	\left(-4T^2 + \Delta^2\right)^4}
	{\Delta ^2 \left(2 T^2+\Delta ^2\right)^2}, \quad \alpha=4.
\end{split}
\end{equation*}
This function can be integrated in the complex plane by considering the residues of the double pole located at $\Delta = -2\sqrt{3} i T$.\footnote{The other double pole would be located at $\Delta = i\epsilon$ had we kept track of the $\epsilon$ terms, and therefore does not have to be taken into account.} The result is the following:
\begin{equation}\label{decayRatePowerLaw}
	\Im\bar\SigmaR(m,T;\vect 0) = -\frac{g^2 }{2 \pi } m  \dm \expp{-2 \sqrt{3} \dm T} =-\frac{ g^2 }{2 \pi }m \dm \expp{-{8 \sqrt{3} \dm}/{H(T)}}
\end{equation}

This expression can be generalized, on the one hand, by considering different values of $\alpha$, and on the other hand by considering particles in motion. Let us first consider different values of $\alpha$. We have found that at this order there is only a non-vanishing contribution to the imaginary part of the self-energy for even values of $\alpha$ larger than two. In the case $\alpha = 6$ the following result is found:
\begin{equation}
	\Im\bar\SigmaR(m,T;\vect 0) =-\frac{\left(3+\sqrt{5}\right)g^2 }{4 \pi }m \dm \expp{-12 \sqrt{5+2 \sqrt{5}}
   \dm T} , \quad \alpha = 6.
\end{equation}
Expressions become increasingly cumbersome with $\alpha$. We have also computed the momentum corrections. For instance, in the case $\alpha=4$, by going to second order in $\vect q^2/m^2$ we find
\begin{multline}\label{powerLawMomentum}
	\Im\SigmaImp([E_\vect q],T;\vect q)= - \frac{g^2 }{2 \pi } m  \dm \expp{-2 \sqrt{3} \dm T}   \\ \times \bigg[1  + \bigg(\frac{3 \dm T \left[2 \dm T \left(8 \dm T+11
   \sqrt{3}\right)-49\right]-182 \sqrt{3}}{75264  \dm T} \bigg)\frac{\vect k^2}{m^2 a^2(T)}\bigg] \\ + O(\vect k^4/m^4), \quad \alpha = 4.
\end{multline}
The correction due to the momentum is small. When $T = 1/\dm$ the numerical value of the factor in round brackets is $-0.00398$, and when $T=0.1/\dm$ its value is $-0.0437$.

\begin{figure}
	\centering\includegraphics[width=.7\textwidth]{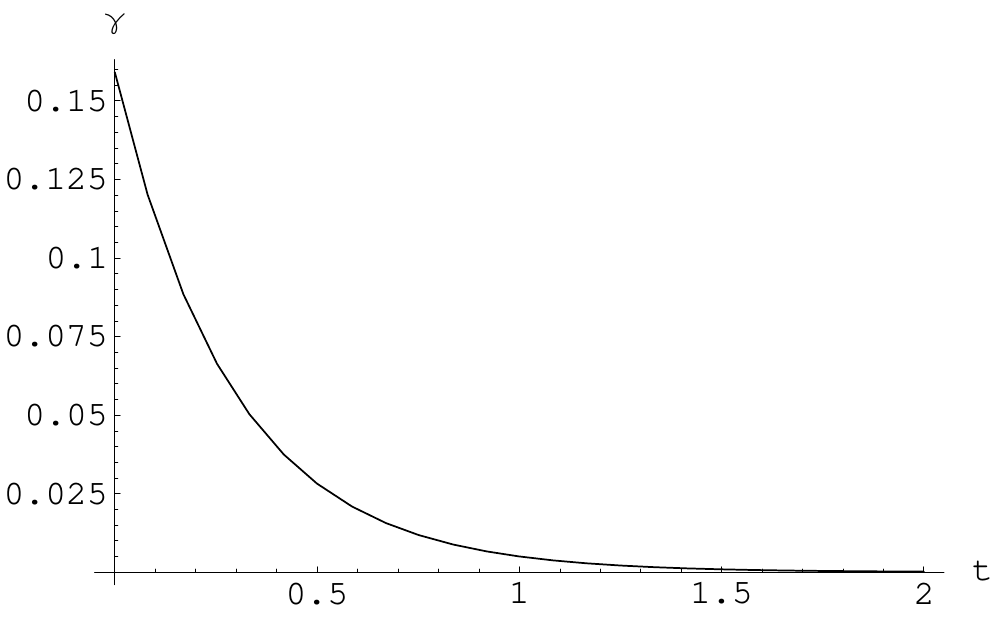}
	\caption{Decay rate of a rest particle in the power-law inflation model. Time is in units of $\dm^{-1}$, and the decay rate is in units of $g^2 \dm$. Adding a non-relativistic amount of momentum does not modify the plot in a noticeable way.}\label{fig:DecayPowerLaw}
\end{figure}

The decay rates can be immediately extracted from the imaginary part of the self-energy following eqs.~\eqref{DecayRates7}. Qualitatively, the behavior is the same as in de Sitter: the vacuum-induced decay rate is exponentially suppressed except when $H \gtrsim \dm$. However, in this case the rates do not have an exact thermal character and depend on time, because the Hubble rate varies during the universe evolution (see fig.~\ref{fig:DecayPowerLaw}).
Unfortunately we have not been able to compute the decay rate in the ultrarelativistic limit, which is the most interesting limit, because even slowly moving particles will become rapidly moving with respect the comoving frame as we go back in time in the universe expansion.

We end this chapter by remarking that the decay rate, derived from the imaginary part of 
the self-energy, has a secular character, as expected. Even small decay rates 
could thus give an important effect when integrated over large periods of time. 
Moreover, dissipation is a generic phenomenon which appears even in the vacuum.
 Therefore, if one considers remote times or large time lapses, 
the dissipative processes will become relevant, specially in situations in which the expansion rate, or the curvature, are large.

	\chapter{Conclusions}

\section{Main results}


Quantum fields have been the main object of interest of this thesis. Despite that, in chapter 2 we began by briefly studying a quantum Brownian motion system (a harmonic oscillator linearly coupled to an environment), with the purpose of analyzing each mode of the field from the perspective of the theory of open quantum systems. In sect.~\ref{sect:OQS} it was found that whenever the field can be treated under a Gaussian approximation each field mode behaves as if it were an open quantum system linearly coupled to an effective environment. 
The equivalence is expressed in eqs.~\eqref{equivalenceQBM} and in table \ref{tbl:equivalence} (page \pageref{tbl:equivalence}), and has been extensively used in this thesis. As a first application, we rederived the well-known result that the imaginary parts of the retarded self-energy corresponds to the net decay rate of an excitation. Using the quantum Brownian motion analogy allowed us to introduce a model-independent derivation.

We next moved to the analysis of the quasiparticle excitations, which were defined as long-lived elementary excitations carrying momentum and energy. Two different approaches were used to analyze their properties: a frequency-based approach, in which the properties of the excitations were deduced from the analysis of the spectral representation, and a real-time approach, wherein the quantum states corresponding to the quasiparticle excitations were explicitly constructed, and the time evolution of the expectation value of the energy in those states was studied. In both approaches it was possible to show that the real and imaginary part of the self-energy determine the energy and the decay rate of the quasiparticles respectively ---see eqs.~\eqref{RGammaBasic}.  Although previous evidences existed, to our knowledge this is the first systematic corroboration that the real part of the self-energy determines the physical energy of the quasiparticles in non-vacuum relativistic field theory. The dynamics of the quasiparticle excitation can be encoded in the form of a generalized dispersion relation, eq.~\eqref{generalizedDispRel}, which needs not be Lorentz-invariant. Moreover, we argued that dissipative phenomena is a generic feature of field theory in general backgrounds.

Several additional interesting points were illustrated by the real-time approach. Using a very simplified model of a quasiparticle creation device, we discussed that quasiparticles were adequately described by the action of the creation operator on the background state for long observation times ---see \Eqref{physicalStateQP}. At the same time, we showed that the quantum state corresponding to the quasiparticle contains actually more than one particle excitation. We argued that this result is actually a statistical effect, due to the fact that sectors with larger occupation numbers are more likely to be excited, because of the Bose-Einstein statistics. We also built the hole state [\Eqref{holeState}], but discussed that it cannot be properly considered a quasiparticle state for bosonic systems.

We ended chapter 3 by comparing our results with other mean-field-based approaches, and argued that the latter were more suited for the study of classical-like configurations in the hydrodynamical limit rather than to the study of the properties of individual quasiparticles. 

In chapter 4 we applied the results of  chapter 3 to the study the modifications to the dispersion relation of a scalar particle propagating in a thermal bath of gravitons. Linearized gravity was treated as an effective field theory. A first interesting outcome is that gravitational interactions do not generate a four-derivative term in the scalar field action, despite the fact that there is no apparent symmetry protecting that term. Anyway, the main results of the chapter are the modified dispersion relations, given by eqs.~\eqref{RelDispT} and \eqref{RelDispHighT}, which can be put in a Lorentz invariant form at low temperatures and/or high momenta (see next section for a discussion of this point). 

In chapter 4 we were not able to properly discuss the dissipative effects because kinematic constraints forbid the appearance of an imaginary part of the self-energy at one loop order. For this reason, in chapter 5 we moved to a different field theory model, consisting of two heavy scalar fields of similar mass interacting with a third massless field, which can be understood as a field theoretic description of a two-level atom. The system was studied in the limit in which the temperature is large enough to induce transitions between the two levels but small enough so that no heavy particles are present in the background. We computed the decay rates of the two fields and verified that they correspond to the transition rates of the internal states of the atom [eqs.~\eqref{RestGamma}]. The same system  was also analyzed in a non-isotropic background, consisting in an extra atom on top of the thermal bath. The results show that the usual interpretation of the self-energy holds even for non-isotropic backgrounds.

We also studied the equations governing the time-evolution of the distribution function of the two atomic states, as a function of the decay and creation rates, both in a standard near-equilibrium situation [eq.~\eqref{DecayThermalEq}] and in an out-of-equilibrium situation [eqs.~\eqref{DecayOutT}]. The extension to the inhomogeneous case led us to the Boltzmann equation \eqref{Boltzmann}. The basic physical assumption was the two-scale separation, also assumed in chapter 3: the time and length scales characteristic of the time evolution of the distribution function were assumed to be much larger than the microscopic interaction scales. 

The results of chapter 3 were extended in chapter 6 to curved backgrounds. We showed that most field theory relations among the propagators valid in flat spacetime can be easily extended to curved spacetimes by rescaling the fields with the metric determinant [eq.~\eqref{rescaling}] and using a local Fourier transform with respect to the Riemann normal coordinates [eq.~\eqref{localFourier}]. When the relevant time and length scales involved are much smaller than the local curvature radius, we recover the expected result that the form and interpretation of the curved spacetime expressions matches their corresponding flat spacetime counterparts. In particular, when this scale separation holds, quasilocal quasiparticles can be introduced, having the same essential properties as the flat quasiparticles.

More appealing results were found in the cosmological case working with the adiabatic approximation. While this approximation still requires the field modes to have frequencies much higher than the expansion rate of the universe, observation times can be arbitrarily large. In this case the evolution of the propagators can be computed in a WKB approximation, in terms of the energy shift and the decay rate ---see eq.~\eqref{InteractingWKB}. Two different physical situations were distinguished. If the interaction timescale is much smaller than the typical expansion rate, energy shift and the decay rate can be extracted from the retarded self-energy in a quasilocal frequency representation [eqs.~\eqref{quasilocaldRGamma}]. If, on the contrary, the interaction timescale is comparable to the expansion rate, then the energy shift and the decay rate are extracted from a kind of frequency transform which takes into account the evolution of the on-shell condition during the interaction time [see \Eqref{GammaRSuperGen}]. We also defined an improved frequency representation, \Eqref{frequencyImproved}, from which the decay rate in the particle rest frame can be extracted. 

These results were applied in chapter 7 to the same physical model analyzed in chapter 5, namely a doublet of massive fields interacting with a third massless field, focusing on the dissipative effects. The massive fields were treated under the adiabatic approximation. We first addressed the propagation in an expanding thermal bath, recovering the expected result that the propagation is governed by the local temperature at each moment of the expansion, provided the temperature was much larger than the expansion rate. When tracing back the particle in the universe, the decay rate of the particles is enhanced  because both the temperature and the momentum of the particle increase.

When the expansion rate of the universe is of the order of the mass gap or larger, vacuum effects become relevant. We found an additional contribution to the imaginary part of the self-energy, and hence to the decay rate.  In the case of de Sitter,  the decay rate corresponds to that of a particle at rest in a flat thermal bath at the de Sitter effective temperature [\Eqref{decayRateDeSitter}]; moreover we verified that the decay rate is de Sitter invariant. In the case of power-law inflation, a qualitatively similar decay rate is found [eqs.~\eqref{decayRatePowerLaw}].

From the structure of the equations leading to the imaginary part of the self-energy [\Eqref{parentEquationDecay}] one can explicitly verify that the imaginary part of the self-energy corresponds to a decay rate. The vacuum contribution can be identified as the lower mass state decaying into the higher mass state with the emission of a massless particle, process which is not forbidden because energy conservation does not hold for energy scales of the order of the expansion rate. 

In this thesis we have analyzed both the flat and the curved spacetime situations, and at the same time have presented both a theoretical framework and explicit applications. We hope that we have at least partially addressed the goal which we stated in the preface: to extract the maximum amount of information on the particle propagation from the two-point correlation functions in general situations. Of course, there are open questions  and more work to be done in the future. 
Let us talk over them in the final sections of this thesis.

\section{Open discussion}

Although many of the relevant points of the thesis have already been discussed in the previous chapters, let us now draw our attention on some relevant aspects.

\index{Quasiparticle}
We start by commenting the subtleties related to the construction of the quasiparticle states, found in chapter 3. Recall that quantum state corresponding to the quasiparticle contains more than one additional particle excitation. As we  already mentioned, this is a statistical effect related to the fact that the background state is not an eigenstate of the number operator. It can be illustrated with  the following toy model. Let $\hat\rho = (|0\rangle\langle 0|+|2\rangle\langle 2|)/2$ be the initial state of a harmonic oscillator. When a ``quasiparticle'' is introduced into the system by the action of the creation operator, the state becomes $\hat\rho' = (|1\rangle\langle 1|+3|3\rangle\langle 3|)/4$. The expected number of particles in the initial and final states is 1 and 5/2 respectively, and therefore the particle number is increased in 3/2, despite the fact that each term of the state is increased with only one particle. Clearly, the reason for this fact is linked to the Bose-Einstein statistics: the higher occupied excited states are more likely to be excited. A similar phenomenon happens if the incoherent mixture is replaced by a coherent superposition of two particle eigenstates.

In chapter 3 we put a great amount of effort in the construction and analysis of the quasiparticle excitations. One could wonder whether this effort was worth, since after all, the same dispersion relation can be obtained from much simpler mean-field-based methods (linear response theory or effective action, see sect.~\ref{sect:Alternative}). We believe that such effort was useful for two different reasons. First, the quasiparticle and mean field approaches provide the same answer to two different physical questions: in the former case one studies the dynamics of individual quasiparticles, and in the latter, the dynamics of the mean fields. Second, the route to the construction and analysis of the quasiparticle properties shed light on many intermediate results which are interesting by themselves. In any case, the equivalence of the answers can be understood in several ways, one of them being the following: loosely speaking, when higher order correlation functions are neglected, the system behaves as if it were effectively linear, and thus the dynamics of all relevant quantities is essentially determined by the solution of the corresponding stochastic problem (see chapter 2). In this context it is not surprising that the dynamics of both the mean field and the quasiparticles is characterized by the same dispersion relation. 

\index{Gaussian approximation}

A key element in our analysis was the Gaussian approximation, which on the one hand allowed having manageable expressions, and on the other hand provided physical interpretations for those expressions, without having to restore to any perturbative expansion in the coupling constant. The introduction of the QBM analogy facilitated the interpretation by providing a dictionary between the field theory and open quantum system languages. The retarded propagator appears in a central position because it determines the spectrum of energies. Recall that the retarded propagator does not depend  on the equivalent linear environment, or, what it is the same, it does not explicitly depend on the occupation numbers ---although in general it depends on the state of the field, since for non-linear systems the spectrum of energies depends on the state.


Let us now comment on a couple of noticeable points found in chapters 4 and 5. First, recall that in the propagation in a thermal graviton bath, even in the vacuum, we did not find any contribution to the higher order derivative term even if that contribution was expected from simple effective field theory arguments. This may indicate that some underlying symmetry is protecting that term, symmetry which we were not able to identify.  Second, recall also the effective dispersion relation at one-loop order was found to be Lorentz-invariant whenever the massive fields are not thermalized.  
We found this effect both in the propagation in a thermal graviton bath and in the three fields model, and it is also present in electrodynamics \cite{DonoghueHolstein83}. The fact that the same phenomenon happens in three different models leads us to think that it might be a generic feature of unexcited massive fields coupled to thermalized massless fields. 

\index{Dispersion relation}

If the modifications to the real part of the dispersion relation have been somewhat daunting, the prospect of generically dissipative dispersion relations is much more encouraging. Recall that from the perspective of second quantization, dissipation does not necessarily mean that the particle decays into  different  products: it may simply mean that the particle changes its momentum.  Let us emphasize that in non-vacuum and non-flat spacetimes dissipation is a generic phenomenon, and moreover it has a secular character: it accumulates over time. For this reason we believe that modifications to the imaginary part of the dispersion relation might be physically more relevant than the modifications to the real part, specially when following the particle over long periods of time. This is in contrast to most approaches studying modified dispersion relations, which usually tend to emphasize real dispersion relations ---see for instance \Eqref{Eq1}.

\index{Decay rate}
Unfortunately, there is no contribution to  the decay rates of particles  in a thermal bath of gravitons at one loop order, because
 kinematic constraints forbid the appearance of dissipative terms. Moreover, ordinary perturbation theory most likely fails for the imaginary part of the propagator in the graviton thermal bath, and this calculation requires considering a Braaten-Pisarski resumed propagator for the gravitons. The motivation for analyzing the quantum gravitational corrections was its universality and its relevance at high energies. However, we expect that the qualitative results for the dissipation in the thermal graviton bath are not much different from the ones we obtained at one loop in the 3-fields model. An expected yet interesting fact is that for ultrarelativistic particles the decay rate becomes relevant whenever the particle energy is large enough to reach the GZK cuttoff. We will come back on this fact later on.

\index{Boltzmann equation}
The main drawback of the second-quantized approach to the analysis of particle propagation, which is the approach we have mostly used in this thesis, is the inability to follow the particle trajectory. Effectively, one studies not the particle itself, but the field mode corresponding to the initial momentum of the particle. This is an essential limitation of the second-quantized approach. In order to follow the particle trajectory at least two other different approaches are possible. On the one hand, a first-quantized description could be introduced provided that the particle number is a conserved quantity. If moreover only one particle is present, the dynamics of that particle can be followed through the equations of motion of its first-quantized density matrix or its first-quantized Wigner function. A different but related description is a statistical description, wherein the dynamics of an ensamble of particles can be studied following the evolution of their distribution function through the Boltzmann equation. The equation of motion of the Wigner function bears some resemblances with the equation of motion of a distribution function, but it is not equivalent because, roughly speaking, the former keeps track of the information on the quantum phases while the latter does not. In chapter 5 we explicitly reconstructed the Boltzmann equation from the information extracted from second-quantized self-energies in out of equilibrium situations. However, we were not able to resconstruct the equation of motion of the Wigner function because quantum phases were lost in the quantum field theory calculation.

\index{Quasiparticle!adiabatic}
In the final chapters of this thesis we dealt with curved spacetimes. At the leading order in the adiabatic quasilocal expansion,  the flat-spacetime expressions are recovered provided the time and length scales are much smaller than the curvature radius. Curved spacetimes are in this sense analogous to flat inhomogeneous backgrounds: curved or inhomogeneous propagation reduce to the corresponding flat or homogeneous results provided the length scale of the curvature or the inhomogeneity is large. We could think of incorporating additional physical effects by going beyond the leading order in the curvature expansion. However, no matter the higher order in the curvature expansion we go, we are always  limited to remain in the quasilocal sector. For this reason, we believe that it is much more promising to adopt an adiabatic expansion, as we did in cosmology, wherein the particle mode can be followed during long periods of time. This approximation is much less restrictive since it only requires large particle energies as compared to the curvature scale, and can accommodate large interaction and observation times. Within the adiabatic expansion additional physical effects not present in flat spacetime are indeed found.

Recall that the analogy between curved spacetimes and flat inhomogeneous backgrounds does not extend to the analysis of the divergent structure of both theories, since the ultraviolet divergent structure in inhomogeneous backgrounds is the same as in Minkowski, whereas the same property does not hold for curved spacetimes.

In curved spacetime the mean field and quasiparticle methods also lead to the same dispersion relations; the interpretation is different in each case, as in Minkowski. However, a noticeable point is that in curved spacetime the mean-field-based methods do not depend on the particle interpetation and can be easily extended to arbitrarily large distances in general situations, by deriving the effective geodesics which determine the dynamics of the mean fields \cite{DrummondHathrell80}.

The formalism presented in chapter 6 was used in chapter 7 to analyze the decay rates of the particles in cosmological backgrounds. One could argue that our results for the decay rates of the particles in cosmological backgrounds can be directly obtained by simpler means without the need of dealing with the framework introduced in chapter 6. While this is obviously true, notice that our goal is not   to simply compute the particle decay rates, but to analyze the dynamics of the particle by studying its two-point correlation functions. For us, the main result is that the particle evolution can be expressed in terms of two semi-local quantities extracted from the retarded self-energy, one of which can be interpreted as a decay rate.  

One of the requeriments of the calculation of the dissipative effects in the adiabatic vacuum is that observation times must be much larger than interaction times. Since interaction times are of the order of the Hubble time $H^{-1}$, interesting results can only be obtained when the Hubble rate is slowly varying, or, in other words, when $\dot H \ll H^2$. This condition can only be achieved in inflationary contexts. Notice that in the power-law inflationary model $\dot H = H^2/\alpha$, so that large values of $\alpha$ should be considered (although for computational simplicity we just studyied the $\alpha=4,6$ cases). Concerning also this inflationary model, recall that no contribution to the decay rate was found for odd values of $\alpha$; this can be probably atributed to some symmetry of this particular case, and most likely does not extend to higher order calculations.

\index{Trans-Planckian problem}
Let us now comment on the possible relevance of the results found for the trans-Plankian problem. Although we obviously cannot reach definite conclusions, some interesting insights can be extracted from our results. First, we should stress the importance of considering interactions and not limiting the analysis to free fields (aspect which had already been much emphasized in the past \cite{Parentani01a,Parentani01b,Parentani02,tHooft96,tHooft06}). Second, we should  also highlight the importance of considering dissipative processes, or, in other words, the importance of considering generalized dispersion relations including imaginary terms. Third, it is often assumed 
that if Lorentz invariance is an exact symmetry of nature there is no such thing as a trans-Planckian problem, since one can always boost to the reference frame where the frequencies are sub-Planckian.  Notice that this argument can only be  applied for free fields, since as soon as interactions are taken into account the effective dispersion relations governing the dynamics of the particle are not in general Lorentz-invariant anymore, even if Lorentz symmetry is an exact symmetry, and therefore the trans-Planckian issue is still a valid question to pose.

More specifically, the analysis of the propagation in flat spacetimes suggests that whenever high energies are reached, the modifications to the real part of the dispersion relation might not be very relevant but, on the contrary, the modifications to the imaginary part could be very significant. Unfortunately in cosmology we have not been able to  compute the high momentum behavior of the decay rate. 

\index{GZK limit}
One could also wonder whether the results found in this thesis may have any observational consequences. In the simplified physical models we have analyzed we have found  imaginary contributions to the dispersion relation which correspond to the GZK limit in the presence of a thermal bath (chapter 5), and a kind of ``dynamical GZK limit'' with an expanding thermal bath (chapter 6). Both effects have observational consequences, but of course these results are not novel and should instead be regarded as a validation of our framework. The possibility of real modifications to the dispersion relation due to gravity, as we have seen in chapter 4, is strongly constrained. More intriguing is the possibility of vacuum contributions to the imaginary part of the dispersion relations. We expect those contributions to be possibly significant for periods in which the Hubble parameter is large, such as in inflation. In any case, still much work would be  needed to go from the physical systems we have analyzed to experimentally relevant physical situations.


\longpage

\section{Future perspectives}

In this final section we mention, with no particular order, several directions in which the work presented in this thesis could be extended. Some of them correspond to topics which are under investigation at the moment of writing these conclusions.

In this thesis we have basically concentrated on the dynamics of the average values of the relevant quantum operators, particularly on the dynamics of the mean value of the Hamiltonian operator, and we have not generally considered quantum fluctuations. In general, the asymptotic formalism of field theory tends to overlook quantum fluctuations. There have been claims \cite{YuFord99a,YuFord99b,BorgmanFord03}, based on semiclassical considerations,  that fluctuations could accumulate over time and, in some circumstances, significantly modify the propagation of particles for large periods of time. Although some general energy conservation arguments seem to preclude those claims, we believe that is is nevertheless interesting to study quantum fluctuations associated to particles and quasiparticles from the perspective of quantum field theory.

A somewhat marginal but nevertheless interesting point of this thesis, and also  related to fluctuations,  is the issue of the particle content of the quasiparticle excitations. While there is nothing fundamental in the fact that more than one additional particle is expected whenever a quasiparticle is introduced in the system, this  phenomenon may lead to interesting physical considerations related, for instance, to energy conservation issues.\footnote{\emph{Note added in the second printing.} We have recently become aware that Unruh and Wald \cite{UnruhWald84} already discussed this kind of phenomena.}

\longpage

A natural extension of the work of chapter 3 would be to generalize the results for non-zero spin fields. The analysis of fermion fields should reveal the emergence of hole excitations as true quasiparticles, in contrast to the bosonic systems, in which we have seen that hole states, although can be constructed, do not have the appropriate quasiparticle properties.

Another direction in which the work of this thesis could be extended is the analysis of quasiparticle interactions. We have implicitly dealt with quasiparticle interactions in many sections of this thesis, when analyzing the interpretation of the imaginary parts of the self-energy and the decay rates, but always relying on heuristic arguments with no proper formal basis. In the same way as the Källén-Lehmann spectral representation can be adapted to non-vacuum situations, the Lehmann-Symanzik-Zimmerman formalism of vacuum field theory could in principle also be extended to non-vacuum situations. Some work has been already done in this direction in the context of thermofield dynamics \cite{NakawakiEtAl89}.

It would also be worth further elaborating on the passage from the second-quantized description to the first-quantized description, investigating on the precise conditions of validity of such a correspondence, and determining the equation of motion for the the first-quantized density matrix or Wigner function from field theory quantities. This would also allow to clarify the regime in which the dispersion relation properly charecterizes the particle propagation.

In chapters 3 and 6 we studied the quasiparticle properties from the direct analysis of the propagators. However, we could also have adopted an indirect approach, based on the analysis of the 2PI effective action, following the work of Calzetta and Hu \cite{CalzettaEtAl88}. Similarly as the dynamics of the mean field can be obtained from the 1PI effective action, the 2PI action determines the dynamics of the propagators. It would be interesting to rederive our results starting from a 2PI formalism. 

Going to more particular aspects, we found that at one loop order the contribution to the imaginary part of the decay rate in a thermal graviton bath was vanishing. It would be interesting to pursue the calculation of the imaginary part by computing the Braaten-Pisarksi resumed propagator.  In order to be sure that the results are physical and not a consequence of the infrared divergences, the gauge should not be fixed, and this way the gauge-independent character of the results could be checked. 

When considering the thermal graviton bath we have used the standard perturbative approach. We could  have alternatively used the large $N$ approach to gravity \cite{Tomboulis77}, wherein one considers a large number $N$ of quantum matter fields coupled to the gravitational field, and singles out one of the fields, which acts as the system of interest. The correlation functions found in the large $N$ expansion coincide with those of stochastic gravity \cite{HuVerdaguer03,HuVerdaguer04}. The  large $N$ expansion allows to systematically consider the backreaction of the fields on the metric perturbations. 

Moving now to curved spacetime, we found that the leading order adiabatic approximation in cosmology allowed for curved spacetime contributions to the self-energy, in contrast to the leading order quasistatic approximation which merely reproduced the flat spacetime results. It would be interesting to generalize the adiabatic particle concept so that it could be applied to more general, non-cosmological, situations. Although we are not aware of such a generalization, its basic ingredient, the WKB approximation has been much used in non-cosmological gravitational contexts (see \eg~ref.~\cite{BroutEtAl95}).

Unfortunately, due to technical difficulties, we have only been able to compute the cosmological contributions to the particle decay rate for non-relativistic particles. However presumably the most interesting case of interest would be that of ultrarelativistic particles. Therefore it is worth pursuing the calculation in this line. 

Finally, we end by mentioning that another interesting extension of the work of this thesis would be to analyze the dynamics of an ensemble of particles subject to the vacuum-induced dissipative effects. The aim would be to determine whether particles generically tend to lower values of the conformal momentum for physically reasonable spacetimes.

\part{Appendices}

\appendix
		
	\selectlanguage{catalan}
\chapter{Resum en català}

\section{Introducció}

L'objectiu bàsic d'aquesta tesi és intentar esbrinar què és allò que es pot aprendre sobre la propagació de les partícules  a través de l'anàlisi de les funcions de correlació de dos punts a teoria de camps, tant en espais plans com en espais corbats.

Pel que fa aquesta tesi, considerarem que les partícules són excitacions elementals amb un cert moment i una certa energia, i un temps de vida mig llarg. Les propietats bàsiques de les partícules es poden condensar en l'anomenada relació de dispersió,
\begin{equation} \label{RelDispCat}
	\mathcal E^2 = R_\vect p^2 - i R_\vect p \Gamma_\vect p,
\end{equation}
que és un lligam entre l'energia generalitzada $\mathcal E$, el moment $\vect p$ i el ritme de decaïment $\Gamma_\vect p$. A l'espai pla i en el buit la relació de dispersió és invariant Lorentz, i es pot obtenir a partir de l'anàlisi dels pols dels propagadors.

L'estudi de les propietats de les partícules i de les seves relacions de dispersió està originalment motivat pel problema transplanckià:  algunes prediccions de física de forats negres (radiació de Hawking) i cosmologia inflacionària depenen de la possibilitat de fer transformacions de Lorentz a energies més altes que l'energia de Planck.  Hom espera que tal vegada la física coneguda no sigui vàlida més enllà de l'escala de Planck, i que en particular la relació de dispersió de les partícules deixi de ser invariant Lorentz a mesura que ens acostem a aquesta escala d'energies. És per això que s'ha estudiat si aquestes prediccions es veurien afectades per modificacions de les relacions de dispersió a altes energies.

La invariància Lorentz de les relacions de dispersió pot ser trencada per efectes físics encara desconeguts quan ens apropem a l'escala de Planck, o, alternativament, pot ser trencada de manera efectiva quan la partícula es propaga en fons no trivials. En efecte, no hi ha cap raó per esperar que la relació de dispersió \eqref{RelDispCat} sigui invariant Lorentz quan s'analitza la propagació en situacions generals, ja sigui la propagació en un espaitemps corbat o en un medi a l'espai pla. De fet, aquest és un efecte ben conegut: per exemple, la velocitat de propagació dels fotons no té per què ser igual a $c$ en presència de camps magnètics \cite{Adler71} o  en espais corbats \cite{DrummondHathrell80}.

En aquesta tesi ens proposem analitzar la propagació en situacions generals des del punt de vista de teoria de camps, tot  estudiant les relacions de dispersió a partir de l'anàlisi de les funcions de correlació a dos punts. En aquest sentit ens preocuparem especialment dels efectes dissipatius, és a dir, de la part imaginària de la relació de dispersió.

\section{Sistemes quàntics oberts i lineals}

Com veurem al següent apartat, una partícula en propagació en un fons arbitrari es pot modelar com un sistema quàntic obert: el mode corresponent a la partícula  relativista és el subsistema obert, i els altres modes del camp, així com qualsevol altre camp que pugui estar en interacció, constitueixen l'ambient. També veurem com, sota algunes hipòtesis addicionals, pot ser tractat com un sistema quàntic lineal. Per tant, és d'interès iniciar aquest resum amb un breu repàs dels sistemes quàntics oberts lineals. Addicionalment això ens permetrà comparar el llenguatge propi de teoria de camps amb el propi de mecànica quàntica de sistemes oberts, contrastar els mètodes pertorbatius amb les solucions exactes, i i\l.lustrar la necessitat del mètode \emph{closed time path} (CTP).

El paradigma d'aquest tipus de sistemes és el model de moviment quàntic brownià (QBM, per les seves inicials en anglès), que consisteix en un osci\l.lador harmònic quàntic (la partícula browniana) en interacció lineal amb un ambient.
En concret, l'acció del sistema tancat es pot descomposar en les parts corresponents al subsistema d'interès, l'ambient i la interacció com
\begin{subequations}
\begin{align}
    S_\mathrm{sys}[q] & = \int \ud{t} \left[ \fud \dot q^2 - \fud     \omega_0^2 q^2 \right], \\
    S_\mathrm{env}[\varphi] & = \int \ud{t} \ud{x}  \left[ \fud(\partial_t
    \varphi)^2-\fud(\partial_x \varphi)^2 \right], \\
    S_{\text{int}}[q,\varphi] & = g \int \ud{t} \ud{x} \delta(x)  \dot q
    \varphi , \label{SIntCat}
\end{align}
\end{subequations}
on $\omega_0$ és la freqüència nua dels sistema i on $g$ és la constant d'acoblament. Considerem que l'osci\l.lador té massa unitat. 

Fem servir un camp unidimensional com a ambient, seguint la ref.~\cite{UnruhZurek89}. Hi ha una representació alternativa
 \cite{CaldeiraLeggett83b} en què l'ambient s'expressa com un conjunt d'osci\l.ladors harmònics :\begin{equation}
    S_{\text{int}}[q,\varphi] = \sqrt{L} \int \ud t \frac{\vd p}{2\pi} g \dot q 
    \varphi_p,
\end{equation}
on $\varphi_p(t)$ és proporcional a la transformada de Fourier espacial del camp
 i $L$ és la longitud de l'eix real.
El model es pot generalitzar fent
\begin{equation}
    S_\text{int}[q,\varphi] = \sqrt{L} \int \ud t  \frac{\ud p}{2\pi} \tilde f(-p) g \dot{q}(t) 
    \varphi_p(t).
\end{equation}
La funció real i parella $\mathcal I(p) := \tilde f(p) \tilde f(-p)$ és la distribució de freqüències, i caracteritza l'acoblament amb l'ambient. El cas estàndard $\mathcal I(p)=1$ correspon a l'anomenat ambient òhmic.

\subsection{Soroll i dissipació}

La matriu densitat reduïda $\hat\rho_\text{s}$ per a un sistema quàntic obert es defineix mitjançant
\begin{equation}
	\hat\rho_\text{s}(t) = \Tr_\text{env} \hat\rho(t),
\end{equation}
on $\hat\rho$ és la matriu densitat del sistema total.
Quan el sistema i l'ambient estan inicialment descorrelacionats ---$\hat{\rho}(t_\mathrm i)=\hat{\rho}_\mathrm{s}(t_\mathrm i)\otimes \hat{\rho}%
_\mathrm{s}(t_\mathrm i)$, on $\hat{\rho}_{s}(t_\mathrm i)$
i $\hat{\rho}_{e}(t_\mathrm i)$---, l'evolució de la matriu densitat es pot expressar com
\begin{equation}
    \rho _\text{s}(q_\mathrm f,q_\mathrm f^{\prime },t_\mathrm f)=\int \ud{q_\mathrm i}\ud{q_\mathrm i^{\prime
    }}J(q_\mathrm f,q_\mathrm f^{\prime },t_\mathrm f;q_\mathrm i,q_\mathrm i^{\prime },t_\mathrm i)\rho
    _{r}(q_\mathrm i,q_\mathrm i^{\prime },t_\mathrm i)\text{,}
\end{equation}
on el propagador $J$ es pot expressar en una representació de camps com
\begin{equation}
    J(q_\mathrm f,q_\mathrm f^{\prime },t_\mathrm f;q_\mathrm i,q_\mathrm i^{\prime
    },t_\mathrm i)=\int\limits_{q(t_\mathrm i)=q_\mathrm i}^{q(t_\mathrm f)=q_\mathrm f}{\cal D}%
    q\int\limits_{q^{\prime }(t_\mathrm i)=q_\mathrm i^{\prime }}^{q^{\prime
    }(t_\mathrm f)=q_\mathrm f^{\prime }}{\cal D}q^{\prime } \expp{i(S[q]-S[q^{\prime
    }]+S_{\mathrm{IF}}[q,q^{\prime }]) }\text{,} 
\end{equation}
on $S_{\mathrm{IF}}[q,q^{\prime }]$ és l'acció d'influència, relacionada amb el funcional d'influència de  Feynman i
Vernon \cite{FeynmanVernon63,FeynmanQMPI} a través de $F[q,q^{\prime
}]=\exp (iS_{\mathrm{IF}}[q,q^{\prime }])$.

Quan la matriu densitat inicial de l'ambient és gaussiana, l'acció d'influència es pot expressar com
%
%
\begin{equation} \label{S_IF2Cat}
\begin{split}
    S_{\mathrm{IF}}[q,q^{\prime}]=
    & \int_{t_\mathrm i}^{t_\mathrm f} \ud{t} \int_{t_\mathrm i}^{t_\mathrm f} \ud{t'} \Delta(t) H(t,t') Q(t')  \\
    & + \frac{i}{2} \int_{t_\mathrm i}^{t_\mathrm f} \ud{t} \int_{t_\mathrm i}^{t_\mathrm f} \ud{t'} \Delta(t) N(t,t')
    \Delta(t'),
\end{split}
\end{equation}
o com
\begin{equation} \label{S_IF3Cat}
\begin{split}
    S_{\mathrm{IF}}[q,q^{\prime}]
    =&- 2\int_{t_\mathrm i}^{t_\mathrm f} \ud{t} \int_{t_\mathrm i}^{t} \ud{t^{\prime
    }}\Delta (t) D(t,t^{\prime }) Q(t^{\prime }) +  \int_{t_i}^{t_f} \ud{t} \delta\omega^2_0 \Delta(t)Q(t) \\ & + \frac{i}{2}\int_{t_\mathrm i}^{t_\mathrm f}\ud{t}%
    \int_{t_\mathrm i}^{t_\mathrm f} \ud{t^{\prime }} \Delta (t) N(t,t^{\prime }) \Delta
    (t^{\prime }).
\end{split}
\end{equation}
Els nuclis $D(t,t')$ i $N(t,t')$ són els nuclis de soroll i dissipació, i  $\delta\omega^2_0$ és una renormalització de la freqüència, formalment infinita. El nucli $H(t,t')$ està relacionat amb el nucli de dissipació mitjançant $H(t,t')=-2\theta(t-t') D(t-t') + \delta \omega_0^2 \delta(t-t')$. Noteu que les parts real i imaginària del nucli $H(t,t')$ segueixen una relació de Kramers-Kronig:
\begin{equation}
  \Re H (\omega) = -2\PV \int \frac{\ud {\omega'}}{2\pi} \frac{
      \Im H(\omega')}{\omega-\omega'}+ \delta\omega^2_0.
\end{equation}

A l'espai de Fourier, el nucli de dissipació es pot expressar com
\begin{subequations}
\begin{equation} \label{DisQBMCat}
    D(\omega) = \frac{i \omega g^2}{2} \mathcal I(\omega).
\end{equation}
Fixeu-vos que és independent de l'estat. En canvi, el nucli de soroll depèn de l'estat, i es pot expressar com
\begin{equation}
	N(\omega)=  {g^2|\omega|}\mathcal I(\omega)\left[\frac12 + n(|\omega|)\right],
\end{equation}
\end{subequations}
on $n(|p|) := \Tr{ ( \hat\rho_\text e \hat a^\dag_p \hat a_p )}$ és el nombre d'ocupació de l'ambient a energia $|p|$.
Els detalls del càlcul es poden trobar a l'apartat \ref{sect:DisNoise}.
Si l'ambient està en un estat tèrmic, llavors els nombres d'ocupació venen donats per $n(|\omega|) = 1/(\expp{|\omega|/T}-1)$ i els nuclis de soroll i dissipació estan relacionats mitjançant el teorema de fluctuació-dissipació:
\begin{equation}\label{FluctDisThCat}
    N(\omega) = - i \sign(\omega) \coth  \left(\frac{|\omega|}{2 
    T}\right)
    D(\omega).
\end{equation}

\subsection{El funcional generador i les funcions de correlació}

Tal com s'explica a l'apèndix~\ref{app:CTP}, el funcional generador CTP es defineix mitjançant
\begin{equation}
    Z[j_1,j_2] = \Tr \left[  \hat \rho\, \widetilde T \expp {- i \int \ud{t} j_2(t)
    \hat q(t) }T \expp {i \int \ud{t} j_1(t)
    \hat q(t)}   \right].
\end{equation}
Per a un sistema gaussià amb condicions de contorn asimptòtiques el funcional generador es pot expressar com
\begin{equation}
\begin{split} \label{ZCTPCat}
    Z[j_1,j_2] =  & \
     \expp{  \fud \int \ud{t_1} \ud{t_2} \ud{t_3}
    \ud{t_4}
     j_\Delta(t_1)
    \Gret(t_1,t_2)N(t_2,t_3)j_\Delta(t_4)\Gret(t_4,t_3) } \\
    & \times
    \expp{  -\int \ud{t_1} \ud{t_2} j_\Delta(t_1) \Gret(t_1,t_2)
    j_\Sigma(t_2)},
\end{split}
\end{equation}
on $j_\Sigma(t) := [j_1(t)+j_2(t)]/2$, $j_\Delta(t) :=
j_1(t) - j_2(t)$ i on $\Gret(t,t')$ és el propagador retardat del nucli
 \begin{equation} \label{KernelLCat}
    L(t,t') = \left( \dert[2]{}{t} + \omega_0^2 \right)
    \delta(t-t') + H(t,t'),
\end{equation}
que coincideix amb el propagador retardat del sistema. 

Diferenciant el propagador retardat obtenim les funcions de correlació. En particular, el propagador retardat es pot escriure com
\begin{subequations}
\begin{equation}
	\GR(\omega) = \frac{-i}{-\omega^2 + \omega_0^2 + H(\omega)},
\end{equation}
i el propagador de Feynman com:
\begin{equation}
\begin{split}
    G_\mathrm F(\omega) 
    &= \frac{- i \left[ - \omega^2 + \omega_0^2  + \Re H(\omega) \right]
    + N(\omega)}{\left[ - \omega^2 + \omega_0^2 + \Re H(\omega)\right]^2 +
    \left[\Im H(\omega)\right]^2}.
\end{split}
\end{equation}
\end{subequations}
	Podeu trobar les expressions per als altres porpagadors a les equacions \eqref{GOtherFourier}.

\subsection{L'aproximació estocàstica}

Així com l'evolució quàntica d'un sistema tancat composat per osci\l.ladors harmònics acoblats linealment està completament determinada per la solució del  corresponent problema clàssic, l'evolució quàntica d'un sistema obert composat per  osci\l.ladors harmònics acoblats linealment està completament determinada per la solució d'un problema estocàstic corresponent, tal com fou notat per Feynman \cite{FeynmanVernon63,FeynmanQMPI}.

Definim una acció efectiva estocàstica donada per
\begin{equation} \label{StochActionCat}
    S_{\text{eff}}[q,q';\xi] = S[q] - S[q'] +  \Re S_{\mathrm{IF}}[q,q']+\int \ud t \xi(t)\left[q'(t)- q(t)\right],
\end{equation}
on $\xi(t)$ és un procés estocàstic gaussià definit pels correladors
\begin{subequations} \label{StochasticCorrCat}
\begin{align}
    \av{ \xi(t) }_\xi &=0, \\
    \av{ \xi(t)\xi(t') }_\xi &=  N(t,t'),
\end{align}
\end{subequations}
on $\av{\cdot}_\xi$ vol dir mitjana estocàstica. La mitjana de l'acció efectiva estocàstica és l'acció efectiva:
\begin{equation}
    \av{ \expp{ {i} S_{\text{eff}}[q,q';\xi]}}_\xi =
    \expp{ i \Gamma[q,q']}.
\end{equation}

Podem obtenir una equació del moviment estocàstica, l'equació de Langevin, fent
\begin{equation}
    \frac{\delta}{\delta q(t)} S_{\text{eff}}[q,q';\xi]
    \Big|_{q'=q} \propto   \ddot q(t) + \int \ud{t'} H(t,t') q(t') +\omega_0^2 q(t)- {\xi(t)} = 0,
\end{equation}
la solució de la qual és
\begin{equation} \label{LangevinSolutionCat}
	q(t) = q_\mathrm h(t;q_0,v_0) - i \int_{-\infty}^{\infty} \ud {s} \Gret(t,s)
    \xi(s),
\end{equation}
on $q_\mathrm h(t;q_0,v_0)$ és una solució homogènia de l'equació de Langevin amb condicions inicials $q_0$ i $v_0$. Mitjançant la solució de l'equació de Langevin podem calcular funcions de correlació estocàstiques per a la variable $q$. Es pot veure \cite{CalzettaRouraVerdaguer03,RouraThesis} que les funcions de correlació estocàstiques corresponen exactament a un subconjunt de les funcions de correlació quàntiques. 

\subsection{Dinàmica de les pertorbacions}

Fins ara hem considerat propietats d'equilibri. Anem ara a considerar la dinàmica de les pertorbacions: considerem que a un cert temps $t_0$ treiem el sistema del seu estat d'equilibri. Assumirem que el sistema és tal que la pertorbació no decau ràpidament, sinó que té una vida mitja llarga. Estem interessats en esbrinar quina és l'energia de la pertorbació i el seu temps de vida mig, fent servir els mètodes estocàstics que acabem de presentar. L'hamiltonià del sistema l'escollirem $\hat H_\text{sys}(t) =   \dot{\hat q}^2(t) /2 + \Omega_1^2 \hat q^2(t)/2$  on $\Omega_1$ és la freqüència efectiva del sistema obert, un paràmetre que voldrem determinar amb algun criteri físic (en general no coincideix amb la freqüència lliure perquè poden haver-hi efectes de renormalització).

L'evolució de les pertorbacions es pot escriure en general com
\begin{equation}
	E(t)=E_0 +  \frac{1}{2}  \av{\dot q_\text h^2(t;q_0,v_0)}_{q_0,v_0} + \fud \Omega_1^2 \av{ q_\text h^2(t;q_0,v_0)}_{q_0,v_0},
\end{equation}
on $\av{ \cdot}_{q_0,v_0}$ indica una mitjana sobre la funció de Wigner de la pertorbació a temps inicial i on $q_\text h(t;q_0,v_0)$ és una solució homogènia de l'equació de Langevin amb condicions inicials $q_0$ i $v_0$. Per a sistemes amb pertorbacions suficientment estables, es pot veure que la solució de l'equació de Langevin es pot escriure en general com
\begin{equation}
	q_\mathrm h(t;q_0,v_0)= q_0 \expp{-\Gamma (t-t_0)/2} \cos{ \Omega(t-t_0)} + \frac{v_0}{\Omega} \expp{-\Gamma(t-t_0)/2} \sin{\Omega(t-t_0)},
\end{equation}
on $\Omega$ està definit a través de l'equació autoconsistent
$
	\Omega^2 :=  \omega_0^2 + \Re H(\Omega)
$
i $\Gamma := - \Im{H(\Omega)} / \Omega$.  Aquesta solució només és vàlida quan $\Omega \gg \Gamma$ i per a temps llargs en comparació amb $\Omega$: $t-t_0 \gg \Omega^{-1}$.

En general, l'energia evoluciona d'acord amb l'equació \eqref{EnergyPertLarge}: si $\Omega_1 \neq \Omega$, hi ha ràpides osci\l.lacions de l'energia amb freqüència $\Omega$, modulades per un decaïment exponencial a un ritme $\Gamma$. Com que físicament no esperem osci\l.lacions de l'energia, establim $\Omega_1 = \Omega$, de manera que  trobem 
\begin{equation}\label{QBMEtCat}
	E(t)=E_0 + \Av{ \frac{\Omega^2 q_0^2+ v^2_0}{2\Omega^2}}_{q_0,v_0} \expp{-\Gamma(t-t_0)}.
\end{equation}
És a dir, podem determinar que l'energia segueix un lent decaïmenta un ritme $\Gamma$. Addicionalment establim que la freqüència efectiva del sistema ve donada per $\Omega$.

\section{Excitacions de quasipartícula en estats generals}

En aquest apartat estudiem les excitations tipus partícula sobre estats generals a l'espai de Minkowski. En particular volem analitzar tota la informació sobre les excitacions elementals que es pot obtenir estudiant les funcions de correlació de dos punts.

\subsection{Partícules en el buit}

En el buit, les propietats de les partícules estables es poden estudiar a partir de la representació de K\"allen-Lehmann. Les funcions de correlació en el buit es poden expressar com 
\begin{subequations}
\begin{align}
	G(p) &= 2\pi \rho(-p^2) \sign(p^0), \label{PauliJordanSpectralCat}\\
	\GR(p) &= \int_0^\infty \frac{-i\rho(s)\,\vd s}{-(p^0+i\epsilon)^2+\vect p^2+s},\label{retardedSpectralCat} \\
	\GF(p) &= \int_0^\infty \frac{-i\rho(s)\,\vd s}{p^2+s-i\epsilon}. \label{FeynmanSpectralCat}
\end{align}
\end{subequations}
on $\rho(-p^2)$ és la funció espectral de buit,
\begin{equation}\label{SpectralFunctionCat}
	\rho(-p^2) := 
\frac{1}{2\pi} \sum_n (2\pi)^4 \delta^{(4)}(p-p_n) |\langle 0 |\phi(0)|n\rangle|^2,
\end{equation}
que verifica les propietats \eqref{PropVacSpectral}.

Si hi ha partícules estables en el buit, aquestes estaran caracteritzats per la seva massa i el seu moment:  $\hat{\vect p }|\vect p\rangle = \vect p |\vect p\rangle$, $\hat{H }|\vect p\rangle = E_\vect p|\vect p\rangle$, amb $E_\vect p^2 = \vect p^2 + \mph^2$. En presència de partícules estables, la representació espectral \eqref{SpectralFunctionCat} es descomposa en el terme corresponent a la partícula estable i la contribució d'estats de moltes partícules:
\begin{subequations}
\begin{equation}
	\rho(-p^2) = Z \delta(p^2 + \mph^2) + \theta(p^2+m_*^2) \sigma(-p^2) ,
\end{equation}
on
\begin{equation}
	Z=\abs{\langle 0 |\phi(0)|\vect q\rangle}^2  
\end{equation}
\end{subequations}
és una constant positiva, $\sigma(-p^2)$ és la contribució de moltes partícules i $m_*$ és la massa mínima dels estats de moltes partícules. En el pla complex $p^2$ el propagador de Feynman té un pol al valor de $p^2$ corresponent a la massa de la partícula estable.
\begin{equation}
	\GF(p) = \frac{ -i Z}{p^2 + m^2 - i\epsilon} + \int_{m_*^2}^\infty \vd s \frac{ -i \sigma(s)}{p^2 + s - i\epsilon}.
\end{equation}
L'expressió corresponent per al propagador retardat es troba a  \eqref{VacuumRetardedAnalytic}.  Vegeu les figures \ref{fig:FeynmanVacuum} i \ref{fig:retardedVacuum} a la pàgina \pageref{fig:FeynmanVacuum} per a una anàlisi de l'estructura analítica a l'espai d'energies complexes.

Estrictament parlant, les partícules inestables no es poden considerar estats asimptòtics de l'espai de Fock, atès que no són estats propis de l'hamiltonià interactuant, sinó que són combinacions d'estats propis  corresponents als productes de decaïment. Per tant, la funció espectral de buit és la corresponent a estats de moltes partícules, i les partícules inestables venen descrites per talls i no pas per pols. Tot i això, quan la funció espectral està molt picada entorn d'una certa energia, es pot veure que la prolongació analítica de la funció de Green es pot caracteritzar aproximadament per un pol:
\begin{equation}\label{decayGFCat}
	\GF(p) = \frac{-iZ}{p^2 + m^2 - i m \gamma},
\end{equation}
on $\gamma$ és el ritme de decaïment. 

Tant per a partícules estables com inestables, la massa i el ritme de decaïment es poden extreure a partir de l'autoenergia retardada $\SigmaR(p)$. La massa física ve donada per la solució d'ordre més baix de la següent equació:
\begin{subequations}
\begin{equation}
	-\mph^2+ m^2 + \Re \Sigma(\mph^2) = 0.
\end{equation}
Per a partícules estables la massa es pot definir exactament, però per a partícules inestables hi ha una incertesa natural associada que ve donada pel ritme de decaïment, que a la vegada es pot trobar a partir de:
\begin{equation}
	\gamma =-\frac{1}{\mph} \Im \Sigma(\mph^2).
\end{equation}
\end{subequations}
Recordem que l'autoenergia està relacionada amb el propagador de Feynman mitjançant
\begin{equation}\label{FeynmanUnstableCat}
	\GF'(p) = \frac{-i}{p^2+m^2 + \Sigma(-p^2) } 
\end{equation}

També és interessant treballar amb la representació temporal de la funció de Green,
\begin{equation}\label{GFTimeDecayCat}
	\GF(t,t';\vect p) \approx \frac{-iZ}{2E_\vect p} \expp{-iE_\vect p|t-t'|}\expp{- \Gamma_\vect p |t-t'|/2},
\end{equation}
on $\Gamma_\vect p = m\gamma/E_\vect p$ és el ritme de decaïment al sistema de referència del laboratori. Val la pena notar que, tot i que experimentalment és molt satisfactòria, hom espera desviacions d'aquesta darrera expressió tant per a temps molt curts com per a temps molt llargs. En tot cas, es pot veure que la representació temporal de la funció de Green està connectada directament amb l'evolució de l'energia en funció del temps,
\begin{subequations}
\begin{equation}
	 E(t,t_{0};\vect p) := \langle \vect p|\hat H_{\vect p}(t)|\vect p\rangle =  E_0 + {4E_\vect p^3} |\GF(t,t_0;\vect p)|^2,
\end{equation}
on $H_\vect p$ és l'hamiltonià del mode en qüestió i $E_0$ és l'energia del buit. A partir d'aquesta darrera expressió i l'eq.~\eqref{GFTimeDecayCat} veiem que $E_\vect p$ és l'energia inicial de la partícula i que $\Gamma_\vect p$ és el ritme de decaïment:
\begin{equation}
	E(t,t_0;\vect p) = E_0 + E_\vect p\expp{-\Gamma_\vect p(t-t_0)}.
\end{equation}
\end{subequations}

\subsection{Propagadors i autoenergies en un estat arbitrari}

En el buit, l'anàlisi del propagador de Feynman és suficient, però no és així amb estats arbitraris. En general cal considerar diferents propagadors, que és convenient organitzar en una matriu $2\times2$, d'acord amb el formalisme CTP. Tenim però una certa llibertat. Podem treballar amb l'anomenada base directa
\begin{subequations}
\begin{equation}
	G_{ab}(x,x') = 
		\begin{pmatrix}
			\GR(x,x') & G_-(x,x') \\
			G_+(x,x') & G_\text{D}(x,x')
 		\end{pmatrix},
\end{equation}
i també amb la base física o de Keldysh:
\begin{equation}\label{KeldyshCat}
	G'_{a'b'}(x,x') = 
		\begin{pmatrix}
			G^{(1)}(x,x') & G_\text R(x,x') \\
			G_\text A(x,x') & 0
 		\end{pmatrix}.
\end{equation}
\end{subequations}
Podeu trobar les definicions dels diferents propagadors al capítol 3 i a l'apèndix \ref{app:GenRel}. Les propietats de les dues bases s'estudien a l'apèndix \ref{app:CTP}. Una altra diferència amb el cas de buit és que els propagadors a l'espai de Fourier estan definits per
\begin{equation} \label{MidPointCat}
    G_{ab}(p;X) = \int \ud[4]{x} \expp{-i p \cdot \Delta}
    G_{ab}(X+\Delta/2,X-\Delta/2),
\end{equation}
i en general contenen una certa dependència espaciotemporal. 

L'autoenergia també té forma matricial $\Sigma^{ab}(x,x')$, i igualment es pot expressar tant en la base física com  en la base de Keldysh.  En general no hi ha una relació senzilla amb els propagadors, sinó que allò que trobem és una relació matricial autoconsistent:
\begin{equation} \label{SelfEnergyGeneralCat}
    G_{ab}(x,x') = G_{ab}^{(0)}(x,x')+
    \int \ud[4]{y} \ud[4]{y'} G_{ac}^{(0)}(x,y) [-i\Sigma^{cd}(y,y')]  G_{db}(y',z).
\end{equation}
Trobem però un cas particular: per motius de causalitat, hi ha una relació directa entre el propagador retardat i l'anomenada autoenergia retardada:
\begin{equation} \label{SelfEnergyGeneralRetardedCat}
    \GR(x,x') = \GR^{(0)}(x,x')+
    \int \ud[4]{y} \ud[4]{y'} \GR^{(0)}(x,y) [-i\SigmaR(y,y')]  \GR(y',z),
\end{equation}
En el cas en què l'estat de fons  és homogeni i estacionari podem introduir una representació de moment i resoldre l'equació anterior:
\begin{equation} 	\label{GSigmaDiagonalCat}
	\GR(p) = \frac{-i}{ p^2 +m^2+\SigmaR(p)-p^0 i\epsilon}.
\end{equation}
Si l'estat no és exactament homogeni, sinó que té inhomogeneïtats a una gran escala $L$, hi haurà correccions a aquesta equació que seran de l'ordre de $p^{-1}L^{-1}$.

Amb estats generals també és possible derivar una relació de Källén-Lehmann. Cal treballar però a l'espai d'energies (i no en l'espai de $p^2$). La funció de Pauli-Jordan fa el paper de funció espectral, i trobem
\begin{equation}
	\GR(p;X) = \int \frac{\vd k^0}{2\pi}   \frac{i G(k^0,\vect p;X)}{p^0 - k^0 + i\epsilon}.
\end{equation}
No hi ha una relació anàloga per al propagador de Feynman. Si l'estat de fons és homogeni i estacionari trobem que la funció de Pauli-Jordan és proporcional a la probabilitat de crear una excitació a un estat d'energia més alta menys la probabilitat de crear una excitació a un estat d'energia més baixa:
\begin{equation} \label{SpectralDetailedCat}
	G(p) = \sum_{\alpha,\beta} \rho_{\alpha}\abs{\langle\alpha|\hat\phi_{\vect p}|\beta\rangle}^{2} \left[ \delta(p^0 + E_{\alpha} - E_{\beta})-\delta(p^0 - E_{\alpha} + E_{\beta})\right].
\end{equation}

\subsection{Interpretació en termes d'un sistema quàntic obert}

Atès que qualsevol teoria de camps està composada per més d'un grau de llibertat (de fet per un nombre infinit d'ells), els camps quàntics es poden analitzar des del punt de vista de la teoria de sistemes quàntics oberts. La divisió entre sistema i ambient és obviament arbitrària, i depèn de les propietats que estem interessats en analitzar. La divisió  que per a nosaltres és més interessant és la següent: el sistema està constituït pels modes del camp amb moments $\vect p$ i $-\vect p$, i l'ambient per la resta de modes del camp, incloent-hi també els modes de qualsevol altre camp que pugui estar en interacció. 

Prenent el model $\lambda\phi^4$ com un exemple, si no hi haguessin divergències, l'acció es podria dividir en l'acció del sistema, acció de l'ambient, acció d'interacció i acció de contratermes de la següent forma:
\begin{subequations}\label{OQSActionCat}
\begin{align}
	S_\text{sys} &= \int \ud t  \left( \dot \phi_\vect p \dot \phi_{-\vect p} -R^2_\vect p \phi_\vect p \phi_{-\vect p} \right),\\
	S_\text{env} &=  \frac{V{\mathcal Z}_\vect p}{2} \int\limits_{\vect q \neq \pm \vect p} \udpi[3]{\vect q} \int \ud t  \left( \dot \phi_\vect q \dot \phi_{-\vect q} -R^2_\vect p \phi_\vect q \phi_{-\vect q} \right),\\
		S_\text{int} &=\frac{ V^2  \lambda \mathcal Z^2_\vect p}{4!} \int \udpi[3]{\vect q} \udpi[3]{\vect q'} \int \ud t
		\phi_{\vect p} \phi_{\vect q} \phi_{\vect q'} \phi_{-\vect p-\vect q -\vect q'}, \\
	S_\text{fc}&=  
	\int \ud t  \left\{ ({\mathcal Z}_\vect p-1)\dot \phi_\vect p \dot \phi_{-\vect p} -\big[{\mathcal Z}_\vect p (\vect p^2+m^2)- R^2_\vect p\big] \phi_\vect p \phi_{-\vect p} \right\},
\end{align}
\end{subequations}
Hem permès un reescalat arbitrari dels camps $\phi_\vect p \to \phi_\vect p/{\mathcal Z}_\vect p$, i hem introduït una renormalització finita de la freqüència dels modes, de manera que la freqüència efectiva és $R_\vect p$. De moment, no especificarem els valors de  ${\mathcal Z}_\vect p$ i  $R^2_\vect p$, que deixarem com a paràmetres lliures. Com que a més a més apareixen divergències en els càlculs, cal permetre també contratermes infinits [vegeu l'equació \eqref{InfCount}].

El funcional generador d'aquest sistema quàntic obert està caracteritzat per una sèrie infinita de funcions de correlació, ja que es tracta d'un sistema quàntic no lineal. Tanmateix, en molts casos hom està interessat en estudiar propietats que tan sols depenen de les funcions de correlació de dos punts. En altres casos  només es té accés a les funcions de correlació de dos punts i s'espera que la part connectada de les funcions de correlació de dos punts sigui petita. En tots aquests casos, el sistema quàntic obert es pot aproximar per un sistema quàntic obert gaussià, caracteritzat per les funció de correlació de dos punts. Si a més a més l'estat és homogeni i isòtrop, la conservació del moment fa que l'únic conjunt de propagadors rellevants sigui $G_{ab}(t,t';\vect p)$. En aquestes condicions, el funcional generador per als dos modes $+\vect p$ i $-\vect p$ es pot expressa simplement com:
\begin{equation} \label{ZDirectBasisCat}
\begin{split}
	Z[j_{a}]&\approx\exp\bigg[-\frac{1}{2!}\int \ud t \ud{t'} j^{a}(t) j^{b}(t) G_{ab}(t,t';\vect p) \bigg].
\end{split}
\end{equation}
D'aquí es pot concloure que, dins de l'aproximació gaussiana, cada conjunt de parells de modes d'un sistema quàntic donat es pot descriure com si fos una partícula quàntica browniana interactuant linealment amb un ambient efectiu. Vegem exactament en què consisteix això.

En termes de la interpretació de sistemes quàntics oberts, podem assignar a cada parell de modes un sistema efectiu, format per a un osci\l.lador harmònic de freqüència $R_\vect p$, interactuant amb un ambient efectiu, caracteritzat per un nucli de dissipació $D(t,t')=-i\Im \SigmaR(t,t';\vect p)$ [$H(t,t')=\SigmaR(t,t';\vect p)$] i per un nucli de soroll $N(t,t')=\SigmaN(t,t';\vect p)$, on $\SigmaR(t,t';\vect p)$ i $\SigmaN(t,t';\vect p)$ són l'autoenergia retardada i l'autoenergia de Hadamard respectivament (vegeu apèndix \ref{app:CTP}). Podem elaborar els detalls del sistema obert: l'acoblament efectiu de l'ambient es pot deduir a partir de l'expressió del nucli de dissipació,
\begin{subequations}
\begin{equation}
	D(\omega) = \frac{ i g^2_\text{lin}}{2} \omega\mathcal I(\omega),
\end{equation}
i l'estat gaussià del sistema efectiu, que està caracteritzat pels nombres d'ocupació $n(|\omega|) = \Tr [\hat\rho_\text{e} \hat a_\omega^\dag \hat a_\omega ] $, es pot deduir a partir de
\begin{equation}
	N(\omega) = g^2_\text{lin} |\omega| \mathcal I(\omega) \left[ \frac{1}{2} + n(|\omega|) \right].
\end{equation}
\end{subequations}
Trobareu més detalls a la taula \ref{tbl:equivalence} a la pàgina \pageref{tbl:equivalence}.

Una primera aplicació de la interpretació com a sistema quàntic obert és l'anàlisi de les autoenergies. Un càlcul senzill en el sistema lineal equivalent mostra que la part imaginària de l'autoenergia és proporcional a la probabilitat que una excitació de l'estat decaigui a l'ambient, menys la probabilitat que l'ambient exciti l'estat,
\begin{equation}
	\Im\SigmaR (\omega) =-\omega[\Gamma_-(\omega) - \Gamma_+(\omega)],
\end{equation}
recuperant així el resultat de Weldon \cite{Weldon83}.

\subsection{Excitacions de quasipartícula en estats genèrics}

La quasipartícula és una generalització del concepte de partícula per a estats diferents del buit. Considerarem que una quasipartícula és una excitació d'un estat quàntic que:
\begin{enumerate}
	\item Té una energia inicial $E$, les fluctuacions de la qual són petites: $\Delta E \ll E$.
	\item Té un moment inicial $\vect p$, les fluctuacions del qual són també petits: $\delta \vect p\ll |E|$.
	\item La seva energia i el seu moment romanen aproximadament constants durant molt de temps. Més precisament, el ritme de decaïment $\Gamma$ ha de ser molt més petit que no la freqüència de de Broglie de la partícula: $\Gamma\ll |E|$.
	\item És elemental, és a dir, no es pot descomposar en la superposició coherent o incoherent de dues pertorbacions que verifiquin les tres propietats anteriors.
\end{enumerate}
Caracteritzarem les quasipartícules per la seva energia $E$, el seu moment $\vect p$ i el ritme de decaïment $\Gamma$; addicionalment, les quasipartícules poden tenir altres nombres quantics com ara l'espín.
Les quasipartícules poden existir fins i tot en sistemes quàntics fortament interactuants. En tot cas, requereixen que les fluctuacions de l'energia i el moment de l'estat original siguin petites per tal de complir amb la condició de la petitesa de les fluctuacions.

\longpage

Les quasipartícules es poden caracteritzar per l'acció de l'operador de creació i anihilació. Podem trobar dos tipus d'excitacions: 
\begin{equation}
	\hat\rho^\pplus _\vect p := \frac{1}{n_\vect p + 1}\, 
\hat a^\dag_\vect p \hat\rho \hat a_\vect p, \qquad
\hat\rho^\pminus _\vect p = \frac{1}{n_\vect p}\, 
\hat a_{-\vect p} \hat\rho \hat a^\dag_{-\vect p}.
\end{equation}
Les primeres corresponen a excitacions d'energia positiva, i les segones a excitacions d'energia negativa (forats). 

Trobem el següent fenomen sorprenent: el valor esperat del nombre de partícules,
\begin{equation}
	\av{\hat N_\vect p}^\pplus = \Tr{\big(\hat\rho^\pplus _\vect p\hat N_\vect p \big)}  = \Tr{\big(\hat\rho^\pplus _\vect p \hat a^\dag_\vect p \hat a_\vect p\big)}  =  \frac{1}{n_\vect p+1} \av{(\hat N_\vect p+1)^2},
\end{equation}
és incrementat en més d'una unitat respecte el seu valor no pertorbat
\begin{equation}
	\av{\hat N_\vect p}^\pplus = n_\vect p + 1 + \frac{\delta n_\vect p^2}{1+n_\vect p} =: n_\vect p + N_\vect p^\pplus,
\end{equation}
on $N^\pplus_\vect p=1 + \delta n_\vect p^2/(1+n_\vect p)$ és el nombre d'excitacions i
	$\delta n_\vect p^2 $
és la dispersió en el nombre de partícules en l'estat de fons. Per a un estat propi del nombre de partícules $\delta n_\vect p^2 =0$ i $N^\pplus_\vect p=1$,  i per a un estat Gaussià $\delta n_\vect p^2 = n_\vect p ( n_\vect p +1)$ i $N^\pplus_\vect p=1 + n_\vect p$. La raó d'aquest fenomen és purament estadística: els components més ocupats de l'estat es veuen afavorits per l'estadística de Bose-Einstein. Pels forats succeeix quelcom similar, però encara més dramàtic: el nombre esperat de partícules augmenta quan treiem una partícula de l'estat de fons. La causa és novament estadística, i en el cas dels forats això ens suggereix que no els podem considerar com a quasipartícules reals.

Un estudi de la dispersió en els moments confirma que els forats no es poden considerar veritables quasipartícules en sistemes bosònics, pel fet que els forats tenen una dispersió comparable amb el valor esperat. En canvi, la dispersió de les quasipartícules d'energia positiva és petita si els nombres d'ocupació de l'estat original no són excessivament elevats. 

El fet que els estats creats a partir dels operadors de creació i anihilació corresponguin als estats físics es pot entendre de la següent forma. Considerem un agent extern caracteritzat per una grau de llibertat $Q$, corresponent a un emissor de quasipartícules. Aquest agent comença en un estat excitat. Per simplicitat, assumim que aquest agent està linealment acoblat amb un parell de modes del camp, $g_Q Q(t) [\phi_\vect p(t)+\phi_{-\vect p}(t)] $, amb una constant d'acoblament petita. El sistema total es troba en un estat inicial $\hat\rho \otimes |1_Q\rangle \langle 1_Q|$. Hom pot veure que per a temps llargs, si l'agent $Q$ es detecta en el seu estat fonamental, i el sistema es troba inicialment en un estat $\hat\rho$ qualsevol, llavors al cap d'un cert temps es trobarà en un estat $\hat\rho_\vect p^\pplus \propto \hat a_\vect p^\dag \hat\rho\,\hat a_\vect p+\hat a_{-\vect p}^\dag \hat\rho\,\hat a_{\vect p}$.

Hem trobat dues formes bàsiques d'analitzar les propietats bàsiques de les quasipartícules: estudiant-ne la representació espectral, o bé estudiant l'evolució del seu estat quàntic. Comencem per la primera de les opcions.

\longpage

Suposem que la funció espectral (propagador de Pauli-Jordan) té la següent forma per a un cert rang d'energies:
\begin{equation}\label{StableQuasiparticleCat} 
	G(\omega,\vect p) \approx \frac{Z_{\vect p}}{2R_{\vect p}}\big[\delta(\omega-R_{\vect p})-\delta(\omega+R_{\vect p})], \quad \text{per a }|\omega|\sim R_{\vect p}.
\end{equation}
Comparant amb l'\Eqref{SpectralDetailedCat} veiem que aquesta funció espectral implica que l'operador camp crea una excitació d'energia $R_\vect p$. Per tant, podem identificar $R_\vect p$ com l'energia de l'excitació. Atès que no hi ha dispersió en l'energia, l'excitació ha de tenir una vida mitja infinita. Aquesta hipòtesi no és realista, ja que els fenòmens dissipatius són genèrics a teoria de camps. Una hipòtesi més realista és suposar que la funció de Pauli-Jordan té una certa dispersió:
\begin{equation}\label{StableQuasiparticleCat2} 
	G(\omega,\vect p) \approx \frac{2Z_\vect p \omega \Gamma_\vect p}{(-\omega^2+R_\vect p^2)^2 + (\omega \Gamma_\vect p)^2}, \quad \text{per a }|\omega|\sim R_{\vect p},
\end{equation}
cosa que ens porta a un propagador retardat de la forma
\begin{equation}
	\GR(\omega,\vect p) \approx \frac{-i Z_\vect p}{-\omega^2 +R^2_{\vect p} - i \omega \Gamma_{\vect p}}.
\end{equation}
Per comparació amb l'expressió del propagador en funció de l'autoenergia obtenim que
\begin{subequations}
\begin{align}	
	R_\vect p &= \Re \SigmaR(R_\vect p,\vect p), \\
	\Gamma_\vect p &= -\frac{1}{R_\vect p}\Re \SigmaR(R_\vect p,\vect p).
\end{align}
\end{subequations}
De l'anàlisi de la part imaginària de l'autoenergia sabem que $\Gamma_\vect p$ correspon al ritme de decaïment net de la pertorbació, i a partir de la representació espectral \eqref{SpectralDetailedCat} obtenim que $R_\vect p$ correspon a l'energia aproximada de la partícula.

\longpage

La segona manera d'estudiar les propietats de les quasipartícules és a través de la seva evolució temporal. El valor esperat de l'hamiltonià en funció del temps ve donat per
\begin{equation}
	E^\pplus(t,t_{0};\vect p) := \av{\hat H_{\vect p}(t)}_\pplus  = \frac{1}{ n_\vect p + 1} \Tr{ \big(\hat a_\vect p^\dag\hat\rho\hat a_\vect p U(t_0,t)\hat H_\vect p U(t,t_{0})   \big) },
\end{equation}
on  $
	H_\vect p =  \dot \phi_\vect p \dot \phi_{-\vect p} + R_\vect p^2  \phi_\vect p \phi_{-\vect p}
$ i
on ja suposem que $R_\vect p$ és la freqüència física del mode. Fent servir l'analogia amb un sistema quàntic obert lineal, i per tant negligint l'efecte de les funcions de quatre punts, trobem que l'evolució de l'energia ve donada per
\begin{equation}\label{TimeEvolQuasi1Cat}
	E^\pplus(t,t_{0};\vect p) \approx E^{(0)}_\vect p+ R_\vect p (n_\vect p +1) \expp{-\Gamma_\vect p (t-t_0)}.
\end{equation}
on $n_\vect p +1$ és el factor estadístic del qual hem parlat abans. A partir d'aquesta equació identifiquem novament $R_\vect p$ com l'energia de la quasipartícula, i $\Gamma_\vect p $ com el seu ritme de decaïment.

Les propietats bàsiques de les quasipartícules es poden condensar adequadament en la forma d'una relació de dispersió, és a dir, una relació entre l'energia i el moment de la quasipartícula. Per a quasipartícules inestables és útil el concepte de la relació de dispersió generalitzada, 
\begin{equation}\label{generalizedDispRelCat}
	\mathcal E^2 = R_\vect p^2 - i R_\vect p \Gamma_\vect p =  m^2 + \vect p^2 + \Sigma(R_\vect p,\vect p)  .
\end{equation}

Les expressions d'aquest apartat les hem deduït per al cas d'estats homogenis, estacionaris i isòtrops. Si els estats no són homogenis o estacionaris les expressions continuen essent vàlides sempre que les escales d'inhomogeneïtat siguin prou grans amb comparació amb les freqüències típiques del sistema. 

\subsection{Mètodes alternatius}

Per acabar aquesta secció esmentem  altres aproximacions que s'han fet servir a la literatura. 

El mètode estàndard per estudiar la dinàmica de les quasipartícules és la teoria de la resposta lineal \cite{Kapusta,LeBellac,FetterWalecka,TheoryOfSolidsIIMIT,Reichl}. La teoria de la resposta lineal estudia el comportament del valors esperats de certs operadors quàntics quan introduïm una pertorbació al sistema. En el context de teoria de camps relativista generalment s'escull com a pertorbació  una corrent del propi camp i s'estudia el valor esperat del camp. El resultat que s'obté és que la pertorbació del camp osci\l.la amb una freqüència que ve donada per $R_\vect p$ i decau amb un ritme $\Gamma_\vect p/2$. Amb la teoria de la resposta lineal hom està estudiant el comportament de pertorbacions macroscòpiques més que no pas el comportament de les quasipartícules individuals. Cal remarcar que per a les quasipartícules que nosaltres hem considerat el valor esperat del camp és zero tota l'estona.

Un altre mètode, equivalent a l'anterior, és l'estudi de la dinàmica del camp mig a partir de l'acció efectiva \cite{Weldon98,DrummondHathrell80}. El resultat que s'obté és que la dinàmica del camp mig $\bar\phi (x)$ ve determinada per l'autoenergia retardada:\begin{equation}\label{EqMotionCat}
       (-\Box_x+m^2)\bar\phi (x)+
     \int \vd[4]{y}  \,
    \Sigma_\mathrm R(x,y) \bar\phi(y)  = 0.
\end{equation}
Novament, la solució d'aquesta equació torna a ser un comportament osci\l.lant de freqüència $R_\vect p$ i constant de decaïment $\Gamma_\vect p/2$. 

\longpage

\section{Propagació en un bany tèrmic de gravitons}

Com a aplicació dels mètodes desenvolupats a l'apartat anterior, considerem la propagació d'una partícula escalar en un bany tèrmic de gravitons. Tractarem gravetat linealitzada com una teoria efectiva. L'objectiu és calcular les correccions a l'autoenergia del camp en un bany tèrmic.

Considerem un camp escalar $\phi$ de massa 
$m$ propagant-se en una espaitemps caracteritzat per una mètrica $g_{\mu\nu}$. L'acció del camp és
\begin{subequations}
\begin{equation}
    S_{\phi,g} = - \int \ud[4]x \sqrt{-g} \left( \fud g^{\mu\nu}
    \partial_\mu \phi\, \partial_\nu \phi + \fud m^2 \phi^2
    \right),
\end{equation}
i l'acció de la mètrica,
\begin{equation}
    S_g =  \frac{2}{\kappa^2} \int \ud[4]x \sqrt{-g}\, R.
\end{equation}
\end{subequations}
on $\kappa = \sqrt{32 \pi G} = \sqrt{32\pi}\, \Lp$ és la constant d'acoblament gravitacional. Si la mètrica és una petita pertorbació de Minkowski, $g_{\mu\nu} =
\eta_{\mu\nu} + \kappa h_{\mu\nu}$, l'acció completa es pot descomposar en
\begin{subequations}
\begin{align}
    S_\phi &=  \int \ud[4]x  \left( - \fud
    \partial_\mu \phi \, \partial^\mu \phi - \fud m^2 \phi^2
    \right), \label{ActPhiClasCat}
    \\
    \begin{split}
    S_h &= \int \ud[4]x \bigg(- \frac12 \partial^\alpha h^{\mu\nu} \partial_\alpha h_{\mu\nu}
    +  \partial_\nu h^{\mu\nu} \partial^{\alpha}
    h_{\mu\alpha}
     \\ &\qquad\qquad- \partial_\mu h \, \partial_\nu h^{\mu\nu}
    + \frac12 \partial^\mu h\, \partial_\mu h \bigg) + O(\kappa),
    \end{split}\label{GravPropCat}\\
    \Sint &=  \int \ud[4]x \left( \frac\kappa2 T^{\mu\nu} h_{\mu\nu} +
    \frac{\kappa^2}{4} U^{\mu\nu\alpha\beta} h_{\mu\nu}
    h_{\alpha\beta} \right)
     +  O(\kappa^3),
    \label{IntTermCat}
\end{align}
on $T_{\mu\nu}$ és el tensor d'energia moment del camp,
\begin{equation}
    T_{\mu\nu}  =  \partial_\mu \phi \partial_\nu \phi - \fud
    \eta_{\mu\nu} \partial_\alpha \phi \partial^\alpha \phi- \fud \eta_{\mu\nu} m^2
    \phi^2,
\end{equation}
\end{subequations}
i on el tensor $U^{\mu\nu\alpha\beta}$ no tindrà rellevància.

Per calcular les  correccions a l'autoenergia cal introduir contratermes a l'acció. L'acció més general per als contratermes compatible amb la simetria fins a ordre $\kappa^2$ és
\begin{equation}
\begin{split}
    S_\mathrm{count} =  &- \int \ud[4]x  \bigg[  \fud (\mathcal Z m_0^2-m^2)
    \phi^2 +  \fud (\mathcal Z -1) (\partial_\mu\phi \partial^\mu\phi + m^2
     \phi^2)
   \\&\qquad+ \frac14 \kappa^2 C_0 \mathcal Z^2
    (\partial_\mu \partial^\mu \phi)^2 \bigg] + O(\kappa^4),
\end{split}
\end{equation}
on $m_0=m/\sqrt{\mathcal Z}+O(\kappa^2)$ és la massa despullada, $\mathcal Z =1+O(\kappa^2)$ és un paràmetre de renormalització del camp i $C_0=C/\mathcal Z^2+O(1)$ és un coeficient de derivades altes. La seva part finita $C$ constitueix una dada addicional del problema. Per acabar, assenyalem que pels gravitons treballem amb el gauge harmònic.

\subsection{Temperatura zero}

A temperatura zero l'autoenergia està relacionada amb el propagador de Feynman a través de 
\begin{equation}\label{FeynmanCat}
    G\TO_\mathrm{F}(p) = \frac{-i}{p^2+m^2+\Sigma\TO(p^2)},
\end{equation}
i recordem que es pot calcular com la suma de tots els diagrames irreductibles d'una partícula.

Cal calcular dos diagrames de Feynman diferents, dibuixats a la figura \ref{fig:FeyDiag} (pàgina \pageref{fig:FeyDiag}). El resultat del càlcul és el següent. L'autoenergia renormalitzada és
\begin{equation}\label{SigmaMCat}
\begin{split}
    \Sigma\TO(p^2) &=
    - \frac{\kappa^2}{(4\pi)^2} \left( \frac{ m^6}{2p^2} + \frac{m^4}{2} \right) \ln\left( 1 +
    \frac{p^2}{m^2}-i\epsilon \right) \\
     &\quad- \frac{\kappa^2}{(4\pi)^2} \left(m^4 + m^2p^2 \right)
    \ln\left(\frac{p^2+m^2}{\mu^2}-i\epsilon\right) \\
    &\quad + C \kappa^2 (p^2+m^2)^2 +
    O(\kappa^4),
\end{split}
\end{equation}
on hem absorbit les divergències a través de 
\begin{subequations}
\begin{align}
    m_0^2 \mathcal Z&= m^2 - C \kappa^2 m^4 + O(\kappa^4)\\
    \mathcal Z &= 1 + 2 C \kappa^2m^2- \frac{\kappa^2m^2}{(4\pi)^2} \left( \frac{1}{\hat\varepsilon} + 2     \right)+ O(\kappa^4),\\
    C_0\mathcal Z^2 &= C + O(\kappa^4).
\end{align}
\end{subequations}
Hem imposat la condició de renormalització on-shell, 
$
    \Sigma\TO(-m^2) = 0.
$
Noteu que el coeficient $C$ no es renormalitza, per la qual cosa hagués estat consistent simplement suposar $C=0$ des del principi, tot i que aquest fet no és d'esperar d'entrada des del punt de vista de l'anàlisi de teories efectives.

L'autoenergia desenvolupa una part imaginària per a $-p^2>m^2$,
\begin{equation} \label{ImSigmaCat}
\begin{split}
    \Im \Sigma\TO (p^2)=\theta(-p^2-m^2)\frac{\kappa^2}{16\pi} \left( \frac{
    m^6}{2p^2}+
    \frac{3m^4}{2} + m^2p^2 \right) \!,
\end{split}
\end{equation}
que quantifica  la probabilitat de que una partícula escalar emeti un gravitó. Noteu que sobre la capa de masses la part imaginària de  l'autoenergia és zero, per la manca d'espai fàsic.

\subsection{Temperatura finita}

Calculem ara les contribucions tèrmiques a l'autoenergia. Comencem pel càlcul de la part real.
Cal considerar els dos diagrames de la figura \ref{fig:FeyDiag}, tenint en compte que ara els propagadors interns són tèrmics i que cal emprar el mètode CTP. Distingint entre els límits de temperatura alta i temperatura baixa, el resultat del càlcul és el següent:
\begin{equation} \label{ReSigmaTCat}
\begin{split}
    &\ \Re \Sigma_\mathrm R (E_\vect p, \vect p) \approx\\
    &\quad    \begin{cases}
         -\tfrac{1}{6} \kappa^2 m^2 T^2
        -   \sqrt{\tfrac{m^5 T^3}{{8\pi^3}}}\kappa^2  \expp{-m/T} \left(
        \tfrac{m^2+2|\vect p|^2}{3m^2+4|\vect p|^2}\right) ,
        & T \ll m,\\[3ex]
         \tfrac{1}{48} \kappa^2 m^2 T^2 \left[  - 11
        + \tfrac{ \sqrt{m^2+ |\vect p|^2}}
     {|\vect p|}
         \ln \left(\tfrac{ 2\sqrt{m^2+ |\vect
       p|^2}+ |\vect p|}{ 2\sqrt{m^2+ |\vect
        p|^2}- |\vect p|}\right)\right] , & T \gg m.
    \end{cases}
\end{split}
\end{equation}
Trobareu els detalls al apartat \ref{sect:scalarReal} i a l'apèndix \ref{app:ABCD}.

En quant la la part imaginària, calculem $\Sigma^{12}(p)$ i $\Sigma^{21}(p)$ i farem servir la relació \eqref{cut}. El resultat és el següent (vegeu l'apartat  \ref{sect:scalarIm}):
\begin{widetext}
\begin{equation} \label{ImSigmaTCat}
\begin{split}
    \Im \Sigma_\mathrm R(p) &= \frac{\kappa^2T m^2(m^2+2p^2)}{32\pi |\vect p|} 
    \ln \left[\frac{
    \sinh \left(\frac{(p^0+|\vect p|)^2+m^2}{4T(p^0+|\vect p|)}\right)
    \sinh \left(\frac{(p^0)^2-|\vect p|^2-m^2}{4T(p^0-|\vect p|)}\right)}
    {\sinh \left(\frac{(p^0-|\vect p|)^2+m^2}{4T(p^0-|\vect p|)}\right)
    \sinh \left(\frac{(p^0)^2-|\vect p|^2-m^2}{4T(p^0+|\vect p|)}\right)}
    \right].
\end{split}
\end{equation}
\end{widetext}
Prenent el límit on-shell d'aquesta equació obtenim:
\begin{equation} \label{ImDOnShellCat}
\begin{split}
    \Im \Sigma_\mathrm R(p) \xrightarrow[p^0 \to E_\vect p]{} -&\ \frac{\kappa^2 m^4T}{32\pi |\vect p|
    }
    \ln \left(\frac{{\sqrt{m^2 + |\vect p|^2}+|\vect p| }}
      {{\sqrt{m^2 + |\vect p|^2}-|\vect p|}}\right),
\end{split}
\end{equation}
que és diferent de zero si $T\neq 0$. Tanmateix, sabem que la part imaginària de l'autoenergia correspon a la probabilitat neta de decaïment d'una partícula. A primer ordre una partícula no pot decaure perquè no hi ha espai fàsic per aquest procés, com es pot comprovar fàcilment: vegeu els diagrames \ref{fig:cut} a la pàgina \pageref{fig:cut}. Es pot veure que de fet el resultat obtingut no és físic, perquè és degut a la presència d'una divergència infraroja en els càlculs intermedis. Regulant aquesta divergència obtenim $\Im \Sigma_\mathrm R(E_\vect p,\vect p)=0$ tal com era d'esperar.

De fet, per a teories gauge sense massa Braaten i Pisarski \cite{Pisarski89,BraatenPisarski90a,BraatenPisarski90b} van argumentar que la teoria de pertorbacions ordinària fallava: en certs casos, una part de les correccions de dos loops és de fet del mateix ordre que les correccions d'un loop. Nosaltres hem comprovat que el càlcul de la part real que hem presentat no es veu afectat. Tanmateix, per calcular la contribució imaginària possiblement sigui necessari considerar la teoria de pertorbacions resumada, que els mateixos Braaten i Pisarski van introduir.

\subsection{Relacions de dispersió}

Recordem que les relacions de dispersió venen donades per l'energia en funció del moment:
\begin{subequations}
\begin{equation} \label{DispRelCat}
     E^2 =R_\vect p^2  = m^2 + |\vect p|^2 + \Re \SigmaR(p^0,\vect p).
\end{equation}
La massa tèrmica és la contribució a les relacions de dispersió a moment zero:
\begin{equation}
    \mth^2 = m^2 + \Re \SigmaR(\mth,\vect0).
\end{equation}
\end{subequations}
La relació de dispersió ve donada per
\begin{subequations}
\begin{equation}\label{RelDispTCat}
\begin{split} 
    E^2 &\approx \mth^2 +|\vect p|^2
        -   \kappa^2 \sqrt{\dfrac{ m^5 T^3
        }{{2\pi^3}}} \expp{-m/T}
        \left( \dfrac{|\vect p|^2}{3m^2+4|\vect p|^2} \right),
\end{split}
\end{equation}
a temperatures baixes i per  
\begin{equation} \label{RelDispHighTCat}
\begin{split}
     E^2 &\approx \mth^2 + |\vect p|^2  +  \frac{\kappa^2 m^2 T^2}{48}
         \left[ \frac{ E_\vect p}{|\vect p|} \ln \left(\frac{ 2E_\vect p+ |\vect p|}{ 2E_\vect p- |\vect p|}\right) - 1 \right],
\end{split}
\end{equation}
\end{subequations}
a temperatures altes. Tant en un cas com en l'altre, adoneu-vos que per al cas de moments molt alts les modificacions a la relació de dispersió es poden reabsorbir a la massa tèrmica.

Noteu també que les modificacions a la relació de dispersió estan exponencial suprimides a temperatures baixes. De fet, només hi ha correccions a la relació de dispersió quan les partícules escalars estan tèrmicament excitades. Aquest mateix fenomen, que podria semblar inesperat, s'esdevé a electrodinàmica i en el model de tres camps que estudiarem a continuació.

\section{Efectes dissipatius en la propagació: un model de tres camps}

En el capítol previ no hem pogut estudiar la part imaginària  de l'autoenergia a causa de les restriccions cinemàtiques. Estudiem ara un model que és qualitativament similar, però que presenta una part imaginària de l'autoenergia ja a un loop. 

Considerem una teoria de camps amb dos camps massius $\phiM$ i $\phim$, de masses $M$ i $m$ respectivament, i un altre camp sense massa $\chi$. L'acció del model és la següent:
\begin{equation}\label{accio3Camps}
\begin{split}
    S &= -\fud \int \ud[4]x \big( \partial_\mu \phiM \partial^\mu \phiM + M^2 \phiM^2 + \partial_\mu \phim \partial^\mu \phim + m^2 \phim^2 \\ &\qquad + \partial_\mu \chi \partial^\mu \chi - 2 \tilde g \phiM \phim \chi
        \big),
\end{split}
\end{equation}
Preferim treballar amb la constant d'acoblament adimensional  $g=\tilde g/m$. En el límit en què la diferència de masses $\dm = M-m$ és molt petita aquesta teoria de camps es pot entendre com un model d'un àtom de dos nivells interactuant amb un camp de radiació.


En el buit l'autoenergia es pot calcular de forma estàndard a partir del diagrama \ref{fig:SigmaVac} a la pàgina \pageref{fig:SigmaVac}. El seu valor és el següent:
\begin{equation}
\begin{split}
   \Sigma\TO(p^2) &=  \frac{g^2 m^2}{(4\pi)^2} \bigg[
     \frac{M^2}{p^2} \ln \left(1+\frac{p^2}{M^2} -i\epsilon \right) \\
	 &\qquad -  \frac{M^2}{m^2} \ln \left(1-\frac{m^2}{M^2} \right) + \ln
     \left(\frac{p^2+M^2}{M^2-m^2}-i\epsilon\right)\bigg].
\end{split}
\end{equation}
L'autoenergia de l'estat excitat es pot trobar intercanviant $m$ i $M$.

El ritme de decaïment és proporcional a la part imaginària de l'autoenergia, tal com hem vist abans. Prenent la part imaginària de l'expressió anterior trobem que per l'estat excitat el ritme de decaïment és $\gamma_-^* = g^2 \dm /(8\pi)$, i per l'estat fonamental és obviament zero: $\gamma_- = 0$.

\subsection{Ritmes de decaïment en banys tèrmics}

Recordem que la part imaginària de l'autoenergia es pot expressar com 
\begin{equation}\label{ImSigmaRGammaCat}
    \Im\SigmaR(E_\vect p, \vect p) = - R_\vect p[\Gamma_-(\vect p) - \Gamma_+(\vect p)],
\end{equation}
on $\Gamma_-(\vect p)$ i  $\Gamma_+(\vect p)$ són respectivament els ritmes de decaïment i creació. Més precisament, $\Gamma_-(\vect p)$ és la probabilitat per unitat de temps que una partícula entrant amb moment $\vect p$ decaigui a qualsevol altre estat, i $\Gamma_+(\vect p)$ és la probabilitat per unitat de temps que un estat de moment $\vect p$ es crei espontàniament. En un bany tèrmic els dos ritmes segueixen una condició de balanç detallat, $\Gamma_+(\vect p) = \expp{- E_\vect p\vect/T} \Gamma_-(\vect p)$.

Per a un conjunt de partícules  la funció de distribució $f(\vect p,t)$, lleugerament separada de l'equilibri tèrmic, evoluciona d'acord amb les equacions \cite{Weldon83}
\begin{subequations}
\begin{equation}\label{DecayThermalEqCat}
    \derp{f(\vect p,t)}{t} = - \Gamma_- (\vect p) f(\vect p,t) + \Gamma_+ (\vect p) [1+f(\vect p,t)].
\end{equation}
Solucionant aquesta equació trobem
\begin{equation}
	f(\vect p, t) = \frac{1}{\expp{E_\vect p/T} -1} - \Delta f_0 (\vect p) \expp{-\Gamma(\vect p)},
\end{equation}
\end{subequations}
on 
\begin{equation} \label{GammaExpCat}
	\Gamma(\vect p)=\Gamma_-(\vect p) - \Gamma_+(\vect p)
\end{equation}
és el ritme d'aproximació a l'equilibri, i és proporcional a la part imaginària de l'autoenergia. 

En el cas que ens ocupa, el ritme de decaïment, calculat a partir de la part imaginària de l'autoenergia, ve donat per
\begin{subequations} \label{RestGammaCat}
\begin{align}
    \Gamma_-^*=\Gamma_-^*(\vect 0)  &=  \frac{g^2}{8\pi} \dm \left[ 1 + n(\Delta m) \right]\\
    \Gamma_-=\Gamma_-(\vect 0) &=  \frac{g^2}{8\pi} \dm \, n(\Delta m) 
\end{align}
\end{subequations}
en el límit en què els àtoms són en repòs. Noteu que aquests ritmes de decaïment també es poden entendre com els ritmes de transició entre els dos estats atòmics. 

Si la temperatura és baixa, els ritmes de decaïment de l'estat fonamental estan exponencialment suprimits. Tanmateix, quan augmentem arbitràriament el moment de la partícula arriba un moment en què el ritme de decaïment augmenta dràsticament: hem arribat al límit GZK. Aquest fenomen és degut al fet que l'energia dels fotons tèrmics augmenta vist des del sistema de referència de la partícula.

\subsection{Ritmes de decaïment i funcions de distribució en casos generals}

En fons generals, l'\Eqref{DecayThermalEqCat} es pot generalitzar a
\begin{subequations} \label{DecayOutTCat}
\begin{align}
    \derp{f(\vect p,t)}{t} &=     - \Gamma_- (\vect p,[f^*],t) f(\vect p,t) + \Gamma_+ (\vect p,[f^*],t) [1+f(\vect p,t)], \\
	\derp{f^*(\vect p,t)}{t} &= - \Gamma^*_- (\vect p,[f],t) f^*(\vect p,t) + \Gamma^*_+ (\vect p,[f],t) [1+f^*(\vect p,t)]
\end{align}
\end{subequations}
Les funcions de distribució $f(\vect p,t)$ i $f^*(\vect p,t)$ corresponen a l'estat fonamental i a l'estat excitat respectivament. El ritme de decaïment $ \Gamma_- (\vect p,[f^*],t)$ està calculat en un fons d'estats excitats $f^*(\vect p,t)$. Definicions anàlogues s'apliquen a les altres quantitats de l'equació \eqref{DecayOutTCat}. Per deduir aquesta equació, hem fet servir hipòtesis físiques raonables, i hem suposat que l'escala de temps d'interacció és molt més petita que l'escala de temps d'evolució de la distribució. Les funcions de distribució estan normalitzades
\begin{equation}
	\int \udpi[3]{\vect p} [ f(\vect p,t) + f^*(\vect p,t) ] = N \frac{(2\pi)^3}{V}.
\end{equation}

Generalitzant l'equació \eqref{DecayOutTCat} per a  distribucions inhomogènies trobem l'equació de Boltzmann:
\begin{subequations} \label{BoltzmannCat}
\begin{align}
    \left[ \derp{}{t} + \vect v \cdot \derp{}{\vect x} \right] f(\vect p,x)&=     - \Gamma_- (\vect p,[f^*],x) f(\vect p,x) + \Gamma_+ (\vect p,[f^*]x) [1+f(\vect p,x)], \\
	\left[ \derp{}{t} + \vect v \cdot \derp{}{\vect x} \right] f^*(\vect p,x)&= - \Gamma^*_- (\vect p,[f],x) f^*(\vect p,x) + \Gamma^*_+ (\vect p,[f],x) [1+f^*(\vect p,x)],
\end{align}
\end{subequations}
on $\vect v$ és la velocitat de grup de les partícules. Perquè aquesta equació sigui vàlida les escales típiques de temps i longitud d'interacció han de ser molt més petites que no pas les escales típiques d'evolució de les funcions.   Tal com hem vist anteriorment, per a funcions de distribució inhomogènies el ritme de decaïment també es pot extreure a partir de l'autoenergia.

Com a exemple particular del càlcul en situacions diferents de l'equilibri tèrmic, hem calculat l'autoenergia de l'estat excitat d'un àtom en presència d'un altre àtom a l'estat fonamenta, fent servir el mètode CTP com de costum. La modificació a la part imaginària de l'autoenergia respecte al cas d'equilibri tèrmic ve donada per
\begin{equation}
\begin{split}
     \Delta\!\Im \SigmaR^* (E_{\vect p},\vect p;{\vect p'}) &=  -\frac{g^2 m^2}{4E_{\vect p'}V} 
	  \bigg[ \frac{2\pi}{2 |\vect {p-p'}| } \delta(E_{\vect p} - E_{\vect p'} - |\vect{p-p'}|)\\ &\qquad\qquad\times[1+n(|\vect{p-p'}|)] \\
    &\qquad- \frac{2\pi}{2 |\vect {p-p'}| } \delta(E_{\vect p} - E_{\vect p'} - |\vect{p-p'}|)n(|\vect{p-p'}|)  \bigg].
\end{split}
\end{equation}
Els dos diagrames de Feynman que corresponen a aquest procés es mostren a la figura \ref{fig:Mdecay} (pàgina \pageref{fig:Mdecay}). Aquest càlcul també i\l.lustra que fins i tot en situacions anisotròpiques la interpretació de la part imaginària de l'autoenergia com un ritme de decaïment és vàlida. 

\section{Excitacions de tipus partícula en espais corbats}

El concepte de partícula en espais corbats ha rebut molta atenció \cite{BirrellDavies,WaldQFT,Fulling,JacobsonQFT}. Tanmateix, la majoria de les anàlisis s'han limitat a l'estudi de partícules no interactuants. En aquesta secció ens centrarem en l'estudi de (quasi)partícules interactuants, tot i que començarem per un repàs de les partícules lliures.

\subsection{Partícules lliures en espais corbats}

Diferents conceptes de partícula existeixen en espais corbats. Recordem que l'operador camp en un espai general es pot descomposar en modes,
\begin{equation}\label{ModeDecompCurvedCat}
	\hat \phi(x)  = \sum_\alpha \big[ \hat a_\alpha   u_\alpha (x) + \hat a_\alpha ^\dag u_\alpha ^*(x) \big].
\end{equation}
Aquesta descomposició és en principi arbitrària. Segons la fem d'una manera o altra trobarem diferents conceptes de partícula.

En primer lloc podem parlar de partícules associades a observadors asimptòtics. Quan en un espaitemps existeix una regió asimptòticament plana és possible descomposar l'operador camp en modes que es redueixen als modes de Minkowski a la regió asimptòtica. Quan existeixen dues regions asimptòtiques podem trobar dues descomposicions diferents, que en general no són equivalents. L'estat que es percep com a buit en una de les regions asimptòtiques és percebut com un estat excitat a l'altra.

Si l'espai no té regions asimptòtiques, però té simetries globals sota una certa classe de transformacions, és possible escollir un conjunt de modes que s'adaptin a aquesta simetria. Per exemple, si l'espai té una simetria sota translacions temporals es possible escollir un conjunt de modes pels quals la seva energia estigui ben definida, o si hi ha simetria sota translacions espacials és possible escollir modes de moment ben definit. En situacions generals, les partícules associades a aquests modes no han de tenir necessàriament res a veure amb les partícules de Minkowski.

En tot cas, pel principi d'equivalència sabem que per regions limitades en comparació amb el radi de curvatura local ha d'existir un concepte de partícula equivalent al de Minkowski. De fet, treballant amb coordenades normals de Riemann al voltant de qualsevol punt és possible definir un espai de moment local i trobar conjunts de modes etiquetats amb aquest moment. Aquesta expansió només és vàlida en regions prou petites, i per a moments prou grans en comparació amb l'escala d'energia de curvatura. A ordre més baix trobem una expressió idèntica a Minkowski, i a següent ordre hi ha correccions amb la curvatura. Les partícules associades a aquests modes les anomenarem partícules quasilocals.

Les partícules quasilocals tenen l'inconvenient que només estan definides en regions prou petites. En cosmologia és possible estendre la definició de partícula quasilocal per a distàncies arbitràriament llargues, fent servir el concepte de partícules adiabàtiques; l'única condició és que l'energia de les partícules sigui prou gran en comparació amb el ritme d'expansió de l'univers. Les partícules adiabàtiques es redueixen a les partícules quasilocals quan la regió considerada és prou petita.

Una altra possibilitat per a considerar les partícules és fer servir un concepte operacional: un determinat estat contindrà partícules si un detector de partícules les detecta. El detector de partícules es pot modelar com un osci\l.lador harmònic linealment acoblat al camp, semblant al model d'emissor de partícules que hem fet servir prèviament. Un resultat important és que la detecció de partícules depèn de la trajectòria del detector. Per a trajectòries inercials, i per osci\l.ladors harmònics sintonitzats a una freqüència prou alta, la probabilitat de detecció és proporcional al nombre de quasipartícules locals presents a l'estat.

\subsection{Camps interactuants en espais corbats}

Considerem ara la situació en què tenim un camp escalar en un espaitemps general en un estat $\hat\rho$. Volem esbrinar quines novetats hi ha respecte al cas de l'espai pla. 

Els propagadors són biescalars que es defineixen exactament com a l'espai pla, i es poden organitzar igualment en les matrius directa o de Keldysh. Les autoenergies es defineixen igualment a través de l'equació
\begin{equation} \label{SelfEnergyGeneralCurvedCat}
\begin{split}
    G_{ab}(x,x') = G_{ab}^{(0)}(x,x') &+
    \int {\vd[4]{z}}\sqrt{-g(z)}\, \vd[4]{z'}\sqrt{-g(z')} \\ &\quad \times G_{ac}^{(0)}(x,z) [-i\Sigma^{cd}(z,z')]  G_{db}(z',x').
\end{split}
\end{equation}
Es convenient treballar amb camps reescalats per un factor $[-g(x)]^{1/4}$ de manera que les expressions esdevenen idèntiques a l'espai pla. En termes dels camps reescalats, el propagador retardat i l'autoenergia retardada verifiquen
\begin{equation} \label{SelfEnergyGeneralRetardedCurvedCat}
    \bar\GR(x,x') = \bar\GR^{(0)}(x,x')+
    \int \ud[4]{z} \ud[4]{z'} \bar\GR^{(0)}(x,z) [-i\bar\SigmaR(z,z')]  \bar\GR(z',x'),
\end{equation}
on la barra indica el reescalament. Totes les propietats que depenen del moment també són aplicables en espais corbats treballant amb coordenades normals i moments locals, mitjançant 
\begin{equation}
	\bar G_{ab}(p;X) = \int \udpi[4]{y} \expp{-ip_a y^a}  \bar G_{ab}(y/2,-y/2),
\end{equation}
on $y$ és un sistema de coordenades normals centrat al voltant del punt $X$.

En principi, l'avaluació pertorbativa dels propagadors té lloc de manera semblant a l'espai pla, fent servir el formalisme CTP. Cal tenir diversos aspectes en compte però:
primer, fins i tot el càlcul dels propagadors lliures sol ser complicat a l'espai corbat; segon, la seva interpretació pot ser diferent, i, tercer, poden aparèixer noves divergències absents a l'espai pla. Tractem ara aquest darrer punt.

A l'espai corbat poden haver-hi divergències ultraviolades que depenguin de la curvatura local de l'espaitemps. En general, cal incloure a l'acció tots aquells termes compatibles amb la simetria, incloent-hi també els termes que depenen de la curvatura. Aquests termes són irrellevants i estan suprimits per potències creixents de la massa de Planck (excepte en el cas de l'acoblament conforme, que és un acoblament marginal). A banda de les divergències ultraviolades, també poden haver-hi divergències infraroges, que són molt més delicades ja que depenen de l'estructura global de la varietat. 

Quan les escales d'energia en què estem interessats són molt més altes que no pas l'escala d'energia de curvatura, la representació de moments de la teoria serà particularment útil. En aquest cas podem trobar expressions anàlogues a l'espai pla com ara
\begin{equation}\label{GSigmaCurvedCat}
	\bar\GR(p;X) = \frac{-i}{p^2 + m^2 + \bar\SigmaR(p;X)} + O(l_\text{int}/L,p^{-2}L^{-2}),
\end{equation} 
on $l_\text{int}$ és l'escala típica d'interacció i $L$ és el radi local de curvatura. En aquesta situació la majoria d'expressions que vam trobar a l'espai pla són també d'aplicació, però tenint en compte que hi haurà correccions com les indicades. En particular continuaràn essent vàlida la representació espectral.

\subsection{Partícules interactuants en espais corbats}

De totes les nocions de partícula que hem esmentat a l'inici d'aquest apartat, aquelles que ens són més útils  són les partícules quasilocals i les partícules adiabàtiques ---recordem que les partícules globals i les partícules asimptòtiques només són aplicables en determinats espaitemps, i que detectors sintonitzats a freqüències prou altes detecten les partícules quasilocals. Resumim ara les propietats de les quasipartícules quasilocals en presència d'interacció.

A l'espai de Fourier local, l'estructura del propagador retardat de les quasipartícules locals és anàloga a la de Minkowski:
\begin{equation}\label{RetardedQLCat}
	\bar\GR(\omega,\vect p;x) \approx \frac{-i Z_\vect p(x)}
	{-\omega^2+R_\vect p^2(x)-i\omega\Gamma_\vect p(x)},\quad 	\omega \sim R_\vect p,
\end{equation}
on $R_\vect p(x)$ i $\Gamma_\vect p(x)$ corresponen respectivament a l'energia i al ritme de decaïment, i es poden obtenir a partir de la part real i imaginària de l'autoenergia respectivament.
De la mateixa manera, l'estat quàntic corresponent a les quasipartícules ve  donat per $\hat a_\vect p^\dag(x) \hat\rho(t_0)\hat a_\vect p(x)$, i la seva energia decau segons
\begin{equation}\label{EnergiaDecau}
	E^\pplus(\tau;\vect p;x) = 
	E_\vect p^{(0)}+ [n_\vect p(x) +1] \expp{-\Gamma_\vect p(x)\tau}.
\end{equation}

En resum, trobem les mateixes expressions que a Minkowski. Cal fer diversos aclariments però. En primer lloc, noteu que hi ha una dependència en el punt de l'espaitemps. En segon lloc, a l'expressió del propagador hi haurà correccions de l'ordre de $R_\vect p^{-2}L^{-2}$ i també de l'ordre de $l_\text{int}/L$, on $l_\text{int}$ és la distància típica d'interacció. Finalment, cal adonar-se que l'aproximació quasilocal només és vàlida per a regions molt més petites que $L$, i per tant l'equació \eqref{EnergiaDecau} només serà vàlida per a temps curts.

\subsection{Partícules interactuants a cosmologia}

A cosmologia, l'anàlisi es simplifica un tant perquè, com sabem, podem etiquetar els modes del camp  pel seu moment conforme, que és una quantitat conservada. El propagador retardat corresponent a aquests modes verifica   
\begin{equation}\label{DysonCat}
	\bar \GR(t,t';\vect k) = \GR^{(0)}(t,t';\vect k) -i 
	\int \ud s \ud {s'}  \bar \GR^{(0)}(t,s;\vect k) 
	\bar\SigmaR(s,s';\vect k) \bar \GR(s',t';\vect k),
\end{equation}
on els modes reescalats són $\bar \phi_\vect k(t) = a^{3/2}(t) \phi_\vect k(t)$. Analitzarem les propietats de les quasipartícules a partir de l'anàlisi dels propagadors.

Per a temps d'observació curts, més petits que no pas el temps típic d'expansió de l'univers, el propagador retardat es pot aproximar per 
\begin{equation}\label{shortimetimeCat}
	\bar G_\text{R}(t,t';\vect k) = \frac{-i}{ R_\vect k(T)} \sin\left[R_\vect k(T)(t-t')\right] \expp{-\Gamma_\vect k(T)(t-t')/2} \theta(t-t'),
\end{equation}
on $T=(t+t')/2$. Com sempre, $R_\vect p(T)$ i $\Gamma_\vect p(T)$ venen determinades per l'autoenergia avaluada sobre la capa de masses:
\begin{subequations}\label{RgammaCat}
\begin{align} 
	R^2_\vect k(t) &=  m^2 + \frac{\vect k^2}{a^2(t)} + \Re\bar\SigmaR\boldsymbol(R_\vect k(t),t;\vect k\boldsymbol),\\
	\Gamma_\vect k(t) &= -\frac{1}{R_\vect k(t)} \Im\bar\SigmaR\boldsymbol(R_\vect k(t),t;\vect k\boldsymbol).
\end{align}
\end{subequations}
L'expressió \eqref{shortimetimeCat} correspon  a la transformada de Fourier de l'equació \eqref{RetardedQLCat}, identificant $\vect k/a(t)= \vect p$ com el moment físic. Com veiem essencialment recuperem l'aproximació quasilocal que hem introduït a l'apartat anterior: el propagador ve determinat pel valor de l'autoenergia al voltant del temps d'observació.

Per a temps d'observació llargs, comparables al temps d'expansió de l'univers, l'expressió \eqref{shortimetimeCat} perd la seva validesa. Vegem en què pot ajudar l'aproximació adiabàtica. L'equació del moviment del propagador retardat ve donada per
\begin{equation} \label{EqRetCat}
\begin{split}
	&\left[ \frac{1}{a^3(t)} \derp{}{t}{} \left(a^3(t) \derp{}{t}{} \right) + m^2 + \xi R(t) + \frac{\vect k^2}{a^2(t)} \right]
	\left[ \frac{\bar\GR(t,t';\vect k)}{a^{3/2}(t) a^{3/2}(t')} \right]  \\
	&\qquad\qquad + \frac{1}{{a^{3/2}(t) a^{3/2}(t')} } \int \ud{s} \bar\SigmaR(t,s;\vect k)  \bar\GR(s,t';\vect k) = \frac{-i}{a^3(t)} \delta(t-t').
\end{split}
\end{equation}
Expandint el terme no local de l'equació com
\begin{equation} \label{expansionCat}
\begin{split}
	N(t,t') &= \int \ud{s} \bar\SigmaR(t,s;\vect k)  \bar\GR(s,t';\vect k)\\ &= \left[ \delta R^2_\vect k(t) + H \Gamma_\vect k(t) \derp{}{t}  + \cdots\right] \bar\GR(t,t';\vect k) ,
\end{split}
\end{equation}
podem trobar una solució WKB a l'equació del moviment de la forma
\begin{equation}\label{InteractingWKBCat}
\begin{split}
	\bar G_\text{R}(t,t';\vect k) &= \frac{-i}{  \sqrt{ R_\vect k(t) R_\vect k(t')}} \sin\left({\int^{t}_{t'} \ud{s} {R_\vect k(s)}} \right) \expp{-{{}\int^{t}_{t'} \ud{t} \Gamma_\vect k(t)/2} }  \theta(t-t').
\end{split}
\end{equation}
Aquesta solució és vàlida sempre que les energies de les partícules siguin prou grans en comparació amb el ritme d'expansió de l'univers.
Introduint l'\Eqref{InteractingWKBCat} a \eqref{EqRetCat} comprovem que  $R_\vect p(t)$ i $\Gamma_\vect p(t)$ venen encara donats per l'\Eqref{RgammaCat}. L'evolució del propagador retardat a cosmologia  és doncs una expressió WKB, on les quantitats que apareixen dins d'aquesta equació poden ser calculades de manera local. Això és així sempre que els temps d'interacció siguin molt més curts que no pas el temps d'expansió de l'univers.

Dins d'aquesta aproximació adiabàtica l'evolució de l'energia de l'estat de quasipartícula ve donada per
\begin{equation} 
	E^\pplus(t,t_{0};\vect p) \approx E^{(0)}_\vect p(t)+ R_\vect p(t)(1+n_\vect p)\expp{-\int_{t_0}^t \ud s \Gamma_\vect p(s)},
\end{equation}
d'on identifiquem $R_\vect p$ com l'energia de les quasipartícules i $\Gamma_\vect p(t)$ com el seu ritme de decaïment.

Fins ara hem suposat temps d'interacció menors que el temps d'expansió de l'univers. Si els temps d'interacció són de l'ordre del temps d'expansió de l'univers, llavors l'evolució del propagador retardat ve també  donada per una solució tipus WKB.
\begin{equation}\label{InteractingWKBLCat}
\begin{split}
	\bar G_\text{R}(t,t';\vect k) &= \frac{-i}{  \sqrt{ R\ret_\vect k(t) R\ret_\vect k(t')}} \sin\left({\int^{t}_{t'} \ud{s} {R\ret_\vect k(s)}} \right) \\ &\quad \times \expp{-{{}\int^{t}_{t'} \ud{t} \Gamma\ret_\vect k(t)/2} }  \theta(t-t'),
\end{split}
\end{equation}
però on ara 
\begin{subequations}
\begin{align}
	(\delta R\ret_\vect k)^2(t) &= \int \ud s  \sqrt{\frac{R\ret_\vect k(t)}{R\ret_\vect k(s)}} \bar\SigmaR(t,s;\vect k) \cos\left({\int^{t}_{s} \ud{s'} {R\ret_\vect k(s')}} \right), \\
	\Gamma\ret_\vect k(t) &= -\int \ud {s} \frac{\bar\SigmaR(t,s;\vect k)}{\sqrt{R\ret_\vect k(t)R\ret_\vect k(s)}}   \sin\left({\int^{t}_{s} \ud{s'} {R\ret_\vect k(s')}} \right).
\end{align}
\end{subequations}
L'\Eqref{InteractingWKBLCat} només té sentit per a temps extremadament llargs, molt més grans que el temps típic d'interacció i evolució. Fent la mitjana per a temps molt més petits que aquest temps trobem que el ritme de decaïment es pot expressar també com
\begin{equation}\label{DecayRateNewCat}
	\Gamma_\vect p(t) = \int \ud \Delta \frac{\Im\SigmaR(t+\tfrac\Delta2,t-\tfrac\Delta2;\vect k)}{\sqrt{E_\vect p(t+\Delta/2)E_\vect p(t-\Delta/2)}} \expp{i\int_{t-\Delta/2}^{t+\Delta/2} \ud s E_\vect k(s)}.
\end{equation}
Podem també definir una versió millorada de la representació de freqüències de l'autoenergia,
\begin{equation}
	\SigmaImp([R_\vect p],t;\vect k) := \int \ud \Delta \bar\SigmaR(t+\Delta/2,t-\Delta/2;\vect k) \expp{i\int_{t-\Delta/2}^{t+\Delta/2} \ud s R_\vect p(s)}.
\end{equation}
La part imaginària d'aquesta autoenergia retardada correspon al ritme de decaïment en el sistema de referència de la partícula.

\section[Propagació cosmològica: un model de tres camps]{Efectes dissipatius en la propagació cosmològica: un model de tres camps}

Apliquem els resultats de l'apartat anterior a un cas concret: l'estudi d'una partícula massiva interactuant amb un camp sense massa en un univers en expansió. Estarem particularment interessats en els efectes dissipatius, que com sabem es poden extreure de la part imaginària de l'autoenergia.
Com hem comentat a la introducció, la interacció pot modificar la propagació de les partícules en els primers estadis d'inflació, i en aquest context els efectes dissipatius poden jugar un paper important 

Farem servir el mateix model de tres camps que hem analitzat prèviament ---vegeu l'equació \eqref{accio3Camps}---, però afegint-hi ara un terme d'acoblament conforme per al camp sense massa. Recordem que aquest model consisteix en dos camps molt massius $\phiM$ i $\phim$, però amb una diferència de masses $\dm$ petita. 
Els propagadors lliures del camp sense massa es poden calcular de manera exacta a causa de l'acoblament conforme; pels camps massius, considerarem que la massa dels camps és molt més gran que no pas el ritme d'expansió de l'univers, i així podrem treballar amb l'aproximació adiabàtica que hem vist a l'apartat anterior.

El nostre objectiu és doncs calcular la part imaginària de l'autoenergia retardada a un loop per al camp menys massiu. Considerarem la possibilitat que el camp sense massa estigui en equilibri tèrmic. Podem aplicar les regles de Feynman al diagrama de la figura \ref{fig:SigmaCurved} (pàgina \pageref{fig:SigmaCurved}), tenint en compte el formalisme CTP. El diagrama el calcularem en la representació temporal, però l'avaluarem en representació de freqüències, fent servir la representació millorada que hem introduït a l'apartat anterior si així fos necessari.

L'aplicació de les regles de Feynman mostra clarament que hi ha dues contribucions diferents a la part imaginària de l'autoenergia (vegeu la figura \ref{fig:ImSigmaCosCurved} a la pàgina \pageref{fig:ImSigmaCosCurved}): l'una és la contribució habitual, que ve donada per l'absorció d'un fotó del bany tèrmic, i l'altra és una nova contribució que correspon a l'emissió d'un fotó acompanyada de l'excitació del camp. Aquesta darrera contribució viola la conservació de l'energia, però l'energia no té per què ser conservada per a  energies de l'ordre del ritme d'expansió.

\subsection{Efectes de temperatura}

Quan la temperatura és molt més alta que el ritme d'expansió, els temps d'interacció seran curts en comparació amb el temps de Hubble, i per tant els propagadors que apareixen a dins dels diagrames de Feynman es poden calcular amb l'aproximació de temps curts. Fent aquesta aproximació trobem que els ritmes de decaïment coincideixen amb els  de l'espai pla, però on ara la temperatura és la temperatura instantània a cada moment de l'expansió. Per exemple, el ritme de decaïment d'una partícula en repòs ve donat per
\begin{equation}
\begin{split}
	\Gamma_\vect 0(t) = -\frac{1}{m} \Im \bar\SigmaR (m,t;\vect 0) &= \frac{g^2}{8\pi} \dm \, 
	n_{\theta/a(t)} (\dm)\, ,
\end{split}
\end{equation}
on $\theta$ és la temperatura conforme.

A la figura \ref{fig:DecayT} a la pàgina \pageref{fig:DecayT} hem representat el ritme de decaïment en funció del factor d'escala. Aquesta figura i\l.lustra que les partícules poc energètiques comencen a decaure tan bon punt la temperatura és de l'ordre de la diferència de masses. Les partícules molt energètiques decauen abans, degut a que en el sistema de referència de la partícula els fotons tèrmics es perceben més energètics a causa de l'efecte Doppler.

Noteu que, tot i que el procés d'interacció té lloc a temps petits, el propagador interactuant es pot calcular per a temps arbitràriament llargs mitjançant l'\Eqref{InteractingWKBCat}. En aquest sentit els efectes dissipatius tenen un caràcter secular, i així efectes petits podrien ser importants per a temps llargs.

\subsection{Efectes de buit}

Considerem ara l'efecte del buit. Recordem que quan el ritme d'expansió és prou gran, hi ha una contribució del primer diagrama de la figura \ref{fig:ImSigmaCosCurved}, contribució que es pot interpretar com la partícula més lleugera excitant-se i emetent un fotó. Aquesta contribució és conseqüència de la no conservació de l'energia per a escales prou petites. Hem analitzat dos models d'expansió diferents: de Sitter i inflació potencial. Noteu que hem fet servir la representació de freqüències millorada perquè el temps típic d'interacció d'aquesta mena de processos és de l'ordre del temps d'expansió de Hubble.

En el cas de de Sitter, $a(t) = a(T) \expp{H(t-T)}$, trobem que el ritme de decaïment en el sistema de referència de la partícula ve donat per
\begin{equation}
	\gamma=\Im\SigmaImp([E_\vect k],T;\vect k) = \frac{g^2}{8\pi}  m \dm\, n_{H/2\pi}(\dm)
\end{equation}
 Recuperem doncs la temperatura efectiva de de Sitter: una partícula comòbil a de Sitter experimenta un bany de radiació tèrmica a temperatura $H/(2\pi)$. El resultat és independent del moment de la partícula a causa de la invariància de de Sitter. El ritme de decaïment al sistema de referència del laboratori ve donat per
\begin{equation}
	\Gamma_\vect k(t) = \frac{g^2}{8\pi}  \dm\, n_{H/2\pi}(\dm) \left( 1 - \frac{\vect k^2}{2 a^2(t) m^2} \right) + O(\vect k^2/m^2).
\end{equation}

En el cas d'inflació potencial, $a(t) = a(T) (t/T)^\alpha$, el ritme de decaïment ve donat per 
\begin{equation}
	\Gamma_\vect 0  = -\frac{g^2 }{2 \pi }   \dm \expp{-2 \sqrt{3} \dm T} =-\frac{ g^2 }{2 \pi }m \dm \expp{-{8 \sqrt{3} \dm}/{H(T)}}
\end{equation}
quan $\alpha=4$ i per 
\begin{equation}
	\Gamma_\vect 0 =-\frac{\left(3+\sqrt{5}\right)g^2 }{4 \pi }m \dm e^{-12 \sqrt{5+2 \sqrt{5}}
   \dm T}
\end{equation}
quan $\alpha=6$. Per a valors imparells de $\alpha$ no hem trobat cap contribució. També hem calculat les correccions de moment [vegeu l'\Eqref{powerLawMomentum}]. En aquest cas trobem una dependència temporal explícita perquè el ritme d'expansió no és constant.

En tot cas, noteu que el comportament qualitatiu és en ambdós casos el mateix: no hi ha contribució de buit al ritme d'expansió a menys que $H \gtrsim \dm$.

\section{Conclusions}

Un dels punts centrals d'aquesta tesi ha estat l'estudi de les excitacions de quasipartícula en situacions generals. Hem emprat dues tècniques bàsiques: l'estudi de la seva representació espectral, d'una banda, i l'anàlisi de l'evolució de l'energia en funció del temps, de l'altra. Amb tots dos mètodes hem mostrat que l'autoenergia determina l'energia física de les partícules i el seu ritme de decaïment, sempre que l'estat del camp sigui de quasiequilibri.

Hem arribat a idèntiques conclusions que amb els mètodes de camp mig, però a diferència d'aquests, en els estats de quasipartícula que hem considerat el valor mig del camp és zero. Un element clau ha estat la introducció de l'aproximació gaussiana, que ha permès simplificar el problema i alhora dotar de significat físic a moltes expressions. Una limitació essencial del nostre mètode de segona quantització és la impossibilitat de seguir la trajectòria de la partícula; tal com hem comentat, per aquesta darrera tasca caldria fer servir una aproximació de primera quantització o estadística.

Hem aplicat aquests mètodes a l'estudi de dos casos particulars: la propagació en un bany tèrmic de gravitons i la propagació en un model de tres camps. En el primer model ens hem centrat en l'estudi de la part real de la relació de dispersió, obtenint que les modificacions solament són rellevants per a temperatures altes i moments petits. En el segon model hem estudiat sobre tot la part imaginària: el ritme de decaïment. Hem estudiat com a partir del ritme de decaïment en situacions generals podem obtenir les equacions de moviment de les funcions de distribució.

Els resultats obtinguts a l'espai pla els hem generalitzat a l'espai corbat. En general, en espaitemps arbitraris i a ordre zero en la curvatura recuperem els resultats de l'espai pla, sempre que l'energia de les quasipartícules sigui prou gran. En el cas particular de cosmologia hem aconseguit anar més enllà, obtenint resultats vàlids per a tot temps treballant amb l'aproximació adiabàtica. Hem distingit diversos casos segons les relacions entre les diverses escales de temps rellevants del problema.

A l'espai corbat la hipòtesi clau és la separació d'escales. Cal suposar que les escales de longitud i temps típiques d'evolució de la partícula són molt més petites que no pas les escales d'evolució de l'espaitemps. A cosmologia aquesta hipòtesi de separació d'escales es tradueix en l'aproximació adiabàtica, que permet, primer, obtenir un concepte de partícula prou ben definit; segon, simplificar tècnicament el problema, i, tercer, obtenir resultats que van més enllà d'aquells de l'espai pla.

 Com a exemple pràctic, hem aplicat els resultats de cosmologia al model de tres camps, centrant-nos en la part imaginària de la relació de dispersió. Per a temperatures altes, la partícula es troba en equilibri tèrmic instantani amb un fons de radiació, la temperatura del qual evoluciona amb l'univers. Per a temperatures baixes trobem una segona contribució, absent a l'espai pla, que correspon al ritme de decaïment generat per les contribucions de buit.
 
Una dels punts que val la pena emfatitzar a la vista dels nostres resultats és la rellevància i universalitat dels termes dissipatius de la relació de dispersió. Atès que els efectes dissipatius tenen un caràcter secular, aquests poden ser més rellevants que no pas la part real, especialment quan estudiem comportaments a temps llargs.

Un altre punt a destacar és la importància de no negligir les interaccions, ja que el comportament qualitatiu pot ser molt diferent respecte al cas lliure en molts casos. En relació amb el problema transplanckià, els nostres resultats apunten a què fins i tot en el cas que la simetria Lorentz sigui una simetria fonamental de la natura, els efectes d'interacció amb l'ambient fan que la qüestió transplanckiana encara sigui  una qüestió adequada a plantejar-se.
 
\selectlanguage{english}
	\chapter{The closed time path method and thermal field theory} \label{app:CTP}

\index{Closed time path method|(}
\index{In-in method|see{closed time path method}}
\index{CTP|see{closed time path method}}

In this appendix we give a brief introduction to the closed time path (CTP) method (also called \emph{in-in} method, in contrast to the conventional \emph{in-out} method) originally proposed by Schwinger \cite{Schwinger61} and Keldysh \cite{Keldysh65}. For further details we address the reader to refs.~\cite{ChouEtAl85,CalzettaHu87,CamposVerdaguer94,Weinberg05}. An approach similar to ours can be found in ref.~\cite{CamposHu98}.

In this appendix we shall also present the real-time approach to thermal field theory as a particular application of the CTP method for thermal states. Further details on the real-time approach to thermal field theory can be found in refs.~\cite{LeBellac,Das,LandsmanWeert87,CamposHu98,CalzettaHu88}. The imaginary-time formalism can be studied in the aforementioned references and additionally in refs.~\cite{Kapusta,FetterWalecka}.

For the purposes of this appendix we shall consider a free or an interacting scalar field $\phi$. For simplicity we shall adopt a flat spacetime notation, although most expressions can be straightforwardly extended to curved spacetime.

\section{The need for a closed time path approach}

The CTP method provides a generic way to deal with a field theory in an arbitrary spacetime background with an arbitrary state $\hat\rho$ for the field. In contrast, the usual \emph{in-out} approach is restricted, at least in its straightforward formulation, to studying field theory the Minkowski vacuum. The most characteristic feature of the CTP method is the doubling of the number of degrees of freedom of the theory. The need for the CTP method and the doubling of the number of degrees of freedom when considering generic situations can be understood as follows: 

Let us consider a 2-point correlation function. In the standard \emph{in-out} approach to quantum field theory, one directly obtains transition elements from the asymptotic \emph{in} vacuum to the asymptotic \emph{out} vacuum: \cite{Peskin,WeinbergQFT}
\begin{equation}
 {}_\text{out}\langle 0 |\hat\phi(t,\vect x)\hat\phi(t',\vect x')|0\rangle_\text{in} 
	=	{}_\text{out}\langle 0 (t) | \hat\phi(\vect x) {U(t,t')} \hat\phi(\vect x') |0(t')\rangle_\text{in}.
\end{equation}
where we recall that according with our conventions the left-hand side expression is written in the Heisenberg picture while the right-hand side is written in the Schrödinger picture. Standard \emph{in-out} Feynman rules are obtained by expanding perturbatively the time evolution operator $U(t,t')$. In the Minkowski spacetime both vacua are equivalent up to an irrelevant phase, so that \emph{in-out} transition elements coincide with true vacuum expectation values. However, when the spacetime is not Minkowski both vacua are not equivalent in general \cite{BirrellDavies}. If in this case one is still interested in considering true expectation values, one has to deal with quantities such that
\begin{equation}
 {}_\text{in}\langle 0 |\hat\phi(t,\vect x)\hat\phi(t',\vect x')|0\rangle_\text{in} 
	=	{}_\text{in}\langle 0(t) | \hat\phi(\vect x) {U(t,t')} \hat\phi(\vect x') U(t',t) |0(t)\rangle_\text{in}
\end{equation}
Notice that two copies of the evolution operator $U(t,t')$ appear, one of them reversed in time. The CTP method deals with this kind of situations, and the two copies of the evolution operator are responsible for the doubling of the number of degrees of freedom. The ``closed time path'' name refers to the fact that, as we shall see, both evolution operators can be considered in a single path integral going forward and backwards in time.

In an even more general situation one is interested in studying not the vacuum but an arbitrary state $\hat\rho$, either in flat or curved spacetime. Then the expectation values are expressed as:
\begin{equation}
 \Tr {\big[\hat \rho \hat\phi(t,\vect x) \hat\phi(t',\vect x')\big]}
	=	 \Tr {\big[\hat \rho(t) \hat\phi(\vect x) U(t,t') \hat\phi(t',\vect x') U(t',t)  \big]}
\end{equation}
Notice that again two copies of the evolution operator appear. The CTP method also allows for these generic averages, which are the ones we will study hereafter.

\section{The CTP generating functional and correlation functions}

\index{Generating functional}

The path-ordered generating functional $Z_{\mathcal C}[j]$ is defined
as
\begin{equation}
    Z_{\mathcal C}[j] := \Tr \left(\hat \rho T_{\mathcal C} \expp{i \int_{\mathcal C} \vd t \int
    \ud[3]{\vect
    x} \hat \phi(x) j(x)}  \right),
\end{equation}
where $\hat \phi(x)$ is the field operator in the Heisenberg
picture, ${\mathcal C}$ is a certain path in the complex $t$
plane, $T_{\mathcal C}$ means time ordering along this path and $j(x)$
is a classical external source.  By functional differentiation of
the generating functional with respect to $\phi$, path-ordered
correlation functions can be obtained:
\begin{equation}
	G_\mathcal C(x,x') := \Tr\big[ \hat\rho T_{\mathcal C} \hat\phi(x) \hat\phi(x')\big] =  - \left. \frac{ \delta^2  Z_{\mathcal C}}{\delta j(x) \delta j(x')}\right|_{j=0}
\end{equation}
Next we introduce a complete basis of eigenstates of the field operator:
\begin{align*}
	\hat\phi(\vect x) |\phi\rangle &= \phi(\vect x) |\phi\rangle, &&\text{(Schrödinger picture)}\\
	\hat\phi(t,\vect x) |\phi,t\rangle &= \phi(t,\vect x) |\phi,t\rangle, && \text{(Heisenberg picture)}
\end{align*}
Introducing the complete basis as a representation of the identity, and working in the Heisenberg picture, the generating functional can be expressed as:
\begin{equation}  \label{PrePathGenFunct}
	Z_\mathcal C[j] = \int \widetilde{\mathrm d} \phi\, \widetilde{\mathrm d} \phi' \langle \phi,\ti|\hat\rho|\phi',\ti\rangle
	\langle \phi',\ti |T_{\mathcal C} \expp{i \int_{\mathcal C} \vd t \int
    \ud[3]{\vect
    x} \hat \phi(x) j(x)}  |\phi,\ti\rangle 
\end{equation}
The functional measures $\widetilde{\mathrm d} \phi$ and $\widetilde{\mathrm d} \phi'$ go over all field configurations of the fields at fixed initial time $t$. If the path $\mathcal C$ begins and ends at the same point $\ti$, then the transition element of the evolution operator can be computed via a path integral:
\begin{equation}\label{PathGenFunct}
	Z_\mathcal C[j] = \int \widetilde{\mathrm d} \phi\, \widetilde{\mathrm d} \phi' \langle \phi,\ti|\hat\rho|\phi',\ti\rangle
	\int_{\varphi(\ti,\vect x)=\phi(\vect x)}^{\varphi(\ti,\vect x)=\phi'(\vect x)} \mathcal D \varphi \expp{i \int_{\mathcal C} \vd t \int
    \ud[3]{\vect x}\{ L[\varphi] +  \varphi(x) j(x)\}},
\end{equation}
where $L[\phi]$ is the Lagrangian density of the scalar field.

\begin{figure}
    \centering
    \includegraphics{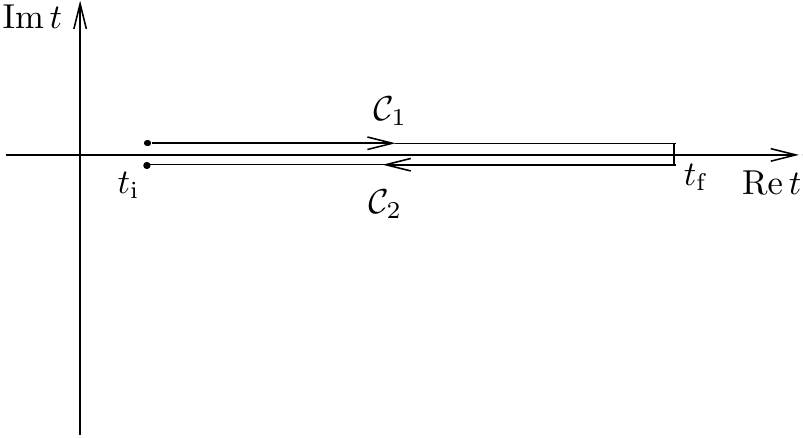}
    \caption{Integration path in the complex-time plane used in the
    CTP method. The forward and backward lines are infinitesimally close to the real axis.}
    \label{fig:CTP}
\end{figure}

\index{Generating functional}
Let us consider the time path depicted in fig.~\ref{fig:CTP}.  If we define $\varphi_{1,2}(t,\vect
x)=\varphi(t,\vect x)$ and $j_{1,2}(t,\vect x)=j(t,\vect x)$ for $t
\in {\mathcal C}_{1,2}$, then the generating functional can be reexpressed as:
\begin{equation}	 \label{CTPGenFunct}
\begin{split}
	Z[j_1,j_2] &= \int \widetilde{\mathrm d} \phi\, \widetilde{\mathrm d} \phi' \widetilde{\mathrm d} \phi'' \langle \phi,\ti|\hat\rho|\phi',\ti\rangle  \\
	&\qquad\times\int_{\varphi_1(\ti,\vect x)=\phi(\vect x)}^{\varphi_1(\tf,\vect x)=\phi''(\vect x)} \mathcal D \varphi_1 
	\expp{i \int \ud[4]{x}\{ L[\varphi_1] +  \varphi_1(x) j_1(x)\}}\\
	&\qquad\times\int_{\varphi_2(\ti,\vect x)=\phi'(\vect x)}^{\varphi_2(\tf,\vect x)=\phi''(\vect x)} \mathcal D \varphi_2 
	\expp{-i\int \ud[4]{x}\{ L[\varphi_2] +  \varphi_2(x) j_2(x)\}}.
\end{split}
\end{equation}
In the following it will prove useful to use a condensed notation where neither the boundary conditions of the path integral nor the integrals over the initial and final states are explicit. With this simplified notation the above equation becomes
\begin{equation}	
\begin{split}
	Z[j_1,j_2] &=  \int \mathcal D \varphi_1 \mathcal D \varphi_2\langle \phi,t|\hat\rho|\phi',t\rangle 
	\expp{i \int \ud[4]{x}\{ L[\varphi_1] - L[\varphi_2] +  \varphi_1(x) j_1(x)-  \varphi_2(x) j_2(x)\}}
\end{split}
\end{equation}
An operator representation is also possible:
\begin{equation}\label{ZCTPOper}
    Z[j_1,j_2] := \Tr 
	\left(\hat \rho \,  
	\widetilde T \expp{-i \int_\ti^\tf \vd t \int
    \ud[3]{\vect
    x} \hat \phi(x) j_2(x)} T \expp{i \int_\ti^\tf \vd t \int \ud[3]{\vect
    x} \hat \phi(x) j_1(x)} \right).
\end{equation}

The choice of the path is a matter of convenience. The one presented here is the simplest and is the one that directly leads to the physically relevant propagators. Other possible paths will be discussed when considering thermal field theory.

\index{Propagator!Feynman}
\index{Propagator!Dyson}
\index{Propagator!Whightman}

By functionally differentiating the generating functional one obtains the Feynman and Dyson propagators and the Whightman functions:
\begin{subequations} \label{CTPPropSet}
\begin{align}
	G_{11}(x,x') &= \GF (x,x') = \Tr{\big[ \hat\rho T \hat\phi(x) \hat\phi(x')\big]} =  - \left. \frac{ \delta^2  Z}{\delta j_1(x) \delta j_1(x')}\right|_{j_1=j_2=0}, \\
	G_{21}(x,x') &= G_+ (x,x') = \Tr\big[ \hat\rho \hat\phi(x) \hat\phi(x')\big] =   \left. \frac{ \delta^2  Z}{\delta j_2(x) \delta j_1(x')}\right|_{j_1=j_2=0} ,\\
	G_{12}(x,x') &= G_- (x,x') = \Tr\big[ \hat\rho \hat\phi(x') \hat\phi(x)\big] =  \left. \frac{ \delta^2  Z}{\delta j_1(x) \delta j_2(x')}\right|_{j_1=j_2=0} ,\\
	G_{22}(x,x') &= G_\text{D} (x,x') = \Tr\big[ \hat\rho \widetilde T \hat\phi(x) \hat\phi(x')\big] =  - \left. \frac{ \delta^2  Z}{\delta j_2(x) \delta j_2(x')}\right|_{j_1=j_2=0}.
\end{align}
\end{subequations}
These four propagators can be conveniently organized in a $2\times2$ matrix:
\begin{equation}  \label{DirectBasisCTP}
	G_{ab}(x,x') = \begin{pmatrix}
				\GF (x,x') & G_-(x,x')\\
				G_+ (x,x') & G_\text{D} (x,x')
	              \end{pmatrix}
\end{equation}
See appendix \ref{app:GenRel} for the properties linking the different propagators.
Higher order correlation functions can be obtained in a similar way.

\index{Keldysh basis}
It is sometimes useful to consider a transformation from the above basis, the direct basis, to the so-called physical or \emph{Keldysh} basis, in which
\begin{equation}
	G'_{a'b'}(x,x')= \begin{pmatrix}
				\Ga (x,x') &  \Gret (x,x') \\
				\Gadv(x,x') & 0 
	               \end{pmatrix},
\end{equation}
where $\Gadv(x,x')$, $\Gret (x,x')$ and $\Ga (x,x')$ are respectively the advanced, retarded propagators and Hadamard functions, defined in appendix \ref{app:GenRel}. Primed indices refer to the Keldysh basis. Both representations are related via the following equation:
\begin{subequations}\label{KeldyshTransQ}
\begin{equation}
	G'_{a'b'} = Q_{a'}^{\phantom{a'}c} \; G_{cd} \; Q_{b'}^{\phantom{b'}d} \qquad
	G_{ab} = (Q^{-1})_{a}^{\phantom{a}c'} \; G'_{c'd'} \; (Q^{-1})_{b}^{\phantom{b}d'}
\end{equation}
or its equivalent matrix version,
\begin{equation}
	G' = Q \, G \,Q^T, \qquad
	G = Q^{-1} G'\, (Q^{-1})^T
\end{equation}
\end{subequations}
with\footnote{There is some freedom in the location of the zero of the Keldysh basis, and therefore slightly different transformation matrices $Q$ are found in the literature.}
\begin{equation}
	Q =\frac{1}{\sqrt{2}} \begin{pmatrix}
				1 & 1\\
				1 & -1
	               \end{pmatrix},\qquad
	Q^{-1} = \frac{1}{\sqrt{2}} \begin{pmatrix}
				1 & 1\\
				1 & -1
	               \end{pmatrix}.
\end{equation}
The matrix relation is based in the non-perturbative relations between the propagators (see chapter 3 ad appendix \ref{app:GenRel}).

\section{Perturbation theory}

At least formally, all the results in the previous section are valid either for interacting or free field theories. However, explicit results can only be obtained when the theory is free and the initial state is Gaussian. In this case the path integrals in \Eqref{CTPGenFunct} can be exactly performed, and one obtains:
\begin{equation}
	Z^{(0)}[j_1,j_2] = Z^{(0)}[0,0]  \expp{ \frac{1}{2} \int \ud[4]{x} \ud[4]{x'} j^a(x) G^{(0)}_{ab}(x,x') j^b(x')}. 
\end{equation}
We recall that lowercase roman indices may acquire the values 1 and 2 are raised and lowered with the ``CTP metric''
$c_{ab}=\mathrm{diag}(1,-1)$. If the initial state is non-Gaussian, there shall be additional contributions in the exponent.

The free propagators $G^{(0)}_{ab}$ are proportional the inverse of the differential operator $c^{ab}(-\partial_{\mu}\partial^\mu + m^2)$,\ie,
\begin{equation}  \label{DirectBasis}
\begin{split}
\begin{pmatrix}
				-\partial_{\mu}\partial^\mu + m^2 & 0\\
				0 & \partial_{\mu}\partial^\mu - m^2
	               \end{pmatrix}
	\begin{pmatrix}
				\GF (x,x') & G_-(x,x')\\
				G_+ (x,x') & G_\text{D} (x,x')
	               \end{pmatrix} \\ = -i 
	\begin{pmatrix}
				\delta^{(4)}(x-x') &0\\
				0 & \delta^{(4)}(x-x')
	               \end{pmatrix}.
\end{split}	               
\end{equation}
In other words, the positive and negative Whightman functions correspond to homogeneous solution of the equation of motion, while the Feynman and Dyson propagators correspond to Green functions. Similarly, the retarded, advanced and Pauli-Jordan propagators are also Green functions of the equation of motion, and the Hadamard function is a homogeneous solution. The more general form for the free propagators is \cite{CalzettaHu87}
\begin{subequations}\label{GeneralFreeProps}
\begin{equation} \label{GeneralFreeProps1}
\begin{split}
         G_{ab}^{(0)}(p)  &=
        \begin{pmatrix}
            \dfrac{-i}{p^2 + m^2 - i \epsilon} & {2\pi \delta(p^2+m^2) \theta(-p^0)}\\
            { 2\pi \delta(p^2+m^2) \theta(p^0)}
             & \dfrac{i}{p^2 + m^2 + i \epsilon}
        \end{pmatrix} \\ &\quad + 2\pi \delta(p^2+m^2) f(p)
        \begin{pmatrix}
            1 & 1 \\
            1 & 1
        \end{pmatrix},
\end{split}
\end{equation}
where $f(p)$ is a positive even function, related to the
mode occupation number $n_\vect p =\av{\hat a^\dag \hat a}$  through:
\begin{equation} \label{GeneralFreeProps2}
	\delta(p^2+m^2) f(p) = \delta(p^2+m^2) [n_\vect p \theta(p^0) + n_{-\vect p}\theta(-p^0)].
\end{equation}
\end{subequations}

Perturbative evaluation of the generating functional can proceed in a similar way to the usual \emph{in-out} case. If the Lagrangian density can be split in its free and interacting parts as $L(x) = L_0(x) + V(x)$, then the  generating functional \eqref{CTPGenFunct} can be reexpressed as
\begin{equation}	
\begin{split}
	Z[j_1,j_2] &=  \int \mathcal D \varphi \mathcal D \varphi'\langle \phi,t|\hat\rho|\phi',t\rangle 
	\expp{i \int \ud[4]{x} \{ V[\varphi_1] -V[\varphi_2] \}  }\\
	&\quad\times\expp{i \int \ud[4]{x}\{ L_0[\varphi_1] - L_0[\varphi_2] +  \varphi_1(x) j_1(x)-  \varphi_2(x) j_2(x)\}}
\end{split}
\end{equation}
Substituting $\phi_1$ and $\phi_2$ by  $-i\delta/\delta j$ and $-i\delta/\delta j'$ respectively,
\begin{equation}	
\begin{split}
	Z[j_1,j_2] &=  \expp{i \int \ud[4]{x} \big\{ V\big[-i\frac{\delta}{\delta j_1}\big] -V\big[-i\frac{\delta}{\delta j_2}\big] \big\}  }
	\int \mathcal D \varphi \mathcal D \varphi'\langle \phi,t|\hat\rho|\phi',t\rangle 
	\\
	&\qquad\times\expp{i \int \ud[4]{x}\{ L_0[\varphi_1] - L_0[\varphi_2] +  \varphi_1(x) j_1(x)-  \varphi_2(x) j_2(x)\}}
\end{split}
\end{equation}
This equation can be rewritten as:
\begin{equation}	
\begin{split}
	Z[j_1,j_2] &=  \expp{i \int \ud[4]{x} \big\{ V\big[-i\frac{\delta}{\delta j_1}\big] -V\big[-i\frac{\delta}{\delta j_2}\big] \big\}  } Z^{(0)}[j_1,j_2]	
\end{split}
\end{equation}
Therefore, the perturbative expansion is very similar to the standard one once we properly take into account the doubling of the number of degrees of freedom and the boundary conditions. 

\index{Feynman rules!in the CTP formalism}
As in the vacuum case, the  perturbative expansion can be organized in terms of Feynman diagrams. The Feynman rules are those of standard scalar field theory, supplemented by the following ones:
\begin{enumerate}
	\item There are two kinds of vertices, type 1 and type 2. 
	\item There are four kinds of (free and interacting) propagators: 11, 12, 21 and 22 [see \Eqref{CTPPropSet}]. Free propagators type $ab$ link a vertex type $a$ with a vertex type $b$.
	\item When computing a Feynman diagram, add an additional minus sign for every type 2 vertex.
	\item Sum over all possible internal vertices of any given Feynman diagram.
\end{enumerate}
These rules can be applied not only to the time-ordered propagator but to any propagator in the direct basis. It is also possible to reexpress the Feynman rules in the Keldysh basis \cite{EijckKobesWeert94}.

The renormalization of the theory proceeds as in the vacuum case, at least for physically reasonable states, since the high energy structure of the theory is not modified. In other words, the counterterms which make the theory finite in the vacuum also make the theory finite in a state $\hat \rho$. However the interpretation of the finite part of the renormalization is different in general (see chapter 3).

\section{The effective action}
\index{Effective action}
\index{Generating functional!connected}

From the generating functional the connected generating functional $W[j,j']$ is defined
\begin{equation}
	W[j,j'] := - i \ln Z[j,j']
\end{equation}
Next we introduce the following objects:
\begin{equation}\label{CTPBarPhi}
	\bar\phi^a(x) = \frac{\delta W}{\delta j_a(x)},
\end{equation}
which must be understood as functionals of $j_1$ and $j_2$ even if this dependence is not explicit. If $j_1=j_2$ both $\phi_1$ and $\phi_2$ give the expectation value of the field under the presence of a classical source $j$. Finally, the effective action $\Gamma[\bar\phi_1,\bar\phi_2]$ is defined as the Legendre transform of the connected generating functional:
\begin{equation}
	\Gamma[\phi_1,\phi_2] :=  W[j_1,j_2] - \int \ud[4] x j_a(x) \bar\phi^a(x).
\end{equation}
In this equation $j_1$ and $j_2$ must be understood as functionals of $\bar\phi_1$ and $\bar\phi_2$, which can be obtained by inverting \Eqref{CTPBarPhi}.

By functionally differentiating the effective action with respect to $\bar\phi_1$ and setting $\bar\phi_2=\bar\phi_1$ the equation of motion for the expectation value of the scalar field is obtained:
\begin{equation}
	\left. \frac{\delta \Gamma}{\delta\bar\phi(x)} \right|_{\bar\phi_2=\bar\phi_1} = j(x).
\end{equation}
In contrast to the conventional \emph{in-out} treatment, the equations of motion obtained from the CTP generating functional are real and causal because they correspond to the dynamics of true expectation values \cite{Jordan86}.

\section{The self-energy} 
\index{Self-energy}

In a generic situation the self-energy is a matrix $\Sigma^{ab}$ defined via the following Dyson equation:
\begin{equation} \label{SelfEnergyGeneralApp}
    G_{ab}(x,x') = G_{ab}^{(0)}(x,x')+
    \int \ud[4]{y} \ud[4]{y'} G_{ac}^{(0)}(x,y) [-i\Sigma^{cd}(y,y')]  G_{db}(y',z),
\end{equation}
Notice that this is a matrix relation which links the different propagator and self-energy components. In general there is no one-to one relation between $G_{11}(x,x')$ and $\Sigma^{11}(x,x')$. 

Similarly to the conventional case, the self-energy component $\Sigma^{ab}$ can be computed as the sum of all one-particle irreducible diagrams (1PI) which go from a vertex $a$ to a vertex $b$. They therefore correspond to amputated diagrams.

The self-energy can also be expressed in the Keldysh basis. Applying the transformation \eqref{KeldyshTransQ} on \Eqref{SelfEnergyGeneralApp} we find this latter equation is also valid in the Keldysh basis,
\begin{equation} 
    G'_{a'b'}(x,x') = G_{a'b'}^{'(0)}(x,x')+
    \int \ud[4]{y} \ud[4]{y'} G_{a'c'}^{'(0)}(x,y) [-i\Sigma'^{c'd'}(y,y')]  G'_{d'b'}(y',z),
\end{equation}
with
\begin{equation}
	\Sigma'_{a'b'}(x,x')= \begin{pmatrix}
				0 &  \SigmaA(x,x') \\
				\SigmaR(x,x') & \SigmaN(x,x') 
	               \end{pmatrix},
\end{equation}
where  $\SigmaA(x,x')$, $\SigmaR(x,x')$ and $\SigmaN(x,x')$ are defined in \Eqref{KeldyshCompDef}.  Both representations of the self-energy are related via the following equations:
\begin{subequations}
\begin{equation}
	\Sigma'^{a'b'} = R^{a'}_{\phantom{a'}c} \; \Sigma^{cd} \; \Sigma^{b'}_{\phantom{b'}d} \qquad
	\Sigma^{ab} = (R^{-1})^{a}_{\phantom{a}c'} \; \Sigma'^{c'd'} \; (R^{-1})^{b}_{\phantom{b}d'}
\end{equation}
or their equivalent matrix version,
\begin{equation}
	\Sigma' = R \, \Sigma \,R^T, \qquad
	\Sigma = R^{-1}  \Sigma' \, (R^{-1})^T
\end{equation}
\end{subequations}
with $R=[Q^T]^{-1}$:
\begin{equation}
	R =\frac{1}{\sqrt{2}} \begin{pmatrix}
				1 & 1\\
				1 & -1
	               \end{pmatrix},\qquad
	R^{-1} = \frac{1}{\sqrt{2}} \begin{pmatrix}
				1 & 1\\
				1 & -1
	               \end{pmatrix}.
\end{equation}

The properties of the self-energy in relation with the propagators are discussed in chapter 3. The different relations among the self-energy components are analyzed in appendix \ref{app:GenRel}.

\section{Thermal field theory}
\index{Thermal field theory}

\index{Closed time path method|)}

Thermal field theory can be seen as a strict particular case of the CTP method, in which the state $\hat\rho$ happens to be 
\begin{equation}
     \hat\rho = \frac{\expp{- \beta\hat
    H}}{\Tr(\expp{- \beta\hat {H}})},
\end{equation}
where $\hat H$ is the Hamiltonian operator of the system. To apply the techniques presented in this chapter the only requeriment is to compute the free thermal propagators. However, it proves convenient to make a slight adpatation of the formalism and modify the complex time path in order to connect with the usual approaches to thermal field theory.

Noticing that the density matrix operator can be seen as the time translation operator in the complex plane, $\expp{-\beta\hat H} = U(t-i\beta,t)$, \Eqref{PrePathGenFunct} can be reexpressed as:
\begin{equation}  
\begin{split}
	Z_\mathcal C[j] = \frac{1}{\Tr(\expp{- \beta\hat {H}})} \int \widetilde{\mathrm d} \phi \langle \phi,\ti-i\beta|T_{\mathcal C} \expp{i \int_{\mathcal C} \vd t \int
    \ud[3]{\vect
    x} \hat \phi(x) j(x)}  |\phi,\ti\rangle ,
\end{split}
\end{equation}
where we have used the completeness relation $\int \widetilde{\mathrm d}\phi'|\phi',\ti\rangle\langle \phi',\ti|=1$. This way we have managed to incorporate the information on the state on the dynamical evolution. Now, if the path $\mathcal C$ starts at $\ti$ and ends at $\ti-i\beta$, a path-integral representation can be introduced:
\begin{equation}
\begin{split}
	Z_\mathcal C[j] &= \frac{1}{\Tr(\expp{- \beta\hat {H}})}
	\int \widetilde{\mathrm d} \phi 
	\int_{\varphi(\ti,\vect x)=\phi(\vect x)}^{\varphi(\ti,\vect x)=\phi(\vect x)} \mathcal D \varphi \expp{i \int_{\mathcal C} \vd t \int
    \ud[3]{\vect x}\{ L[\varphi] +  \varphi(x) j(x)\}},
\end{split}
\end{equation}
which can be rewritten in a more compact form as
\begin{equation}
\begin{split}
	Z_\mathcal C[j] &= N
	\int \mathcal D \varphi \expp{i \int_{\mathcal C} \vd t \int
    \ud[3]{\vect x}\{ L[\varphi] +  \varphi(x) j(x)\}},
\end{split}
\end{equation}
where $N$ is a normalization constant, and the boundary conditions $\phi(t_\mathrm
i,\vect x) = \phi(t_\mathrm i-i\beta,\vect x)$ are assumed.

\index{Thermal field theory!real time}
\index{Thermal field theory!imaginary time}

\begin{figure}
    \centering
    \includegraphics{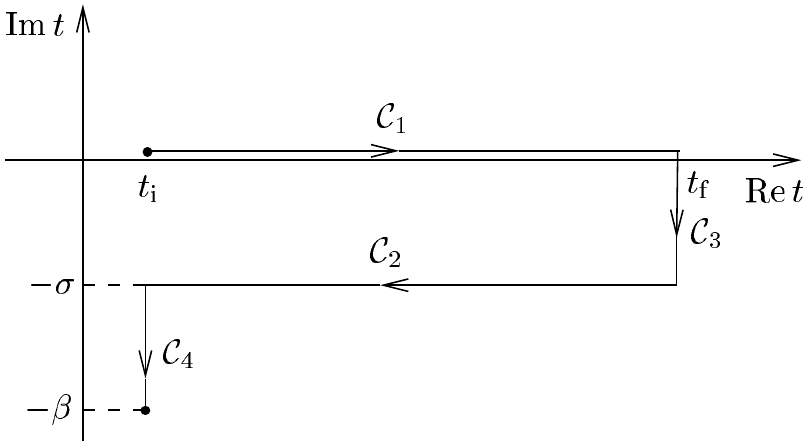}
    \caption{Integration contour in the complex-time plane used in the
    real-time approach to thermal field theory. The choice
    $\sigma=0^+$ makes the formalism analogous to the CTP method.}
    \label{fig:ThermalPath}
\end{figure}

Different elections for the path ${\mathcal C}$ lead to different approaches to
thermal field theory: a straight line from $t_\mathrm i$ to
$t_\mathrm i - i\beta$ leads to the imaginary-time formalism, and
the contour shown in Fig.~\ref{fig:ThermalPath} leads to the real-time
formalism. By choosing $\sigma= 0^+$ the real-time formalism is virtually identical to
the CTP method, since the  the path
along ${\mathcal C}_3$ and ${\mathcal C}_4$ can be neglected if we are interested in real-time correlation functions and the boundary
conditions of the path integral are properly taken into account.

\index{Thermal field theory!thermofield dynamics}

The imaginary-time formalism avoids the doubling of the number of degrees of freedom by considering a path consisting in a single straight line. However, correlation functions obtained from the imaginary-time generating functional are only defined at discrete imaginary frequencies, and an analytic continuation to real time has to be introduced in order to obtain physical observables. For instance, the imaginary-time 2-point propagator is analytically continued to the retarded propagator. In contrast, in the real-time formalism the number of degrees of freedom is doubled, but no analytic continuation is needed. For completeness let us mention that besides the real- and imaginary-time approaches to field theory there also exists the so-called thermofield dynamics approach, which is a real-time operational formalism which aims to closely reproduce the zero-temperature methods by introducing the notion of ``thermal vacuum''. A doubling in the number of degrees of freedom also appears naturally in this case. Thermofield dynamics can be shown to correspond to the real-time formalism with $\sigma=\beta/2$. 

In this thesis we work in the real-time formalism with $\sigma=0^+$, which is the case that we shall explore hereafter and which corresponds to the CTP formalism. Therefore, all expressions derived in the previous sections of this chapter also apply here. In the following, we shall explore additional properties particular of the thermal case.

 If, as before, we define $\phi_{1,2}(t,\vect
x)=\phi(t,\vect x)$ and $j_{1,2}(t,\vect x)=j(t,\vect x)$ for $t
\in {\mathcal C}_{1,2}$ the contribution to the generating functional can also be split in two pieces $Z_C[j] = Z_{34} Z[j_1,j_2]$, where $Z_{34}$ is the constant contribution coming from the segments $3$ and $4$ and where 
\begin{equation}
\begin{split}
    Z[j_1,j_2]= \int \uD \phi_1 \uD \phi_2 & \expp{i\int \ud[4]x
    \left\{ {\mathcal L}[\phi_1(x)] + j_1(x) \phi_1(x)\right\}}
    \\ \times & \expp{ -i\int \ud[4]x \left\{
      {\mathcal L}[\phi_2(x)] + j_2(x)
    \phi_2(x) \right\}}.
\end{split}
\end{equation}
The boundary conditions of these path integrals are the following: $\phi_1(\tf)=\phi_2(\tf)$, $\phi_1(\ti)=\phi_2(\ti-i\beta)$. 

Again, correlation functions may be obtained by functional differentiation [see \Eqref{CTPPropSet}]. The boundary conditions of the path integral impose the following relation between the Whightman functions:
\begin{equation}\label{KMS}
	G_-(t,\vect x;t',\vect x') = G_+(t-i\beta,\vect x;t',\vect x'),
\end{equation}
which is known as the Kubo-Martin-Schwinger relation (KMS), and is further explored in appendix \ref{app:GenRel}.

Free propagators are given by
\begin{equation} \label{FreeGT}
\begin{split}
         G_{ab}^{(0)}(p)  &=
        \begin{pmatrix}
            \dfrac{-i}{p^2 + m^2 - i \epsilon} & {2\pi \delta(p^2+m^2) \theta(-p^0)}\\
            { 2\pi \delta(p^2+m^2) \theta(p^0)}
             & \dfrac{i}{p^2 + m^2 + i \epsilon}
        \end{pmatrix} \\ &\quad + 2\pi \delta(p^2+m^2) {n(|p^0|)}
        \begin{pmatrix}
            1 & 1 \\
            1 & 1
        \end{pmatrix},
\end{split}
\end{equation}
where $n(E)$ is the Bose-Einstein distribution function:
\begin{equation}
    n(E) = \frac{1}{1-\expp{\beta E}}.
\end{equation}
The retarded propagator and Hadamard functions are given by
\begin{equation}
	\GR^{(0)}(p) = \frac{-i}{p^2 + m^2 - i \epsilon p^0}, \quad (G^{(1)})^{(0)}(p) = 2\pi \delta(p^2+m^2) [1+2n(|p^0|)].
\end{equation}
As expected, the retarded propagator coincides with its vacuum expression.

Thermal contributions to Feynman diagrams are always finite in the
ultraviolet regime because the Bose-Einstein function acts as a
soft cutoff for momenta larger than the temperature $T$. The
counterterms which renormalize the theory at zero temperature also
renormalize the theory at finite temperature. Note also that the
thermal part of the propagator, which breaks the Lorentz symmetry
through an explicit dependence on $p^0$, is always on shell.

In the thermal case, instead of working with the four propagators of the direct basis, or the three propagators of the Keldysh basis, one can take profit of the KMS relation and reorganize perturbation theory in a way such
that just the retarded and the advanced propagators are involved
\cite{EijckKobesWeert94}. In this case one has to consider how
Feynman rules are transformed when working with the
retarded/advanced basis.

	\chapter{Relations between the different 2-point propagators}\label{app:GenRel}

In this appendix we summarize the non-perturbative relations between the
different 2-point propagators that are employed throughout the thesis.  To simplify the notation we consider the propagators of an interacting quantum mechanical degree of freedom $\hat q$. However, exactly the same relations apply for the propagators of a scalar field $\hat \phi(\vect x)$. Both the time and frequency representations are discussed.

\section{Definition of the 2-point propagators}

\index{Propagator!Feynman}
\index{Propagator!Dyson}
\index{Propagator!Whightman}
\index{Propagator!retarded}
\index{Propagator!advanced}
\index{Propagator!Pauli-Jordan}
\index{Propagator!Hadamard}
Most definitions of the propagators are already introduced in chapters 2, 3 and appendix \ref{app:CTP}, but we repeat them here for  completeness. The Feynman and Dyson propagators, and the positive and negative Whightman functions are defined, respectively as, as:
\begin{subequations}
\begin{align}
	\GF(t,t') &:= \av{ T \hat q(t) \hat q(t)},\\
	G_\text{D}(t,t') &:= \av{ \widetilde T \hat q(t) \hat q(t')},\\
	G_+(t,t') &:= \av{ \hat q(t) \hat q(t') },\\
	G_-(t,t') &:= \av{ \hat q(t) \hat q(t') }.
\end{align}
\end{subequations}
The average is done with respect to an arbitrary state $\hat\rho$. These propagators form the CTP direct basis (appendix \ref{app:CTP}):
The Pauli-Jordan and Hadamard propagators are respectively given by:
\begin{subequations}
\begin{align}
	G(t,t') &:= \av{ [ \hat q(t) ,\hat q(t')]},\\
	G^{(1)}(t,t') &:= \av{ \{ \hat q(t) ,\hat q(t')\}}.
\end{align}
\end{subequations}
Notice that for free fields the Pauli-Jordan and Hadamard propagators are closely related to the dissipation and noise kernels introduced in chapter 2. 
Finally, the retarded and advanced propagators are
\begin{subequations}
\begin{align}
	\Gret(t,t') &:= \theta(t-t') \av{ [\hat q(t), \hat q(t')] } ,\\
	G_\text{A}(t,t') &:= \theta(t'-t) \av{ [\hat q(t), \hat q(t')] } .
\end{align}
\end{subequations}
The advanced and retarded propagators, together with the Pauli-Jordan function form the Keldysh basis in the CTP formalism (see page \pageref{DirectBasis} in appendix \ref{app:CTP}).

Notice that the above definitions may differ by factors of $i$ and $1/2$ from those used in other references.

\index{Propagator!Relation between the different}

\section{General relations valid for any state}

The relations developed in this section are completely general and follow directly from the definitions of the propagators.

\subsection{Time representation}

From the definitions of the propagators one trivially deduces that the Pauli-Jordan propagator is anti-symmetric under the exchange of $t$ and $t'$ and that the Hadamard propagator is symmetric. Similarly, from the hermiticity properties of the operator $\hat q$ it can be shown that the Pauli-Jordan, retarded and advanced propagators are purely imaginary and that the Hadamard propagator is purely real. Other propagators are complex and have both real and imaginary parts. The above statements can be summarized as:
\begin{subequations}
\begin{align}
	G(t,t') = -G(t',t) &\in i \mathbf R,\\
	G^{(1)}(t,t') = G^{(1)}(t',t) &\in  \mathbf R,\\
	\Gret(t,t'), G_\text{A}(t,t') &\in i\mathbf R,\\
	\GF(t,t'),G_\text{D}(t,t'),G_+(t,t'),G_-(t,t') &\in \mathbf C.
\end{align}
\end{subequations}
Symmetry properties also imply:
\begin{subequations}
\begin{align}
	G_\text{D}(t,t') & = \GF^*(t,t'),\\
	G_{-}(t,t') & = G^*_+(t,t') = G_+(t',t), \\
	G_\text{A}(t,t') &= - \Gret(t',t).
\end{align}
\end{subequations}

As a direct consequence of the definitions it can be trivially shown that:
\begin{subequations}
\begin{align}
	\Gret(t,t') &= G(t,t') \theta(t-t'), \label{GretG} \\
	\Gadv(t,t') &= G(t,t') \theta(t'-t), \label{GadvG} \\
	\Gc(t,t')  &=  G_+(t,t') - G_-(t,t') ,\\
	\Ga(t,t')  &=  G_+(t,t') + G_-(t,t') ,\\
	\GF(t,t') &= \theta(t-t') G_+(t,t') + \theta(t'-t) G_-(t,t'), \\
	\GD(t,t') &= \theta(t'-t) G_+(t,t') + \theta(t-t') G_-(t,t').
\end{align}
\end{subequations}

The following relations give the
decomposition of the complex propagators, which form the direct basis, in its real and imaginary parts:
\begin{subequations}\label{Rel1Time}
\begin{align}
    G_\mathrm F(t,t') &= \fud \big[ \Ga(t,t') + \Gc(t,t') \sign(t-t') \big]  , \\
    G_\mathrm D(t,t') &= \fud \big[ \Ga(t,t') - \Gc(t,t') \sign(t-t') \big] , 	\\
    G_{+}(t,t') &=  \fud \big[ \Ga(t,t') +  \Gc(t,t') \big],\\
    G_{-}(t,t') &= \fud \big[ \Ga(t,t') -  \Gc(t,t') \big].
\end{align}
\end{subequations}
These four propagators are connected through
\begin{equation}
	\GF(t,t') + G_\text{D}(t,t') = G_+(t,t) + G_-(t,t'),
\end{equation}
The propagators forming the Keldysh basis (namely, the retarded and advanced propagators and Hadamard functions) can be expressed in terms of the above as:
\begin{subequations}
\begin{align}
	\Gret(t,t') &= \GF(t,t') - G_- (t,t') = G_+(t,t') - \GD(t,t') \\
	\Gadv(t,t') &= \GF(t,t') - G_+ (t,t') = G_-(t,t') - \GD(t,t')\\
	\Ga(t,t') &=  G_+(t,t') + G_-(t,t') = \GF(t,t') + \GD(t,t')
\end{align}
\end{subequations}

The full set of propagators is determined by one complex quantity (\eg\ the Whightman functions, the Feynman propagator) or by one real and one imaginary quantity (\eg\ the Pauli-Jordan or retarded propagator and the Hadamard function).

\subsection{Frequency representation}

To go to the frequency representation, it will be useful to recall that the Fourier transform of a real even function is even and real and that the Fourier transform of a real odd function is odd and purely imaginary. Also we recall the following Fourier transforms:
\[
	\theta(t-t') \to \frac{i}{\omega+i\epsilon}, \qquad \sign(t-t') \to \PV \frac{2i}{\omega}.
\]
In the generic case the Fourier-transformed propagators are defined as [see also \Eqref{MidPoint} chapter 3]
\begin{equation} \label{MidTime}
	G(\omega,T)= \int \ud \Delta\expp{i\Delta\omega} G(T+\Delta/2,T-\Delta/2).
\end{equation}
It will be also useful to recall that $1/(z \pm
i\epsilon) = \PV (1/z) \mp i\pi \epsilon$ and $\theta(z) =
[1+\sign(z)]/2$.

Differently to the time representation, the Hadamard function is real and the Feynman and Dyson propagators acquire an imaginary part in the frequency representation: 
\begin{subequations}
\begin{align}
	G(\omega,T) = -G(-\omega,T) &\in \mathbf R,\\
	G^{(1)}(\omega,T) = G^{(1)}(-\omega,T) &\in  \mathbf R,\\
	G_+(\omega,T),G_-(\omega,T) &\in \mathbf R,\\
	\GF(\omega,T),G_\text{D}(\omega,T),\Gret(\omega,T),G_\text{A}(\omega,T) &\in \mathbf C.
\end{align}
\end{subequations}
The symmetry properties also imply:
\begin{subequations}
\begin{align}
	G_\text{D}(\omega,T) & = \GF^*(\omega,T),\\
	G_{-}(\omega,T) & = G_+(-\omega,T), \\
	G_\text{A}(\omega,T) &= \Gret^*(\omega,T)= - \Gret(-\omega,T).
\end{align}
\end{subequations}

Eqs.~\eqref{GretG} and \eqref{GadvG} in the frequency space read:
\begin{subequations}
\begin{align}
	\Gret(\omega,T) &= \int \frac{\vd \omega'}{2\pi} \frac{iG(\omega',T)}{\omega'-\omega+i\epsilon},\\
	\Gadv(\omega,T) &= \int \frac{\vd \omega'}{2\pi} \frac{iG(\omega',T)}{\omega'-\omega-i\epsilon}.
\end{align}
\end{subequations}
The inverse relations are:
\begin{equation}
	G(\omega,T) = 2 \Re \Gret(\omega,T) = -2\Re\Gadv(\omega,T).
\end{equation}

The following relations give the
decomposition of the complex propagators in its real and imaginary parts:
\begin{subequations}\label{Rel2Time}
\begin{align}
\Gret(\omega,T) &= \fud  G(\omega,T) + i \PV \int \frac{\vd \omega'}{2\pi} \frac{\Gc(\omega',T)}{\omega'-\omega}\\
G_\text{A}(\omega,T) &= \fud  G(\omega,T) - i \PV \int \frac{\vd \omega'}{2\pi} \frac{\Gc(\omega',T)}{\omega'-\omega} \\
    G_\mathrm F(\omega,T) &= \fud  \Ga(\omega,T) + i \PV \int \frac{\vd \omega'}{2\pi} \frac{\Gc(\omega',T)}{\omega'-\omega}   , \\
    G_\mathrm D(\omega,T) &= \fud  \Ga(\omega,T) - i \PV \int \frac{\vd \omega'}{2\pi} \frac{\Gc(\omega',T)}{\omega'-\omega}  .
\end{align}
\end{subequations}

Relations which do not involve time-ordering properties are identical in the frequency representation and will not be repeated here.

In the frequency representation the full set of propagators is determined by two real quantities, the Pauli-Jordan propagator and the Hadamard function, which can be regarded as the basic building blocks of the propagator set.

\index{Thermal field theory}
\section{Thermal and vacuum states}

If the state of the system is thermal, \ie, 
\begin{equation}
	\hat\rho = \frac{\expp{-\beta \hat H}}{\Tr{ \expp{-\beta \hat H}}},
\end{equation}
then the Kubo-Martin-Schiwnger (KMS) condition is verified
(see appendix \ref{app:CTP}):
\begin{equation}
	G_+(t-i\beta,t') = G_-(t,t').
\end{equation}
The KMS condition is equivalent to the following expression, which connects the Hadamard function and the Pauli-Jordan propagator:
\begin{equation}\label{KMS-CTP}
	G^{(1)}(\omega) = \coth{\left(\frac{\beta \omega}{2}\right)} G(\omega).
\end{equation}
The fluctuation-dissipation theorem [see \Eqref{FluctDisTh} in section 2] can be understood as a particular application of this last equation.  Yet another alternative expression of the KMS condition is:
\begin{equation}
	\GF(\omega) + G_\text{D}(\omega) = \expp{-\beta\omega} G_+(\omega) + \expp{\beta \omega} G_-(\omega).
\end{equation}
Therefore, for thermal states the Pauli-Jordan propagator alone determines the properties of the full set of propagators.
\index{Fluctuation-dissipation theorem}


In the  zero temperature case (\ie, in the case of being in the vacuum) the KMS condition implies
\begin{equation}
	\Ga(\omega) = \sign(\omega) G(\omega).
\end{equation}
In this case the Hadamard function and the Pauli-Jordan propagator essentially coincide, and new relations among the propagators emerge:
\begin{subequations}\label{Rel2Fourier}
\begin{align}
    G_\mathrm F(\omega) &= \int \frac{\ud {\omega'}}{2\pi} \frac{ i\Gc(\omega')}{\omega' - \omega + i \epsilon  \omega}, \\
    G_\mathrm D(\omega) &= \int \frac{\ud {\omega'}}{2\pi} \frac{ i\Gc(\omega')}{\omega' - \omega - i \epsilon  \omega}, \\
    G_{+}(\omega) &= 2i  \theta(\omega) \Gc(\omega), \\
G_{-}(\omega) &= 2i  \theta(-\omega) \Gc(\omega).
\end{align}
\end{subequations}

\section{Linear systems}

For linear systems (for instance, the QBM model considered in chapter 2) the Pauli-Jordan
propagator is state-independent since the commutator is
proportional to the identity operator (it is a c-number):
\[
    [\hat q(t),\hat q(t')] = [\hat q f(t) + \hat p g(t),\hat q f(t') + \hat p
    g(t')] = i[f(t)g(t') - g(t)f(t')] \hat 1,
\]
In this derivation we have used the fact that the position operators in the Heisenberg
picture can be expressed in the linear case as $\hat q(t) = f(t)
\hat q + g(t) \hat p$, where $\hat q$ and $\hat p$ are the
operators in the Schrödinger representation. (Recall that the Heisenberg equations of motion for
the operators coincide with the equations of motion for the
classical variables; $\hat q$ and $\hat p$ play the role of
initial conditions.) We have also used that $[\hat q,\hat p]=i$. In chapters 2 and 3 we give alternative proofs of the state independence of those propagators.

Thus, for free or linearly coupled systems the Pauli-Jordan, retarded and advanced propagators are independent of the state $\hat\rho$.

\section{Relations between the self-energy components}

The self-energy components $\Sigma^{ab}$ (see appendix \ref{app:CTP}) are not independent, but obey similar relations to these discussed in the previous sections of this appendix. There are two ways to understand those relations. First, the self-energy matrix is linked to the propagator matrix via the Schwinger-Dyson equation \eqref{SelfEnergyGeneral}, and therefore the relations between propagators translates into relations between the self-energy components. Second, one should recall (see section 2 and appendix \ref{app:CTP}) that the self-energy components actually correspond to amputated propagators, so that they follow essentially the same relations as ordinary propagators, bearing in mind that there is an additional $i$ factor  of the self-energy and that Feynman rules in the CTP formalism imply an additional minus sign for every type-2 vertex.

In general, the self-energy components satisfy:;
\begin{subequations} \label{RelSigma}
\begin{gather}
    \Sigma^{11}(\omega,T) =- (\Sigma^{22})^*(\omega,T), \qquad \Sigma^{12}(\omega,T)=\Sigma^{21}(-\omega,T)\\
    \Sigma^{11}(\omega,T) + \Sigma^{22}(\omega,T) =-  \Sigma^{12}(\omega,T) -
    \Sigma^{21}(\omega,T).
\end{gather}
\end{subequations}
Concerning the reality properties, $\Sigma^{12}(\omega,T)$ and $\Sigma^{21}(\omega,T)$ are purely imaginary, and $\Sigma^{11}(\omega,T)$ and $\Sigma^{22}(\omega,T)$ have both real and imaginary part. The retarded and advanced self-energies are related by complex conjugation: $\SigmaA(\omega,T)=\SigmaR^*(\omega,T)$.

The retarded, advanced and Hadamard self-energies are defined by:
\begin{subequations} \label{KeldyshCompDef}
\begin{align}
	\SigmaR(\omega,T) &:= \Sigma^{11}(\omega,T) + \Sigma^{12} (\omega,T) = - \Sigma^{21}(\omega,T) - \Sigma^{22}(\omega,T) \\
	\SigmaA(\omega,T) &:= \Sigma^{11}(\omega,T) + \Sigma^{21} (\omega,T) = - \Sigma^{12}(\omega,T) - \Sigma^{22}(\omega,T)\\
	\SigmaN(\omega,T) &:= -\Sigma^{21}(\omega,T) - \Sigma^{12}(\omega,T) = \Sigma^{11}(\omega,T) + \Sigma^{22}(\omega,T).
\end{align}
\end{subequations}
Moreover, $\SigmaN(\omega,T)$ is purely imaginary, and $\SigmaR(\omega,T)$ and $\SigmaA(\omega,T)$ have both real and imaginary parts.

The following equation, which is a consequence of the KMS condition, is only verified if the state is
thermal:
\begin{equation}
    \Sigma^{11}(\omega) + \Sigma^{22}(\omega) = -\expp{\beta \omega} \Sigma^{12}(\omega,T) - \expp{-\beta \omega}
    \Sigma^{21}(\omega).
\end{equation}
Thus, in the thermal case all the components of the self-energy can be determined from the
knowledge of just one of them. 

\index{Self-energy!retarded}

As explained in chapter 3, a particularly useful combination is the retarded self-energy $\SigmaR(\omega,T)$, which is directly connected to the retarded propagator. The imaginary part of the retarded self-energy is of particular interest:
\begin{equation}
    \Im \Sigma_\mathrm R(\omega,T) = \frac{i}{2} [ \Sigma^{21}(\omega,T) -
    \Sigma^{12}(\omega,T)   ].
\end{equation}
This equation can be easily derived from
\begin{equation} \label{cut}
    \Im \Sigma^{11}(\omega,T) = \frac{i}{2} [ \Sigma^{12}(\omega,T) +
    \Sigma^{21}(\omega,T)
    ],
\end{equation}
which in turn is derived from relations \eqref{RelSigma}. 
The real part can be expressed as
\begin{equation}
	\Re \Sigma_\mathrm R(\omega,T) = \Re \Sigma^{11}(\omega,T) = \fud\big[\Sigma^{11}(\omega,T)- \Sigma^{22}(\omega,T) \big].
\end{equation}
For a thermal state, $\Sigma_\mathrm R(\omega)$ is related to
$\Sigma^{11}(\omega)$ through:
\begin{equation}
    \SigmaR(\omega)  = \Re \Sigma^{11}(\omega) + \tanh
    \left( \frac{\beta \omega}{2} \right) \Im \Sigma^{11}(\omega) \label{SigmaR11}.
\end{equation}
and the Hadamard self-energy is connected to the imaginary part of the self-energy as follows:
\begin{equation}
	\SigmaN(\omega) = -2i \Im \SigmaR(\omega) \coth \left( \frac{\omega \beta}{2} \right).
\end{equation}

	\chapter{Gaussian states}\label{app:Gaussian}
\index{Gaussian state}

In the thesis we make extensive use of the Gaussian states. In the following we shall give a brief definition and a description of the aspects relevant to us without entering into details. Gaussian states are of crucial importance in quantum optics. See \eg\ refs.~\cite{WallsMilburn,GardinerZoller} for a more complete description.

In this appendix $N$ will always represent the correct normalization constant, which can always be computed with the Gaussian integration formula if desired.

\section{Single harmonic oscillator}

In general, Gaussian states are those whose density matrices have  a Gaussian form in a coordinate representation, or equivalently those whose Wigner functions have a Gaussian form. Given a single quantum mechanical degree of freedom, the more general Gaussian state is given by the following density matrix \cite{RouraThesis}:
\begin{equation}
	\rho(q,q') = N \exp{\big[-A (q-\bar q)^2 - A^*(q'-\bar q)^2 +2B(q-\bar q)(q'-\bar q)\big]}
\end{equation}
where $B$ and $\bar q$ are real, with $B$ satisfying $B < \Re A$ (so that the function can be normalized). Gaussian states can also be equivalently characterized by the following Wigner function [see \Eqref{Wigner}]:
\begin{equation}
	W(q,p) = N \exp{\left( \frac{ -C(p-\bar p)^2 - D(q-\bar p)^2 + 2E(q-\bar q)(p-\bar p)}{2(CD-E^2)}\right)},
\end{equation}
where $\bar q,\bar p,C,D,E$ are real, and where the relation $CD - E^2 \geq \Omega^2/4$,  has to be verified for the above Wigner function to represent the state of a harmonic oscillator of frequency $\Omega$  (the inequality is saturated for pure states) \cite{DodonovEtAl94}. The quantities $\bar q$ and $\bar p$ correspond to the expectation values of the position and the momentum respectively, $\bar q=\av{\hat q}$ and $\bar p = \av{\hat p}$.
In a operator representation, in terms of the creation and destruction operators,
\begin{equation}
	\hat q = \frac{\hat a + \hat a^\dag}{\sqrt{2\Omega}}, 
	\qquad
	\hat p = \frac{\hat a - \hat a^\dag}{i\sqrt{2\Omega}},
\end{equation}
a generic Gaussian state with $\av{\hat p}=0$ and $\av{\hat q}=0$ (or equivalently with $\av{\hat a}=\av{\hat a^\dag}=0$) can be represented as:
\begin{equation}
	\hat\rho = N \exp{\big( -F \hat a^\dag \hat a + G \hat a \hat a + G^* \hat a^\dag \hat a^\dag \big)},
\end{equation}
where $F$ is real. In this thesis we have exclusively used Gaussian states with zero mean (many times with making explicit mention).

Any Gaussian state can be obtained by applying an unitary transformation to the density matrix of a thermal state,
\begin{equation}
	\hat\rho_\beta = N \exp{\Big[-\beta \Omega  \Big( \hat a^\dag \hat a + \frac{1}{2} \Big)\Big]}.
\end{equation}
with $\beta=1/T$, $T$ being the temperature. The unitary operation is the product of displacement, squeeze and rotation operators \cite{Adam94}.

Stationary states are those which conmute with the Hamiltonian,
\begin{equation}
	[\hat \rho,\hat H] = 0.
\end{equation}
The stationary state of a harmonic oscillator with
\begin{equation}
	\hat H = \frac{1}{2} \hat p^2 + \frac{1}{2} \Omega^2 \hat q^2 =
	\Omega \Big( \hat a^\dag \hat a + \frac{1}{2} \Big)
\end{equation}
must be of the form $\hat\rho = N \exp{(-F  \hat a^\dag \hat a)}$, and therefore corresponds to a thermal state. 

\section{Reduced 2-mode sector of a scalar field}

For a scalar field, decomposed in modes $\phi_\vect p$, the most general state for the two-mode system $\pm \vect p$ which is translation invariant, \ie, which conmutes with the momentum operator,\footnote{Do not confuse the  physical momentum operator $\hat{\vect p}$, with the canonical momentum operator $\hat\pi_\vect p$, conjugate of the field operator.}
\begin{equation}
	[\hat\rho_\text{r},\hat{\vect p}]=0, \qquad \hat{\vect p}= \vect p \big(\hat a^\dag_\vect p \hat a_\vect p  - \hat a^\dag_{-\vect p} \hat a_{-\vect p}\big).
\end{equation}
and which verifies $\av{\hat a_{\pm \vect p}}=\av{\hat a^\dag_{\pm \vect p}}=0$ is given by
\begin{equation}
	\hat\rho_\text{r} = N \exp{\big[
	-F \big( \hat a_\vect p^\dag \hat a_\vect p 
	+ \hat a_{-\vect p}^\dag \hat a_{-\vect p} \big)
	+ 2G \hat a_\vect p \hat a_{-\vect p} 
	+ 2G^* \hat a^\dag_\vect p \hat a^\dag_{-\vect p} \big]}.
\end{equation}
If we further impose stationarity with respect to the reduced Hamiltonian,
\begin{equation}
	[\hat\rho_\text{r},\hat{H_\text{s}}]=0, \quad \hat{H_\text{s}}= \hat\pi_\vect p \hat\pi_{-\vect p} + R^2_\vect p \hat\phi_\vect p \hat\phi_{-\vect p}  =R^2_\vect p \big(\hat a^\dag_\vect p \hat a_\vect p  + \hat a^\dag_{-\vect p} \hat a_{-\vect p} + 1\big),
\end{equation}
the most general Gaussian state corresponds to a factorized thermal state:
\begin{equation}
	\hat\rho_\text{r} = N \exp{\big[
	-F \big( \hat a_\vect p^\dag \hat a_\vect p 
	+ \hat a_{-\vect p}^\dag \hat a_{-\vect p} \big) \big]} =  N \exp{\big(
	-F  \hat a_\vect p^\dag \hat a_\vect p \big) } \exp{\big(
	-F  \hat a_{-\vect p}^\dag \hat a_{-\vect p} \big) }.
\end{equation}

\index{Wick theorem}
\index{Gaussian system}
\section{Gaussian systems and the Wick theorem}

A quantum mechanical or field theory system is Gaussian if its generating functional is Gaussian [see eq.~\eqref{ZCTPOper}]. For a single degree of freedom this means (assuming vanishing expectation values for the position operator)
\begin{equation} \label{AppGenF}
\begin{split}
	Z[j_{a}]&=\exp\bigg[-\frac{1}{2}\int \ud t \ud{t'} j^{a}(t)  G_{ab}(t,t')j^{b}(t) \bigg]
\end{split}
\end{equation}
See \Eqref{ZCTPGaussian} for the equivalent expression for a mode-decomposed field theory. Gaussian systems correspond either to Gaussian states following quadratic equations of motion (as in chapter 2), or alternatively to an approximation for general states following general equations of motion (as in chapter 3).

For Gaussian systems, according to the Wick theorem, the $n$-point correlation functions can be reduced to the two-point correlation functions. Instead of trying to give a general formulation of the Wick theorem let us simply show some particular applications. The general Wick theorem is studied among others in refs.~\cite{FetterWalecka} and \cite{LeBellac}. 

Whenever the system is Gaussian, any time-ordered four-point correlation function can be expressed in terms of two-point correlation functions (we  assume that the expectation value of the field operators vanishes):
\begin{equation}
\begin{split}
	\av{T\hat q(t_1) \hat q(t_2)\hat q(t_3)\hat q(t_4)}
	&= \av{T\hat q(t_1) \hat q(t_2)}\av{T\hat q(t_3)\hat q(t_4)} \\ &\quad +
	\av{T\hat q(t_1) \hat q(t_3)}\av{T\hat q(t_2)\hat q(t_4)}
	\\ &\quad +
	\av{T\hat q(t_1) \hat q(t_4)}\av{T\hat q(t_2)\hat q(t_3)}
\end{split}
\end{equation}
If the correlation function is a mixture of time- and antitime-ordered expressions, the equivalent expression goes as follows
\begin{equation} \label{WickMixed}
\begin{split}
	\av{T\hat q(t_1) \hat q(t_2)\widetilde T\hat q(t_3)\hat q(t_4)}
	&= \av{T\hat q(t_1) \hat q(t_2)}\av{\widetilde T\hat q(t_3)\hat q(t_4)} \\ 
	&\quad + \av{ \hat q(t_1)\hat q(t_3)}  \av{\hat q(t_2) \hat q(t_3)}\\
	&\quad  + \av{\hat q(t_1) \hat q(t_4)}\av{ \hat q(t_2)\hat q(t_3)} \\
\end{split}
\end{equation}
These expressions can be demonstrated by taking derivatives on the Gaussian generating functional \eqref{AppGenF}.

In turn, any of the two-point correlators can be expressed as a function of the creation and annihilation operators:
\begin{equation}
\begin{split}
	\av{\hat q(t_1) \hat q(t_2)} = \frac{1}{2\Omega}\Big(
	\av{\hat a(t_1) \hat a(t_2)} &+ 
	\av{\hat a^\dag (t_1) \hat a (t_2) } \\ &+
	\av{\hat a (t_1) \hat a^\dag  (t_2)} + 
	\av{\hat a^\dag(t_1) \hat a^\dag (t_2)} \Big).
\end{split}
\end{equation}
If the state is stationary, only those terms having a creation and an annihilation term survive:
\begin{equation}
\begin{split}
	\av{\hat q(t_1) \hat q(t_2)} = \frac{1}{2\Omega}\left(
	\av{\hat a^\dag(t_1) \hat a(t_2)} + 
	\av{\hat a (t_1) \hat a^\dag  (t_2)}  \right).
\end{split}
\end{equation}
The Wick theorem can also be directly applied to the creation and annihilation operators.

In field theory, if the state is translation-invariant, momentum conservation can simplify the application of the Wick theorem. For instance, if $\vect k \neq \vect q$,
\begin{equation}
\begin{split}
	\av{T\hat \phi_\vect k(t_1) \hat \phi_{-\vect q}(t_2)\hat \phi_\vect q(t_3)\hat \phi_{-\vect k}(t_4)}
	&= \av{T\hat \phi_\vect k(t_1) \hat \phi_{-\vect k}(t_4)}
	+ \av{T\hat \phi_\vect q(t_2) \hat \phi_{-\vect q}(t_3)}.
\end{split}
\end{equation}
The two-point correlators can be also expressed as a function of the creation and annihilation operators:
\begin{equation}
\begin{split}
	\av{\hat \phi_\vect k (t_1) \hat \phi_{-\vect k}(t_2)} =  \frac{1}{2R_\vect p} &\Big(
	\av{\hat a_\vect k(t_1) \hat a_{-\vect k}(t_2)} + 
	\av{\hat a^\dag_{-\vect k} (t_1) \hat a_{-\vect k} (t_2) } \\ &+
	\av{\hat a_{\vect k} (t_1) \hat a^\dag_\vect k  (t_2)} + 
	\av{\hat a^\dag_{-\vect k}(t_1) \hat a^\dag_\vect k (t_2)} \Big).
\end{split}
\end{equation}
If the state is stationary, only those terms having a creation and an annihilation operator survive:
\begin{equation}
\begin{split}
	\av{\hat \phi_\vect k (t_1) \hat \phi_{-\vect k}(t_2)} =  \frac{1}{2R_\vect p} &\Big(
	\av{\hat a^\dag_{-\vect k} (t_1) \hat a_{-\vect k} (t_2) } +
	\av{\hat a_{\vect k} (t_1) \hat a^\dag_\vect k  (t_2)}  \Big).
\end{split}
\end{equation}

	\chapter{Feynman rules of a scalar field in a graviton background} \label{app:FeyRules}
\index{Feynman rules!scalar field in a graviton background}

At zero temperature the free propagator for the scalar field is
\begin{equation}
    G_\mathrm F^{(0,{\scriptscriptstyle T=0})}(p)=\frac{-i}{p^2+m^2-i\epsilon},
\end{equation}
and the free propagator for the gravitons in the harmonic gauge is
\cite{Donoghue94a,Donoghue94b,Veltman76}
\begin{equation}
    (\Delta_\mathrm F)\TO_{\mu\nu\alpha\beta}(p) = \frac{-i \mathcal P_{\mu\nu\alpha\beta}}{p^2-i\epsilon},
\end{equation}
where
\begin{equation}
    \mathcal P_{\mu\nu\alpha\beta} = \fud \left(
    \eta_{\mu\alpha}\eta_{\nu\beta} + \eta_{\mu\beta}\eta_{\nu\alpha}
    - \frac{2}{d-2} \eta_{\mu\nu}\eta_{\alpha\beta}
    \right)
\end{equation}
in $d$ spacetime dimensions. In four dimensions, 
\begin{equation}\mathcal
P_{\mu\nu\alpha\beta} =  \left(
    \eta_{\mu\alpha}\eta_{\nu\beta} + \eta_{\mu\beta}\eta_{\nu\alpha}
    - \eta_{\mu\nu}\eta_{\alpha\beta}
    \right)/2.
\end{equation}

\begin{figure}
    \centering
    \hfill
    \includegraphics{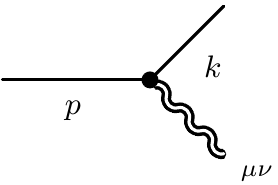}
    \hfill
    \includegraphics{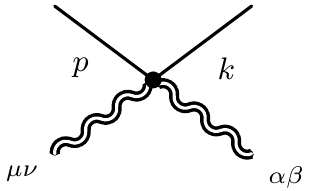}
    \hfill\mbox{}
    \caption{On the left, the two-scalar--one-graviton vertex, denoted
    by $\tau_{\mu\nu}(p)$. On the right, the two-scalar--two-graviton vertex,
    denoted by $V_{\mu\nu\alpha\beta}(p)$.}
    \label{fig:Vert}
\end{figure}

The two-scalar--one-graviton vertex, shown in fig.~\ref{fig:Vert},
is given by \cite{Donoghue94b}
\begin{equation}
    \tau_{\mu\nu}(p,k) = \frac{i \kappa}{2} [ p_\mu k_\nu + p_\nu k_\mu
    -
    (p\cdot k) \eta_{\mu\nu} - m^2 \eta_{\mu\nu} ],
\end{equation}
and the two-scalar--two-graviton vertex, also shown in
fig.~\ref{fig:Vert}, is given by \cite{Donoghue94b}
\begin{equation}
\begin{split}
    V_{\mu\nu\alpha\beta}(p,k) = - &\ i\kappa^2  \bigg[
    I_{\mu\nu\rho\lambda}
    I^\lambda_{\phantom\lambda\sigma\alpha\beta} \left(p^\rho k^\sigma
    + p^\sigma k^\rho\right) \\
    &+\frac12 \left(\eta_{\mu\nu} I_{\alpha\beta\rho\sigma} +
    \eta_{\alpha\beta} I_{\mu\nu\rho\sigma} \right) p^\rho
    k^\sigma \\
    &-\frac12 \left(I_{\mu\nu\alpha\beta} - \fud
    \eta_{\mu\nu}\eta_{\alpha\beta}\right)(p\cdot k + m^2) \bigg],
\end{split}
\end{equation}
where
\begin{equation*}
    I_{\mu\nu\alpha\beta} = \fud \left(
    \eta_{\mu\alpha}\eta_{\nu\beta} +
    \eta_{\mu\beta}\eta_{\nu\alpha}\right).
\end{equation*}

At finite temperature the free propagators for the scalar field
are given by eq.~\eqref{FreeGT} and the free propagators for the
gravitons are \[ (\Delta_{ab})_{\mu\nu\alpha\beta}(p)=\mathcal
P_{\mu\nu\alpha\beta} G^{(0)}_{ab}(p)|_{m=0}. \] Vertices type 1 are
described by $\tau_{\mu\nu}$ and $V_{\mu\nu\alpha\beta}$ and
vertices type 2 are described by $-\tau_{\mu\nu}$ and
$-V_{\mu\nu\alpha\beta}$.
	\addtocontents{toc}{\bigskip}
	\chapter{Technical details of some calculations}

\section[The Hadamard function and the self-energy (chapter 3)]{The Hadamard function and the self-energy (chapter 3)}\label{app:SelfEnergyHadamard}

The aim is to deduce the simplified relation between the Hadamard propagator and the self-energy, \Eqref{SelfEnergyHadamardSimp}, from \Eqref{SelfEnergyHadamard}. To simplify the expressions, we will use a matrix-like notation for the different propagators, omitting the space-time dependence and the space-time integrals. With this notation, \Eqref{SelfEnergyHadamardSimp} becomes
\begin{equation}
	\GN = \GN^{(0)}-i 
	  \GN^{(0)} \SigmaA \GA
	-i\GR^{(0)}\SigmaR \GN
	-i \GR^{(0)}\SigmaN \GA,
\end{equation}
or
\begin{equation}\label{SelfEnergyAbr}
	\left( 1+i\GR^{(0)} \SigmaR\right)\GN = \GN^{(0)}-i 
	  \GN^{(0)} \SigmaA \GA
	-i \GR^{(0)}\SigmaN \GA.
\end{equation}
Taking into account the relation between the retarded self-energy and the retarded propagator,
\begin{equation}
	\GR = \GR^{(0)} - i \GR^{(0)} \SigmaR \GR,
\end{equation}
and introducing the inverse retarded propagator 
\begin{equation}
	\int \ud[4]{y} \GR^{-1}(x,y) \GR(y,x')  =\int \ud[4]{y} \GR(x,y) \GR^{-1}(y,x') = \delta^{(4)}(x-x'),
\end{equation}
or simply $\GR \GR^{-1} =\GR^{-1} \GR=1$, we can write \Eqref{SelfEnergyAbr} as
\begin{equation}
	\GR^{(0)} \GR^{-1} \GN = (\GN)^{(0)}-i 
	  (\GN)^{(0)} \SigmaA \GA
	-i \GR^{(0)}\SigmaN \GA.
\end{equation}
Acting with the operator $\partial_{\mu}\partial^\mu + m^2$ on both sides of this equation, the first and second term on the right hand side vanish and we have
\begin{equation}
	 \GR^{-1} \GN = 
	-i \SigmaN \GA.
\end{equation}
Multiplying by the retarded propagator on the left we recover  \Eqref{SelfEnergyHadamardSimp}:
\begin{equation}
	  \GN = -i\GR \SigmaN \GA.
\end{equation}

Notice that we have assumed that the retarded propagator is the unique inverse of $\GR^{-1}$ and viceversa. This need not be the case, as it is highlighted by the fact that the differential operator $\partial_{\mu}\partial^\mu + m^2$ has different inverse kernels, among them the free retarded and advanced propagators. Therefore, the validity of \Eqref{SelfEnergyHadamardSimp} is subject to the retarded propagator being the unique inverse of a kernel. Without entering into mathematical details, it can be argued that for systems exhibiting sufficiently dissipative dynamics, and with asymptotic boundary conditions, such uniqueness is indeed guaranteed. 

\section[Integration by residues of $I(t,t_0; p)$ (chapter 3)]{Integration by residues of $I(t,t_0;\vect p)$ (chapter 3)} \label{app:Integral}

In chapter 3, when computing the time evolution of the propagator, the following integral appears [see eqs.~\eqref{IntegralI} and \eqref{IntegralI2}]:
\begin{equation}
 I(t,t_0;\vect p) =  \frac{1}{2R_\vect p}\int \frac{\vd \omega}{2\pi} \expp{-i\omega(t-t_0)} \frac{ |\omega| (\omega+ R_\vect p)^2  \Gamma_\vect p \left[ n_\vect p + \theta(\omega)\right]  }{\left(-\omega^2+R_\vect p^2\right)^2 + \left( \omega \Gamma_\vect p\right)^2}
\end{equation}
Apparently, this integral cannot be evaluated with complex plane techniques since the integrand contains the factor and $|\omega|$, which is non-analytic with the usual prescription $|\omega| = \sqrt{\omega \omega^*}$, and the factor $\theta(\omega)$, which in principle is defined in the real axis. Let us do a more careful analysis. 

We begin by extending the problematic terms to the complex plane in the following way:
\[
	\theta(\omega) \to \theta(\Re \omega), \qquad
	|\omega| \to \omega \sign(\Re \omega).
\]
Those terms continue to be non-analytic, but only in a branch cut located in the imaginary axis. Therefore, with this prescription the integrand is analytic everywhere in the complex plane except in the branch cut and in the poles. Notice that the branch cut is rather special, since the function is continuous across the branch cut.

Second, notice that the following related contour integrals are well-defined for $t>t_0$ and can be computed by residues:
\begin{align*}
	I_3(t,t_0;\vect p) &=  \frac{1}{2R_\vect p}\oint\limits_{C_3} \frac{\vd \omega}{2\pi} \expp{-i\omega(t-t_0)} \frac{ -\omega (\omega+ R_\vect p)^2  \Gamma_\vect p  n_\vect p  }{\left(-\omega^2+R_\vect p^2\right)^2 + \left( \omega \Gamma_\vect p\right)^2}, \\
	I_4(t,t_0;\vect p) &=  \frac{1}{2R_\vect p}\oint\limits_{C_4}  \frac{\vd \omega}{2\pi} \expp{-i\omega(t-t_0)} \frac{ \omega (\omega+ R_\vect p)^2  \Gamma_\vect p ( n_\vect p +1)  }{\left(-\omega^2+R_\vect p^2\right)^2 + \left( \omega \Gamma_\vect p\right)^2},
\end{align*}
where $C_3$ and $C_4$ are the closed anticlockwise paths at the boundaries of the left lower and right lower quadrants respectively

Third, note that with the above prescription $I(t,t_0;\vect p) = I_3(t,t_0;\vect p) + I_4(t,t_0;\vect p)$, since the path at infinity does not contribute, and the contribution from the path in the imaginary axis cancels (because the function is continuous at the branch singularity). Therefore, in practice, the integral $I(t,t_0;\vect p)$ can be computed by residues as if there were no branch cut singularity.

Let us now compute the integrals $I_3(t,t_0;\vect p)$ and $I_4(t,t_0;\vect p)$. We start by $I_4(t,t_0;\vect p)$. There is a pole in the lower right quadrant at $\omega \approx R_\vect p - i \Gamma_\vect p/2$. Neglecting $\Gamma_\vect p$ in front of $R_\vect p$ we find
\begin{equation*}
	I_4(t,t_0;\vect p) \approx R_\vect p (1+n_\vect p) \expp{-iR_\vect p(t-t_0)} \expp{-\Gamma_\vect p/2}.
\end{equation*}
With respect to the integral $I_4(t,t_0;\vect p)$, the contribution from the pole in the lower left quadrant is very suppressed because of the factor $(\omega+R_\vect p)^2$ in the numerator. Therefore $I_3(t,t_0;\vect p) \ll I_4(t,t_0;\vect p)$. Given all this we obtain the final result:
\begin{equation}
	I_4(t,t_0;\vect p) \approx R_\vect p (1+n_\vect p) \expp{-iR_\vect p(t-t_0)} \expp{-\Gamma_\vect p/2}.
\end{equation}

\section{Integrals in dimensional regularization (chapter 4)}\label{app:Int}

In Eq.~\eqref{SigmaDecInt} we need the following integrals:
\begin{gather}
I(p) = \mu^\varepsilon\int \frac{\vd[d] k}{(2\pi)^d}
     \frac{1}{[k^2+m^2-i\epsilon][(p-k)^2-i\epsilon]}, \\
I_\mu(p) = \mu^\varepsilon\int \frac{\vd[d] k}{(2\pi)^d}
    \frac{k_\mu}{[k^2+m^2-i\epsilon][(p-k)^2-i\epsilon]},\\
I_{\mu\nu}(p) = \mu^\varepsilon \int \frac{\vd[d] k}{(2\pi)^d}
\frac{k_\mu
    k_\nu}{[k^2+m^2-i\epsilon][(p-k)^2-i\epsilon]}.
\end{gather}
The result in arbitrary $d$ dimensions after series expansion in
$\varepsilon = 4-d$ is
\begin{widetext}
\begin{align}
\begin{split}
    I &= \frac{i}{(4\pi)^2} \left[ \frac{1}{\hat\varepsilon} +  2
    - \frac{m^2}{p^2} \ln \left(1+\frac{p^2}{m^2} -i\epsilon \right) \right.  \\   &\qquad\left.- \ln \left(\frac{p^2+m^2}{\mu^2}-i\epsilon\right)\right] +
    O(\varepsilon),
\end{split}\\
\begin{split}
  I_\mu(p)&= \frac{i p_\mu}{2(4\pi)^2} \bigg[ \frac{1}{\hat\varepsilon} +  2 -
    \frac{m^2}{p^2}
    + \frac{m^4}{p^4} \ln \bigg(1+\frac{p^2}{m^2} -i\epsilon\bigg)
    \\ &\qquad-  \ln \left(\frac{p^2+m^2}{\mu^2}-i\epsilon\right)\bigg] +
    O(\varepsilon),
\end{split}\\
\begin{split}
    I_{\mu\nu}(p) &= - \frac{\eta_{\mu\nu}}{2(4\pi)^2} \bigg[ \frac{m^2}{2\hat\varepsilon} + \frac{p^2}{6\hat\varepsilon}+ \frac{m^4}{6p^2} + \frac{7m^2}{6}
    +
    \frac{4p^2}{9}\\
      &\qquad - \left( \frac{m^6}{6p^4} + \frac{m^4}{2p^2} \right) \ln \left( 1 + \frac{p^2}{m^2}-i\epsilon \right)  \\&\qquad-\left(\frac{m^2}{2}+ \frac{p^2}{6}\right)\ln \left( \frac{p^2+m^2}{\mu^2} - i\epsilon \right) \bigg]
    \\
    &\quad+ \frac{p_\mu p_\nu}{(4\pi)^2}
    \bigg[ \frac{1}{3\hat\varepsilon}  - \frac{m^6}{3p^6}
    \ln\left(1+\frac{p^2}{m^2}-i\epsilon
    \right) \\ \qquad& -\frac13  \ln\left(\frac{p^2+m^2}{\mu^2}-i\epsilon\right)
     \frac{m^4}{3p^4} - \frac{m^2}{6p^2} + \frac{13}{18} \bigg] +
     O(\varepsilon),
\end{split}
\end{align}
where
\[
\frac{1}{\hat\varepsilon} = \frac{2}{\varepsilon} - \gamma +
    \ln 4\pi.
\]
\end{widetext}

\section[{Computation of $A(p)$, $B(p)$, $C(p^2)$ and
$D(p)$ (chapter 4)}]{Computation of $A(p)$, $B(p)$, $C(p^2)$ and
$D(p)$ (chapter 4)}\label{app:ABCD}

In this appendix we  compute the integrals $A(p)$, $B(p)$,
$C(p^2)$ and $D(p)$ which appear in the calculation of the
self-energy at finite temperature; see eqs.~\eqref{A}--\eqref{C}
and \eqref{D}.

Let us start by computing the integral $A(p)$, defined in
\Eqref{A}. The Dirac delta can be expanded as
\begin{equation*}
    \delta\boldsymbol((p-k)^2\boldsymbol) = \delta(q^2)
    = \frac{1}{2|\vect q|} \left[ \delta(-q^0+ |\vect q|) +
    \delta(q^0+|\vect{q}|) \right],
\end{equation*}
where we have introduced the new variable $q=p-k$ and where
$p=(p^0,\vect p)$ and $q=(q^0,\vect q)$. Introducing now spherical
coordinates $(\phi,\theta)$ in the three spatial dimensions, with
$\theta$ being the angle between $\vect p$ and $\vect q$, and
integrating with respect to $q^0$ with the aid of the delta
function we get
\begin{widetext}
\begin{equation*}
\begin{split}
    A(p) &= \int_0^\infty \frac{n(Q) Q\mathrm d Q}{2(2\pi)^2} \int_{-1}^1 \ud x  \left[ \PV\frac{
     g_2(p^2,0,-p^0 Q +  P Q x)}{-(p^0)^2+P^2+2p^0 Q -  2P Q x+m^2} \right. \\ &\qquad+ \left. \PV\frac{
     g_2(p^2,0,p^0 Q +  P Q x)}{-(p^0)^2+P^2-2p^0 Q - 2 P Q x+m^2} \right],
\end{split}
\end{equation*}
where $Q=|\vect q|$, $P=|\vect p|$, $x=\cos \theta$,
\begin{equation}
\begin{split}
    g_2(p^2,q^2,p\cdot q)&=g_1\boldsymbol(p^2,(p-q)^2,p\cdot(p-q)\boldsymbol)\\
    &= -2m^4 - 2 m^2 p^2 + p^4 + 2 m^2 (p \cdot q) - 2 p^2 (p \cdot q)+
      p^2 q^2,
\end{split}
\end{equation}
and we have performed the trivial angular integration over $\phi$.
We now integrate with respect to $x$ to get
\begin{equation} \label{PreB}
\begin{split}
    A(p) &= \int_0^\infty \frac{n(Q) \mathrm d Q}{16\pi^2P} 
    \Bigg[
    8PQ[-m^2-(p^0)^2+P^2] + m^2\{m^2+2[(p^0)^2-P^2]\} \\  &\qquad\times\ln
    \left(
    \tfrac{\left[m^2 - (p^0-P+2Q)(p^0+P)\right]\left[m^2 -
    (p^0-P)(p^0+P-2Q)\right]}{\left[m^2 - (p^0-P-2Q)(p^0+P)\right]\left[m^2 -
    (p^0-P)(p^0+P+2Q)\right]}\right) \Bigg].
\end{split}
\end{equation}
\end{widetext}
The result of this integral cannot be given in closed analytic
form, in general. However in chapter 4 we are mainly interested
in its on-shell value $p^0 = E_\vect p = \sqrt{m^2+P^2}$, and in
this limit the integral can be computed exactly at any temperature
since the logarithmic term in \Eqref{PreB} vanishes. In this case
the value of the integral is given by
\begin{equation} \label{AOnShell}
    A(E_\vect p,\vect p) = - \frac{ m^2}{\pi^2} \int_0^\infty \ud Q n(Q)
    Q = - \frac{1}{6}m^2T^2,
\end{equation}
where we used that
\begin{equation*}
    \int_0^\infty \ud Q n(Q) Q = \frac{\pi^2 T^2}{6}.
\end{equation*}

We now proceed with the computation of $B(p)$, defined in
\Eqref{B}. Repeating similar steps as in the previous integral we
get
\begin{equation*}
\begin{split}
    B(p) &= \int_0^\infty \frac{n(E_\vect k) K^2\mathrm d K}{2(2\pi)^2E_\vect k} \int_{-1}^1 \ud x  \left[ \PV\frac{
     g_1(p^2,0,-p^0 E_\vect k +  P K x)}{-(p^0)^2+P^2+2p^0 E_\vect k - 2 P K x} \right. \\ &\qquad+ \left. \PV\frac{
     g_1(p^2,0,p^0 E_\vect k +  P K x)}{-(p^0)^2+P^2-2p^0 E_\vect k - 2 P K x}
     \right],
\end{split}
\end{equation*}
where we recall that $E_\vect k=\sqrt{m^2+K^2}$. The integral with
respect to $x$ can be analytically performed; the result is a very
large and cumbersome expression, which we shall not reproduce
here. The resulting expression cannot be integrated again in a
closed analytic form. However, we may find particular expressions
valid at low and high temperatures. As in the case of previous
integral, we will restrict to the on shell results.

At low temperatures, only those momenta $\vect k$ whose
corresponding energies are at most of the order of the
temperature, $E_\vect k \lesssim T$, contribute significantly to
the integral because of the presence of the thermal factor
$n(E_\vect k)$, which acts as a soft cutoff. Hence, low
temperature also implies low-energy and low momentum. Therefore,
in the low-temperature approximation we may retain only the
leading term in a $K$ expansion:
\begin{equation*}
\begin{split}
    B(E_\vect p, \vect p) &=  \frac{m^2+2P^2}{\pi^2(3m^2+4P^2)} 
    \int_0^\infty \ud K n(E_\vect k) \left[ K^2 +
    O(K^3) \right].
\end{split}
\end{equation*}
Taking into account that for low temperatures
\begin{equation*}
    n(E_\vect k) \approx
    \expp{-m/T}\expp{-K^2/(2mT)}
\end{equation*}
and that
\begin{equation*}
    \int_0^\infty \ud K \expp{-K^2/(2mT)} K^2 = \sqrt{\frac\pi2}
    \,
    (mT)^{3/2},
\end{equation*}
we find the following expression for $B(p)$ at low temperature:
\begin{equation}
     B(E_\vect p,\vect p) \approx \sqrt{\frac{m^5 T^3}{2\pi^3}}
     \left(\frac{m^2+2P^2}{3m^2+4P^2}\right)\expp{-m/T}.
\end{equation}
We have not made precise the exact meaning of the
``low-temperature'' approximation employed above. In principle,
this approximation would require the temperature $T$ to be much
smaller than any relevant quantity with dimensions of energy that
could be formed by a combination of $m$ and $P$. However, a
detailed analysis of the expressions shows that the condition $m
\gg T$ is sufficient to guarantee the validity of the result.

Let us now proceed to the calculation of $B(p)$ in the high
temperature regime. Since $B(p)$ would be divergent if no thermal
cutoff were present, at high temperatures the leading contribution
to the integral is given by those momenta close to the temperature
$T$. Thus, as a first approximation, we can retain only the leading
term in a $1/K$ expansion:
\begin{equation}
\begin{split}
    B(E_\vect p, \vect p) &=
    \bigg[ \frac{m^2 \sqrt{m^2+ P^2}}{4\pi^2P} \ln \left(\frac{ 2\sqrt{m^2+ P^2}- P}{ 2\sqrt{m^2+ P^2}+ P}\right)  + \frac{3}{4\pi^2 } m^2 \bigg] \\ &\qquad \times \int_0^\infty \ud K
    n(E_\vect k) \left[ K + O(1/K^0) \right]
\end{split}
\end{equation}
Since the leading contribution to the integral is given in the
ultrarelativistic regime, we can approximate the energy by the
momentum in the Bose-Einstein function, $n(E_\vect k)\approx
n(K)$. With this approximation we find
\begin{equation}\label{BAppendix}
\begin{split}
    B(E_\vect p, \vect p) &=
    \frac{m^2 T^2 E_{\vect p}}{24P} \ln \left(\frac{ 2E_{\vect p}- P}{ 2E_{\vect P}+ P}\right)  + \frac{1}{8} m^2 T^2.
\end{split}
\end{equation}
Analogously to the low-temperature case, the high temperature
approximation would \emph{a priori} require the temperature $T$ to
be much higher than $m$, $P$ and any relevant energy scale formed
by combination of these two. Again, one can show that the
condition $T\gg m$ is sufficient to guarantee the validity of
\Eqref{BAppendix}.

We now move to the integral $C(p^2)$, defined in \Eqref{C}. Its
evaluation is straightforward:
\begin{equation}
    C(p^2) = g_3(p^2) \int_0^\infty \frac{K \vd K}{2\pi^2} n(K) =
    \frac{T^2}{12} (10m^2+4p^2).
\end{equation}
We only need its on shell value $C(-m^2) = T^2 m^2 / 6$.

Let us now consider the integral $D(p)$, defined in \Eqref{D}. We
start by introducing the variable $q = p -k$:
\begin{widetext}
\begin{equation*}
    \begin{split}
    D(p) = \int \frac{\mathrm d q^0 \, \mathrm d^3 \vect q}{(2\pi)^2} &\ \, F(p^0,q^0)\, g_2\boldsymbol(-(p^0)^2+|\vect p|^2,-(q^0)^2+\vect q^2,-p^0 q^0+\vect p \cdot \vect q \boldsymbol)
      \\ &\times
      \delta\boldsymbol( - (q^0)^2 + \vect q^2 \boldsymbol)\, \delta\boldsymbol(-(p^0-q^0)^2+(\vect p - \vect q)^2
    +m^2\boldsymbol),
    \end{split}
\end{equation*}
where $p=(p^0,\vect p)$ and $q=(q^0,\vect q)$. Next we expand the
first delta function and integrate over $q^0$:
\begin{equation} \label{DpNice}
    \begin{split}
    D(p) &= \int \udpi[3]{\vect q} \frac{2\pi}{2 |\vect q|} F(p^0,|\vect q|)  g_2\boldsymbol(|\vect p|^2-(p^0)^2,0,\vect p \cdot \vect q - p^0 |\vect q|
    \boldsymbol)\\ &\qquad\times
    \delta\boldsymbol(-(p^0-|\vect q|)^2+(\vect p - \vect q)^2
    +m^2\boldsymbol)\\
    &\quad+\int \udpi[3]{\vect q} \frac{2\pi}{2 |\vect q|} F(p^0,-|\vect q|)\, g_2\boldsymbol(|\vect p|^2-(p^0)^2,0,\vect p \cdot \vect q + p^0 |\vect q| \boldsymbol) \\ &\qquad\times  \delta\boldsymbol(-(p^0+|\vect q|)^2+(\vect p - \vect q)^2
    +m^2\boldsymbol).
    \end{split}
\end{equation}
We now introduce spherical coordinates over $\vect q$ and expand
the second delta function:
\begin{equation*}
    \begin{split}
    D(p) &= \frac{1}{8\pi P} \int_{-1}^1 \ud x \int_0^\infty \ud Q F(p^0,Q) \, g_2\boldsymbol(P^2-(p^0)^2,0,Q P x - Q p^0\boldsymbol) \\ &\qquad\times \delta\boldsymbol( x - (p^2+ 2 p^0 Q + m^2
    )/(2PQ)\boldsymbol)\\
    &\quad+\frac{1}{8\pi P} \int_{-1}^1 \ud x \int_0^\infty \ud Q F(p^0,-Q) \, g_2\boldsymbol(P^2-(p^0)^2,0,Q P x + Q p^0\boldsymbol) \\ &\qquad\times  \delta\boldsymbol( x - (p^2- 2 p^0 Q + m^2
    )/(2PQ)\boldsymbol).
    \end{split}
\end{equation*}
\end{widetext}
  where $Q = |\vect q|$, $P=|\vect p|$ and $x=\cos
\theta$. We have already performed the trivial angular integration
over $\phi$. Integrating with respect to $x$ with the aid of the
delta function we get
\begin{equation}
    D(p) = \frac{g_2\boldsymbol(p^2,0,(p^2+m^2)/2\boldsymbol)}{8\pi P} \left| \int_{Q_1}^{Q_2}  \ud Q
    F(p^0,Q)\right|
\end{equation}
with
\begin{equation*}
    Q_2=\frac{(p^0)^2-P^2-m^2}{2(p^0-P)}, \qquad
    Q_1=\frac{(p^0)^2-P^2-m^2}{2(p^0+P)},
\end{equation*}
which can be finally arranged as
\begin{equation}
    D(p) = \frac{-m^2(m^2+2p^2)}{8\pi P} \left| \int_{Q_1}^{Q_2}  \ud Q
    F(p^0,Q)\right|.
\end{equation}
Recall that $p^2 = -(p^0)^2+P^2$.

\section{Free propagators in a 1-particle state background (chapter 5)}\label{app:FreeOne}

In this section we compute the free propagators for the field $\phim$ in the background state characterized by the presence of a 1-particle excitation with momentum ${\vect p'}$. This state will be denoted by $|{\vect p'}\rangle$. We need the full set of CTP propagators, \ie,
\begin{equation}
    G_{ab}(x,x';{\vect p'})=
    \begin{pmatrix}
        \langle {\vect p'} | T\hat \phi_m(x) \hat \phi_m(x') | {\vect p'} \rangle &
        \langle {\vect p'} | \hat \phi_m(x) \hat \phi_m(x') | {\vect p'} \rangle  \\
        \langle {\vect p'} | \hat \phi_m(x') \hat \phi_m(x') | {\vect p'} \rangle &
        \langle {\vect p'} | \widetilde T\hat \phi_m(x) \hat \phi_m(x') | {\vect p'} \rangle .
    \end{pmatrix}
\end{equation} From now on we drop the subindex $m$ to simplify the notation.

Let us now compute a Whightman function in the background state $|\vect
p'\rangle$:
\begin{equation*}
\begin{split}
    G^{(0)}_{21} &(t,\vect x ;t',\vect x';{\vect p'}) = \langle {\vect p'} |  \hat\phi (t,x)
\hat\phi(t',x') |   {\vect p'} \rangle =  \langle 0 | \hat
a_{\vect p'}  \hat\phi     (t,x) \hat\phi(t',x') \hat a^\dag_{\vect p'}|0 \rangle \\
        &=  \frac{V}{2 \sqrt{E_\vect kE_{\vect k'} }} \int \udpi[3]{\vect k} \udpi[3]{ \vect k'}
    \expp{-iE_\vect p t + i \vect p \cdot \vect x}
    \expp{iE_\vect {k'} t' - i \vect {k'} \cdot \vect x'}
    \langle 0 | \hat a_{\vect p'} \hat a_\vect k \hat a^\dag_\vect {k'} \hat a_{\vect p'}^\dag |0 \rangle
     \\
    &\quad +\frac{V}{2 \sqrt{E_\vect kE_{\vect k'} }}  \int \udpi[3]{\vect k} \udpi[3]{\vect k'}
    \expp{iE_\vect k t - i \vect k \cdot \vect x}
    \expp{-iE_\vect {k'} t' + i \vect {k'} \cdot \vect x'}
    \langle 0 | \hat a_{\vect p'} \hat a^\dag_\vect k \hat a_\vect {k'} \hat a_{\vect p'}^\dag |0 \rangle,
\end{split}
\end{equation*}
where we have used the expansion of the field operator in terms of creation and annihilation operators:
\begin{equation}
    \hat\phi(t,x)= \sqrt{V} \int \udpi[3]{\vect k} \frac{1}{2E_\vect k} \left(  \hat a_\vect k \expp{-iE_\vect k t + i \vect k \cdot \vect x}  +
    \hat a^\dag_\vect k \expp{iE_\vect k t - i \vect k \cdot \vect x} \right).
\end{equation}
Next, we evaluate explicitly the different contractions of the creation and annihilation operators:
\begin{equation*}
\begin{split}
    G^{(0)}_{21}(t,\vect x ;t',\vect x';{\vect p'}) &= \int \udpiE{3}{k} \expp{-iE_\vect k(t-t') + i \vect k \cdot (\vect x-\vect x')} \\
    &\quad+ \frac{(2\pi)^3}{2E_{\vect p'}V} \left[ \expp{-iE_\vect k (t-t') + i \vect k \cdot (\vect x-\vect x')}
    + \expp{iE_\vect k (t-t') - i \vect k \cdot (\vect x-\vect x')}\right],
\end{split}
\end{equation*}
which in the Fourier representation becomes
\begin{equation*}
\begin{split}
    G^{(0)}_{21}(k;{\vect p'}) =  \frac{2 \pi}{2E_\vect k} \delta(k^0 - E_\vect k) &+ \frac{(2 \pi)^4V}{2E_{\vect p'} } \bigg[
    \delta^{(3)}(\vect k - {\vect p'})   \delta(k^0 - E_\vect k) \\ &\qquad +
    \delta^{(3)}(\vect k + {\vect p'})   \delta(k^0 + E_\vect k) \bigg].
\end{split}
\end{equation*}
Using the properties of the delta functions, the result can be further reorganized as
\begin{equation}
\begin{split}
    G^{(0)}_{21}(k;{\vect p'}) = 2\pi \delta(k^2+m^2) &\ \bigg[ \theta(k^0) + \frac{(2\pi)^3}{V} \delta^{(3)}(\vect k - {\vect p'})  \theta(k^0) \\ &+
     \frac{(2\pi)^3}{V} \delta^{(3)}(\vect k + {\vect p'})  \theta(-k^0) \bigg].
\end{split}
\end{equation}
Repeating a similar calculation for the Feynman propagator we find:
\begin{equation}
\begin{split}
    G^{(0)}_{11}(k;{\vect p'}) = \frac{-i}{p^2+m^2-i\epsilon}
    + 2\pi&\ \delta(k^2+m^2) \frac{(2\pi)^3}{V}  \big[  \delta^{(3)}(\vect k - {\vect p'})  \theta(k^0) \\ &\qquad +
    \delta^{(3)}(\vect k + {\vect p'})  \theta(-k^0) \big].
\end{split}
\end{equation}
The complete matrix of CTP propagators is given by
\begin{equation} \label{FreeGT2}
\begin{split}
         &\ G_{ab}^{(0)}(k;{\vect p'})  =
        \begin{pmatrix}
            \dfrac{-i}{k^2 + m^2 - i \epsilon} & {2\pi \delta(k^2+m^2) \theta(-k^0)}\\
            { 2\pi \delta(p^2+m^2) \theta(k^0)}
             & \dfrac{i}{k^2 + m^2 + i \epsilon}
        \end{pmatrix} \\ &\quad + \delta(k^2+m^2) \frac{(2\pi)^4}{V} \left[  \delta^{(3)}(\vect k - {\vect p'})  \theta(k^0) +
    \delta^{(3)}(\vect k + {\vect p'})  \theta(-k^0) \right]
        \begin{pmatrix}
            1 & 1 \\
            1 & 1
        \end{pmatrix}.
\end{split}
\end{equation}

Alternatively, we could have simply used relations \eqref{GeneralFreeProps} with $n_\vect k =(2\pi)^3 \delta^{(3)}(\vect k - {\vect p'}) /V$.

\section{Adiabatic approximation for the Hadamard function (chapter 6)}\label{app:adiabaticHadamard}

In chapter 6 we introduced the following adiabatic approximation for the Hadamard function [see \Eqref{adiabaticHadamard}].
\begin{equation}\label{adiabaticHadamard3}
	G^{(1)}(t,t';\vect k) = \frac{ 1+2 n_\vect k }{\sqrt{R_\vect p(t)R_\vect p(t')}} \cos\left({\int^{t}_{t'} \ud{s} {R_\vect k(s)}} \right) \expp{-\left|\int^{t}_{t'} \ud{s} \Gamma_\vect k(s)/2\right| }..
\end{equation}
We will not attempt to demonstrate this expression by constructive methods; we will simply look for the differential equation obeyed by the Hadamard function, and verify that the above is an adiabatic solution to that equation.

First, we introduce the reescaled time $u=tH$. Then, following \Eqref{operatorRetardedAdiabatic}
\begin{subequations}
\begin{equation} 
	\left[ H^2 \derp[2]{}{u} + H \Gamma_\vect k(u) \derp{}{u} + R_\vect k^2(u)\right]
	 \bar\GR(u,u';\vect k)  = -i H \delta(u-u')
\end{equation}
and the equivalent expression for the advanced propagator,
\begin{equation}
	\left[ H^2 \derp[2]{}{u} - H \Gamma_\vect k(u) \derp{}{u} + R_\vect k^2(u)\right]
	 \bar\GA(u,u';\vect k)  = i H \delta(u-u')
\end{equation}
\end{subequations}
we obtain, from \eqref{SelfEnergyHadamard2},
\begin{multline*}
	\left[ H^2 \derp[2]{}{u} + H \Gamma_\vect k(u) \derp{}{u} + R_\vect k^2(u)\right]
	\left[ H^2 \derp[2]{}{u'} - H \Gamma_\vect k(u') \derp{}{u'} + R_\vect k^2(u')\right]
	 \\ \times \bar G^{(1)}(u,u';\vect k)   = - i \SigmaN(u,u';\vect k)
\end{multline*}
Using the  equivalent expression of \eqref{SigmaNRPole} in the time representation,
\begin{equation}
	\SigmaN(u,u';\vect k) = 2 iH R_\vect k(u)	\Gamma_\vect k(u) (1+2n_\vect k) \delta(u-u'),
\end{equation}
we get
\begin{multline}
	\left[ H^2 \derp[2]{}{u} + H \Gamma_\vect k(u) \derp{}{u} + R_\vect k^2(u)\right] \left[ H^2 \derp[2]{}{u'} - H \Gamma_\vect k(u') \derp{}{u'} + R_\vect k^2(u')\right]
	 \\ \times \bar G^{(1)}(u,u';\vect k)  = 2 H R_\vect k(u)	\Gamma_\vect k(u)  (1+2n_\vect k) \delta(u-u').
\end{multline}
This is the key equation of this appendix: it gives the differential equation obeyed by the interacting Hadamard propagator in cosmology.

It can be now verified that the Hadamard function given in \Eqref{adiabaticHadamard} verifies the above equation, following similar steps as in subsect.~\ref{sect:adiabaticParticles}. It proves convenient to analyze the regular and singular contributions to the differential equation separately. It is easy to check that the regular contributions cancel to first adiabatic order; some more work is needed to verify the cancellation of the singular terms.


\backmatter

	\bibliographystyle{tesi}

\printindex

\end{document}